\documentclass[11pt]{report}
\usepackage[utf8]{inputenc}
\usepackage{graphicx}
\usepackage[justification=centering]{caption}
\usepackage{subcaption}
\usepackage{float}
\usepackage[width=150mm,top=35mm,bottom=27mm,bindingoffset=6mm]{geometry}
\usepackage{fancyhdr}
\usepackage{setspace}
\usepackage{amssymb, amsmath, amsfonts, amsthm, amscd, bm, helvet, mathrsfs, mathtools, tocloft, xcolor}

\AtBeginDocument{
  \setlength{\abovedisplayskip}{9pt plus 3pt minus 5pt}
  \setlength{\belowdisplayskip}{9pt plus 3pt minus 5pt}
  \setlength{\abovedisplayshortskip}{0pt plus 3pt}
  \setlength{\belowdisplayshortskip}{6pt plus 3pt minus 3pt}
}
\usepackage{etoolbox}

\usepackage{needspace}

\usepackage{enumitem}

\usepackage[T1]{fontenc}
\usepackage{lmodern}

\usepackage{tikz}
\usetikzlibrary{cd}
\usetikzlibrary{matrix,calc}
\usetikzlibrary{decorations.markings,shapes.geometric,shapes.misc}
\usetikzlibrary{decorations.pathreplacing,positioning}
\usetikzlibrary{arrows}
\usetikzlibrary{hobby}
\usetikzlibrary{patterns, fadings}

\usepackage{graphicx, verbatim}

\usepackage[all]{xy}
\usepackage{relsize}
\usepackage{bbold}

\usepackage[style=alphabetic, sorting=nyt, backref=true, citestyle=alphabetic, maxnames=99,
  minnames=1,
  maxcitenames=99,
  mincitenames=1]{biblatex}

\DefineBibliographyStrings{english}{
  backrefpage  = {$\rightarrow$~p.},
  backrefpages = {$\rightarrow$~pp.},
}

\renewbibmacro*{pageref}{%
  \iflistundef{pageref}
    {}
    {\newline%
       \ifnumgreater{\value{pageref}}{1}
         {\bibstring{backrefpages}\ppspace}
         {\bibstring{backrefpage}\ppspace}%
       \printlist[pageref][-\value{listtotal}]{pageref}}}

\usepackage{xstring}

\DeclareFieldFormat{doi}{}

\newcommand{\formaturl}[1]{%
  \IfBeginWith{#1}{https://www.}%
    {\StrGobbleLeft{#1}{12}[\tempurl]\href{#1}{\tempurl}}%
    {\IfBeginWith{#1}{https://}%
      {\StrGobbleLeft{#1}{8}[\tempurl]\href{#1}{\tempurl}}%
      {\href{#1}{#1}}%
    }%
}

\DeclareFieldFormat{url}{%
  \newline\small%
  \formaturl{#1}%
  \nopunct%
}
\DeclareFieldFormat{eprint}{\newline\small\href{https://arxiv.org/abs/#1}{arXiv:#1\iffieldundef{eprintclass}{}{\addspace\thefield{eprintclass}}}\nopunct}
\DeclareFieldFormat{journaltitle}{#1}  % Remove italics from journal names
\DeclareFieldFormat[article,periodical]{journaltitle}{#1}
\DeclareFieldFormat[incollection]{booktitle}{#1}  % Remove italics from booktitle in incollection

\DeclareFieldFormat[article,periodical]{number}{}
\DeclareFieldFormat{number}{}
\DeclareFieldFormat[article,periodical]{volume}{\textbf{#1}}
\DeclareFieldFormat{volume}{\textbf{#1}}

\DeclareFieldFormat{pages}{#1}

\AtEveryBibitem{\clearfield{number}}

\renewbibmacro*{volume+number+eid}{%
  \printfield{volume}%
}

\renewbibmacro*{journal+issuetitle}{%
  \usebibmacro{journal}%
  \setunit*{\addspace}%
  \usebibmacro{volume+number+eid}%
  \setunit{\addcomma\space}%
  \usebibmacro{issue+date}%
  \setunit{\addcolon\space}%
  \usebibmacro{issue}%
  \newunit%
}

\DeclareBibliographyDriver{article}{%
  \usebibmacro{bibindex}%
  \usebibmacro{begentry}%
  \usebibmacro{author/translator+others}%
  \setunit{\printdelim{nametitledelim}}\newblock
  \usebibmacro{title}%
  \newunit
  \printlist{language}%
  \newunit\newblock
  \usebibmacro{byauthor}%
  \newunit\newblock
  \usebibmacro{bytranslator+others}%
  \newunit\newblock
  \printfield{version}%
  \newunit\newblock
  \usebibmacro{journal}%
  \setunit*{\addspace}%
  \printfield{volume}%
  \setunit{\addcomma\space}%
  \printfield{pages}%
  \setunit{\addspace}%
  \printtext[parens]{\usebibmacro{date}}%
  \newunit\newblock
  \usebibmacro{doi+eprint+url}%
  \newunit\newblock
  \usebibmacro{addendum+pubstate}%
  \setunit{\bibpagerefpunct}\newblock
  \usebibmacro{pageref}%
  \newunit\newblock
  \iftoggle{bbx:related}
    {\usebibmacro{related:init}%
     \usebibmacro{related}}
    {}%
  \usebibmacro{finentry}%
}

\DeclareFieldFormat[book]{edition}{#1 edition}

\DeclareBibliographyDriver{book}{%
  \usebibmacro{bibindex}%
  \usebibmacro{begentry}%
  \usebibmacro{author/editor+others/translator+others}%
  \setunit{\printdelim{nametitledelim}}\newblock
  \usebibmacro{maintitle+title}%
  \iffieldundef{note}
    {\newunit\newblock}
    {\setunit*{\addperiod\space}%
     \printfield{note}%
     \newunit\newblock
    }%
  \printlist{language}%
  \newunit\newblock
  \usebibmacro{byauthor}%
  \newunit\newblock
  \usebibmacro{byeditor+others}%
  \newunit\newblock
  \usebibmacro{publisher+location+date}%
  \setunit{\addspace}%
  \printtext[parens]{\usebibmacro{date}}%
  \newunit\newblock
  \usebibmacro{doi+eprint+url}%
  \newunit\newblock
  \usebibmacro{addendum+pubstate}%
  \setunit{\bibpagerefpunct}\newblock
  \usebibmacro{pageref}%
  \newunit\newblock
  \iftoggle{bbx:related}
    {\usebibmacro{related:init}%
     \usebibmacro{related}}
    {}%
  \usebibmacro{finentry}%
}

\DeclareBibliographyDriver{incollection}{%
  \usebibmacro{bibindex}%
  \usebibmacro{begentry}%
  \usebibmacro{author/translator+others}%
  \setunit{\printdelim{nametitledelim}}\newblock
  \usebibmacro{title}%
  \newunit
  \printlist{language}%
  \newunit\newblock
  \usebibmacro{byauthor}%
  \newunit\newblock
  \usebibmacro{in:}%
  \usebibmacro{maintitle+booktitle}%
  \setunit{\addperiod\space}%
  \printfield{series}%
  \setunit{\addspace}%
  \printfield{volume}%
  \setunit{\addcomma\space}%
  \printfield{pages}%
  \setunit{\addperiod\space}%
  \printfield{note}%
  \setunit{\addperiod\space}%
  \usebibmacro{publisher+location+date}%
  \setunit{\addspace}%
  \printtext[parens]{\usebibmacro{date}}%
  \addperiod
  \newunit\newblock
  \usebibmacro{doi+eprint+url}%
  \newunit\newblock
  \usebibmacro{addendum+pubstate}%
  \setunit{\bibpagerefpunct}\newblock
  \usebibmacro{pageref}%
  \newunit\newblock
  \iftoggle{bbx:related}
    {\usebibmacro{related:init}%
     \usebibmacro{related}}
    {}%
  \usebibmacro{finentry}%
}

\renewbibmacro*{publisher+location+date}{%
  \printlist{location}%
  \iflistundef{publisher}
    {\setunit*{\addcomma\space}}
    {\setunit*{\addcolon\space}}%
  \printlist{publisher}%
  \setunit*{\addspace}%
}

\DeclareBibliographyDriver{inbook}{%
  \usebibmacro{bibindex}%
  \usebibmacro{begentry}%
  \usebibmacro{author/translator+others}%
  \setunit{\addperiod\newline}%
  \printfield[emph]{chapter}%
  \setunit{\addperiod\newline}%
  \printfield[plaintext]{title}%
  \setunit{\addperiod\space}%
  \printlist{publisher}%
  \setunit{\addspace}%
  \printtext[parens]{\printfield{year}}%
  \setunit{\addperiod}%
  \newunit\newblock
  \usebibmacro{doi+eprint+url}%
  \newunit\newblock
  \usebibmacro{addendum+pubstate}%
  \setunit{\bibpagerefpunct}\newblock
  \usebibmacro{pageref}%
  \newunit\newblock
  \iftoggle{bbx:related}
    {\usebibmacro{related:init}%
     \usebibmacro{related}}
    {}%
  \usebibmacro{finentry}%
}

\DeclareBibliographyDriver{misc}{%
  \usebibmacro{bibindex}%
  \usebibmacro{begentry}%
  \usebibmacro{author/editor+others/translator+others}%
  \setunit{\printdelim{nametitledelim}}\newblock
  \usebibmacro{title}%
  \newunit
  \printlist{language}%
  \newunit\newblock
  \usebibmacro{byauthor}%
  \newunit\newblock
  \usebibmacro{byeditor+others}%
  \newunit\newblock
  \printfield{howpublished}%
  \newunit\newblock
  \printfield{type}%
  \newunit
  \printfield{version}%
  \newunit
  \printfield{note}%
  \newunit\newblock
  \printlist{organization}%
  \newunit\newblock
  \printlist{location}%
  \setunit{\addspace}%
  \printtext[parens]{\usebibmacro{date}}\addperiod
  \newunit\newblock
  \usebibmacro{doi+eprint+url}%
  \newunit\newblock
  \usebibmacro{addendum+pubstate}%
  \setunit{\bibpagerefpunct}\newblock
  \usebibmacro{pageref}%
  \newunit\newblock
  \iftoggle{bbx:related}
    {\usebibmacro{related:init}%
     \usebibmacro{related}}
    {}%
  \usebibmacro{finentry}%
}

\DeclareSourcemap{
  \maps[datatype=bibtex]{
    \map{
      \step[fieldsource=doi, final]
      \step[fieldset=url, null]
      \step[fieldsource=doi]
      \step[fieldset=url, origfieldval]
      \step[fieldsource=url, match=\regexp{(.+)}, replace={https://doi.org/$1}]
    }
  }
}

\renewbibmacro*{doi+eprint+url}{%
  \iftoggle{bbx:doi}
    {\printfield{doi}}
    {}%
  \newunit\newblock
  \iftoggle{bbx:url}
    {\usebibmacro{url+urldate}}
    {}%
  \newunit\newblock
  \iftoggle{bbx:eprint}
    {\usebibmacro{eprint}}
    {}}

\DeclareFieldFormat[article,incollection,inproceedings,patent,thesis,unpublished]{title}{\emph{#1}}
\DeclareFieldFormat[article,incollection,inproceedings,patent,thesis,unpublished]{citetitle}{\emph{#1}}
\DeclareFieldFormat[inbook]{title}{#1}  % Plain text for inbook title
\DeclareFieldFormat{plaintext}{#1}      % Plain text format
\DeclareFieldFormat{emph}{\emph{#1}}    % Emphasized format for chapter

\DeclareDelimFormat{titlepunct}{\addspace}

\renewbibmacro*{author}{%
  \ifboolexpr{
    test \ifuseauthor
    and
    not test {\ifnameundef{author}}
  }
    {\printnames{author}%
     \setunit{\addspace}}
    {}}

\renewbibmacro*{title}{%
  \ifboolexpr{
    test {\iffieldundef{title}}
    and
    test {\iffieldundef{subtitle}}
  }
    {}
    {\newline%
     \printtext[title]{%
       \printfield[titlecase]{title}%
       \setunit{\subtitlepunct}%
       \printfield[titlecase]{subtitle}}%
     \nopunct%
     \newline}}

\DeclareFieldFormat{titlecase:title}{\emph{#1}}
\DeclareFieldFormat{titlecase:subtitle}{\emph{#1}}

\renewbibmacro{in:}{}

\AtEveryBibitem{\clearfield{issn}}
\AtEveryBibitem{\clearfield{isbn}}

\allowdisplaybreaks[1]

\usepackage{pict2e}

\usepackage[pdfencoding=auto,psdextra,hidelinks]{hyperref}

\hypersetup{
     colorlinks=true,         % false: boxed links; true: colored links,false is default
     linkcolor=blue,
     urlcolor=purple,
     citecolor=violet
}

\usepackage{xurl} % better URL line breaking than breakurl
\appto\bibsetup{\sloppy}

\AtEveryBibitem{\needspace{6\baselineskip}}

\makeatletter
\DeclareFontFamily{OMX}{MnSymbolE}{}
\DeclareSymbolFont{MnLargeSymbols}{OMX}{MnSymbolE}{m}{n}
\SetSymbolFont{MnLargeSymbols}{bold}{OMX}{MnSymbolE}{b}{n}
\DeclareFontShape{OMX}{MnSymbolE}{m}{n}{
    <-6>  MnSymbolE5
   <6-7>  MnSymbolE6
   <7-8>  MnSymbolE7
   <8-9>  MnSymbolE8
   <9-10> MnSymbolE9
  <10-12> MnSymbolE10
  <12->   MnSymbolE12
}{}
\DeclareFontShape{OMX}{MnSymbolE}{b}{n}{
    <-6>  MnSymbolE-Bold5
   <6-7>  MnSymbolE-Bold6
   <7-8>  MnSymbolE-Bold7
   <8-9>  MnSymbolE-Bold8
   <9-10> MnSymbolE-Bold9
  <10-12> MnSymbolE-Bold10
  <12->   MnSymbolE-Bold12
}{}

\DeclareMathAlphabet{\mathscrbf}{OMS}{mdugm}{b}{n}

\let\llangle\@undefined
\let\rrangle\@undefined
\DeclareMathDelimiter{\llangle}{\mathopen}%
                     {MnLargeSymbols}{'164}{MnLargeSymbols}{'164}
\DeclareMathDelimiter{\rrangle}{\mathclose}%
                     {MnLargeSymbols}{'171}{MnLargeSymbols}{'171}
\makeatother

\theoremstyle{plain}
\newtheorem{theorem}{Theorem}[section]
\newtheorem*{theorem*}{Theorem}
\newtheorem{proposition}[theorem]{Proposition}
\newtheorem*{proposition*}{Proposition}
\newtheorem{lemma}[theorem]{Lemma}
\newtheorem{corollary}[theorem]{Corollary}

\theoremstyle{definition}
\newtheorem{definition}[theorem]{Definition}

\newtheorem{remarkx}[theorem]{Remark}

\newenvironment{remark}
  {\pushQED{\qed}\remarkx}
  {\popQED\endremarkx}

\DeclareMathOperator{\Res}{Res}
\DeclareMathOperator{\Aut}{Aut}

\DeclareMathOperator{\Ad}{Ad}
\DeclareMathOperator{\ad}{ad}

\DeclareMathOperator{\Tr}{Tr}

\newcommand{\longhookrightarrow}{\lhook\joinrel\relbar\joinrel\rightarrow}

\newcommand{\SimTo}{\xrightarrow{\raisebox{-0.3 em}{\smash{\ensuremath{\text{\tiny$\cong$}}}}}}

\newcommand{\lda}{\lambda}

\newcommand{\g}{\mathfrak{g}}
\newcommand{\Lg}{\widetilde{\mathfrak{g}}}

\newcommand{\LG}{\widetilde{G}}

\newcommand{\CP}{\mathbb{P}^1}

\newcommand{\Lag}{\mathscr{L}}

\newcommand{\sgm}{\sigma}
\newcommand{\omg}{\omega}

\newcommand{\Q}{\mathcal{Q}}

\newcommand{\ti}[1]{_{\mathbf{\underline{#1}}}}

\newcommand{\sfE}{\mathsf{E}}
\newcommand{\sfH}{\mathsf{H}}

\newcommand{\sfp}{\mathsf{p}}
\newcommand{\sfq}{\mathsf{q}}

\def\sl{\mathfrak{sl}}
\def\Id{\textup{Id}}
\def\h{\mathfrak{h}}

\def\s{\mathfrak{s}}

\def\d{\mathrm{d}}

\def\1{\bm{1}}

\newcommand{\lau}[1]{(\kern-.2em( #1 )\kern-.2em)}%{(( #1 ))}

\newcommand{\parder}[2]{\frac{\partial #1}{\partial #2}}
\newcommand{\ie}{{\it i.e.}\ }

\definecolor{myPurple}{rgb}{0.5,0.1,0.6}
\definecolor{myOrange}{rgb}{1.0,0.5,0.0}
\definecolor{myRed}{rgb}{1.0,0.0,0.0}
\definecolor{myGreen}{rgb}{0.0,0.5,0.0}
\definecolor{LatexBlue}{rgb}{0.211765,0.227451,0.666667}
\definecolor{myBlue}{rgb}{0.0,0.0,1.0}
\definecolor{myBlack}{rgb}{0.0,0.0,0.0}
\definecolor{myGray}{rgb}{0.3,0.3,0.3}

\def\be{\begin{equation}}
\def\ee{\end{equation}}
\def\bea{\begin{eqnarray}}
\def\eea{\end{eqnarray}}

\def\A{\mathcal{A}}

\def\CC{\mathbb{C}}
\def\CP{\mathbb{C}P^1}

\def\RR{\mathbb{R}}

\def\ZZ{\mathbb{Z}}

\def\O{\mathcal{O}}
\def\P{\mathcal{P}}

\def\M{\mathcal M}

\def\sfE{\mathsf{E}}

\def\1{\bm{1}}

\title{Lie dialgebras, gauge theory, and Lagrangian multiforms for integrable models}
\author{Anup Anand Singh}
\date{September 2025}

\begin{document}
\pagenumbering{roman} 

\begin{titlepage}
    \begin{center}
        \vspace*{0.7cm}
        
        \Huge
        \textbf{Lie dialgebras, gauge theory,}\\ \textbf{and Lagrangian multiforms for}\\\textbf{integrable models}
        
        \vspace{1.4cm}
        \LARGE
    
        \textbf{Anup Anand Singh}

        \vspace{2cm}

        \includegraphics[width=0.25\textwidth]{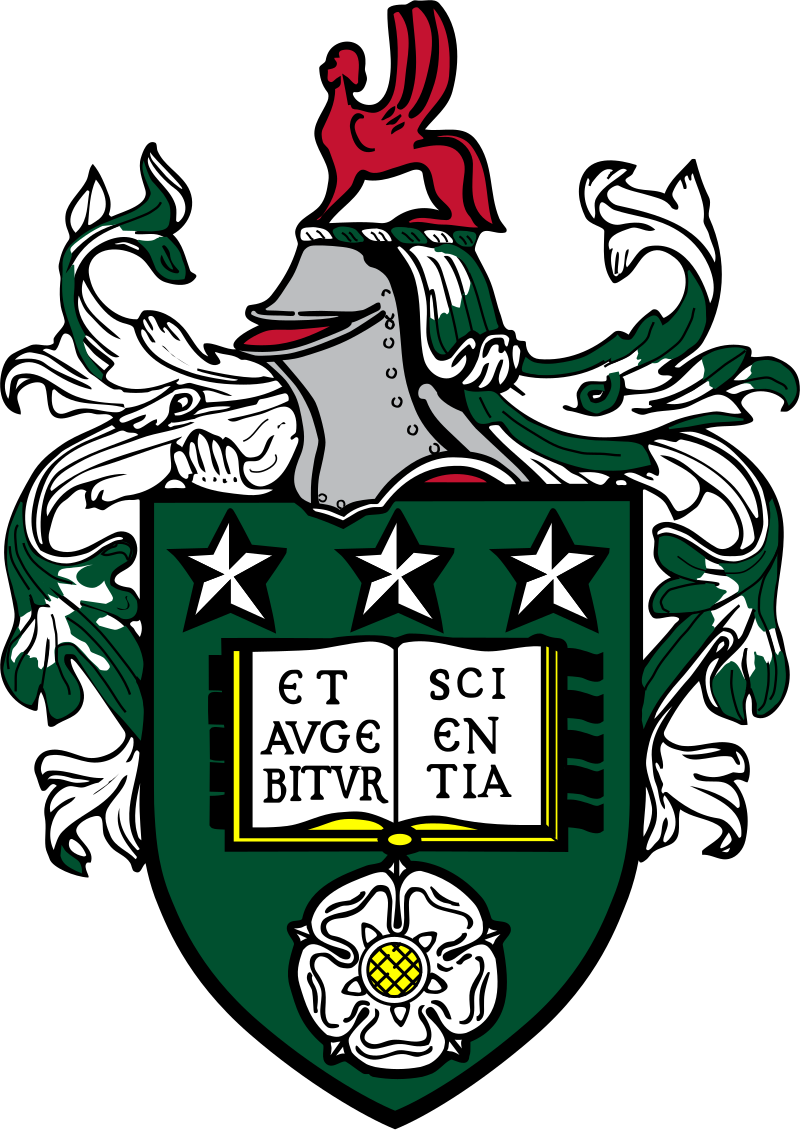}
        
        \vfill
        
        \Large
        Submitted in accordance with the requirements\\
        for the degree of Doctor of Philosophy
        
        \vspace{1.5cm}

        \textbf{School of Mathematics}\\
        \textbf{University of Leeds}
        
        \vspace{0.4cm}

        September 2025
        
    \end{center}
    
\end{titlepage}

\newpage
\thispagestyle{empty}
\mbox{}
\newpage

\cleardoublepage
\thispagestyle{empty}
\vspace*{0.2\textheight}
\begin{center}
\Large{\emph{For Leo}}
\end{center}

\newpage
\thispagestyle{empty}
\mbox{}
\newpage

\addtocontents{toc}{\protect\setlength{\cftbeforechapskip}{0pt}}

\chapter*{Acknowledgements}
I am infinitely grateful to my supervisor Vincent Caudrelier, whom I can only thank for his immense patience, the many hours we spent puzzling at the board, those delightful conversations over lunch, and above all, his boundless enthusiasm for mathematical physics, which has deepened my own love for the subject. 

I am deeply indebted to Frank Nijhoff, who gave me my first introduction to Lagrangian multiforms and without whose support pursuing this work would not have been possible. Heartfelt thanks are also due to Derek Harland, for his invaluable guidance as a supervisor and his crucial perspectives on the many problems in geometry.

I am extremely thankful to my examiners Ben Hoare and Oleg Chalykh for taking the time to read this thesis and suggesting edits that have enhanced the presentation.

Over the last two years, it has been a great pleasure and privilege to work with Benoît Vicedo. Our collaborations have been pivotal in shaping my understanding of integrable systems, for which I will always be grateful. I also feel fortunate to have had the opportunity to collaborate with Marta Dell'Atti during the early part of my doctoral work --- learning from her was a great joy!

I also owe the wonderful experience I have had at Leeds to all the other members of the \emph{Integrable Systems} group: Farrokh Atai, Allan Fordy, Alexander Mikhailov, and Taras Skrypnyk. And to those from the broader mathematical community at Leeds with whom I was lucky to have crossed paths: Adam Aldridge, Aurélie Astoul, Cas Chaudhuri, Ilaria Colazzo, Dylan Crook, João Faria Martins, Francesca Fedele, Ravil Gabdurakhmanov, Tathagata Ghosh, Lluís Hernández-Navarro, Benjamin Lambert, Matthew Lawrence, Yang Lu, Paul Martin, Benjamin Morris, Ruheyan Nuermaimaiti, Jonathan Schilhan, and Luca Seemungal. I am thankful for all the different ways they have contributed to this journey.

Thank you to Asaf Karagila for his mentorship and an unfailing supply of caffeine over the last three years. To Luigi Appolloni and Anna Guseva for keeping the aforementioned supply of caffeine indeed unfailing, and for the many interesting conversations. To Linden Disney-Hogg for providing his insightful perspective on questions of all sorts and being there to co-found and run the \emph{Mathematical Physics at Leeds Seminar Series} with me --- there could not have been a better collaborator! To Ibrahim Mohammed for being a great friend, both on and off the tennis court. To Emine Y{\i}ld{\i}r{\i}m for always looking out for me and sharing countless pieces of valuable advice. And to Matteo Spadetto for always being there and filling my time at Leeds with countless stimulating conversations about life, the universe, and everything. Finally, I need to thank Shashank Pratap Singh from across the pond who, like the years before, continued to provide the deepest of mathematical insights and the silliest of reasons to laugh.

My doctoral work was funded by the School of Mathematics EPSRC Doctoral Training Partnership Studentship (Project Reference Number 2704447). I am also grateful for the additional financial support from the School of Mathematics, which made my participation in various scientific events possible. To this end, I would also like to thank the school administration for their support in handling practical arrangements.

It is only fitting that I close by expressing my gratitude to Chanchal Singh and Arun Singh, my parents, with whom this story began, and who never stopped believing in me --- thank you for everything!
\addcontentsline{toc}{chapter}{Acknowledgements}

\cleardoublepage
\thispagestyle{empty}
\vspace*{0.5\textheight}
\emph{Suddenly, without warning, his feverish activity was interrupted and was replaced by a kind of fascination. He spent several days as if he were bewitched, softly repeating to himself a string of fearful conjectures without giving credit to his own understanding. Finally, one Tuesday in December, at lunchtime, all at once he released the whole weight of his torment. The children would remember for the rest of their lives the august solemnity with which their father, devastated by his prolonged vigil and by the wrath of his imagination, revealed his discovery to them:}

\emph{`The earth is round, like an orange.'}

\vspace{0.7cm}

\begin{flushright}        
Gabriel García Márquez\\
\emph{One Hundred Years of Solitude}\\
\emph{(Cien años de soledad)}\\
Translated from the Spanish by Gregory Rabassa
\end{flushright}

\newpage
\thispagestyle{empty}
\mbox{}
\newpage

\chapter*{Declaration}
I confirm that the work submitted is my own, except where work which has formed part of jointly authored publications has been included. These publications are listed below.

\begin{description}[labelwidth=5em, leftmargin=!, align=left]
    \item[\normalfont{[CDS24]}] V.~Caudrelier, M.~Dell'Atti, and A.A.~Singh\\
    \textit{Lagrangian multiforms on coadjoint orbits for finite-dimensional integrable systems}\\
    \href{https://doi.org/10.1007/s11005-023-01766-9}{Lett. Math. Phys. \textbf{114}, 34 (2024)}\\
    \href{https://arxiv.org/abs/2307.07339}{arXiv:2307.07339 [math-ph]}

    \item[\normalfont{[CSV24]}] 
    V.~Caudrelier, A.A.~Singh, and B.~Vicedo\\
    \textit{Lagrangian multiform for cyclotomic Gaudin models}\\
    \href{https://doi.org/10.3842/SIGMA.2024.100}{SIGMA \textbf{20}, 100 (2024)}\\
    \href{https://arxiv.org/abs/2405.12837}{arXiv:2405.12837 [math-ph]}

    \item[\normalfont{[CHSV25]}]
    V.~Caudrelier, D.~Harland, A.A.~Singh, and B.~Vicedo\\
    \textit{The $3$d mixed BF Lagrangian $1$-form: a variational formulation of Hitchin's integrable system}\\
    \href{https://doi.org/10.1007/s00220-025-05535-8}{Commun. Math. Phys. \textbf{407}, 40 (2026)}\\
    \href{https://arxiv.org/abs/2509.05127}{arXiv:2509.05127 [math-ph]}
    
\end{description}
The contributions of the authors to the above publications are as follows. 

\begin{itemize}

  \item In the paper \cite{CDS}, my supervisor Dr~Vincent Caudrelier and I developed the general framework for Lagrangian one-forms on coadjoint orbits in collaboration with Dr~Marta Dell'Atti, then at the University of Portsmouth. The result connecting the closure relation for Lagrangian one-forms to the involutivity of Hamiltonians was proved by Dr~Caudrelier. Lagrangian one-forms for two different realisations of the open Toda chain were obtained by Dr~Caudrelier and Dr~Dell'Atti, while a Lagrangian one-form for the rational Gaudin hierarchy was derived by me with help from Dr~Caudrelier in resolving certain issues. 
  
  Content from \cite{CDS} appears in Chapters \ref{chap:dialgebra-background}, \ref{chap:lm-background}, \ref{chap:lm-orbits} and \ref{chap:orbitexamples}.
  
  \item The paper \cite{CSV}, which generalises the framework of \cite{CDS} to obtain a Lagrangian one-form for the cyclotomic Gaudin hierarchy, resulted from my collaboration with my supervisor Dr~Vincent Caudrelier and Dr~Benoît Vicedo from the University of York. A key insight for the underlying algebraic setup was provided by Dr~Vicedo. The construction of Lagrangian one-forms for the hierarchies of the periodic Toda chain, the discrete self-trapping (DST) model, and the coupled Toda--DST system, and the explicit derivations of the corresponding Euler--Lagrange equations were done by me with guidance from Dr~Caudrelier and Dr~Vicedo in overcoming conceptual difficulties. 
  
  Content from \cite{CSV} appears in Chapters \ref{chap:dialgebra-background} and \ref{chap:orbitexamples}.
  
  \item The paper \cite{CHSV} is based on my joint work with my supervisors Dr~Vincent Caudrelier and Dr~Derek Harland, and our collaborator Dr~Benoît Vicedo from the University of York. The concept of gauged Lagrangian one-forms was primarily developed by Dr~Caudrelier and Dr~Harland. The application of this formalism to obtain a Lagrangian one-form for $3$d mixed BF theory coupled with the so-called type A and type B defects, which provides a variational description of Hitchin systems, was done by Dr~Caudrelier, Dr~Harland, and Dr~Vicedo. With this Lagrangian one-form as the starting point, an initial derivation of a unifying Lagrangian one-form for a hierarchy of Lax equations describing Hitchin systems in terms of meromorphic Lax matrices was done by me, and later improved upon by Dr~Caudrelier, Dr~Harland, and Dr~Vicedo. Explicit Lagrangian one-forms for rational and elliptic Gaudin hierarchies were obtained as special cases of this unifying Lagrangian one-form by me with help from Dr~Caudrelier and Dr~Harland in resolving conceptual issues.
  
  Content from \cite{CHSV} appears in Chapters \ref{chap:lm-background}, \ref{chap:bf-hitchin}, \ref{chap:lax-hitchin} and \ref{chap:hitchinexamples}.

\end{itemize}

I confirm that appropriate credit has been given within the thesis where reference has been made to the work of others.
\addcontentsline{toc}{chapter}{Declaration}

\chapter*{Abstract}
Lagrangian multiforms provide a variational framework for describing integrable hierarchies. This thesis presents two approaches for systematically constructing Lagrangian one-forms, which cover the case of finite-dimensional integrable hierarchies, thus addressing one of the central open problems in the theory of Lagrangian multiforms.

The first approach, based on the theory of Lie dialgebras, incorporates into Lagrangian one-forms the notion of the classical $r$-matrix and produces Lagrangian one-forms living on coadjoint orbits. We prove an important structural result relating the closure relation for Lagrangian one-forms to the Poisson involutivity of Hamiltonians and the double zero on Euler--Lagrange equations. As applications of this approach, we obtain explicit Lagrangian one-forms for the hierarchies of the open Toda chain and the non-cyclotomic and cyclotomic rational Gaudin models, as well as the periodic Toda chain and the discrete self-trapping model as realisations of the cyclotomic Gaudin model. The versatility of this approach is further demonstrated by coupling the periodic Toda chain with the discrete self-trapping model and obtaining a Lagrangian one-form for the corresponding hierarchy.

In the second approach, we extend the notion of Lagrangian one-forms to the setting of gauge theories and derive a variational formulation of the Hitchin system associated with a compact Riemann surface of arbitrary genus. We show that this description corresponds to a Lagrangian one-form for classical $3$d holomorphic-topological BF theory coupled with so-called type A and type B defects. Notably, this establishes an explicit connection between $3$d holomorphic-topological BF theory and the Hitchin system at the classical level. Further, we obtain a unifying action for a hierarchy of Lax equations describing the Hitchin system in terms of meromorphic Lax matrices. The cases of genus zero and one are treated in greater detail, leading to explicit Lagrangian one-forms for the hierarchies of the rational and the elliptic Gaudin models, respectively, and of the elliptic spin Calogero--Moser model as a special subcase of the latter.
\addcontentsline{toc}{chapter}{Abstract}

\newpage
\thispagestyle{empty}
\mbox{}
\newpage

\chapter*{Nomenclature}
For the reader's convenience, here is a list of abbreviations that are used repeatedly throughout this thesis.

\begin{itemize}

    \item \textbf{AKS}: Adler--Kostant--Symes (scheme)
    
    \item \textbf{CYBE}: Classical Yang--Baxter Equation

    \item \textbf{DST}: Discrete Self-Trapping (model)

    \item \textbf{HT}: Holomorphic-Topological (gauge theory)
    
    \item \textbf{IFTs}: Integrable Field Theories

    \item \textbf{mCYBE}: modified Classical Yang--Baxter Equation
    
\end{itemize}
\addcontentsline{toc}{chapter}{Nomenclature}

\newpage
\thispagestyle{empty}
\mbox{}
\newpage

\addtocontents{toc}{\protect\setlength{\cftbeforechapskip}{10pt}}

\tableofcontents

\newpage
\thispagestyle{empty}
\mbox{}
\newpage

\setcounter{chapter}{-1}

\chapter{Introduction}
\pagenumbering{arabic} 
Our quest to understand the perplexing complexity of nature and develop a unifying picture of its underlying principles has led to remarkable discoveries. Integrable models --- the central theme of this thesis --- sit right at the heart of this quest. They have long offered valuable insights by serving as theoretical laboratories for studying nature using exact techniques.

One can trace back the study of integrable systems to attempts to find exact solutions to Newton's equations of motion, made soon after Newton's dynamical laws were introduced over three centuries ago. The Kepler two-body problem, historically one of the first integrable systems to be studied, was solved by Newton himself.

Two different but closely related pictures soon evolved from Newtonian mechanics: the \emph{Hamiltonian} and the \emph{Lagrangian} frameworks. Both approaches provide useful reformulations of Newtonian mechanics and form the foundation of much of modern physics. However, when it comes to integrable systems, the \emph{traditional} Lagrangian framework has serious limitations. As we explain later in this chapter, this framework is ill-equipped for detecting and encoding integrability purely variationally. It is therefore not surprising that the modern era of research in integrable systems --- which started with the discovery of the \emph{classical inverse scattering method} \cite{GGKM} by Gardner, Greene, Kruskal and Miura --- has been primarily driven by the Hamiltonian formalism. A natural question arises: how does one address this inadequacy of Lagrangians? After all, they have otherwise been central to the formulation of the fundamental theories of nature.

However, the motivations behind this question go beyond this somewhat philosophical consideration. The Hamiltonian approach, despite having led to significant advances in our understanding of integrability, has its own shortcomings. For instance, the language of Hamiltonians is ill-suited for studying Lorentz-invariant theories and fully discrete systems, two important classes that play foundational roles in physics.

Thus, the need for a purely variational approach to integrability is also prompted by the possibility of using the advantages afforded by Lagrangian techniques to tackle problems where the Hamiltonian approach encounters limitations. This work, aimed at constructing a systematic Lagrangian framework for integrable models, is an attempt in this direction.

\section{Integrability in the Hamiltonian framework}\label{sec:hamintegrability-intro} 

Let us start by talking about integrable systems in the language of Hamiltonians where we have a definition of integrability through the work of Liouville. A Hamiltonian system with phase space of dimension $2n$ is said to be \emph{integrable in the sense of Liouville} or \emph{Liouville integrable}, if it possesses $n$ independent conserved quantities, say $H_i$, $i \geq 1$, that are in involution, that is,
\begin{equation}\label{eq:invol}
  \{H_i, H_j\} = 0,
\end{equation}
for every $i,j \geq 1$, with respect to some Poisson bracket on the phase space. The Liouville--Arnol'd theorem \cite{Li, Ar} then states that the equations of motion of the system can be solved by quadrature: this implies that the system is \emph{solvable} or \emph{integrable}.\footnote{One can extend this notion of integrability to Hamiltonian field theories as well: an integrable Hamiltonian field theory should possess infinitely many conserved charges in involution. We do not discuss the case of integrable field theories here, the focus of this thesis being finite-dimensional integrable systems.}

The field of integrable systems was to lie dormant for more than a century after the work of Liouville. This changed with the pioneering papers of Gardner, Greene, Kruskal and Miura \cite{GGKM}, and Zakharov and Shabat \cite{ZS} on the inverse scattering method. Shortly after, Ablowitz, Kaup, Newell and Segur \cite{AKNS} generalised this method and introduced the notion of an \emph{integrable hierarchy}. This notion stems very naturally from the definition of Liouville integrability we discussed above: each of the quantities $H_i$ in \eqref{eq:invol} can be used as a Hamiltonian to define a time flow with respect to the Poisson bracket. We can impose all these flows simultaneously on the phase space, as a result of which we obtain an entire collection of equations of motion referred to as an integrable hierarchy. 

The notion of an integrable hierarchy has played a crucial role in the study of integrable systems: it is often useful to treat an integrable model not on its own but as a part of the entire hierarchy it lives in. Studying the hierarchy as a whole then reveals more structure and properties of the initial models. For instance, the nonlinear Schr\"odinger (NLS) equation and the modified Korteweg-de Vries (mKdV) equation were originally studied as integrable field theories (IFTs) in their own right, only for it to be discovered later that they both naturally fit into the Ablowitz--Kaup--Newell--Segur (AKNS) hierarchy \cite{AKNS}. These ideas grew further and led to interesting developments in different directions, bringing in valuable algebraic and geometric perspectives to the subject of integrability. Some of these developments of relevance to this thesis will be discussed in Chapter \ref{chap:dialgebra-background}. For now, let us return to the notion of integrable hierarchies but in the framework of Lagrangians.

\section{Integrability in the Lagrangian framework}\label{sec:lagintegrability-intro} 
A scalar Lagrangian function only describes an individual model and cannot capture the notion of commuting (Hamiltonian) flows, or its discrete analogue known as multidimensional consistency \cite{BS, N1}. Given the crucial role the least action principle and Lagrangians play in describing the fundamental theories of nature, it is worthwhile addressing this limitation of the traditional Lagrangian formalism in encoding the notion of integrable hierarchies.

The obstacle to studying integrable hierarchies within a variational framework was overcome for the first time within the discrete setup in \cite{LN} by introducing \emph{Lagrangian multiforms}. A generalised action and a variational principle, involving a Lagrangian multiform, were proposed to capture the notion of multidimensional consistency purely variationally. Since its introduction in the work \cite{LN} of Lobb and Nijhoff in 2009, this idea has been developed in several directions, covering various realms of integrable systems: 
\begin{itemize}
    \item discrete and continuous finite-dimensional systems\\
    \cite{YKLN, Su, PS, V2}
    \item discrete systems in $2$ and $3$ dimensions\\
    \cite{LNQ, LN2, BobS, XNL, BollS, ALN, BollPS, BollPS2, LN3, RicV1, RicV2, NZ, N4}
    \item IFTs in $1+1$ and $2+1$ dimensions\\
    \cite{XNL, Su2, SuV, SNC1, SNC2, CS1, PV, CS2, V2, SNC3, N2, CStV, FNR}
    \item semi-discrete systems\\
    \cite{XNL, SV, N4}
\end{itemize}
Relations between discrete and continuous multiforms were explored in \cite{V3, V}, while the concept was extended to non-commuting flows in \cite{CNSV, CH}. More recently, connections between the multiform framework and the theory of multidimensional dispersionless integrable systems were established in \cite{FerVer}. 

This thesis is concerned with the case of \emph{continuous Lagrangian one-forms} which cover \emph{continuous finite-dimensional integrable systems}. We only present a brief introduction to the main ideas\footnote{The interested reader is also referred to \cite[Chapter 12]{HJN} for a comprehensive introduction to these ideas in the discrete setup.} here, reserving a detailed discussion of the necessary background on Lagrangian one-forms for later in Chapter \ref{chap:lm-background}.

Unlike a traditional Lagrangian, a Lagrangian multiform is a differential $d$-form which is integrated over a $d$-dimensional hypersurface in a so-called multitime of dimension greater than $d$ to yield an action functional depending not only on the field configurations but also on the hypersurface. One then postulates a principle of least action, which must be valid for any hypersurface embedded in the multitime space, together with the so-called \emph{closure relation} which provides a variational analogue of the Poisson involutivity of the Hamiltonians of a hierarchy.\footnote{Some early works on this theme are based on the framework of the so-called pluri-Lagrangian systems, which differ from Lagrangian multiforms in that they do not require the closure relation to be satisfied.} The generalised variational principle produces equations\footnote{Classifying all possible Lagrangian multiforms along these lines would amount to classifying all integrable hierarchies, which lends a further appeal to this framework.} that come in two flavours:
\begin{itemize}
    \item \emph{Standard Euler--Lagrange equations} associated with each of the coefficients of the Lagrangian multiform, which form a collection of Lagrangian densities
    \item \emph{Corner or structure equations} on the Lagrangian coefficients themselves which select possible models and ensure the compatibility of the various equations of motion imposed on a common set of fields
\end{itemize}

In practice, it is a non-trivial task to obtain all the Lagrangian coefficients of a multiform which produce compatible equations of motion. Beyond brute-force calculations to solve the corner equations~\cite{SNC1}, several works have used the idea of variational symmetries to achieve this goal \cite{PS,PV,SNC2}. This produces an algorithm to construct the Lagrangian coefficients one after the other from a given initial Lagrangian. Although perfectly fine in theory, this can become quickly unmanageable in practice, and usually formulas for only a few Lagrangian coefficients are obtained.\footnote{This also has the disadvantage of singling out some independent variables in the hierarchy which then appear as the so-called ``alien derivatives'' in the higher Lagrangian coefficients. See, for instance, \cite{V3, V} for discussions on alien derivatives.} Developing a systematic framework for the construction of Lagrangian multiforms has therefore been an open problem in general. 

This thesis is based on three joint works, \cite{CDS}, \cite{CSV}, and \cite{CHSV}, that provide a solution to this problem for the case of Lagrangian one-forms. We achieve this by introducing \emph{geometric Lagrangian one-forms} for large classes of finite-dimensional integrable hierarchies and developing two approaches for their construction.

Crucially, the framework of geometric Lagrangian one-forms is formulated in \emph{phase space} in contrast to most works on Lagrangian multiforms so far which are instead based in \emph{coordinate space}. As we will see later in this thesis, this has the advantage of placing various techniques from symplectic geometry at our disposal and of allowing for direct connections with more traditional features of integrability to be made. In connection with our approach, we also note the recent work \cite{CH} where a new variational principle for Lagrangian one-forms, termed the \emph{univariational principle}, was introduced by Caudrelier and Harland.

Apart from providing solutions to various problems in the framework of Lagrangian multiforms, this thesis also presents the first instance of the merging of this framework with that of \emph{mixed holomorphic-topological (HT) gauge theories}, another recent development in the study of integrable systems that has seen immense activity in the last few years.

In the next section, we present an overview of these results and the content of this thesis.

\section{Structure of the thesis}
This thesis has four parts. 

\begin{itemize}
  \item In Part \ref{part:background}, we start by providing the necessary background on aspects from the Hamiltonian framework for integrability: in particular, we discuss the notions of Lax pairs and classical $r$-matrices and the theory of Lie dialgebras in Chapter \ref{chap:dialgebra-background}. Then, in Chapter \ref{chap:lm-background}, we present the theory of Lagrangian multiforms in detail, focusing on the case of finite-dimensional integrable systems. These two chapters are intended to provide the required background for the work presented in the rest of the thesis. Together with older results by other authors, they also contain some novel content from the three works this thesis is based on, namely \cite{CDS}, \cite{CSV}, and \cite{CHSV}. 

  \item Part \ref{part:coadjointmultiform} is devoted to the framework of Lagrangian multiforms on coadjoint orbits we introduced in \cite{CDS} and generalised further in \cite{CSV}. The general framework is the content of Chapter \ref{chap:lm-orbits}, where we also discuss several main results for the geometric Lagrangian one-form. This framework resolves the open problem of a systematic construction of Lagrangian one-forms for a large class of integrable hierarchies that fall within the Lie dialgebra setup. Then, in Chapter \ref{chap:orbitexamples}, we illustrate the construction of geometric Lagrangian one-forms for some well-known integrable models and, in the process, fill multiple gaps in the landscape of Lagrangian multiforms.

  \item We then switch from the algebraic approach of \cite{CDS} and \cite{CSV} to a gauge-theoretic one in Part \ref{part:gaugemultiform}. This is based on the joint work \cite{CHSV} and deals with casting a large class of integrable systems, called Hitchin systems, into the framework of Lagrangian multiforms. This variational setting naturally produces a multiform version of the action of $3$d mixed BF theory with defects, a lower-dimensional analogue of the celebrated $4$d semi-holomorphic Chern--Simons theory. In Chapter \ref{chap:bf-hitchin}, we give details on the required geometric setup and describe the construction of a Lagrangian one-form for the Hitchin system. Then, in Chapter \ref{chap:lax-hitchin}, we obtain a reduced Lagrangian one-form that provides a variational formulation of the Lax description of the Hitchin system. Finally, in Chapter \ref{chap:hitchinexamples}, we focus on the cases of genus 0 and 1 to produce explicit geometric Lagrangian one-forms that describe some well-known rational and elliptic integrable models.   
  
  \item We conclude this thesis by discussing some perspectives and open questions in Chapter \ref{chap:perspectives} in Part \ref{part:conclusion}.

\end{itemize}

\part{Background}\label{part:background}
\thispagestyle{empty}

\newpage
\thispagestyle{empty}
\mbox{}
\newpage

\chapter{Classical \texorpdfstring{$\boldsymbol{r}$}{r}-matrices and Lie dialgebras}\label{chap:dialgebra-background}
This chapter is devoted to some fundamental concepts in the Hamiltonian framework for finite-dimensional integrable systems. We present a basic introduction to the Lax formulation and the notion of classical $r$-matrices in Section \ref{sec:lax-rmatrix-background}. Then, in Section \ref{sec:dialgebra-background}, which is adapted from \cite{CDS} and \cite{CSV}, we review the key results from the theory of Lie dialgebras that form the foundation of the work presented in Part \ref{part:coadjointmultiform} of this thesis. Proofs have been omitted as this chapter is intended to serve as a quick overview of these topics. There are many wonderful references that can be referred to for further details. See, for instance, \cite{RSTS, Aud, BBT, STS}.

\section{Lax pairs and classical \texorpdfstring{$r$}{r}-matrices}\label{sec:lax-rmatrix-background}

We begin with the notion of \emph{Lax pairs} which arose out of the discovery made by Lax in \cite{La} that equations of motion of some dynamical systems can be expressed as a \emph{Lax equation}
\begin{equation}\label{eq:laxequation}
    \frac{\d L}{\d t} = [M, L].
\end{equation} 
Here $L$ and $M$ belong to some Lie algebra $\g$ with Lie bracket $[\,,\,]$, and form a Lax pair, a pair of globally defined maps from the phase space of the dynamical system to $\g$ such that its equations of motion are equivalent to \eqref{eq:laxequation}. Crucially, the Lax matrix $L$ contains all information on initial data and $M$ is a function of $L$. Further, Lax matrices depend analytically on an auxiliary parameter $\lda$, called the \emph{spectral parameter}, for many systems.\footnote{Except for the open Toda chain, presented in Sections \ref{Flaschka} and \ref{Toda_pq}, this will be the case for all systems that appear in this thesis.} The underlying Lie algebra in such cases is a \emph{loop algebra} $\g \otimes \mathbb{C}[\lda, \lda^{-1}]$ with $\g$ being a matrix Lie algebra.

In general, the Lax formulation of a system (if it exists) is not unique. Nonetheless, the notion of Lax pairs provides a powerful general principle that allows one to associate a linear operator, namely a Lax matrix, with a nonlinear equation, so that the eigenvalues of the linear operator are conserved quantities of the nonlinear equation.

Let us emphasise, however, that the flow described by the Lax equation \eqref{eq:laxequation} is not integrable in general. Recall from Section \ref{sec:hamintegrability-intro} that for a Hamiltonian system on a $2n$-dimensional phase space to be Liouville integrable, it is necessary (and sufficient) for it to have $n$ independent conserved quantities \emph{in involution}. This additional requirement for these quantities to be in involution leads to an important result first established in the remarkable work \cite{BV} of Babelon and Viallet: any Hamiltonian system that is Liouville integrable admits a Lax representation, at least locally at generic points in phase space. The Lax representation further leads to the notion of a \emph{classical $r$-matrix} which we shortly define. First, let us introduce some notation we use in the discussion to follow. Denoting by $\sfE_{i}$ a basis of the Lie algebra $\g$, we write 
\begin{equation}
    L(\lda) = \sum_i L_i(\lda) \sfE_i
\end{equation}
where the components $L_i$ are functions on the phase space, and
\begin{equation}
    L_{\ti{1}}(\lda) = L(\lda) \otimes \1 = \sum_i L_i(\lda) (\sfE_i \otimes \1), \qquad L_{\ti{2}}(\mu) = \1 \otimes L(\mu) = \sum_i L_i(\mu) (\1 \otimes \sfE_i)
\end{equation}
with the indices on $L_{\ti{1}}$ and $L_{\ti{2}}$ denoting that $L$ sits in the first and the second factors in the tensor product respectively. Similarly, we write
\begin{equation}
    r_{\ti{12}}(\lda, \mu) = \sum_{ij} r_{i,j}(\lda, \mu) \sfE_i \otimes \sfE_j, \qquad r_{\ti{21}}(\mu, \lda) = \sum_{ij} r_{i,j}(\mu, \lda) \sfE_j \otimes \sfE_i
\end{equation}
for $r$ living in the tensor product $\g \otimes \g$. Further, we define
\begin{equation}
    \{L_{\ti{1}}(\lda), L_{\ti{2}}(\mu)\} = \sum_{ij} \{L_i(\lda), L_j(\mu)\} \sfE_i \otimes \sfE_j.
\end{equation}
Then, we have
\begin{theorem}
    The eigenvalues of $L(\lda)$ are in involution if and only if there exists a function $r_{\ti{12}}(\lda, \mu) \in \g \otimes \g$ defined on the phase space such that
    \begin{equation}\label{eq:lax-poisson}
        \{L_{\ti{1}}(\lda), L_{\ti{2}}(\mu)\} = [r_{\ti{12}}(\lda, \mu), L_{\ti{1}}(\lda)] - [r_{\ti{21}}(\mu, \lda), L_{\ti{2}}(\mu)].
    \end{equation}
\end{theorem}
The object $r_{\ti{12}}$ is called a classical $r$-matrix and has played a central role in the study of integrable systems from both algebraic and geometric perspectives since it first appeared in the works of Sklyanin \cite{Sk1, Sk2}. Famously, it led to the discovery of the quadratic Poisson brackets on Poisson--Lie groups \cite{Sk3}, and also played a central role in relating the Hamiltonian structure of integrable systems with the Riemann--Hilbert factorisation problem \cite{STS2}, which, in turn, provides a setup for obtaining explicit solutions.

\begin{remark}\label{rem:nonultralocal}
The case when $r$ is skew-symmetric, that is $r_{\ti{12}}(\lda, \mu) = - r_{\ti{21}}(\mu, \lda)$, is in general better understood compared to the non-skew-symmetric case. This presents a serious limitation when dealing with integrable field theories (IFTs) in particular. The quantisation of \emph{non-ultralocal classical integrable field theories} --- a class of IFTs whose underlying Poisson structure is governed by non-skew-symmetric classical $r$-matrices --- has been a long-standing open problem. The case of non-ultralocal IFTs incorporates the important family of integrable $\sgm$-models which includes the well-known principal chiral model and the Wess--Zumino--Witten model. At present, the most successful approach for quantising IFTs and establishing their integrability at the quantum level is the quantum inverse scattering method \cite{TaF, KuS}. However, this approach suffers from the drawback of only being applicable to the case of ultralocal IFTs, which correspond to the skew-symmetric case, leaving the problem of a systematic quantisation of non-ultralocal IFTs open. As this thesis only deals with the finite-dimensional case, we do not discuss this further and refer the interested reader to \cite{FR, STSSev, DMV} for some interesting works on this topic.
\end{remark}

When $r_{\ti{12}}$ is non-dynamical, the iteration of \eqref{eq:lax-poisson} together with the Jacobi identity leads to the following condition:
\begin{equation}
    [[r_{\ti{12}}(\lda, \mu), r_{\ti{13}}(\lda, \nu)] + [r_{\ti{12}}(\lda, \mu), r_{\ti{23}}(\mu, \nu)] + [r_{\ti{32}}(\nu, \mu), r_{\ti{13}}(\lda, \nu)], L_{\ti{1}}(\lda)] = 0.
\end{equation}
For a general $L(\lda)$, the condition gives the celebrated \emph{classical Yang--Baxter equation}
\begin{equation}\label{eq:cybe}
    [r_{\ti{12}}(\lda, \mu), r_{\ti{13}}(\lda, \nu)] + [r_{\ti{12}}(\lda, \mu), r_{\ti{23}}(\mu, \nu)] + [r_{\ti{32}}(\nu, \mu), r_{\ti{13}}(\lda, \nu)] = 0.
\end{equation}
This is the general version of the CYBE applicable to non-skew-symmetric matrices. For a simple Lie algebra $\g$, the CYBE has large classes of solutions, in both the spectral parameter-independent and spectral parameter-dependent cases. These solutions for the skew-symmetric case, under an additional
nondegeneracy condition, were classified by Belavin and Drinfel'd \cite{BD1,BD2}. In \cite{D}, Drinfel'd also provided a geometric interpretation of the CYBE which eventually led to the beautiful theory of Poisson--Lie groups. Also see, for instance, \cite{STS3}.

\begin{remark}
    As mentioned, \eqref{eq:cybe} corresponds to the case of a non-dynamical classical $r$-matrix. However, this is not always the case. The dynamical analogue of \eqref{eq:cybe}, called the \emph{classical dynamical Yang--Baxter equation}, and its quantum counterpart are important objects of study with interesting connections to conformal field theory and statistical mechanics. The models we deal with using an algebraic approach in Part \ref{part:coadjointmultiform} all belong to the case where the underlying $r$-matrix is non-dynamical, so we do not discuss the dynamical case in detail here. The interested reader is referred to \cite{EtS} for a review of developments related to the classical dynamical Yang--Baxter equation.     
\end{remark}

\section{The theory of Lie dialgebras}\label{sec:dialgebra-background}

Having provided an introduction to the Lax formulation of integrable systems and the notion of classical $r$-matrices, we now turn to discussing the theory of Lie dialgebras, which underlies the framework of Lagrangian multiforms on coadjoint orbits \cite{CDS, CSV} to be covered in Part \ref{part:coadjointmultiform} of this thesis. For simplicity, we only work with complex matrix Lie algebras and their corresponding connected, simply connected complex Lie groups throughout this section.

The notion of Lie dialgebras, introduced by Semenov-Tian-Shansky \cite{STS2}, is associated with the factorisation aspects of classical $r$-matrices and provides a useful setup for studying integrable systems.\footnote{Some earlier works --- for instance, \cite{STS2, KS} --- refer to Lie dialgebras as \emph{double Lie algebras} instead. We follow \cite{STS} in referring to them by their newer name.} It is important to note that Lie dialgebras are different from the perhaps better-known \emph{Lie bialgebras} appearing in Drinfel'd's theory of Poisson--Lie groups. Relations and differences between these two structures can be found in \cite{STS} and \cite{KS}, for instance.

\subsection{The general setup}

Let $\g$ be a matrix Lie algebra, with matrix Lie group $G$, and $\g^*$ its dual space. We have the usual (co)adjoint actions for all $X,Y\in\g$, $\xi\in\g^*$, and $g\in G$,
\begin{subequations}
\begin{align}
 {\rm ad}_X\cdot Y&=[X,Y],\qquad ({\rm ad}^*_X\cdot \xi)(Y)=-\xi({\rm ad}_X\cdot Y)=-\xi([X,Y]),\\[1ex]
 \text{Ad}_{g}\cdot X&=g\,X\,g^{-1},\qquad   \text{Ad}^*_g\cdot \xi(X)=\xi(\text{Ad}_{g^{-1}}\cdot X).
\end{align}
\end{subequations}
Further, the Lie bracket on $\g$ allows us to endow the dual space $\g^*$ with a Poisson structure through the Lie--Poisson bracket\footnote{Interestingly, the Lie--Poisson bracket was first discovered by Lie \cite{Lie} in the late nineteenth century but not given much attention to till the works of Kirillov \cite{Kir2}, Kostant \cite{Kos2}, and Souriau \cite{Sou} in the second half of the following century.} defined by
\begin{equation}
\label{def_Lie_Poisson}
    \{f,g\}(\xi)= (\xi,[\nabla f(\xi),\,\nabla g(\xi)]),\qquad f,g\in C^\infty(\g^*),
\end{equation}
where we introduced the convenient notation $(\,~,~)$ for the natural pairing between $\g^*$ and $\g$: $\xi(X)=(\xi,X)$. The gradient $\nabla f(\xi)$ is the element of $\g$ defined from the differential $\delta f(\xi)$ by using the pairing
\begin{equation}
    \delta f(\xi)(\eta)=\lim_{\epsilon\to 0}\frac{f(\xi+\epsilon \eta)-f(\xi)}{\epsilon}=(\eta,\nabla f(\xi)).
\end{equation}
The Lie--Poisson bracket is degenerate in general, and the $\text{Ad}^*$-invariant functions on $\g^*$ are the Casimir functions. Every maximal connected submanifold of $\mathfrak g^*$ on which all Casimir functions are constant inherits a nondegenerate Poisson structure, and is therefore a symplectic manifold. Since the Hamiltonian vector fields of linear functions generate the infinitesimal coadjoint action, these maximal connected level sets are exactly the coadjoint orbits of $G$ in $\g^*$. We thus obtain a symplectic foliation of $\mathfrak g^*$ into coadjoint orbits, and the restriction of the Poisson structure to each orbit gives rise to the Kostant--Kirillov--Souriau symplectic form. The interested reader is referred to \cite{Sou, Sou2} for a comprehensive introduction to these ideas and a detailed treatment of classical mechanics in the framework of symplectic geometry.

\begin{remark}
    Since $G$ acts transitively on any coadjoint orbit of $G$ in $\g^*$, it preserves the Kostant--Kirillov--Souriau symplectic form on the coadjoint orbit. A coadjoint orbit is therefore an example of a \emph{homogeneous} symplectic manifold, that is, a symplectic manifold with a transitive Lie group action that preserves the symplectic form. We refer the reader to \cite[Chapter 1]{Kir3} for a discussion on the geometry of coadjoint orbits.
\end{remark}

Now, let $R:\g\to\g$ be a linear map. It is a solution of the {\it modified} classical Yang--Baxter equation (mCYBE) if it satisfies
\begin{equation}
\label{mCYBE}
[R(X),R(Y)]-R\left([R(X),Y]+[X,R(Y)]\right)=-[X,Y],
\end{equation}
for all $X, Y \in \g$. We will refer to a solution $R$ of \eqref{mCYBE} as a (classical) $r$-matrix, in relation to the fact that with $R$ one can associate a classical $r$-matrix $r\in\g\otimes \g$ when $\g$ is equipped with a nondegenerate ad-invariant symmetric bilinear form $\langle\,~,~\rangle$. For instance, this is given by the Killing form when $\g$ is a finite-dimensional semisimple Lie algebra. A well-known example of an $r$-matrix arises in the case where $\g$ admits a direct sum decomposition\footnote{Following \cite{STS}, we use $\dotplus$ to denote a direct sum of vector spaces, and $\oplus$ to denote a direct sum of Lie algebras throughout this work.} as a vector space into two Lie subalgebras
\begin{equation}
	\g=\g_+\dotplus\g_-.
\end{equation}
Then, $R=P_+-P_-$ is a solution of \eqref{mCYBE}, where $P_\pm$ is the projector on $\g_\pm$ along $\g_\mp$. 

Given a solution $R$ of the mCYBE, we can define on the vector space $\g$ a second Lie bracket
\begin{equation}
\label{def_R_bracket}
	[X,Y]_R=\frac{1}{2}\left([R(X),Y]+[X,R(Y)]\right),
\end{equation}
and denote the corresponding Lie algebra by $\g_R$. Note that $\g$ and $\g_R$, being the same as vector spaces, have the same dual space $\g^*$. We have an adjoint action of $\g_R$ on itself and a coadjoint action of $\g_R$ on $\g^*$:
\begin{subequations}
\begin{align}
&	{\rm \text{ad}}^R_X\cdot Y=[X,Y]_R,\\
 &({\rm \text{ad}^*}^R_X\cdot \xi)(Y)=-(\xi,\,{\rm \text{ad}}^R_X\cdot Y)=-(\xi,\,[X,Y]_R),
\end{align}
\end{subequations}
for all $X, Y \in \g$ and $\xi \in \g^*$.

We refer to the pair $(\g, R)$ as a \emph{Lie dialgebra}. The algebraic significance of the mCYBE and of the second Lie bracket $[\,~,~]_R$ is given by the following results which lead to essential factorisation properties underlying integrable systems. The key objects are the maps
\begin{equation}
    R_\pm=\frac{1}{2}\left(R\pm {\rm Id}\right).
\end{equation}
\begin{proposition}
	Let $\g_\pm={\rm Im}\,R_\pm$. Then,
	\begin{enumerate}
		\item $R_\pm:\g_R\to\g$ are Lie algebra homomorphisms:
		\begin{equation}
			R_\pm\left([X,Y]_R\right)=\left[R_\pm(X),R_\pm(Y)\right].
		\end{equation} 
  In particular, $\g_\pm\subset\g$ are Lie subalgebras of $\g$.
	
	\item The mapping $i_R:\g_R\to \g_+\oplus\g_-$, $i_R(X)=(R_+(X),R_-(X))$ is a Lie algebra embedding. Thus, $\widetilde{\g}_R={\rm Im}\,i_R$ is a Lie subalgebra of $\g_+ \oplus \g_-$ and isomorphic to $\g_R$.

\item The composition of the maps
 \begin{equation}
   i_R:  \g_R ~{\to}~ \g_+\oplus \g_-,\quad X\mapsto(R_+(X),R_-(X)),
 \end{equation}
 followed by
  \begin{equation}
a:\g_+\oplus \g_-~{\to}~\g,\quad (X_+,X_-)\mapsto X_+-X_-,
 \end{equation}
provides a unique decomposition of any element $X\in\g$ as $X=R_+(X)-R_-(X)$.
	\end{enumerate}
\end{proposition}
\noindent
Note that $R_+-R_-={\rm Id}$ and
\begin{equation}
[X,Y]_R=	R_+\left([X,Y]_R\right)-R_-\left([X,Y]_R\right)=\left[R_+(X),R_+(Y)\right]-\left[R_-(X),R_-(Y)\right].
\end{equation}
We can express the actions of $\g_R$ in terms of those of $\g$. For convenience, we write $X_\pm=R_\pm(X)$ for $X\in\g$. Then,
\begin{subequations}
\begin{align}
&{\rm ad}^R_X\cdot Y=\frac{1}{2}{\rm ad}_{R(X)}\cdot Y+\frac{1}{2}{\rm ad}_{X}\cdot R(Y)={\rm ad}_{X_+}\cdot Y_+ -{\rm ad}_{X_-}\cdot Y_-,\\
\label{relation_ad_adR}
&{\rm \text{ad}^*}^R_X\cdot \xi=\frac{1}{2}{\rm \text{ad}}^*_{R(X)}\cdot\xi+\frac{1}{2}R^*({\rm \text{ad}}^*_{X}\cdot\xi)=R_+^*({\rm \text{ad}}^*_{X_+}\cdot\xi)-R_-^*({\rm \text{ad}}^*_{X_-}\cdot\xi),
\end{align}
\end{subequations}
where the adjoint $A^*:\g^*\to\g^*$ of a linear map $A:\g\to\g$ is defined by $(A^*(\xi),X)=(\xi,A(X))$.

The application of this framework to integrable systems relies on the interplay between the two Lie--Poisson brackets one can define on $\g^*$. Indeed, having a second Lie bracket, we can repeat the definition \eqref{def_Lie_Poisson} to obtain
\begin{equation}
\label{Lie_PB}
    \{f,g\}_R(\xi)= (\xi,[\nabla f(\xi),\,\nabla g(\xi)]_R).
\end{equation}
Exactly like the earlier case of the space $\g^*$ endowed with the Lie--Poisson bracket $\{\,~,~\}$ defined by \eqref{def_Lie_Poisson}, we have a symplectic foliation of the space $\g^*$ equipped with $\{\,~,~\}_R$ into coadjoint orbits of $G_R$, the Lie group of $\g_R$, in $\g^*$. The restriction to a coadjoint orbit gives rise to the symplectic form which we denote by $\omega_R$. It is the interplay between these two structures that provides integrable systems whose equations of motion take the form of a Lax equation. For this last part, one needs one more ingredient: an $\text{Ad}$-invariant nondegenerate bilinear symmetric form $\langle\,~,~\rangle$ on $\g$. It allows us to identify $\g^*$ with $\g$ and the coadjoint actions with the adjoint actions.
Specifically, one has 
\begin{theorem}
The $\textup{Ad}^*\!$-invariant functions on $\g^*$ are in involution with respect to $\{\,~,~\}_R$. The equation of motion 
\begin{equation}
    \frac{\d}{\d t}L=\{L,H\}_R
\end{equation}
induced by an $\textup{Ad}^*\!$-invariant function $H$ on $\g^*$ takes the following equivalent forms, for an arbitrary $L\in\g^*$,
\begin{equation}
\label{Lax2}
   \frac{\d}{\d t}L=\textup{ad}^{R*}_{\nabla H(L)}\cdot L=\frac{1}{2}\,\textup{ad}^*_{R\nabla H(L)}\cdot L=\textup{ad}^*_{R_\pm \nabla H(L)}\cdot L.
\end{equation}
When there is an $\textup{Ad}$-invariant nondegenerate bilinear form $\langle\,~,~\rangle$ on $\g$ so that we can identify $\g^*$ with $\g$ and $\textup{ad}^*\!$ with $\textup{ad}$, the last equation takes the desired form of a Lax equation for $L\in\g$,
\begin{equation}
  \frac{\d}{\d t}L=[M_\pm,L],\qquad M_\pm=R_\pm \nabla H(L).
\end{equation}
\end{theorem}
The proof can be found in \cite{BBT}, for instance, and we only elaborate on certain points that we need here. The crucial point is to exploit the $\text{Ad}^*$-invariance of the function $H$ defining the time flow. The latter means 
 that the following property holds
\begin{equation} 
\label{ad_invariance_property}
    \text{ad}^*_{\nabla H(\xi)} \cdot\xi = 0, \quad \Leftrightarrow \quad (\xi,\, [\nabla H(\xi),X])=0, \qquad \forall \xi \in \mathfrak{g}^*,\quad \forall X\in\g.  
\end{equation}
Thus, for any two $\text{Ad}^*$-invariant functions $H_1$ and $H_2$,
\begin{equation}\label{involutivity}
    \begin{split}
    \{H_1,H_2\}_R(\xi)&= (\xi,[\nabla H_1(\xi),\,\nabla H_2(\xi)]_R)\\
    &=\frac{1}{2}(\xi,[R\nabla H_1(\xi),\,\nabla H_2(\xi)]+[\nabla H_1(\xi),\,R\nabla H_2(\xi)])=0.
    \end{split}
\end{equation}
For any function $f$ on $\g^*$, the time evolution associated with the $\text{Ad}^*$-invariant $H$ with respect to the Poisson bracket $\{\,~,~\}_R$ is defined by
\begin{equation}
    \frac{\d}{\d t}f(L)=\{f,H\}_R(L) , 
\end{equation}
\ie
\begin{equation}
    \begin{split}
    \left(\frac{\d}{\d t}L,\, \nabla f(L)  \right)&= (L,[\nabla f(L),\,\nabla H(L)]_R)=-\frac{1}{2}\,(L,[R\nabla H(L),\,\nabla f(L)])\\
    &=(\text{ad}^{R*}_{ \nabla H(L)}\cdot L,\,\nabla f(L))=\frac{1}{2}(\text{ad}^{*}_{R\nabla H(L)}\cdot L,\,\nabla f(L)).
    \end{split}
\end{equation}
Finally, in view of \eqref{def_R_bracket} and \eqref{ad_invariance_property}, we have 
\begin{equation}
    \text{ad}^{*}_{R\nabla H(L)}\cdot L=2\,\text{ad}^{*}_{R_\pm \nabla H(L)}\cdot L,
\end{equation}
thus establishing the various equivalent forms of the equations in \eqref{Lax2} (by restricting $f$ to be any of the coordinate functions on $\g^*$).

The involutivity property \eqref{involutivity} ensures that we can define compatible time flows associated with a family of $\text{Ad}^*$-invariant Hamiltonian functions $H_k$, $k=1,\dots,N$.
If one can supply enough such independent functions, or work on a coadjoint orbit of low-enough dimension, one obtains an integrable system described by an integrable hierarchy of equations in Lax form (again using the identification provided by $\langle\,~,~\rangle$)
\begin{equation}
\label{Lax_system}
    \partial_{t^k}L=[R_\pm \nabla H_k(L),L],\qquad k=1,\dots,N.
\end{equation}
The typical example of an invariant function $H_k$ is given by $H_k=\frac{1}{k+1}\Tr(L^k)$.

For our purposes, the Lie groups associated with $\g$ and $\g_R$ will be important. We introduce $G$ and $G_R$ as the (connected, simply connected) Lie groups defined for $\g$ and $\g_R$ respectively. For simplicity, we only think of matrix groups in this work. Only in special circumstances are $G$ and $G_R$ diffeomorphic. In general, this is only true in a neighbourhood of the identity where the crucial difference between the two groups lies in their multiplications induced by $[\,~,~]$ and $[\,~,~]_R$ respectively. The homomorphisms $R_\pm$ give rise to Lie group homomorphisms (which we denote by the same symbols) and we obtain a factorisation at the group level. With $g=\text{e}^X$, $X\in\g$, we have 
\begin{equation}
    R_\pm\, g=\text{e}^{R_\pm X}.
\end{equation}
Specifically, let $G_\pm=R_\pm(G_R)$ be the subgroups of $G$ corresponding to $\g_\pm$. The composition of the maps 
 \begin{equation}
   i_R:  G_R\,\to\,G_+ \times G_-,\quad g\mapsto(R_+(g),R_-(g)),
 \end{equation}
 followed by
  \begin{equation}
m:G_+\times G_- \,\to\,G,\quad (g_+,g_-)\mapsto g_+ \, g_-^{-1},
 \end{equation}
allows us to factorise {\it uniquely} an arbitrary element $g\in G$ (sufficiently close to the identity) as 
\begin{equation}
    g=g_+ \, g_-^{-1},\quad (g_+,g_-)\in \widetilde{G}_R={\rm Im}\,i_R.
\end{equation}
An element $g\in G_R$ can be identified with its image $(g_+,g_-)\in \widetilde{G}_R\subseteq G_+\times G_-$ and the multiplication $\cdot_R$ in $G_R$ is most easily visualised using the homomorphism property
\begin{equation}
    i_R( g\cdot_R h)=i_R( g)*i_R( h)=(g_+\,h_+,\,g_-\,h_-)
\end{equation}
where $*$ is the direct product group structure of $G_+\times G_-$. This is usually shortened to
\begin{equation}
  g\cdot_R h= (g_+\,h_+,\,g_-\,h_-).
\end{equation}
The group $G_R$ acts on $\g_R$ by the adjoint action and on $\g^*$ via the coadjoint action
\begin{subequations}
\begin{align}
     &\text{Ad}^R_g\cdot X =g\cdot_R X\cdot_R g^{-1},\qquad  \forall \, X\in\g_R,\quad g\in G_R,\\[1ex]
    &\text{Ad}^{R*}_g\cdot \xi(X) =(\xi,\,\text{Ad}^R_{g^{-1}}\cdot X),\qquad  \forall \,g\in G_R,\xi\in\g^*,\quad X\in\g_R.
\end{align}
\end{subequations}
\begin{remark}
\label{rem_assoc}
    When using the notation $g\cdot_R X\cdot_R g^{-1}$ for the adjoint action, we assume that $\cdot_R$ is an associative product on the matrix Lie algebra and its Lie group. One possible case in which this assumption can be made occurs when $\g$ is an associative algebra and $R$ is a solution of the associative Yang--Baxter equation (AYBE), \begin{equation}
        R(X)\, R(Y)-R(R(X)\, Y+ X\, R(Y))+ X\, Y=0.
    \end{equation}
    This implies that $X\cdot_R Y=\frac{1}{2}\left(R(X)\, Y+X\, R(Y) \right)$ defines a second associative product on $\g$ and allows us to view $[X,Y]_R$ as the commutator $X\cdot_R Y-Y\cdot_R X$, in direct analogy with $[X,Y]=X\, Y-Y\, X$, see \cite{STS2}. However, we should emphasise that the AYBE is only a sufficient (and not necessary condition) for $\cdot_R$ to be an associative product. In the rest of the thesis, we will assume that $\cdot_R$ is such an associative product (and not that $R$ is necessarily a solution of the AYBE) in order to use its consequences, for instance, $[X,Y]_R=X\cdot_R Y-Y\cdot_R X$.
\end{remark}

The following relations are most useful in the practical calculations of the examples discussed below. With $g_\pm=R_\pm\,g$, $X_\pm=R_\pm\,X$, $g\in G_R$, $X\in \g_R$,
\begin{subequations}
\begin{align}
   \text{Ad}^R_g\cdot X&=g_+\, X_+\,g_+^{-1}-g_-\, X_-\,g_-^{-1},\\[1ex]
\label{Rcoadjoint_action} 
\text{Ad}^{R*}_g\cdot \xi&=R_+^*(\text{Ad}^*_{g_+} \xi)-R_-^*(\text{Ad}^*_{g_-}\xi),\qquad  \forall \xi\in\g^*.
\end{align}
\end{subequations}
Thus, the dual space $\g^*$ hosts two coadjoint actions of $G$ and $G_R$, as it does with the two coadjoint actions of the Lie algebras $\g$ and $\g_R$. 

The last main result of this framework is known as the {\it factorisation theorem}, see, for instance, \cite{STS,BBT,AVV}.
\begin{theorem}
 Consider the system of compatible equations with the given initial condition 
\begin{equation}
\label{system_Lax}
 \partial_{t^k}L=\textup{ad}^*_{R_\pm \nabla H_k(L)}\cdot L,\qquad k=1,\dots,N,
\qquad L(0,\dots,0)=L_0\in\g^*.
\end{equation} 
Denote $(t^1,\dots,t^N)={\bf t}$ for conciseness. Let $g_\pm({\bf t})$ be the smooth curves in $G_\pm$ which solves the factorisation problem
    \begin{equation}
    \label{factorisation}
     \text{e}^{-\sum_{k=1}^N t^k\nabla H_k(L_0)}=g_+({\bf t})^{-1}\,g_-({\bf t}),\qquad g_\pm({\bf 0})=e.
    \end{equation}
    Then, the solution to the initial-value problem \eqref{system_Lax} is given by
    \begin{equation}
    \label{solution_L}
        L({\bf t})=\textup{Ad}^*_{g_+({\bf t})}\cdot L_0=\textup{Ad}^*_{g_-({\bf t})}\cdot L_0,
    \end{equation}
    and $g_{\pm}({\bf t})$ satisfy 
     \begin{equation}
          \label{split}
  \partial_{t^k} g_\pm({\bf t}) =R_\pm \nabla H_k(L({\bf t}))\,g_\pm({\bf t}).
          \end{equation}
\end{theorem}
This result shows that the solution lies at the intersection of coadjoint orbits of $G$ and $G_R$. Combined with the fact that the coadjoint orbits provide the natural symplectic manifolds associated with the corresponding Lie--Poisson bracket, this means that the natural arena to define our phase space, \ie where $L$ lives, is a coadjoint orbit of $G_R$ in $\g^*$,
\begin{equation} \label{eq:coadj_orbit}
    {\cal O}_\Lambda=\{\text{Ad}^{R*}_{\varphi}\cdot \Lambda;\varphi\in G_R\},\qquad \text{for some}\,\,\Lambda\in\g^*. 
\end{equation}

\begin{remark}
The essential results of the theory of Lie dialgebra discussed above extend to the infinite-dimensional setting, for instance, to the case of loop algebras\footnote{There are several subtleties related to duals in infinite dimensions and completions which we do not detail out here for keeping the discussion concise.}. As discussed earlier, infinite-dimensional Lie algebras are relevant when one needs Lax matrices with spectral parameters. This is typically the case for integrable field theories but it can also be required for some finite-dimensional systems such as the closed Toda chain or Gaudin models. We will discuss the extension of the Lie dialgebra construction to this infinite-dimensional setting later in this chapter and present it via the examples of non-cyclotomic and cyclotomic Gaudin models in Sections \ref{sec:noncycloGaudin} and \ref{sec:cycloGaudin} respectively. 
\end{remark}

\subsection{The Adler--Kostant--Symes scheme}\label{sec:AKS}

A special case of the Lie dialgebra setup arises when $\g$ admits a direct sum decomposition (as a vector space) into two Lie subalgebras
\begin{equation}
	\g=\g_+\dotplus\g_-,
\end{equation}
and we take $R=P_+-P_-$, where $P_\pm$ is the projector on $\g_\pm$ along $\g_\mp$. The decomposition of $\g$ induces the decomposition 
\begin{equation}
	\g^*=\g_+^*\dotplus\g_-^*.
\end{equation}
Using a nondegenerate ad-invariant bilinear form on $\g$, we can identify $\g_\pm^*$ with $\g_\mp^\perp$, where $\g_\mp^\perp$ denotes the orthogonal complement of $\g_\mp$. We fix $\Lambda$ to be in $\g_-^*$ and consider the coadjoint orbit of elements $L=\text{Ad}^{R*}_{\varphi} \cdot \Lambda$. As a result, only the subgroup $G_-$ in $G_R\simeq G_+\times G_-$ plays a role since $L=\text{Ad}^{R*}_{\varphi} \cdot \Lambda=-R_-^*(\text{Ad}^*_{\varphi_-}\cdot \Lambda)$ and the coadjoint orbit ${\cal O}_\Lambda$ lies in $\g_-^*$. This is the historic setup of the so-called Adler--Kostant--Symes (AKS) scheme \cite{A, K, Sy} which can be used to formulate the open Toda chain in Flaschka coordinates. $R$ is {\it not} skew-symmetric in this case. We will present this example in Section \ref{Flaschka} where details on our Lagrangian multiform for this model will be given.

We conclude this chapter by presenting an infinite-dimensional generalisation\footnote{In the rest of this chapter, we use the terms ``Lie groups'' and ``manifolds'' simply in analogy with the finite-dimensional case. The particular instances of this infinite-dimensional generalisation that appear in this thesis (in Chapter \ref{chap:orbitexamples}) all correspond to the cases of loop algebras and loop groups which are well-understood. The interested reader is referred to \cite{PreSeg, Kac} for detailed discussions on various aspects of infinite-dimensional Lie algebras and Lie groups.} of the AKS scheme that will be required for the case of the cyclotomic Gaudin model dealt with in Section \ref{sec:cycloGaudin}. In the case of this model, as well as its non-cyclotomic version in Section \ref{sec:noncycloGaudin}, we need to work with infinite-dimensional Lie algebras to cast them in the framework of Lie dialgebras. So, for clarity of exposition, we denote by $\mathbb{g}$ the infinite-dimensional Lie algebra of interest and reserve the notation $\g$ to denote finite-dimensional Lie algebras. 

Let $\mathbb{g}$ be the infinite-dimensional Lie algebra of interest with Lie group $\mathbb{G}$, which we assume is equipped with a nondegenerate invariant bilinear pairing 
\begin{equation} \label{pairing V g}
    \langle \cdot, \cdot \rangle : V \times \mathbb{g} \longrightarrow \CC
\end{equation}
    with some representation $V$ of $\mathbb{G}$. 
The AKS scheme, as detailed above, can then be applied more generally by using $V$ as a model for the dual space $\mathbb{g}^\ast$ of $\mathbb{g}$. We will denote the representation of $\mathbb{G}$ by $\Ad^\ast : \mathbb{G} \times V \to V$, $(\varphi, \xi) \mapsto \Ad^\ast_\varphi \xi$ and the corresponding Lie algebra representation by $\ad^\ast : \mathbb{g} \times V \to V$, $(X, \xi) \mapsto \ad^\ast_X \xi$, and refer to these as the coadjoint representations. 
When $\mathbb{g}$ and $V$ are subspaces of a common ambient Lie algebra and the coadjoint action $\ad^\ast : \mathbb{g} \times V \to V$ is given by a Lie bracket in this ambient Lie algebra, as will be the case for us in Section \ref{sec:cycloGaudin} below, we obtain Lax equations on the symplectic leaves of the Poisson manifold $\big( \mathbb{g}^\ast, \{ ,\, \}_R \big)$ given by the coadjoint orbits
\begin{equation} \label{eq:coadj_orbit-aks}
    {\cal O}_\Lambda=\{\Ad^{R\ast}_{\varphi}  \Lambda;\varphi\in \mathbb{G}_R\}, \quad \text{for}\,\, \Lambda \in \mathbb{g}^\ast,
\end{equation}
as in the general case and provide natural phase spaces for integrable systems.

\begin{remark}
    Note that if $\mathbb{g}$ were finite-dimensional, then a good choice of representation space $V$ would be the algebraic dual $\mathbb{g}'$ of $\mathbb{g}$, in which case \eqref{pairing V g} is given by the canonical pairing. In the infinite-dimensional setting we are considering, the algebraic dual is too big. When $\mathbb{g}$ is equipped with a nondegenerate invariant symmetric bilinear form $(\cdot, \cdot) : \mathbb{g} \times \mathbb{g} \to \CC$, then one possible replacement for the algebraic dual is afforded by the smooth dual $(\mathbb{g}, \cdot) \subset \mathbb{g}'$, which is canonically isomorphic to $\mathbb{g}$ itself. However, in the setting required for Gaudin models, discussed in Chapter \ref{chap:orbitexamples}, it turns out that the smooth dual will not be the appropriate notion of dual space, which is why we need the above more general working definition for the dual space $\mathbb{g}^\ast$.
\end{remark}

\begin{remark}
The case where $R$ is not obtained from a decomposition into two subalgebras but rather from a decomposition into nilpotent and Cartan subalgebras can also be accommodated within the framework of Lie dialgebras in a straightforward manner. Interestingly, as we illustrate in Section \ref{Toda_pq}, it can also be used to describe the same open Toda chain obtained from the AKS scheme, as presented in Section \ref{Flaschka}. The underlying algebraic structures are very different, though. In particular, $R$ is skew-symmetric in this case, while it is not in the AKS formulation, showing that the same Toda chain can arise from two distinct constructions. 
\end{remark}

We will employ the results from the theory of Lie dialgebras we have discussed here in Part \ref{part:coadjointmultiform}, where they will provide the ingredients for the construction of geometric Lagrangian one-forms living on coadjoint orbits. But before we do that, we present the necessary details on Lagrangian one-forms in the next chapter, which will serve as the foundation for the rest of the thesis.

\chapter{Lagrangian one-forms}\label{chap:lm-background}
This chapter provides the necessary background on Lagrangian multiforms. We restrict the discussion to the case of Lagrangian one-forms, which covers finite-dimensional integrable systems.

We start by reviewing the position-space formulation of Lagrangian one-forms and the so-called bivariational principle in Section \ref{sec:positionlm-bivarprin}. Then, in Section \ref{sec: univar principle CH}, we describe the setup of phase-space or geometric Lagrangian one-forms --- first introduced in \cite{CDS} --- followed by a discussion of the univariational principle of \cite{CH}. In Section \ref{sec: group actions}, we explain how to incorporate a symmetry in this setup, represented by the free action of a connected Lie group $G$ on a manifold $M$. This allows us to recast the relation between Noether charges and moment maps in the context of Lagrangian one-forms. Then, in Section \ref{gauged_univariational_principle}, we introduce the gauged univariational principle, explaining how it describes the symplectic reduction procedure traditionally presented in Hamiltonian terms purely in the variational language of Lagrangian one-forms.

The content of this chapter is adapted from \cite{CDS} and \cite{CHSV}. Here and in everything that follows, we always use the summation convention according to which repeated upstairs and downstairs indices of any kind are summed over.

\section{Lagrangian one-forms on position space}\label{sec:positionlm-bivarprin}

As we briefly discussed in Section \ref{sec:lagintegrability-intro}, the original approach to Lagrangian multiforms was based on a position-space formulation, and has so far been the one used most predominantly in subsequent works. Here we present a quick overview of this formulation for the case of Lagrangian one-forms in the continuous setup and describe how it captures the notion of an integrable hierarchy.

The central object in the position-space formulation is a \emph{position-space Lagrangian one-form} 
\begin{equation}
\label{def_Lag_multiform}
    \Lag[q]= \Lag_i[q] \, \d t^i, \quad i = 1, \ldots, n,
\end{equation}
where $\Lag_i$ denotes the Lagrangian coefficient corresponding to the time $t^i$ and $q$ denotes generic configuration coordinates. For instance, $q$ could be a position vector in $\RR^d$ for some $d$ or an element of a (matrix) Lie group. The notation $\Lag[q]$ and $\Lag_k[q]$ mean that these quantities depend on $q$ and a finite number of derivatives of $q$ with respect to the times $t^1,\dots,t^n$. Here we restrict ourselves to only the case of first derivatives and simply write $\Lag_i$ for the Lagrangian coefficients.
The traditional action is replaced by a \emph{multiform action}
\begin{equation}\label{eq:multiform-action}
    S[q,\Gamma]=\int_\Gamma \Lag[q]
\end{equation}
where $\Gamma$ is a curve in the multitime $\RR^n$ with (time) coordinates $t^1,\dots,t^n$. The multiform action \eqref{eq:multiform-action} is a functional of both the configuration coordinates $q$ and an arbitrary curve $\Gamma$. One then applies to this action a generalised variational principle, referred to as the \emph{bivariational principle}.\footnote{The term \emph{bivariational principle} was used for the first time relatively recently in \cite{CH} to contrast it with the \emph{univariational principle}, discussed here in Section \ref{sec: univar principle CH}, that combines the two-step procedure of the former into one unified step.} This is done in two steps. First, we vary over the degrees of freedom for an arbitrary choice of the curve $\Gamma$ and require criticality of the action. This leads to the following \emph{multitime} Euler--Lagrange equations:
\begin{subequations}
\begin{align}
    \label{simple_multitime_EL1}
  \frac{\partial \Lag_i}{\partial q}-\partial_{t^i}  \frac{\partial \Lag_i}{\partial q_{t^i}} &= 0,\\[1ex]
  \label{simple_multitime_EL2}
  \frac{\partial \Lag_i}{\partial q_{t^j}} &= 0,\qquad j \neq i,\\[1ex]
  \label{simple_multitime_EL3}
  \frac{\partial \Lag_i}{\partial q_{t^i}} &= \frac{\partial \Lag_j}{\partial q_{t^j}},\qquad i,j = 1,\dots,n.
\end{align}
\end{subequations}
Note that \eqref{simple_multitime_EL1} is simply the \emph{standard Euler--Lagrange equation} for each $\Lag_i$ describing the dynamics of the degrees of freedom. Equation \eqref{simple_multitime_EL2} states that the Lagrangian coefficient $\Lag_i$ cannot depend on the velocities $q_{t^j}$ for $j \neq i$. The last equation \eqref{simple_multitime_EL3} requires that the conjugate momentum to $q$ be the same with respect to all times $t^i$. Equations \eqref{simple_multitime_EL2} and \eqref{simple_multitime_EL3} provide conditions that the Lagrangian coefficients $\Lag_i$ must satisfy, and are often referred to as \emph{corner equations}, a term borrowed from the discrete setup. This is where they first appeared in the context of discrete-time Calogero--Moser model \cite{YKLN}. However, we should emphasise that, in general, equation \eqref{simple_multitime_EL2} is an identity only on shell, and in some cases, it also provides non-trivial equations of motion.

Finally, as a separate second step, we vary over the curve $\Gamma$ and require criticality of the action on solutions of the multitime Euler--Lagrange equations \eqref{simple_multitime_EL1}--\eqref{simple_multitime_EL3}. This gives the so-called \emph{closure relation} for Lagrangian one-forms
\begin{equation}
\label{closure}
    \d \Lag[q]=0~~\Leftrightarrow ~~\partial_{t^i}\Lag_j-\partial_{t^j}\Lag_i=0,
\end{equation}
on solutions of \eqref{simple_multitime_EL1}--\eqref{simple_multitime_EL3}.
Crucially, this is the variational equivalent of the Poisson involutivity of Hamiltonians, the Liouville criterion for integrability. The relation between the closure relation for Lagrangian one-forms and the Poisson involutivity of Hamiltonians will be the subject of Section \ref{sec:closure-KK}.

\section{Lagrangian one-forms on phase space} \label{sec: univar principle CH}

The objects of interest in this thesis are Lagrangian one-forms formulated on phase space, as opposed to the position-space Lagrangian one-forms discussed in Section \ref{sec:positionlm-bivarprin}. Throughout this work, we refer to these phase-space Lagrangian one-forms as \emph{geometric Lagrangian one-forms} in line with the terminology used in the literature, in particular to do with \emph{geometric actions}, which our Lagrangian one-forms share a structural similarity with. Geometric actions can be traced back (at least) to works \cite{AFS, AS, W, ANPZ} on the path integral quantisation of coadjoint orbits based on ideas coming from the orbit method \cite{Kir1, Kir2} due to Kirillov.\footnote{Geometric actions have been studied in connection with integrable hierarchies in \cite{NP}. They have also appeared in the study of matrix models and Fermi fluids \cite{DMW1, DMW2}, and of gravity in two, three, and four dimensions. See, for instance, \cite{MNW, BGS, MR, BNR}.}

We now discuss the framework of geometric Lagrangian one-forms, together with a new generalised variational principle introduced in \cite{CH}, referred to as the \emph{univariational principle}. 

We start by describing the phase space for our Lagrangian one-form. Let $M$ be an $m$-dimensional manifold with coordinates $q^\mu$, for $\mu = 1, \ldots, m$. The cotangent bundle $T^\ast M$ is parameterised by coordinates $(q^\mu, p_\mu)$ where $p_\mu$, for $\mu = 1, \ldots, m$, are dual coordinates along the fibres. The tautological one-form $\alpha$ and symplectic form $\omega$ on $T^\ast M$ are respectively given by 
\begin{equation}
    \alpha=p_\mu \d q^\mu \qquad \text{and} \qquad \omega = \d p_\mu\wedge \d q^\mu.
\end{equation}
The Hamiltonian vector field $\mathcal{X}_f$ associated with any given function $f$ on $T^\ast M$ is defined by the property $\mathcal{X}_f \lrcorner \omega = \d f$ and given explicitly by
\begin{equation} \label{Ham vec def}
\mathcal{X}_f = \frac{\partial f}{\partial q^\mu} \frac{\partial}{\partial p_\mu} - \frac{\partial f}{\partial p_\mu} \frac{\partial}{\partial q^\mu} .
\end{equation}
The Poisson bracket for any pair of functions $f$ and $g$ is then given by 
\begin{equation}
\{ f, g \} = - \mathcal{X}_f g = \mathcal{X}_g f.
\end{equation}

Let us now introduce the geometric Lagrangian one-form on $T^\ast M\times\RR^n$,
\begin{equation}\label{L}
\Lag = \alpha - H_i \d t^i = p_\mu \d q^\mu - H_i \d t^i,
\end{equation}
where $H_1,\ldots, H_n$ are $n \in \ZZ_{\geq 1}$ real functions on $T^\ast M$ and $t^i$ are Cartesian coordinates on $\RR^n$. Note that in \eqref{L}, the phase space variables $p_\mu$, $q^\mu$ and the time variables $t^i$ are treated on an equal footing.\footnote{When writing the geometric Lagrangian one-form \eqref{L}, we do not assume any relation between the dimension of the manifold $M$ and the number of time variables $t^i$ here. The case $n=1$ corresponds to an ordinary Hamiltonian system while the case $n = m$ corresponds to that of a Liouville integrable system.} The action associated to a parameterised curve $\gamma:(0,1)\to T^\ast M\times \RR^n$ can therefore be expressed as
\begin{equation} \label{pre ungauged action}
S_0[\gamma] = \int_0^1\gamma^\ast \Lag=\int_0^1\left(p_\mu \frac{\d q^\mu}{\d s} - H_i(p,q) \frac{\d t^i}{\d s}\right)\d s.
\end{equation}

Let us start by applying the standard principle of least action to $S_0[\gamma]$: we seek a curve $\gamma:(0,1)\to T^\ast M\times \RR^n$ that is a critical point of this action, that is, under an arbitrary variation $\delta \gamma(s)=(\delta p_\mu(s), \delta q^\mu(s), \delta t^i(s))$ satisfying the boundary conditions $0=\lim_{s\to0,1}\delta q(s)=\lim_{s\to0,1}\delta t(s)$, we require
\begin{align}
0 = \delta S_0[\gamma]&=\int_0^1\left(\delta p_\mu \frac{\d q^\mu}{\d s} + p_\mu \frac{\d \delta q^\mu}{\d s}- \left( \frac{\partial H_i}{\partial p_\mu} \delta p_\mu+ \frac{\partial H_i}{\partial q^\mu} \delta q^\mu \right)\frac{\d t^i}{\d s}-H_i \frac{\d \delta t^i}{\d s}\right) \d s\nonumber\\
& =\int_0^1\left(\delta p_\mu \left(\frac{\d q^\mu}{\d s}-\frac{\partial H_i}{\partial p_\mu}\frac{\d t^i}{\d s} \right) - \delta q^\mu\left(\frac{\d p_\mu}{\d s}+ \frac{\partial H_i}{\partial q^\mu}\frac{\d t^i}{\d s}\right) + \frac{\d H_i}{\d s}\delta t^i\right) \d s,
\end{align}
where in the second line we have integrated by parts and used the boundary conditions. This leads to the Euler--Lagrange equations
\begin{equation} \label{EL equations ungauged}
\gamma'\lrcorner \d \Lag = 0 \quad \Longleftrightarrow \quad \frac{\d q^\mu}{\d s}=\frac{\partial H_i}{\partial p_\mu} \frac{\d t^i}{\d s},~~
\frac{\d p_\mu}{\d s}=-\frac{\partial H_i}{\partial q^\mu} \frac{\d t^i}{\d s},~~\frac{\d H_i}{\d s}=0
\end{equation}
where we have denoted by $\gamma'$ the tangent vector to the curve $\gamma$.

The first two Euler--Lagrange equations imply that locally along the curve, we must always have $\frac{\d t^i}{\d s} \neq 0$, for some $i \in \{ 1, \ldots, n \}$, since otherwise we would have $\gamma'=0$. We can then look for solutions of \eqref{EL equations ungauged} such that $\frac{\d t^i}{\d s} \neq 0$ for a fixed $i$. Up to a change of variable, this amounts to working with the parameter $s = t^i$ along the curve $\gamma$. However, this leads to an issue: this inevitably singles out the corresponding Hamiltonian $H_i$ as our ``preferred'' Hamiltonian.

To avoid this, rather than looking for individual curves $\gamma : (0,1) \to T^\ast M \times \RR^n$ which are critical points of the action \eqref{pre ungauged action}, we replace the curve $\gamma$ by an {\it immersion}
\begin{equation} \label{map Sigma}
\Sigma : \RR^n \longrightarrow T^\ast M \times \RR^n , \quad (u^j) \longmapsto \big( p_\mu(u), q^\mu(u), t^i(u) \big),
\end{equation}
\ie a map $\Sigma$ such that its tangent map 
\begin{equation}
    \d_u\Sigma:T_u\RR^n\to T_{\Sigma(u)}(T^\ast M \times \RR^n)
\end{equation}
is injective for every $u\in\RR^n$. In particular, $\Sigma$ is then a (local) diffeomorphism onto its image in $T^\ast M \times \RR^n$. We require $\Sigma$ to satisfy boundary conditions that fix the functions $\lim_{\lVert u \rVert\to\infty}q^\mu(u)$ and $\lim_{\lVert u \rVert\to\infty}t^i(u)$ of $S^{n-1}$ at spatial infinity. Next, we introduce a curve
\begin{equation} \label{Gamma def}
\Gamma : (0,1) \longrightarrow \RR^n , \quad s \longmapsto \big( t^i(s) \big)
\end{equation}
such that $\lim_{s\to0,1} \lVert \Gamma(s) \rVert =\infty$. Evaluating the action \eqref{pre ungauged action} on the composite map
\begin{equation*}
\begin{tikzcd}
\gamma \coloneqq \Sigma \circ \Gamma : (0,1) \arrow[r,"\Gamma"] & \RR^n \arrow[r,"\Sigma"] & T^\ast M \times \RR^n
\end{tikzcd}
\end{equation*}
leads to a family of multiform actions labelled by $\Gamma$, namely
\begin{equation} \label{ungauged action}
S_\Gamma[\Sigma] \coloneqq S_0[\Sigma \circ \Gamma] = \int_0^1\left(p_\mu \frac{\partial q^\mu}{\partial u^j} - H_i(p,q) \frac{\partial t^i}{\partial u^j}\right) \frac{\d u^j}{\d s} \d s.
\end{equation}
We treat this family of actions as depending only on the map $\Sigma$ in \eqref{map Sigma}, and the curve $\Gamma$ as a parameter labelling the family. This distinction between the roles of $\Sigma$ and $\Gamma$, reflected in the notation $S_\Gamma[\Sigma]$, sits at the heart of the \emph{univariational principle} which we can now formulate as seeking an immersion $\Sigma$ which is simultaneously a critical point of the family of actions $S_\Gamma[\Sigma]$ for all curves $\Gamma$ in $\RR^n$. It can equivalently be stated as the following single-step procedure: 
\begin{center}
Find $\Sigma$ such that $\delta_\Sigma S_\Gamma[\Sigma] = 0$ for all curves $\Gamma$ in $\RR^n$.    
\end{center}
As shown in \cite{CH}, the univariational principle implies that the map $\Sigma$ in \eqref{map Sigma} can be written as $\Sigma = (f_\Sigma, \Id_{\RR^n})$ for some map $f_\Sigma : \RR^n \to T^\ast M$, $(t^i)\mapsto \big( p_\mu(t), q^\mu(t)\big)$. This is equivalent to the system of equations
\begin{subequations} \label{univar EL eq}
\begin{align}
\label{univar EL eq a} &\frac{\partial q^\mu}{\partial t^i}=\frac{\partial H_i}{\partial p_\mu} ,\quad 
\frac{\partial p_\mu}{\partial t^i}=-\frac{\partial H_i}{\partial q^\mu},\qquad i=1,\dots,n ,\quad \mu = 1, \ldots, m ,\\
\label{univar EL eq b} &\frac{\partial H_i}{\partial p_\mu}\frac{\partial H_j}{\partial q^\mu}-\frac{\partial H_i}{\partial q^\mu}\frac{\partial H_j}{\partial p_\mu}=\{H_i,H_j\}=0, \qquad i,j=1,\dots,n.    
\end{align}
\end{subequations}
First, we have \eqref{univar EL eq a} which implies that the time coordinate $t^i$ parameterises the Hamiltonian flow of $H_i$ for each $i=1,\ldots, n$. Additionally, as a result of having included the times $t^i$ among the dynamical variables in the map $\Sigma$, we also have \eqref{univar EL eq b}, which tells us that the univariational principle already encodes the closure relation through the variation of $\Sigma$. In particular, the univariational principle admits solutions if and only if the Hamiltonian functions $H_1, \ldots, H_n$ pairwise Poisson commute. Therefore, the univariational principle applied to the Lagrangian one-form \eqref{L} encodes multitime Hamiltonian mechanics, which includes the case of Liouville integrable systems when $n = m$.

\begin{remark}
    The univariational principle is a natural generalisation of the well-known {\it extended phase space} approach to classical mechanics to the multitime case. The case of a single time $t$, \ie $n=1$, in our setup so that $\Gamma:(0,1)\to \RR$, $s\mapsto t$ is a (re)parametrisation of time, corresponds to the original extended phase space approach. The application of this idea to the Lagrangian formalism can be found, for instance, in \cite{Sou,Sou2}, but the idea itself is likely much older. 
\end{remark}

\begin{remark}
As mentioned earlier, the univariational principle encapsulates the two-step procedure of the bivariational principle into a single step. In the bivariational approach, one views $\Sigma$ as $(f_\Sigma, \Id_{\RR^n})$ for some map $f_\Sigma : \RR^n \to T^\ast M$, $(t^i)\mapsto \big( p_\mu(t), q^\mu(t)\big)$ {\it from the outset} so that $S_0[\Sigma \circ \Gamma]$ in \eqref{ungauged action} becomes the action 
 \begin{equation}
S[f_\Sigma , \Gamma]\coloneqq S_0[f_\Sigma \circ \Gamma] = \int_0^1\left(p_\mu \frac{\partial q^\mu}{\partial t^j} - H_j(p,q) \right) \frac{\d t^j}{\d s} \d s.
\end{equation}
The price to pay in this approach is that one has to vary with respect to both $f_\Sigma$ {\it and} $\Gamma$ to obtain the full set of equations \eqref{univar EL eq}. We do not discuss the relation between the two approaches in any further detail and refer the interested reader to \cite{CH}.
\end{remark}

\section{Group actions and symmetry} \label{sec: group actions}

Let us now suppose that $M$ admits a free right action of a connected Lie group $G$, $\rho : G \times M \to M$. Let $X_a$ be a basis of the Lie algebra $\g$ of $G$ and let $X_a^\mu(q)\parder{}{q^\mu}$ denote the corresponding fundamental vector fields generating the action of $\g$ on $M$. The right action of $G$ lifts to $T^\ast M$, which we also denote by $\rho : G \times T^\ast M \to T^\ast M$, and the corresponding action of $\g$ is generated by vector fields
\begin{equation}\label{action coords}
    X_a^\sharp = X_a^\mu(q)\parder{}{q^\mu} - p_\nu \parder{X_a^\nu}{q^\mu} \parder{}{p_\mu}
\end{equation}
satisfying $[X_a^\sharp,X_b^\sharp]=f_{ab}{}^c X_c^\sharp$, where $f_{ab}{}^c$ denotes the structure constants of $\g$.
We further lift this to a right action $\rho : G \times T^\ast M \times \RR^n \to T^\ast M \times \RR^n$ of $G$ on $T^\ast M \times \RR^n$ by letting $G$ act trivially on $\RR^n$ so that the infinitesimal action of $\g$ is still generated by the same vector fields \eqref{action coords}.

In what follows, we will consider only the corresponding left action of the group $G$ on $T^\ast M \times \RR^n$, given by $G \times T^\ast M \times \RR^n \to T^\ast M \times \RR^n$, $(g, x) \mapsto g \cdot x \coloneqq \rho_{g^{-1}}(x)$ and which is infinitesimally generated by the vector fields $\delta_{X_a}\coloneqq - X_a^\sharp$, for $a = 1, \ldots, \dim \g$.

\begin{proposition}\label{prop_global_symmetry}
The action \eqref{ungauged action} is invariant under the infinitesimal action of $G$ on $T^\ast M \times \RR^n$ generated by \eqref{action coords} if and only if each $H_i$ is invariant under this infinitesimal group action, \ie
\begin{equation} \label{Hi invariant}
\mathcal{L}_{X^\sharp_a}H_i=0, \qquad i=1,\dots,n,\quad a=1,\dots, {\rm dim}~\g.
\end{equation}
The Noether charges associated with this global $G$ symmetry are then given by $\mu_a(p,q) = - p_\nu X_a^\nu(q)$.
\end{proposition}
\begin{proof}
The variation of any given map $\Sigma : \RR^n \to T^\ast M\times \RR^n$, $(u^j) \mapsto \big( p_\mu(u), q^\mu(u), t^i(u) \big)$ under the infinitesimal left action of $G$ on $T^\ast M \times \RR^n$ generated by the vector fields \eqref{action coords} is given by $\delta_X \Sigma(u) = - X^\sharp \big( \Sigma(u) \big)$, or more explicitly in components
\begin{subequations} \label{inf var p q t}
\begin{align}
    \delta_X q^\mu(u) &= - \lambda^a X_a^\mu(q(u))\\
    \delta_X p_\mu(u) &= \lambda^a p_\nu(u)\parder{X_a^\nu}{q^\mu}(q(u))\\
    \delta_X t^i(u) &= 0
\end{align}
\end{subequations}
for arbitrary $X=\lambda^a X_a \in\g$. Noting that
\begin{equation} \label{delta p dq}
\delta_X \bigg( p_\mu \frac{\partial q^\mu}{\partial u^j} \bigg) = \lambda^a p_\nu \parder{X_a^\nu}{q^\mu} \frac{\partial q^\mu}{\partial u^j} - \lambda^a p_\nu \frac{\partial X_a^\nu}{\partial u^j} = 0,
\end{equation}
where the last step is by the chain rule for the function $X_a^\nu(q(u))$, the corresponding variation of the action $S_\Gamma[\Sigma]$ in \eqref{ungauged action} is then
\begin{equation} \label{inf_invariance_S}
\delta_X S_\Gamma[\Sigma] = \lambda^a\int_0^1 \bigg(X_a^\mu\frac{\partial H_i}{\partial q^\mu}-p_\nu\parder{X_a^\nu}{q^\mu}\frac{\partial H_i}{\partial p_\mu}\bigg) \frac{\d t^i}{\d s} \d s
= \lambda^a\int_0^1 \big( \mathcal L_{X^\sharp_a} H_i \big) \frac{\d t^i}{\d s} \d s,
\end{equation}
where the second step is by definition \eqref{action coords} of $X^\sharp_a$. Now the resulting expression in \eqref{inf_invariance_S} should be zero for any $\lambda^a$, any map $\Sigma$ and any curve $\Gamma$ from which \eqref{Hi invariant} follows.

The expression $\mu_a = - p_\nu X^\nu_a$ for the Noether charge associated with this global symmetry can be obtained by using the standard trick of promoting the constant parameters $\lambda^a \in \CC$ to functions $\lambda^a : \RR^n \to \CC$. This leads to an additional term $\frac{\partial \lambda^a}{\partial u^j} \mu_a$ on the right-hand side of \eqref{delta p dq} so that the variation of the action now reads $\delta_X S_\Gamma[\Sigma] = \int_0^1 \frac{\d \lambda^a}{\d s} \mu_a \d s$, as required.
\end{proof}

In the Hamiltonian formalism, the Noether charges $\mu_a$ from Proposition \ref{prop_global_symmetry} are encoded in a moment map $\mu : T^\ast M \to \g^\ast$ such that
\begin{equation}\label{moment_map}
\langle\mu(p,q),X_a\rangle=\mu_a(p,q)=-p_\mu X^\nu_a(q),
\end{equation}
where $\langle~,~\rangle : \g^\ast \times \g \to \CC$ denotes the canonical pairing.  In general, a moment map is required to satisfy the equations
\begin{equation} \label{moment map eqs}
\langle\d \mu, X_a\rangle = X_a^\sharp \lrcorner \omega \quad\text{and}\quad \mathcal{L}_{X_a^\sharp} \mu + \mathrm{ad}^\ast_{X_a} \mu = 0.
\end{equation}
For us, the first of these follows by a direct calculation from \eqref{action coords}, \eqref{moment_map} and $\omega=\d p_\mu\wedge \d q^\mu$.  The second is proved as follows.  First we note that $\langle\mu,X_b\rangle=-X_b^\sharp\lrcorner\alpha$ and that $\mathcal{L}_{X_a^\sharp}\alpha=0$.  So
\begin{equation}
\langle\mathcal{L}_{X_a^\sharp}\mu,X_b\rangle=-\mathcal{L}_{X_a^\sharp}(X_b^\sharp\lrcorner\alpha)=-[X_a^\sharp,X_b^\sharp]\lrcorner\alpha = \langle\mu,[X_a,X_b]\rangle=-\langle\mathrm{ad}^\ast_{X_a} \mu,X_b\rangle,
\end{equation}
where in the second last step we used the fact that $[X_a^\sharp,X_b^\sharp] = [X_a,X_b]^\sharp$. Hence, the second equation of \eqref{moment map eqs} holds.

\section{Gauged Lagrangian one-forms}\label{gauged_univariational_principle}

When a physical system is invariant under the action of a Lie group $G$, such as in the context of Section \ref{sec: group actions}, this symmetry can be used to reduce the number of degrees of freedom. Indeed, since the components of the moment map $\mu : T^\ast M \to \g^\ast$ are preserved under all the Hamiltonian flows $\mathcal X_{H_i}$, \ie $\mathcal X_{H_i} \mu_a = 0$, one may consistently restrict to the zero-level set $\mu^{-1}(0) \subset T^\ast M$ of this moment map by imposing the constraint $\mu(p,q) = 0$. Furthermore, by the $G$-equivariance property of the moment map, the zero-level set $\mu^{-1}(0)$ is invariant under the action of $G$ so that we may further reduce the number of degrees of freedom by working on the quotient space $\mu^{-1}(0)/G$.

In order to implement the above symplectic reduction procedure in the variational setting, and thereby construct a Lagrangian one-form and action for the reduced system, we can impose the constraint $\mu(p,q) = 0$ in $\g^\ast$ by introducing a $\g$-valued Lagrange multiplier.
Specifically, since we want to gauge the action of $G$ on the map $\Sigma : \RR^n \to T^\ast M \times \RR^n$ described infinitesimally in \eqref{inf var p q t},
we introduce a $\g$-valued Lagrange multiplier one-form on $\RR^n$ which we denote by
\begin{equation} \label{cal A def}
\mathcal A =\mathcal A_j \d u^j = \mathcal A^a_j X_a \d u^j \in \Omega^1(\RR^n, \g) .
\end{equation}

Using the canonical pairing $\langle~,~\rangle : \g^\ast \times \g \to \CC$ we can combine \eqref{cal A def} with the moment map $\mu : T^\ast M \to \g^\ast$, which we view as a map $\mu : T^\ast M \times \RR^n \to \g^\ast$ that is constant along $\RR^n$, to obtain a one-form $\langle\Sigma^\ast \mu, \mathcal A\rangle \in \Omega^1(\RR^n)$. The gauging procedure then simply consists in adding this term to the pullback $\Sigma^\ast \Lag \in \Omega^1(\RR^n)$ of the Lagrangian one-form $\Lag$. We obtain a family of actions for a map $\Sigma : \RR^n \to T^\ast M \times \RR^n$ and a one-form $\mathcal A \in \Omega^1(\RR^n, \g)$, parameterised by curves $\Gamma : (0,1) \to \RR^n$,
\begin{equation}\label{uni_gauged action}
S_\Gamma[\Sigma, {\cal A}] = \int_0^1 \Gamma^\ast \Big( \Sigma^\ast \Lag + \langle\Sigma^\ast \mu, {\cal A}\rangle \Big) .
\end{equation}
The constraint $\mu(p,q) = 0$ will now be enforced dynamically through the equations of motion for $\mathcal A$, see Section \ref{sec: gauged univar princ} below. Moreover, the effect of adding this new term to the Lagrangian is to promote the $G$-symmetry we started with, from Section \ref{sec: group actions}, to a gauge symmetry as we now show.

\subsection{Gauging a symmetry in a Lagrangian one-form} \label{sec: gauging symmetry}

We consider local gauge transformations parameterised by smooth maps $g: \RR^n \to G$, which act pointwise on the map $\Sigma : \RR^n \to T^\ast M \times \RR^n$ and on the Lagrange multiplier $\mathcal A \in \Omega^1(\RR^n, \g)$ as a gauge transformation, i.e.
\begin{subequations} \label{gauge_transfo_A}
\begin{align}
\label{gauge_transfo_A a} \Sigma(u) &\longmapsto g(u) \cdot \Sigma(u) ,\\
\label{gauge_transfo_A b} \mathcal A &\longmapsto g \,\mathcal A \,g^{-1}- \d_{\RR^n} g \,g^{-1} ,
\end{align}
\end{subequations}
where $\d_{\RR^n}$ denotes the de Rham differential on $\RR^n$.
Due to its transformation property \eqref{gauge_transfo_A b}, we will henceforth often refer to the Lagrange multiplier one-form $\mathcal A \in \Omega^1(\RR^n, \g)$ as a gauge field.

\begin{proposition}\label{prop_local_symmetry}
The action $S_\Gamma[\Sigma,\mathcal A]$ is invariant under an infinitesimal version of the $G$-valued local gauge transformation in \eqref{gauge_transfo_A} if and only if the functions $H_i$ are $G$-invariant, i.e.
\begin{equation}\label{moment map eqs 2}
\mathcal{L}_{X^\sharp_a}H_i=0, \qquad i=1,\dots,n, \quad a=1,\dots, {\rm dim}~\g.
\end{equation}
\end{proposition}
\begin{proof}
Let $\Sigma:\RR^n \to T^\ast M\times \RR^n$, $(u^j) \mapsto \big( p_\mu(u), q^\mu(u), t^i(u) \big)$ be a given map as in the proof of Proposition \ref{prop_global_symmetry}. The action \eqref{uni_gauged action} for this $\Sigma$ and any curve $\Gamma : (0,1) \to \RR^n$ then explicitly reads
\begin{equation} \label{uni_gauged action explicit}
S_\Gamma[\Sigma,\mathcal A] = \int_0^1 \bigg(p_\mu \frac{\partial q^\mu}{\partial u^j} -H_i(p,q)\frac{\partial t^i}{\partial u^j} + \mu_a(p,q) \mathcal A^a_j \bigg) \frac{\d u^j}{\d s} \d s,
\end{equation}
where $p_\mu$, $q^\mu$, $t^i$ and $A^a_j$ all depend on $s$ through their dependence on the functions $u^j(s)$.
Consider an infinitesimal gauge transformation parameterised by a $\g$-valued function $X:\RR^n \to\g$, which we write in components as $X = \lambda^a(u) X_a$. The variations of the components of $\Sigma$ and $\mathcal A$ under the infinitesimal left action of $G$ on $T^\ast M \times \RR^n$ then read (cf. the variations \eqref{inf var p q t} where $\lambda^a$ was constant):
\begin{subequations}
\begin{align}
    \delta_X q^\mu(u) &= - \lambda^a(u)X_a^\mu\big(q(u)\big)\\
    \delta_X p_\mu(u) &= \lambda^a(u) p_\nu(u)\parder{X_a^\nu}{q^\mu}\big(q(u)\big)\\
    \delta_X t^i(u) &= 0 \\
    \delta_X \mathcal A^a_j(u) &= - \frac{\partial \lambda^a}{\partial u^j} - f_{bc}{}^a \mathcal A^b_j (u) \lambda^c(u).
\end{align}
\end{subequations}
By the exact same computation as in the proof of Proposition \ref{prop_global_symmetry}, where in the end of that proof $\lambda^a$ was already treated as a function $\lambda^a : \RR^n \to \CC$, the corresponding variation of $S_\Gamma[\Sigma, \mathcal A]$ reads
\begin{equation} \label{variation action gauged}
\delta_X S_\Gamma[\Sigma, \mathcal A] = \int_0^1\left[ \lambda^a \big( \mathcal L_{X^\sharp_a} H_i \big) \frac{\partial t^i}{\partial u^j} - \lambda^a \mathcal A^b_j \Big(\mathcal L_{X^\sharp_a}\mu_b -f_{ab}{}^c\mu_c\Big) \right] \frac{\d u^j}{\d s} \d s.
\end{equation}
We note, in particular, that the term $- \mu_a \frac{\partial \lambda^a}{\partial u^j}$ appearing in the variation $\mu_a \delta_X A^a_j$ cancels with the term $- p_\nu X^\nu_a \frac{\partial \lambda^a}{\partial u^j}$ from the variation \eqref{delta p dq} when $\lambda^a$ is not constant.
The right-hand side of \eqref{variation action gauged} needs to vanish for all functions $\lambda^a$, $\mathcal A^a_i$ and all maps $\Sigma$ and curves $\Gamma$, but the last bracketed term vanishes by the second relation in \eqref{moment map eqs} since
\begin{equation}
    \langle\mathcal{L}_{X^\sharp_a}\mu, X_b\rangle+ \langle\mathrm{ad}^\ast_{X_a}\mu, X_b\rangle = \mathcal{L}_{X^\sharp_a}\mu_b - \langle\mu, [X_a,X_b]\rangle = \mathcal{L}_{X^\sharp_a}\mu_b -f_{ab}{}^c\mu_c .
\end{equation}
Therefore, gauge invariance is equivalent to the condition \eqref{moment map eqs 2}.
\end{proof}

By combining Propositions \ref{prop_global_symmetry} and \ref{prop_local_symmetry} we see that the gauged action $S_\Gamma[\Sigma, \mathcal A]$ in \eqref{uni_gauged action} is invariant under the gauge group $C^\infty(\RR^n, G)$ if and only if the original action $S_\Gamma[\Sigma]$ in \eqref{ungauged action} is invariant under the Lie group $G$.

\subsection{Gauged univariational principle} \label{sec: gauged univar princ}

Recall from Section \ref{sec: univar principle CH} that the univariational principle seeks an immersion $\Sigma : \RR^n \to T^\ast M\times \RR^n$ such that for all curves $\Gamma : (0,1) \to \RR^n$ we have $\delta_\Sigma S_\Gamma[\Sigma] = 0$.
We introduce the gauged version in the following definition. To do so, note that, using the infinitesimal right action of $G$ on $T^\ast M\times\RR^n$, we obtain the linear map
\begin{equation}
\A^\sharp_u :T_u\RR^n\to T_{\Sigma(u)}(T^\ast M\times \RR^n), \quad \parder{}{u^i}\mapsto \A_i^aX_a^\sharp,
\end{equation}
where we recall that $X_a^\sharp$ is given in \eqref{action coords}. 
\begin{definition}
The \emph{gauged univariational principle} seeks a map
\begin{equation*}
\Sigma : \RR^n \longrightarrow T^\ast M\times \RR^n
\end{equation*}
and a gauge field ${\cal A} \in \Omega^1(\RR^n, \g)$ such that the linear map $\d_u\Sigma-\A^\sharp_u$ is injective for every $u \in \RR^n$,
and such that
the pair $(\Sigma, {\cal A})$ is simultaneously a critical point of the family of actions $S_\Gamma[\Sigma, {\cal A}]$ for all curves $\Gamma : (0,1) \to \RR^n$, namely such that
\begin{equation*}
\delta_\Sigma S_\Gamma[\Sigma, \mathcal A] = 0 \quad \text{and} \quad \delta_{\mathcal A} S_\Gamma[\Sigma, \mathcal A] = 0 .
\end{equation*}
\end{definition}

\begin{theorem} \label{thm: gauged univar}
The gauged univariational principle applied to the gauge invariant action $S_\Gamma[\Sigma, \mathcal A]$ in \eqref{uni_gauged action} gives rise to the following set of equations:
\begin{subequations} \label{EL gauged univar thm}
\begin{align} \label{A variation}
\mu(p,q) &= 0,\\
\label{univariational eqs1}
\parder{q^\mu}{t^i} - \frac{\partial H_i}{\partial p_\mu} &= \widetilde{\mathcal A}^a_i X_a^\mu, \\
\label{univariational eqs2}
\parder{p_\mu}{t^i}+ \frac{\partial H_i}{\partial q^\mu} &=- \widetilde{\mathcal A}^a_i p_\nu \parder{X_a^\nu}{q^\mu}, \\
\label{univariational eqs3} \{H_i,H_j\} &= 0 ,
\end{align}
\end{subequations}
where the composition $\pi_{\RR^n} \circ \Sigma : \RR^n \to \RR^n$, $(u^j) \mapsto \big( t^i(u) \big)$ with the projection $\pi_{\RR^n} : T^\ast M \times \RR^n \to \RR^n$, $( p_\mu, q^\mu,t^i )\mapsto (t^i)$ to the second factor is a (local) diffeomorphism and is used to define $\widetilde{\mathcal A}^a_i = \frac{\partial u^j}{\partial t^i} \mathcal A^a_j$.

Moreover, the $\g$-valued connection $\d_{\RR^n} + \mathcal A$ is flat, i.e. it satisfies the zero-curvature equation $F = \d_{\RR^n}\mathcal A + \frac 12 [\mathcal A, \mathcal A] = 0$ or in components
\begin{equation}
\label{flatness}
    F_{ij}^a \coloneqq \parder{\widetilde{\mathcal A}_j^a}{t^i} - \parder{\widetilde{\mathcal A}_i^a}{t^j}+f_{bc}{}^a \widetilde{\mathcal A}_i^b \widetilde{\mathcal A}_j^c=0,\qquad a=1,\dots,{\rm dim}~\g, \quad i,j=1,\dots,n.
\end{equation}
\end{theorem}
\begin{proof}
Let $\Sigma : \RR^n \to T^\ast M\times \RR^n$, $(u^j) \mapsto \big( p_\mu(u), q^\mu(u), t^i(u) \big)$ and $\Gamma : (0,1) \to \RR^n$ be arbitrary maps. The action \eqref{uni_gauged action} for these can be written as in \eqref{uni_gauged action explicit}.

We first perform a variation of the action \eqref{uni_gauged action explicit} with respect to the functions $\mathcal A^a_j : \RR^n \to \RR$
\begin{equation}
\delta_{\mathcal A} S_\Gamma[\Sigma, \mathcal A] = \int_0^1 \mu_a(p, q) \delta \mathcal A^a_j\big( u(s) \big) \frac{\d u^j}{\d s} \d s.
\end{equation}
In order for this to vanish for arbitrary variations $\delta \mathcal A^a_j$ we must have
\begin{equation} \label{EL constraint gauged univar}
\mu_a(p, q) = 0,
\end{equation}
which is the expected constraint equation $\mu = 0$.

Next, consider the variation of the action \eqref{uni_gauged action explicit} with respect to the map $\Sigma : \RR^n \to T^\ast M \times \RR^n$. Under an arbitrary variation $\delta \Sigma(u) = \big( \delta p_\mu(u), \delta q^\mu(u), \delta t^i(u) \big)$ satisfying the boundary conditions $0=\lim_{\lVert u \rVert\to\infty} \delta q^\mu(u)=\lim_{\lVert u \rVert\to\infty} \delta t^i(u)$, the action varies as
\begin{multline} \label{delta gamma gauged action}
\delta_\Sigma S_\Gamma[\Sigma, \mathcal A] =\int_0^1\frac{\d}{\d s}\bigg[p_\mu\delta q^\mu-H_i\delta t^i\bigg]\d s + \int_0^1\bigg[ \bigg( \frac{\partial q^\mu}{\partial u^j} - \frac{\partial t^i}{\partial u^j} \frac{\partial H_i}{\partial p_\mu} + \frac{\partial \mu_a}{\partial p_\mu} \mathcal A^a_j \bigg) \, \delta p_\mu \\
- \bigg( \frac{\partial p_\mu}{\partial u^j} + \frac{\partial t^i}{\partial u^j} \frac{\partial H_i}{\partial q^\mu} - \frac{\partial \mu_a}{\partial q^\mu} \mathcal A^a_j \bigg) \, \delta q^\mu 
+ \bigg( \frac{\partial H_i}{\partial p_\mu} \frac{\partial p_\mu}{\partial u^j} + \frac{\partial H_i}{\partial q^\mu} \frac{\partial q^\mu}{\partial u^j} \bigg) \, \delta t^i \bigg] \frac{\d u^j}{\d s} \d s.
\end{multline}
The total derivative term on the right-hand side of this equation vanishes due to the boundary conditions on $\delta\Sigma$ and the boundary condition $\lim_{s\to 0,1} \lVert \Gamma(s) \rVert = \infty$ on $\Gamma$. We now want the above variation to vanish for all $\delta \Sigma$ and all curves $\Gamma$, and hence for arbitrary $\frac{\d u^j}{\d s}$, which is equivalent to the system of equations:
\begin{subequations} \label{eom gauged action}
\begin{align}
\label{eom gauged action a} \frac{\partial q^\mu}{\partial u^j} &= \frac{\partial t^i}{\partial u^j} \frac{\partial H_i}{\partial p_\mu} - \frac{\partial \mu_a}{\partial p_\mu} \mathcal A^a_j ,\\
\label{eom gauged action b} \frac{\partial p_\mu}{\partial u^j} &= - \frac{\partial t^i}{\partial u^j} \frac{\partial H_i}{\partial q^\mu} + \frac{\partial \mu_a}{\partial q^\mu} \mathcal A^a_j ,\\
\label{eom gauged action c} \frac{\partial H_i}{\partial p_\mu} \frac{\partial p_\mu}{\partial u^j} + \frac{\partial H_i}{\partial q^\mu} \frac{\partial q^\mu}{\partial u^j} &= 0 .
\end{align}
\end{subequations}

Next, we claim that these equations of motion imply that the Jacobian matrix $\big( \frac{\partial t^j}{\partial u^j} \big)$ must be invertible, thus ensuring that the composition $\pi_{\RR^n} \circ \Sigma : \RR^n \to \RR^n$ is a (local) diffeomorphism. To show this we adapt the similar argument \cite[\S2.2]{CH} to the present case with gauge field.  
The differential of $\Sigma$ at $u\in\RR^n$ reads
\begin{equation}
\d_u\Sigma:T_u\RR^n\to T_{\Sigma(u)}(T^\ast M\times \RR^n), \quad \parder{}{u^i}\mapsto \parder{p_\mu}{u^i}\parder{}{p_\mu} + \parder{q^\mu}{u^i}\parder{}{q^\mu} + \parder{t^j}{u^i}\parder{}{t^j},
\end{equation}
and therefore proving the claim is equivalent to showing that the tangent map of the composition $\pi_{\RR^n} \circ\Sigma:\RR^n\to\RR^n$, $u^j\mapsto t^i(u)$ at every $u \in \RR^n$, namely
\begin{equation}
\d_u (\pi_{\RR^n} \circ \Sigma)= \d_{\Sigma(u)} \pi_{\RR^n} \circ \d_u \Sigma:T_u \RR^n\to T_{\pi_{\RR^n}(\Sigma(u))}\RR^n,\quad \parder{}{u^i}\mapsto \parder{t^j}{u^i}\parder{}{t^j} ,
\end{equation}
is injective when the univariational equations \eqref{eom gauged action a}--\eqref{eom gauged action c} hold. To do this we will use the fact that the univariational equations \eqref{eom gauged action a}--\eqref{eom gauged action c} can equivalently be written as
\begin{equation}
\label{compact_univ_eqs}
\d_u\Sigma\left(\parder{}{u^i}\right)\lrcorner \d\Lag -\A^\sharp_u\left(\parder{}{u^i}\right)\lrcorner\omega=0\quad\forall i,
\end{equation}
in which
\begin{equation}
\label{form_dL}
\d\Lag = \omega-\d H_i\wedge \d t^i = \d p_\mu\wedge \d q^\mu - \parder{H_i}{p_\mu}\d p_\mu\wedge \d t^i - \parder{H_i}{q^\mu}\d  q^\mu\wedge \d t^i.
\end{equation}

So, let $V\in\ker \big( \d_{\Sigma(u)} \pi_{\RR^n} \circ \d_u \Sigma \big)$. On the one hand, \eqref{compact_univ_eqs} and \eqref{form_dL} imply
\begin{equation}
    0= (\d_u \Sigma-\A^\sharp_u)(V)\lrcorner\omega - [\d_u \Sigma(V)\lrcorner \d H_i]\d t^i+[\d_u \Sigma(V)\lrcorner \d t^i]\d H_i.
\end{equation}
On the other hand, $\d_u \Sigma(V)\lrcorner \d t^i=(\d_{\Sigma(u)} \pi_{\RR^n} \circ \d_u \Sigma)(V)\lrcorner \d u^i=0$ since $V\in\ker \big( \d_{\Sigma(u)} \pi_{\RR^n} \circ \d_u \Sigma \big)$.
Thus, we are left with
\begin{equation}
(\d_u \Sigma-\A^\sharp_u)(V)\lrcorner\omega - [\d_u \Sigma(V)\lrcorner \d H_i]\d t^i=0
\end{equation}
and each term must individually vanish since the first belongs to $T^*(T^\ast M)$ and the second belongs to $T^*\RR^n$. In particular, since $\omega$ is nondegenerate, we must have $(\d_u \Sigma-\A^\sharp_u)(V)=0$, from which it follows that $V=0$ since $\d_u \Sigma-\A^\sharp_u$ is injective. 

As a consequence, performing a suitable coordinate transformation on $\RR^n$, we may bring $\Sigma$ to the canonical form $(f_\Sigma, \Id_{\RR^n})$ for some function $f_\Sigma : \RR^n \to T^\ast M$, $t^i \mapsto \big( p_\mu(t), q^\mu(t) \big)$.
We may then rewrite \eqref{eom gauged action} as
\begin{subequations} \label{EL eqs gauged univar}
\begin{align}
\label{EL eqs gauged univar a} \frac{\partial q^\mu}{\partial t^i} &= \frac{\partial H_i}{\partial p_\mu} - \frac{\partial \mu_a}{\partial p_\mu} \widetilde{\mathcal A}^a_i, \\
\label{EL eqs gauged univar b} \frac{\partial p_\mu}{\partial t^i} &= - \frac{\partial H_i}{\partial q^\mu} + \frac{\partial \mu_a}{\partial q^\mu} \widetilde{\mathcal A}^a_i, \\
\label{EL eqs gauged univar c} \frac{\partial H_i}{\partial p_\mu} \frac{\partial p_\mu}{\partial t^j} + \frac{\partial H_i}{\partial q^\mu} \frac{\partial q^\mu}{\partial t^j} &= 0,
\end{align}
\end{subequations}
where $\widetilde{\mathcal A}^a_i \coloneqq \frac{\partial u^j}{\partial t^i} \mathcal A^a_j$ is defined using the inverse $\big( \frac{\partial u^j}{\partial t^i} \big)$ of the matrix $\big( \frac{\partial t^i}{\partial u^j} \big)$.
Substituting \eqref{EL eqs gauged univar a} and \eqref{EL eqs gauged univar b} into the left-hand side of \eqref{EL eqs gauged univar c} we obtain
\begin{equation*}
\{ H_i, H_j \} = \{ H_i, \mu_a \} \widetilde{\mathcal A}^a_j.
\end{equation*}
The right-hand side vanishes using Proposition \ref{prop_global_symmetry} since $\{ H_i, \mu_a \} = \mathcal L_{X^\sharp_a} H_i = 0$, where the first equality follows from the definitions \eqref{action coords}, \eqref{Ham vec def} and \eqref{moment_map}. It then follows that $\{ H_i, H_j \} = 0$, which completes the derivation of the Euler--Lagrange equations \eqref{EL gauged univar thm}.

The relation \eqref{flatness} is a consistency condition between the equations \eqref{EL eqs gauged univar a}--\eqref{EL eqs gauged univar b} under the assumption that $S_\Gamma[\Sigma, \mathcal A]$ is gauge invariant (that is, the Hamiltonians are invariant).
To see this, consider the identities
\begin{equation*}
0 = \frac{\partial}{\partial t^i} \frac{\partial q^\mu}{\partial t^j} - \frac{\partial}{\partial t^j} \frac{\partial q^\mu}{\partial t^i} ,\quad
0 = \frac{\partial}{\partial t^i} \frac{\partial p_\mu}{\partial t^j} - \frac{\partial}{\partial t^j} \frac{\partial p_\mu}{\partial t^i} .
\end{equation*}
Evaluating the right-hand sides using \eqref{univariational eqs1} and \eqref{univariational eqs2} eventually leads to
\begin{equation}\label{FX}
    0=F_{ij}^a X^\sharp_a - \widetilde{\mathcal A}_i^a[X^\sharp_a,\mathcal X_{H_j}] + \widetilde{\mathcal A}_j^a[X^\sharp_a,\mathcal X_{H_i}] + [\mathcal X_{H_i},\mathcal X_{H_j}],
\end{equation}
where $F^a_{ij}$ are the components of the curvature of $\mathcal A$ defined in the statement of the theorem.
The second and third terms in \eqref{FX} vanish because $[X^\sharp_a,\mathcal X_{H_i}] = \mathcal X_{X^\sharp_a H_i}$ and $X^\sharp_a H_i = \mathcal{L}_{X^\sharp_a} H_i=0$ where the last step uses the condition \eqref{moment map eqs 2} from Proposition \ref{prop_local_symmetry} since we are assuming that $S_\Gamma[\Sigma, \mathcal A]$ is gauge invariant. The fourth term in \eqref{FX} vanishes because $[\mathcal X_{H_i}, \mathcal X_{H_j}] = \mathcal X_{\{H_i, H_j\}} = 0$ using the fact that $\{H_i,H_j\}=0$, as we have already established. Therefore, $F_{ij}^a X^\sharp_a=0$. Since the action of $G$ is free, the tangent vectors $X^\sharp_a$ are linearly independent at each point $(p_\mu,q^\mu)$ of $T^\ast M$, and so we deduce the relation \eqref{flatness}, as required.
\end{proof}

The equation \eqref{A variation} together with \eqref{univariational eqs1}--\eqref{univariational eqs2} for any $i = 1, \ldots, n$ represent Hamiltonian flow equations for a constrained Hamiltonian system. Indeed, the effect of the first equation \eqref{A variation} is to impose the set of constraints $\mu^a(p,q) = 0$ on the phase space $T^\ast M$, restricting the dynamics to the submanifold $\mu^{-1}(0) \subset T^\ast M$. On the other hand, the equations \eqref{univariational eqs1}--\eqref{univariational eqs2} written in the form \eqref{EL eqs gauged univar a}--\eqref{EL eqs gauged univar b} describe, for each $i=1,\ldots, n$, the flow equation for the time-dependent Hamiltonian
\begin{equation}
\widetilde{H}_i(p,q, t) \coloneqq H_i(p,q) - \mu_a(p,q) \widetilde{\mathcal A}^a_i(t) .
\end{equation}
This flow is a linear combination of the Hamiltonian flow $\mathcal X_{H_i}$ of the unconstrained system and an arbitrary time-dependent linear combination of the flows $\mathcal X_{\mu_a}$, for $a = 1, \ldots, \dim \g$, along the orbits of the $G$-action, which implements the quotienting by $G$ to $\mu^{-1}(0)/G$. The last equation \eqref{univariational eqs3} encodes the commutativity of the flows for different $i=1,\ldots, n$ and has been obtained here in our variational setting. 

Crucially, the commutativity of the flows is related to the closure relation in the theory of Lagrangian multiforms that we briefly talked about earlier in this chapter. This relation will be discussed in further detail in Section \ref{sec:closure-KK} in Part \ref{part:coadjointmultiform}. More generally, this next part of the thesis will deal with the first of our two approaches for the construction of geometric Lagrangian one-forms, which employs the theory of Lie dialgebras to obtain Lagrangian one-forms for a large class of finite-dimensional integrable systems.

\part{Lagrangian multiforms on coadjoint orbits}\label{part:coadjointmultiform}
\thispagestyle{empty}

\newpage
\thispagestyle{empty}
\mbox{}
\newpage

\chapter{From Lie dialgebras to Lagrangian multiforms}\label{chap:lm-orbits}
This part of the thesis is concerned with the framework of Lagrangian multiforms on coadjoint orbits introduced in \cite{CDS}. As we discussed in Section \ref{sec:lagintegrability-intro}, the primary motivation behind \cite{CDS} has been to address the absence of a systematic framework for constructing Lagrangian multiforms. It draws and expands upon the insight of \cite{CS2,CStV} to incorporate into Lagrangian multiforms the notions of the classical $r$-matrix and the ``compounding'' of hierarchies following \cite{N3}.\footnote{Another precursor to this construction is the short but remarkable paper \cite{ZM} where a Lagrangian (but not a multiform) was introduced to provide a variational description of zero-curvature equations corresponding to a Lax pair consisting of rational functions of the spectral parameter with distinct simple poles.}

In this chapter, which draws content from \cite[Sections 3-4]{CDS}, we show how the theory of Lie dialgebras, discussed in Section \ref{sec:dialgebra-background}, can be used to achieve this for a large class of finite-dimensional integrable systems. In Section \ref{sec:gen-lm}, we introduce \emph{geometric Lagrangian one-forms} living on coadjoint orbits and present the first main result of this chapter: Theorem \ref{Th_multi_EL}. Then, in Section \ref{sec:closure-KK}, we obtain a structural result, namely Theorem \ref{prop_double_zero}, for geometric Lagrangian one-forms that brings together aspects from the framework of Lagrangian multiforms and the Hamiltonian framework for integrability. Section \ref{reduction} deals with another main result. We recast our results from Sections \ref{sec:gen-lm} and \ref{sec:closure-KK} in the context of reduction from free motion on the cotangent bundle of a Lie group and produce a Lagrangian one-form on a general coadjoint orbit. We show how to recover the case of a Lagrangian one-form associated with a Lie dialgebra described in Section \ref{sec:gen-lm}.

\section{Geometric Lagrangian one-forms on coadjoint orbits}\label{sec:gen-lm}

Let us start by recalling from the discussion on Lie dialgebras in  Section \ref{sec:dialgebra-background} that a coadjoint orbit of $G_R$ in $\g^*$,
\begin{equation}
    {\cal O}_\Lambda=\{\text{Ad}^{R*}_{\varphi}\cdot \Lambda;\varphi\in G_R\}, \qquad \text{for some}~\Lambda\in\g^*,
\end{equation}
provides a natural description of the phase space of an integrable system. With this as the underlying setup, we now introduce a Lagrangian one-form, which we refer to as the geometric Lagrangian one-form, on this coadjoint orbit,
\begin{equation}
\label{our_Lag}
  \Lag[\varphi] = \Lag_k \, \d t^k ={\cal K}[\varphi]-{\cal H}[\varphi]
\end{equation}
with kinetic part
\begin{equation}
\label{kin_part}
  {\cal K}[\varphi] =  \left( L,\,\partial_{t^k}\varphi \cdot_R \varphi^{-1}  \right) \, \d t^k , \qquad   L = \text{Ad}^{R*}_{\varphi} \cdot \Lambda,~~\varphi \in G_R, 
\end{equation}
and potential part
\begin{equation}
\label{pot_part}
{\cal H}[\varphi]= H_k(L) \, \d t^k ,
\end{equation}
where $k = 1, \ldots, N$. We remind the reader that we always use the summation convention according to which repeated upstairs and downstairs indices of any kind are summed over. The field $\varphi\in G_R$ contains the dynamical degrees of freedom and, as we will see, the Euler--Lagrange equation will take a natural form when expressed in terms of $L = \text{Ad}^{R*}_{\varphi} \cdot \Lambda$. $\Lambda$ is a fixed non-dynamical element of $\g^*$ which defines ${\cal O}_\Lambda$, the phase space of the model. Each Lagrangian $\Lag_k$ in the Lagrangian multiform has a structure comparable to the familiar Lagrangian $p\dot{q}-H$ in classical mechanics.
The potential part is expressed in terms of $\text{Ad}^*$-invariant functions $H_k \in C^{\infty}(\mathfrak{g}^*)$, and we suppose we have $N$ of them.\footnote{At this stage, we do not necessarily have that $N$ is exactly half of the dimension of ${\cal O}_\Lambda$. As in the AKS scheme, this needs to be addressed in specific cases by choosing a coadjoint orbit of appropriate dimension to ensure Liouville integrability. We will not worry about this for now as our construction follows through anyway.}

\begin{remark}
An important ingredient in producing equations of motion in Lax form from the coadjoint orbit construction of Section \ref{sec:dialgebra-background} is an $\text{Ad}$-invariant nondegenerate bilinear symmetric form $\langle\,~,~\rangle$ on $\g$ to identify $\g^*$ with $\g$ and the coadjoint action with the adjoint action. However, we write the geometric Lagrangian one-form using the pairing $(~,~)$, an element $\Lambda\in\g^*$ and functions $H_k$ on $\g^*$, since it is less confusing to do so when deriving results in the general case here and in the examples in Chapter \ref{chap:orbitexamples}. It must be emphasised that ultimately we \emph{always} use the bilinear form $\langle\,~,~\rangle$ to make all the identifications and obtain equations in Lax form.
\end{remark}        
We now formulate our first main result.
\begin{theorem}
\label{Th_multi_EL}
The geometric Lagrangian one-form \eqref{our_Lag} satisfies the corner equations of the multitime Euler--Lagrange equations. The standard Euler--Lagrange equations associated with the Lagrangian coefficients $\Lag_k$ take the form of compatible Lax equations
        \begin{equation}
\label{EL_Lax}
    \partial_{t^k}L=[R_\pm \nabla H_k(L),L],\qquad k=1,\dots,N.
\end{equation}
The closure relation holds: on solutions of \eqref{EL_Lax}, we have 
\begin{equation}
    \partial_{t^k}\Lag_j-\partial_{t^j}\Lag_k=0,\qquad j,k=1,\dots,N.
\end{equation}
\end{theorem}
\begin{proof} It is clear that each $\Lag_k$ does not depend on $\partial_{t^\ell} \varphi$, for $\ell\neq k$, so the first corner equation is satisfied. To see that the second corner equation holds, it is convenient to introduce local coordinates $\phi_\alpha$, $\alpha=1,\dots,M$, on the group $G_R$. The only source of dependence on velocities is in the kinetic term of $\Lag_k$. Now, 
\begin{equation} \label{kinetic_in_coords}
\begin{split}
  \left( \text{Ad}^{R*}_{\varphi} \cdot \Lambda,\,\partial_{t^k}\varphi \cdot_R \varphi^{-1}  \right)
  &=\left(  \Lambda,\,\text{Ad}^{R}_{\varphi^{-1}}\cdot \left(\partial_{t^k}\varphi \cdot_R \varphi^{-1}\right)\right) \\[1ex]
&=  \left(  \Lambda,\,\varphi^{-1} \cdot_R  \partial_{t^k}\varphi \right) \\[1ex]  
&= \sum_{\alpha=1}^M\left(  \Lambda,\,\varphi^{-1} \cdot_R  \frac{\partial \varphi}{\partial \phi_\alpha}    \right)\partial_{t^k}\phi_\alpha \equiv \sum_{\alpha=1}^M  \pi_\alpha\, \partial_{t^k}\phi_\alpha 
\end{split} 
\end{equation}
where we have introduced the momentum
\begin{equation}
    \pi_\alpha=\left(  \Lambda,\,\varphi^{-1} \cdot_R  \frac{\partial \varphi}{\partial \phi_\alpha}    \right)
\end{equation}
conjugate to the field $\phi_\alpha$. Thus, 
\begin{equation}
    \frac{\partial \Lag_k}{\partial \left(\partial_{t^k}\phi_\alpha\right)}=\pi_\alpha
\end{equation}
is independent of $k$.
The remainder of the multitime Euler Lagrange equations consists of the standard Euler--Lagrange equations for each $\Lag_k$. We compute 
\begin{equation}
\begin{split}
    \delta \Lag_k  &= \left( \delta L,\,\partial_{t^k}\varphi \cdot_R \varphi^{-1}  \right) +  \left( L,\,\delta\,(\partial_{t^k}\varphi \cdot_R \varphi^{-1})  \right)-\delta H_k(L)  ,
\end{split}
\end{equation}
with\footnote{More rigorously, the notation $\delta L$ denotes the tangent vector to ${\cal O}_\Lambda$ at the point $L$ induced by the element $X\in\g_R$ which we write more suggestively as $\delta \varphi\cdot_R\varphi^{-1}$. The latter notation is closer to the more familiar one in variational calculus using matrix-valued fields.}
\begin{equation}
    \delta L=\text{ad}^{R*}_{\delta \varphi\cdot_R\varphi^{-1}}\cdot L
\end{equation}
and
\begin{equation}
\begin{split}
\delta H_k(L)&=\left(\delta L,\nabla H_k(L) \right)\\
&=-\left( L,\,\left[ \delta \varphi\cdot_R\varphi^{-1},\nabla H_k(L)\right]_R \right)\\[1ex]
&=\frac{1}{2}\left( L,\,\left[R\nabla  H_k(L),\, \delta \varphi\cdot_R\varphi^{-1}\right] \right)=-\frac{1}{2}\left(\text{ad}^*_{R\nabla  H_k(L)}\cdot L,\,\delta \varphi\cdot_R\varphi^{-1} \right).
\end{split}
\end{equation}
So, 
\begin{equation}
\begin{split}
    \delta \Lag_k &= \left( \text{ad}^{R*}_{\delta \varphi\cdot_R\varphi^{-1}}\cdot L,\,\partial_{t^k}\varphi \cdot_R \varphi^{-1}  \right) +  \left( L,\,\delta\,(\partial_{t^k}\varphi) \cdot_R \varphi^{-1}  \right)\\[1ex] 
    &\quad- \left( L,\,\partial_{t^k}\varphi \cdot_R \varphi^{-1} \cdot_R \delta \varphi\cdot_R \varphi^{-1}  \right)+\frac{1}{2}\left(\text{ad}^*_{RdH_k(L)}\cdot L,\,\delta \varphi\cdot_R\varphi^{-1} \right) \\[1ex]
    &= \left( \text{ad}^{R*}_{\delta \varphi\cdot_R\varphi^{-1}}\cdot L,\,\partial_{t^k}\varphi \cdot_R \varphi^{-1}  \right) - \left( \partial_{t^k}L,\,\delta\,\varphi \cdot_R \varphi^{-1} \right)\\
    &\quad+ \left( L,\,\delta \varphi\cdot_R \varphi^{-1} \cdot_R \partial_{t^k}\varphi \cdot_R \varphi^{-1}  \right)\\[1ex]
    &\quad + \partial_{t^k}\left( L,\,\delta\,\varphi \cdot_R \varphi^{-1}  \right)- \left( L,\,\partial_{t^k}\varphi \cdot_R \varphi^{-1} \cdot_R \delta \varphi\cdot_R \varphi^{-1}  \right)\\
    &\quad+\frac{1}{2}\left(\text{ad}^*_{R\nabla H_k(L)}\cdot L,\,\delta \varphi\cdot_R\varphi^{-1} \right)\\[1ex]
    &= \left( \text{ad}^{R*}_{\delta \varphi\cdot_R\varphi^{-1}}\cdot L,\,\partial_{t^k}\varphi \cdot_R \varphi^{-1}  \right) - \left( \text{ad}^{R*}_{\partial_{t^k}\varphi\cdot_R\varphi^{-1}}\cdot L,\,\delta\,\varphi \cdot_R \varphi^{-1} \right) \\[1ex]
    &\quad+ \left( L,\,\left[\delta \varphi\cdot_R \varphi^{-1} ,\,\partial_{t^k}\varphi \cdot_R \varphi^{-1} \right]_R \right)+\frac{1}{2}\left(\text{ad}^*_{R\nabla H_k(L)}\cdot L,\,\delta \varphi\cdot_R\varphi^{-1} \right)\\
    &\quad+\partial_{t^k}\left( L,\,\delta\,\varphi \cdot_R \varphi^{-1}  \right).
\end{split}
\end{equation}
The first and third terms cancel each other. In the second term, we recognise 
\begin{equation}
\text{ad}^{R*}_{\partial_{t^k}\varphi\cdot_R\varphi^{-1}}\cdot L=\partial_{t^k} L.
\end{equation}
Hence, 
\begin{equation}
\label{delta_L}
    \delta \Lag_k=
    \left( -\partial_{t^k} L+\frac{1}{2}\text{ad}^*_{R\nabla H_k(L)}\cdot L,\,\delta\,\varphi \cdot_R \varphi^{-1} \right) +\partial_{t^k}\left( L,\,\delta\,\varphi \cdot_R \varphi^{-1}  \right)  
\end{equation}
and we obtain the Euler--Lagrange equation for each $\Lag_k$ as 
\begin{equation}
  \partial_{t^k} L=  \frac{1}{2}\, \text{ad}^*_{R\nabla H_k(L)}\cdot L.
\end{equation}
Now recall that $  \frac{1}{2}\, \text{ad}^*_{R\nabla H_k(L)}\cdot L= \text{ad}^*_{R_\pm\nabla H_k(L)}\cdot L$ and that, with $\g$ being equipped with an $\text{Ad}$-invariant nondegenerate bilinear form, $\text{ad}^*_{R_\pm\nabla H_k(L)}\cdot L$ is identified with $[R_\pm\nabla H_k(L), L]$. Thus, we have obtained \eqref{EL_Lax} variationally as desired.
That this set of equations is compatible follows from the commutativity of the flows which is a consequence of the mCYBE and the $\text{Ad}$-invariance of $H_k$ as we now show. Going back to having $L\in\g^*$ and evaluating its derivatives on a fixed but arbitrary $X\in\g$, we have
\begin{equation}
\begin{split}
    (\partial_{t^k}\partial_{t^j}L)(X)&=-\frac{1}{2}\partial_{t^k}\left( L,\, [R\nabla H_j(L),X] \right)\\
    &=\frac{1}{4}\left( L,\,[R\nabla H_k(L),\, [R\nabla H_j(L),\,X]] \right)-\frac{1}{4}\left( L,\,[R [R\nabla H_k(L),\,\nabla H_j(L)],\,X] \right).
\end{split}
\end{equation}
Hence, using the Jacobi identity, we get
\begin{equation}
\begin{split}
    ([\partial_{t^k},\,\partial_{t^j}]L)(X)&=\frac{1}{4}\left( L,\,[[R\nabla H_k(L), R\nabla H_j(L)], X]\right)\\
    &\quad-\frac{1}{4}\left(L,\,R ([R\nabla H_k(L), \nabla H_j(L)]+[\nabla H_k(L), R\nabla H_j(L)]), X] \right)\\
    &=-\frac{1}{4}\left( L,\,[[\nabla H_k(L), \nabla H_j(L)], X]\right)=0,
\end{split}
\end{equation}
where we use the mCYBE in the second equality and property \eqref{ad_invariance_property} in the last step.
We now establish the closure relation, \ie $\d \Lag=0$ on shell. It turns out that the kinetic and potential contributions vanish separately. 
We have 
\begin{equation}
    \partial_{t^j} \Lag_k-\partial_{t^k} \Lag_j = \partial_{t^j} \left(L,\,\partial_{t^k}\varphi \cdot_R \varphi^{-1}  \right)-\partial_{t^k} \left(L,\,\partial_{t^j}\varphi \cdot_R \varphi^{-1}  \right) -\partial_{t^j}H_k(L)+\partial_{t^k}H_j(L).
\end{equation}
Now, using \eqref{ad_invariance_property}, we find
\begin{equation}\label{dtjhkl}
\partial_{t^j}H_k(L)=\left(\partial_{t^j} L ,\, \nabla H_k(L)\right)=-\frac{1}{2}\,\left( L ,\, \left[R\nabla H_j(L), \nabla H_k(L)\right]\right)=0.
\end{equation}
Thus, it is a direct consequence of the $\text{Ad}^*$-invariance of $H$ that the potential contribution to $\d \Lag$ is zero on shell. We are now left with just the kinetic terms which can be rewritten as
\begin{equation}
\begin{split}
  &\left(\partial_{t^j} L,\,\partial_{t^k}\varphi \cdot_R \varphi^{-1}  \right)- \left(\partial_{t^k} L,\,\partial_{t^j}\varphi \cdot_R \varphi^{-1}  \right)+ \left( L,\,\partial_{t^j}(\partial_{t^k}\varphi \cdot_R \varphi^{-1} ) \right)\\
  &\quad-\left( L,\,\partial_{t^k}(\partial_{t^j}\varphi \cdot_R \varphi^{-1} ) \right)\\
    &= \left(\partial_{t^j} L,\,\partial_{t^k}\varphi \cdot_R \varphi^{-1}  \right)- \left(\partial_{t^k} L,\,\partial_{t^j}\varphi \cdot_R \varphi^{-1}  \right)\\
    &\quad+ \left( L,\,\partial_{t^j}\partial_{t^k}\varphi \cdot_R \varphi^{-1}\,- \,\partial_{t^k}\partial_{t^j}\varphi \cdot_R \varphi^{-1}\right) + \left( L,\,\partial_{t^k}\varphi \cdot_R \partial_{t^j}\varphi^{-1}\,-\,\partial_{t^j}\varphi \cdot_R \partial_{t^k}\varphi^{-1}\right).
\end{split}
\end{equation}
From the commutativity of flows, we have $\partial_{t^j}\partial_{t^k}\varphi - \partial_{t^k}\partial_{t^j}\varphi = 0$, which leaves us with
\begin{eqnarray*}
\left(\partial_{t^j} L,\,\partial_{t^k}\varphi \cdot_R \varphi^{-1}  \right)- \left(\partial_{t^k} L,\,\partial_{t^j}\varphi \cdot_R \varphi^{-1}  \right)+ \left( L,\,\partial_{t^k}\varphi \cdot_R \partial_{t^j}\varphi^{-1}\,-\,\partial_{t^j}\varphi \cdot_R \partial_{t^k}\varphi^{-1}\right).
\end{eqnarray*}
The on-shell relation
\begin{equation}
\partial_{t^j} L= \frac{1}{2}\, \text{ad}^*_{R\nabla H_j(L)}\cdot L
\end{equation}
allows us to express the first term as
\begin{equation}
\begin{split}
\left(\partial_{t^j} L,\,\partial_{t^k}\varphi \cdot_R \varphi^{-1}\right) &= \frac{1}{2}  \left(\text{ad}^*_{R\,\nabla H_j(L)}\cdot L,\,\partial_{t^k}\varphi \cdot_R \varphi^{-1}\right)\\
&= -\frac{1}{2}  \left(\text{ad}^*_{\partial_{t^k}\varphi \cdot_R \varphi^{-1}}\cdot L,R\,\nabla H_j(L)\right).
\end{split}
\end{equation}
Since
\begin{equation}
\begin{split}
\left(\text{ad}^{R*}_{\partial_{t^k}\varphi \cdot_R \varphi^{-1}}\cdot L,\nabla H_j(L)\right) &= \frac{1}{2} \left(\text{ad}^*_{\partial_{t^k}\varphi \cdot_R \varphi^{-1}}\cdot L,R\nabla H_j(L)\right)\\
&\quad+ \frac{1}{2} \left(\text{ad}^*_{R\,\partial_{t^k}\varphi \cdot_R \varphi^{-1}}\cdot L,\nabla H_j(L)\right)
\end{split}
\end{equation}
and
\begin{equation}
\left(\text{ad}^*_{R\,\partial_{t^k}\varphi \cdot_R \varphi^{-1}}\cdot L,\nabla H_j(L)\right)=-\left(\text{ad}^*_{\nabla H_j(L)}\cdot L,R\partial_{t^k}\varphi \cdot_R \varphi^{-1}\right) = 0,
\end{equation}
we have a further simplification to
\begin{equation}
\begin{split}
    \left(\partial_{t^j} L,\,\partial_{t^k}\varphi \cdot_R \varphi^{-1}\right) &= -  \left(\text{ad}^{R*}_{\partial_{t^k}\varphi \cdot_R \varphi^{-1}}\cdot L,\nabla H_j(L)\right)\\[1ex]
    &= - \left(\partial_{t^k}L,\nabla H_j(L)\right)\\[1ex]
    &= -\,\partial_{t^k}H_j(L) = 0,
\end{split}
\end{equation}
where we have used the result from \eqref{dtjhkl} (with $k\leftrightarrow j$).
Similarly, we have for the second term
\begin{equation}
     \left(\partial_{t^k} L,\,\partial_{t^j}\varphi \cdot_R \varphi^{-1}\right) = -\,\partial_{t^j}H_k(L) = 0.
\end{equation}
For the last remaining term, we have
\begin{equation}
\begin{split}
    &\left( L,\,\partial_{t^k}\varphi \cdot_R \partial_{t^j}\varphi^{-1}\,-\,\partial_{t^j}\varphi \cdot_R \partial_{t^k}\varphi^{-1}\right) \\[1ex]
   &\qquad\qquad=  \left( L,\,-\partial_{t^k}\varphi \cdot_R \varphi^{-1} \cdot_R \partial_{t^j}\varphi \cdot_R \varphi^{-1}\,+\,\partial_{t^j}\varphi \cdot_R \varphi^{-1} \cdot_R \partial_{t^k}\varphi \cdot_R \varphi^{-1}\right)\\[1ex]
     &\qquad\qquad= \left( L,\, \left[ \,\partial_{t^j}\varphi \cdot_R \varphi^{-1},\, \partial_{t^k}\varphi \cdot_R \varphi^{-1}\, \right]_R\right)\\[1ex]
     &\qquad\qquad= -\left(\text{ad}^{R*}_{\partial_{t^j}\varphi \cdot_R \varphi^{-1}}\cdot L,\,\partial_{t^k}\varphi \cdot_R \varphi^{-1}\right)\\[1ex]
     &\qquad\qquad= - \left(\partial_{t^j} L,\,\partial_{t^k}\varphi \cdot_R \varphi^{-1}\right)= \,\partial_{t^k}H_j(L) = 0.
\end{split}
\end{equation}
\end{proof}
It is worth noting that the properties of the geometric Lagrangian one-form heavily rely on the mCYBE for $R$. It is at the heart of the commutativity of the flows and the closure relation. The connection between the closure relation and the CYBE was first identified and established in \cite{CStV} in the context of integrable field theories. Here it is established in the finite-dimensional context and related to Lie dialgebras.

\section{Closure relation and Hamiltonians in involution}\label{sec:closure-KK}

In this section, we obtain a structural result which brings together Lagrangian multiforms and essential Hamiltonian aspects of integrable systems. It will be convenient and clearer to work with local coordinates $\phi_\alpha$, $\alpha=1,\dots,M$, on the group $G_R$, as we did in \eqref{kinetic_in_coords}. Then, the geometric Lagrangian one-form can be written in the form 
\begin{equation} 
\label{eq:kin_term_canonical_coord}
  \Lag[\varphi] =\left( \,\sum_{\alpha=1}^M  \pi_\alpha\, \partial_{t^k}\phi_\alpha - H_k\right) \d t^k\\  
\end{equation}
where we recall that the momentum $\pi_\alpha$ is defined by
\begin{equation}\label{eq:mom-loccoord}
    \pi_\alpha=\left(  \Lambda,\,\varphi^{-1} \cdot_R  \frac{\partial \varphi}{\partial \phi_\alpha}    \right).
\end{equation}
Each Lagrangian $\Lag_k$ in the multiform has the structure $p\dot{q}-H$ of a Lagrangian in phase space, 
\begin{equation}
    \Lag_k=\sum_{\alpha=1}^M  \pi_\alpha\, \partial_{t^k}\phi_\alpha - H_k,
\end{equation}
and yields its Euler--Lagrange equations from the variation
\begin{equation}
   \delta \Lag_k=\sum_{\beta=1}^M\left(\sum_{\alpha=1}^M \left(\frac{\partial \pi_\alpha}{\partial \varphi_\beta}-\frac{\partial \pi_\beta}{\partial \varphi_\alpha} \right)\, \partial_{t^k}\phi_\alpha - \frac{\partial H_k}{\partial \phi_\beta}\right)\delta\phi_\beta+\partial_{t^k}\left(\sum_{\alpha=1}^M \pi_\alpha\delta \phi_\alpha \right).
\end{equation}
This is of course consistent with the general result of the previous section. Using \eqref{delta_L} and \eqref{eq:mom-loccoord}, we can write
\begin{equation}
\label{link}
\left( -\partial_{t^k} L+\frac{1}{2}\,\text{ad}^*_{R\nabla H_k(L)}\cdot L,\,\delta\,\varphi \cdot_R \varphi^{-1} \right)=    \sum_{\beta=1}^M\left(\sum_{\alpha=1}^M \Omega_{\alpha\beta}\, \partial_{t^k}\phi_\alpha - \frac{\partial H_k}{\partial \phi_\beta}\right)\delta\phi_\beta
\end{equation}
and 
\begin{equation}
\left( L,\,\delta\,\varphi \cdot_R \varphi^{-1}  \right)=\left(\sum_{\alpha=1}^M \pi_\alpha\delta \phi_\alpha \right).
\end{equation}
Thus, we have natural coordinate versions of key components of the theory. In particular, let us denote by $\theta_R$ the vertical one-form
\begin{equation}
    \theta_R=-\sum_{\alpha=1}^M \pi_\alpha\delta \phi_\alpha=-\sum_{\alpha=1}^M \left(  \Lambda,\,\varphi^{-1} \cdot_R  \frac{\partial \varphi}{\partial \phi_\alpha}    \right)\delta \phi_\alpha=-\left(  \Lambda,\,\varphi^{-1} \cdot_R  \delta \varphi\right),
\end{equation}
and let us introduce the vertical $2$-form
\begin{equation}
    \Omega_R=\sum_{\alpha<\beta} \Omega_{\alpha\beta}\,\delta \phi_\alpha\wedge \delta \phi_\beta,\qquad \Omega_{\alpha\beta}=\frac{\partial \pi_\alpha}{\partial \phi_\beta}-\frac{\partial \pi_\beta}{\partial \phi_\alpha}.
\end{equation}
Observe the important relation
\begin{equation}
\label{exact_form}
    \Omega_R=\delta \theta_R.
\end{equation}
The form $\Omega_R$ is the pullback to the group $G_R$ by the map
\begin{equation}
\begin{split}
    \chi:~&G_R \to {\cal O}_\Lambda \\[1ex] 
    &\varphi \mapsto \text{Ad}^{R*}_{\varphi} \cdot \Lambda
\end{split} 
\end{equation}
of the Kostant--Kirillov--Souriau symplectic form $\omega_{R}$ on the coadjoint orbit through $\Lambda\in\g^*$. We recall here that we consider the coadjoint action of the group $G_R$, not the group $G$. Relation \eqref{exact_form} is the well-known fact \cite{Ar, CandS} that this pullback is an exact form. The expression $\varphi^{-1} \cdot_R  \delta \varphi$ appearing in $\theta_R$ can be interpreted as the Maurer--Cartan form on $G_R$. The structure of our Lagrangian coefficients, in particular their kinetic part, is now elucidated in terms of fundamental objects associated with $G_R$ and its coadjoint orbits in $\g^*$.

It is known that the map $\chi$ is a submersion\footnote{We suppose that we are in a situation where this holds, for instance, excluding the trivial case where the orbit is reduced to a point and assuming that the $G_R$ action is proper.}. Also, a coadjoint orbit is always even-dimensional as it admits the nondegenerate symplectic form $\omega_{R}$. Let us introduce local coordinates $\xi_m$, $m=1,\dots, 2p$, on ${\cal O}_\Lambda$ ($2p\le M$). The tangent map $\chi_*$ is represented locally by the $2p\times M$ matrix $\left(\frac{\partial \xi_m}{\partial \phi_\alpha}\right)$. From now on, summation over repeated indices is understood. The pushforward of the vector fields $\frac{\partial }{\partial \phi_\alpha}$ on $G_R$
is given by
\begin{equation}
    \chi_*\!\left(\frac{\partial }{\partial \phi_\alpha}\right) =\frac{\partial \xi_m}{\partial \phi_\alpha}\,\frac{\partial }{\partial \xi_m}
\end{equation}
and the pullback of the differential one-forms $\delta \xi_m$ on ${\cal O}_\Lambda$ reads
\begin{equation}
    \chi^*(\delta \xi_m)=\frac{\partial \xi_m}{\partial \phi_\alpha}\, \delta \phi_\alpha.
\end{equation}
If we write for the Kostant--Kirillov--Souriau form
\begin{equation}
    \omega_R=\omega_{mn}\,\delta \xi_m\wedge \delta \xi_n , 
\end{equation}
then we have the following relation with the coefficients of its pullback $\Omega_R=\chi^*(\omega_R)$,
\begin{equation}
    \Omega_{\alpha\beta}= \frac{\partial \xi_m}{\partial \phi_\alpha}\,\frac{\partial \xi_n}{\partial \phi_\beta}\,  \omega_{mn} . 
\end{equation}
In view of \eqref{link}, it remains to introduce the Euler--Lagrange vertical one-forms on $G_R$
\be
EL_k\equiv EL_k^\beta \delta \phi_\beta \equiv\left(\Omega_{\alpha\beta}\,\partial_{t^k}\phi_\alpha -\frac{\partial  H_k}{\partial \phi_\beta}\right) \delta \phi_\beta.
\ee
This is the pullback of the following vertical one-form on ${\cal O}_\Lambda$,
\be
EL_k=\chi_*(\Upsilon_k)= \Upsilon_k^n \chi_*(\delta \xi_n)=\left(\sum_{m}\omega_{mn} \partial_{t^k}\xi_m -\frac{\partial  H_k}{\partial \xi_n}\right)\chi_*(\delta \xi_n)
\ee
with the relation
\be
EL_k^\beta= \Upsilon_k^n \frac{\partial \xi_n}{\partial \phi_\beta}.
\ee
Since $\chi$ is a submersion, the matrix $\left(\frac{\partial \xi_m}{\partial \phi_\alpha}\right)$ has maximal rank $2p$, so the Euler--Lagrange equations $EL_k^\beta=0$ imply the equations $\Upsilon_k^n=0$ (and vice versa). This is of course just the confirmation in the present coordinate notations of the result we obtained previously that the (multitime) Euler--Lagrange equations from the geometric Lagrangian one-form produce Lax equations naturally living on coadjoint orbits of $G_R$. 

As a consequence, whenever we say that an equality holds ``on shell'', we mean that it holds modulo $EL_k^\beta=0$ or equivalently $\Upsilon_k^n=0$. We can take advantage of this in the following way. $\Omega_R$ is the pullback of the Kostant--Kirillov--Souriau form $\omega_R$ on the coadjoint orbit ${\cal O}_R$. The latter is nondegenerate and therefore induces a Poisson bracket 
with bivector
\begin{equation}
    P_R=\sum_{m<n}P_{mn}\frac{\partial }{\partial \xi_m}\wedge \frac{\partial }{\partial \xi_n},\qquad P_{mn}\, \omega_{nr}=\delta_{mr}.
\end{equation}
The corresponding Poisson bracket on ${\cal O}_\Lambda$ is known (see, for instance, \cite[Chapter 14]{BBT}) to be the restriction of the Lie--Poisson bracket \eqref{Lie_PB} on $\g^*$:
\begin{equation}
    \{f,g\}_R(\xi)=\Big(\xi,\left[\nabla f(\xi), \nabla g(\xi)\right]_R\Big).
\end{equation}
In other words, when $f$, $g$ are restricted to ${\cal O}_\Lambda$, we have 
\be
\label{def_PB}
\{f, g\}_R=P_{mn}\frac{\partial f}{\partial \xi_m}\frac{\partial g}{\partial \xi_n}.
\ee
With these notions introduced, we see that the Euler--Lagrange equations $\Upsilon_k^n=0$ take the form
\begin{equation}
\label{ELk}    \sum_{m}\omega_{mn}\,\partial_{t^k}\xi_m =\frac{\partial  H_k}{\partial \xi_n},
\end{equation}
and can be written in Hamiltonian form 
\be
\label{syst_eqs_q}
\partial_{t^k}\xi_m=P_{mn}\frac{\partial H_k}{\partial \xi_n}=\{\xi_m, H_k\}_R.
\ee
The system of simultaneous equations \eqref{syst_eqs_q} on the $\xi_m$ admits a solution (at least locally) if and only if the flows are compatible, \ie $[\partial_{t^k},\partial_{t^\ell}]=0$. For an arbitrary function $f$, this means 
\begin{equation}
[\partial_{t^k},\partial_{t^\ell}]f=\{\{H_k, H_\ell\}_R,f\}_R=0.
\end{equation}
The stronger condition $\{ H_k, H_\ell\}_R=0$ is the familiar Hamiltonian criterion for integrability (together with a sufficient number of independent such functions $H_k$, of course).

After these preliminary steps, we are now ready to state our second main result and its corollary, the significance of which will be discussed after the proofs.
\begin{theorem}\label{prop_double_zero}
The following {\bf identity} holds
	\be
\frac{\partial \Lag_k}{\partial t^\ell}-\frac{\partial \Lag_\ell}{\partial t^k}+\Upsilon_k^m\,P_{mn}\,\Upsilon_\ell^n=\{H_k, H_\ell\}_R.
	\ee    
\end{theorem}
\begin{proof}
    The proof is by direct computation.
    \begin{equation}
    \begin{split}
        \frac{\partial \Lag_k}{\partial t^\ell}-\frac{\partial \Lag_\ell}{\partial t^k}&=\left(\frac{\partial \pi_\alpha}{\partial \phi_\beta}-\frac{\partial \pi_\beta}{\partial \phi_\alpha}\right)\partial_{t^\ell}\phi_\beta\,\partial_{t^k} \phi_\alpha-\frac{\partial H_k}{\partial \phi_\beta}\partial_{t^\ell}\phi_\beta+\frac{\partial H_\ell}{\partial\phi_\alpha}\partial_{t^k}\phi_\alpha\\
    &=\left(\Omega_{\alpha\beta} \,\partial_{t^k} \phi_\alpha-\frac{\partial H_k}{\partial \phi_\beta}\right)\partial_{t^\ell}\phi_\beta+\frac{\partial H_\ell}{\partial \xi_m}\frac{\partial \xi_m}{\partial \phi_\alpha}\partial_{t^k}\phi_\alpha\\
    &=\left(\omega_{mn} \,\partial_{t^k} \xi_m-\frac{\partial H_k}{\partial \xi_n}\right)\partial_{t^\ell}\xi_n+\frac{\partial H_\ell}{\partial \xi_m}\partial_{t^k}\xi_m\\
    &=\left(\omega_{mn} \,\partial_{t^k} \xi_m-\frac{\partial H_k}{\partial \xi_n}\right)P_{nr}\left( \omega_{rs}\,\partial_{t^\ell}\xi_s+\frac{\partial H_\ell}{\partial \xi_r}-\frac{\partial H_\ell}{\partial \xi_r}  \right)+\frac{\partial H_\ell}{\partial \xi_m}\partial_{t^k}\xi_m\\
    &=-\Upsilon_k^n\,P_{nr}\,\Upsilon_\ell^r+\frac{\partial H_k}{\partial \xi_n}\,P_{nr} \, \frac{\partial H_\ell}{\partial \xi_r},
    \end{split}
    \end{equation}
    hence the result.
\end{proof}
\begin{corollary}
\label{coro}
The closure relation for the Lagrangian multiform $\Lag$ is equivalent to the involutivity of the Hamiltonians $H_k$ with respect to the Lie--Poisson $R$-bracket $\{\,~,~\}_R$.
\end{corollary}
\begin{proof}
    The closure relation 
    requires that {\it on shell}, we have 
    \be
\d \Lag= \left(\frac{\partial \Lag_\ell}{\partial t^k}-\frac{\partial \Lag_k}{\partial t^\ell}\right)\d t^k\wedge dt^\ell=0, \quad k<\ell.
\ee
From the previous theorem, on shell we have 
	\be
\frac{\partial \Lag_k}{\partial t^\ell}-\frac{\partial \Lag_\ell}{\partial t^k}=\{H_k, H_\ell\}_R,
	\ee    
hence the result.
\end{proof}
\begin{remark}
    The connection between the closure relation for Lagrangian one-forms and the involutivity of Hamiltonians was first discussed in \cite{Su}. The content of our Corollary establishes this result for all Lagrangian one-forms in the class that we have introduced in this paper. They include any system describable by the coadjoint orbit and $r$-matrix methods of Lie dialgebras. An extension of the connection between closure and involutivity to the field theory context (Lagrangian two-forms) was discussed in \cite{V2}. 
\end{remark}

\begin{remark}
    The content of the theorem sheds fundamental light on the link between the closure relation and the involutivity of the Hamiltonian as it establishes an {\it off-shell} identity which clearly shows the interplay between the coefficients of $\d \Lag$, the Euler--Lagrange equations, the Poisson tensor on the coadjoint orbit, and the Poisson bracket of the Hamiltonians related to our Lagrangian coefficients. A particular point is that it shows in the present general setting that $\d \Lag$ has a so-called ``double zero'' on the equation of motion. This idea was introduced in \cite{SNC2} and developed in \cite{SNC3} as an important ingredient of Lagrangian multiform theory. However, the relation to Hamiltonians in involution was not noticed there. The status of the ``double zero'' term $\Upsilon_k^n\,P_{nr}\,\Upsilon_\ell^r$ is now clearly identified as well as its relation to the Euler--Lagrange equations. This term is the off-shell element linking the Hamiltonian integrability criterion $\{H_k,H_\ell\}_R=0$ and the integrability criterion advocated in Lagrangian multiform theory: the closure relation $\d \Lag=0$ on shell. 
\end{remark}

\section{Geometric Lagrangian one-forms from reduction}\label{reduction}

It is well-known that many integrable systems arise from Hamiltonian reduction on the cotangent bundle of a Lie group $A$ following the intuitive idea that the more intricate dynamics of the integrable system of interest on the reduced phase space comes from the simplest ``free'' dynamics on the cotangent bundle. In this section, we show how one can construct a general Lagrangian multiform on a coadjoint orbit by a Lagrangian analogue of the procedure of Hamiltonian reduction. The Lagrangian multiform of Section \ref{sec:gen-lm} is recovered as a special case. 

We follow mainly the exposition and ideas in \cite[Lectures 1-2]{STS} to summarise the key notions. By fixing a left trivialisation of $T^*A$ we can parametrise it with $(\alpha,a)\in\mathfrak{a}^*\times A$ where $\mathfrak{a}^*$ is the dual of the Lie algebra $\mathfrak{a}$ of $A$. The canonical symplectic form $\Omega$ is exact and derives from the canonical one-form $\theta$
\begin{equation}
    \Omega=\delta \theta \qquad \text{with}~~\theta=(\alpha,a^{-1}\delta a).
\end{equation}
The cotangent lifts to $T^*A$ of the action of $A$ on itself by left and right translations read
\begin{equation}
    \lambda_b:(\alpha,a)\mapsto(\alpha,b\,a),\quad \rho_b:(\alpha,a)\mapsto(\text{Ad}_b^* \cdot\alpha,a\,b^{-1}), \qquad b\in A.
\end{equation}
The canonical one-form and hence the symplectic form are invariant under these actions. The corresponding moment maps are given by
\begin{equation}
    \mu_{\ell}(\alpha,a)=\text{Ad}^*_{a}\cdot \alpha ,~~\mu_{r}(\alpha,a)=-\alpha.
\end{equation}
In applications to integrable systems, one usually consider the case where only Lie subgroups $A_+$ and $A_-$ of $A$ act by left and right translations. In this case, the moment maps are the restriction of the above moment maps to $\mathfrak{a}_\pm$, the Lie algebras of $A_\pm$. Thus, they are elements of $\mathfrak{a}^*_\pm$, and we denote them by
\begin{equation}
\label{projected_moments}
\mu_\ell(\alpha,a)=\Pi_{\mathfrak{a}^*_+}\left(\text{Ad}^*_{a}\cdot \alpha \right) ,~~\mu_{r}(\alpha,a)=-\Pi_{\mathfrak{a}^*_-}\alpha.
\end{equation}
In the special case where $A_+$ is the trivial group and $A_-=A$, it is known that the quotient Poisson manifold $T^*A/A$ is isomorphic to $\mathfrak{a}^*$ equipped with the Lie--Poisson bracket. See, for instance, \cite[Proposition 1.24]{STS}.

Since our emphasis is on the Lagrangian formalism, let us describe the translation of the above situation into this framework. We consider the following Lagrangian on $T^*A$ 
\begin{equation}
    \Lag^0=\left(\alpha,a^{-1}\frac{\d a}{\d t} \right).
\end{equation}
The importance of $\Lag^0$ is that the Cartan form arising from its variation is precisely the canonical one-form on $T^*A$. Indeed, we have
\begin{equation}
\begin{split}
    \delta \Lag^0&=\left(\delta \alpha,a^{-1}\frac{\d a}{\d t} \right)-\left(\alpha,a^{-1}\delta a a^{-1}\frac{\d a}{\d t} \right)+\left( \alpha,a^{-1}\delta\frac{\d a}{\d t} \right)\\
    &=\left(\delta \alpha,a^{-1}\frac{\d a}{\d t} \right)-\left(\alpha,a^{-1}\delta a a^{-1}\frac{\d a}{\d t} \right)-\left(\frac{\d \alpha}{\d t},a^{-1}\delta a \right)\\
    &\quad+\left(\alpha,a^{-1}\frac{\d a}{\d t} a^{-1}\delta a  \right)+\frac{\d}{\d t}\left(\alpha,a^{-1}\delta a \right)\\
    &=\left(\delta \alpha,a^{-1}\frac{\d a}{\d t} \right)-\left(\frac{\d}{\d t}\left(\text{Ad}^*_a\cdot \alpha\right),\delta a a^{-1}\right)+\frac{\d}{\d t}\left(\alpha,a^{-1}\delta a \right)
\end{split}
\end{equation}
In the last term, we recognise that the Cartan form is $\theta$ (up to a conventional sign). Also, we see that this Lagrangian yields trivial equations of motion
\begin{equation}
\label{free_eqs}
    a^{-1}\frac{\d a}{\d t}=0,~~\frac{\d}{\d t}\left(\text{Ad}^*_a\cdot \alpha\right)=0\quad \Leftrightarrow \quad  \frac{\d a}{\d t}=0,~~\frac{\d \alpha}{\d t}=0.
\end{equation}
The Lagrangian $\Lag^0$ is invariant under the global transformations $(\alpha,a)\mapsto(\alpha,b\,a)$ and $(\alpha,a)\mapsto(\text{Ad}_b^*\cdot \alpha,a\,b^{-1})$ where $b\in A$ is constant. The conserved currents produced by Noether's theorem are the moment maps $\mu_{\ell,r}$. It is immediate from \eqref{free_eqs} that they are indeed conserved currents. The symmetry group $A\times A$ of this free theory is too large to produce systems of interest. One easy way to reduce the symmetry group to $A_+\times A_-=\{e\}\times A$ acting by right translations only is to include a potential term where the potential function depends on $(\alpha,a)$ only through $\mu_\ell$:
\begin{equation}
    \Lag=\left(\alpha,a^{-1}\frac{\d a}{\d t} \right)-H\left(-\mu_\ell(\alpha,a)\right)=\left(\alpha,a^{-1}\frac{\d a}{\d t} \right)-H\left(\text{Ad}^*_{a}\cdot \alpha\right),
\end{equation}
where $H$ is a function on $\mathfrak{a}^*$. By Noether's theorem, we expect that $\mu_r=-\alpha$ is still a conserved current. Indeed, a computation analogous to that in the proof of Theorem \eqref{Th_multi_EL} gives 
\begin{equation}
\begin{split}
    \delta \Lag
    &=\left(\delta \alpha,a^{-1}\frac{\d a}{\d t} -\text{Ad}_{a^{-1}}\cdot\nabla H\left(\text{Ad}^*_{a}\cdot \alpha\right)\right)\\
    &-\left(\frac{\d}{\d t}\left(\text{Ad}^*_a\cdot \alpha \right)-\text{ad}^*_{\nabla H\left(\text{Ad}^*_{a}\cdot \alpha\right)}\cdot \left(\text{Ad}^*_{a}\cdot\alpha\right),\delta a a^{-1}\right)+\frac{\d}{\d t}\left(\alpha,a^{-1}\delta a \right).
\end{split}
\end{equation}
Thus, the equations of motion read
\begin{equation}
    \frac{\d a}{\d t}\,a^{-1} -\nabla H\left(\text{Ad}^*_{a}\cdot \alpha\right)=0,~~
\frac{\d}{\d t}\left(\text{Ad}^*_a\cdot \alpha \right)-\text{ad}^*_{\nabla H\left(\text{Ad}^*_{a}\cdot \alpha\right)}\cdot \left(\text{Ad}^*_{a}\cdot\alpha\right)=0,
\end{equation}
or equivalently,
\begin{equation}
    \frac{\d a}{\d t}\,a^{-1} -\nabla H\left(\text{Ad}^*_{a}\cdot \alpha\right)=0,~~
\frac{\d}{\d t}\alpha=0.
\end{equation}
The analogue of fixing the moment map $\mu_r=-\alpha$ to some fixed value $-\Lambda\in\mathfrak{a}^*$ in the Hamiltonian reduction approach consists of ``integrating out degrees of freedom'' by solving $\frac{\d}{\d t}\alpha=0$ to $\alpha=\Lambda\in\mathfrak{a}^*$, and inserting back into the Lagrangian to get the effective Lagrangian of the reduced model. This yields 
\begin{equation}
\label{eff_Lag}
    \Lag_{\text{eff}}=\left(\Lambda,a^{-1}\frac{\d a}{\d t} \right)-H\left(\text{Ad}^*_{a}\cdot \Lambda\right).
\end{equation}
This Lagrangian describes a system on the coadjoint orbit of $\Lambda\in\mathfrak{a}^*$ under $A$. At this stage, if we equip $\mathfrak{a}$ with a nondegenerate symmetric bilinear form to identify $\mathfrak{a}^*$ with $\mathfrak{a}$, we obtain as before that the equations of motion take the Lax form for $L=\text{Ad}_{a}^*\cdot \Lambda$
\begin{equation}
    \frac{\d L}{\d t}=\left[\nabla H(L),\, L\right].
\end{equation}
Note that we have not assumed anything special about the function $H$, so strictly speaking there is no notion of integrability at this stage, only that the equations for the system under consideration are written in Lax form. 

Applying this construction to the case $A=G_R$, $\mathfrak{a}=\g_R$, we see that each of our Lagrangians $\Lag_k$ in \eqref{our_Lag} is of the form of $\Lag_{\text{eff}}$. Of course, in that case, each function $H_k$ was assumed to have the additional property of being invariant under the coadjoint action of $G$ so that the closure relation, the Lagrangian criterion for integrability\footnote{This is modulo the requirement of having a sufficient number of such independent functions, as always.}, was valid. However, let us go back to the general situation above and suppose we form a Lagrangian one-form by assembling $N$ effective Lagrangians of the form \eqref{eff_Lag} with $N$ independent (arbitrary smooth) functions $H_k$ defined on $\mathfrak{a}^*$ (or possibly only on ${\cal O}_\Lambda$):
\begin{equation}
\label{general_form_A}
  \Lag = \Lag_k \, \d t^k = \left(\left(\Lambda,a^{-1}\partial_{t^k}a \right)-H_k\left(\text{Ad}^*_{a}\cdot \Lambda\right) \right)\, \d t^k.
\end{equation}
The arguments of Section \ref{sec:closure-KK} can be repeated verbatim and lead to the same conclusion as in Theorem \ref{prop_double_zero}.
\begin{theorem}
\label{gen_th}
The following {\bf identity} holds
	\be
 \label{general_off_shell}
\frac{\partial \Lag_k}{\partial t^\ell}-\frac{\partial \Lag_\ell}{\partial t^k}+\Upsilon_k^m\,P_{mn}\,\Upsilon_\ell^n=\{H_k, H_\ell\}_{\mathfrak{a}^*},
	\ee    
 where $\{~,~\}_{\mathfrak{a}^*}$ is the Lie--Poisson bracket on $\mathfrak{a}^*$ and $P_{mn}$ the corresponding Poisson tensor on ${\cal O}_\Lambda$.
\end{theorem}
Note that so far we have not assumed anything on the functions $H_k$. The exact analogue of Corollary \ref{coro} follows from Theorem \ref{gen_th} in the present context. We stress its importance in this general situation: if we can solve for $\Lag_k$ such that the multitime Euler--Lagrange equations and the closure relation hold for the one-form \eqref{general_form_A} then it qualifies as a Lagrangian multiform and \eqref{general_off_shell} implies that the corresponding functions $H_k$ are in involution. Conversely, if we use functions $H_k$ that are in involution with respect to $\{~,~\}_{\mathfrak{a}^*}$ then the one-form \eqref{general_form_A} satisfies the closure relation and is a Lagrangian multiform. In Section \ref{sec:gen-lm}, we used the latter point of view in a special situation: we took advantage of the Lie dialgebra construction which uses $\text{Ad}^*_G$-invariant functions to produce functions in involution with respect to $\{~,~\}_R$ on the dual of $\g_R$. In the present section, the above results imply the stronger statement that one can associate a Lagrangian multiform with any family of Hamiltonians in involution on any coadjoint orbit of a Lie group.

The perspective of the reverse procedure consisting of solving the multitime Euler--Lagrange equations and the closure relations to produce (new?) integrable systems and Hamiltonians in involution is tantalising. However, it is far from clear whether such a philosophy is more (or less) promising than the established integrable system classification tools such as symmetry analysis or classification of the solutions of the (modified) classical Yang--Baxter equation.  

It is instructive to recover our Lagrangians $\Lag_k$ as effective Lagrangians via a slightly different, but related, mechanism. Suppose now that we reduce the symmetry group to $A_+\times A_-$ such that, at least locally, any element $a\in A$ factorise uniquely as $a=a_+^{-1}a_-$, $a_\pm\in A_\pm$. At the Lie algebra level, we have a unique decomposition $X=X_+-X_-$, $X_\pm\in \mathfrak{a}_\pm$ and by duality $\alpha=\Pi_{\mathfrak{a}^*_+}\alpha-\Pi_{\mathfrak{a}^*_-}\alpha$. Identifying $a$ with $(a_+,a_-)$, we see that the action of $A_+\times A_-$ on $a$ amounts to an action by right translations $(a_+,a_-)\mapsto(a_+b_+^{-1},a_-b_-^{-1})$. Thus,
\begin{equation}
    s:T^*A\to \mathfrak{a}^*,~~(\alpha,a)\mapsto \text{Ad}^*_{a_-}\cdot \alpha
\end{equation}
is invariant under the action of $A_+\times A_-$. Equipped with this, as before, it is easy to introduce a Lagrangian which has $A_+\times A_-$ as symmetry group 
\begin{equation}
    \Lag=\left(\alpha,a^{-1}\frac{\d a}{\d t} \right)-H\left(-s(\alpha,a)\right),
\end{equation}
where $H$ is a function on $\mathfrak{a}^*$. By Noether's theorem we expect that $\mu_{\ell,r}$ in \eqref{projected_moments} are conserved currents.
The direct verification from the Euler--Lagrange equations follows by noticing that 
\begin{equation}
\left(\alpha,a^{-1}\frac{\d a}{\d t} \right)=-\left(s,\frac{\d a_+}{\d t}a_+^{-1}-\frac{\d a_-}{\d t}a_-^{-1} \right).
\end{equation}
Therefore, 
\begin{equation}
\begin{split}
    \delta \Lag=&-\left(\delta s,\frac{\d a_+}{\d t}a_+^{-1}-\frac{\d a_-}{\d t} a_-^{-1}-\nabla H(-s)\right)+\left( \frac{\d}{\d t}\left( \text{Ad}^*_{a_+^{-1}}\cdot s\right),\,\delta a_+a_+^{-1} \right)\\[1ex]
    &-\left( \frac{\d}{\d t}\left( \text{Ad}^*_{a_-^{-1}}\cdot s\right),\,\delta a_-a_-^{-1} \right) - \frac{\d}{\d t}\left(s,\delta a_+a_+^{-1}-\delta a_-a_-^{-1} \right)
\end{split}
\end{equation} 
giving 
\begin{equation}
\begin{cases}
    \dfrac{d}{\d t}\,\Pi_{\mathfrak{a}^*_+}\left(\text{Ad}^*_{a_+^{-1}}\cdot s \right)=\dfrac{d}{\d t}\,\Pi_{\mathfrak{a}^*_+}\left(\text{Ad}^*_{a}\cdot \alpha \right)=\dfrac{d}{\d t}\,\mu_\ell=0,\\[2ex]
    \dfrac{d}{\d t}\,\Pi_{\mathfrak{a}^*_-}\left(\text{Ad}^*_{a_-^{-1}}\cdot s \right)=\dfrac{d}{\d t}\,\Pi_{\mathfrak{a}^*_-} \alpha =-\dfrac{d}{\d t}\,\mu_r=0,\\[2ex]
    \dfrac{\d a_+}{\d t}\,a_+^{-1}-\dfrac{\d a_-}{\d t}\,a_-^{-1} -\nabla H(-s)=0.
\end{cases}
\end{equation}
    The first two equations are indeed the conservation of the Noether currents, as expected. We use them to integrate out the corresponding degrees of freedom. Namely, we set 
    \begin{equation}
        \mu_\ell=-\Lambda_+\in\mathfrak{a}_+,~~\mu_r=\Lambda_-\in\mathfrak{a}_-,
    \end{equation}
    and substitute back into the Lagrangian. To obtain the effective Lagrangian, note that 
    \begin{equation}
        s=\Pi_{\mathfrak{a}^*_+}\left(\text{Ad}^*_{{a_+}}\cdot \mu_\ell \right)+\Pi_{\mathfrak{a}^*_-}\left(\text{Ad}^*_{{a_-}}\cdot \mu_r \right).
    \end{equation}
    This can be seen by the following computation\footnote{This computation is the generalisation to the present context of the analogous argument used in \cite{FG} where our $s$ corresponds to their $X$.}. For any $X\in\mathfrak{a}$,
\begin{equation}
\begin{split}
    \left(s,X\right)=\left(s,X_+-X_-\right)=&\left(\text{Ad}^*_{a_+^{-1}}\cdot s,\text{Ad}_{a_+^{-1}}\cdot X_+\right)-\left(\text{Ad}^*_{a_-^{-1}}\cdot s,\text{Ad}_{a_-^{-1}}\cdot X_-\right)\\
    =&\left(\mu_\ell,\text{Ad}_{a_+^{-1}}\cdot X_+\right)+\left(\mu_r,\text{Ad}_{a_-^{-1}}\cdot X_-\right)\\
=&\left(\Pi_{\mathfrak{a}^*_+}\!\left(\text{Ad}^*_{{a_+}}\cdot \mu_\ell \right)+\Pi_{\mathfrak{a}^*_-}\!\left(\text{Ad}^*_{{a_-}}\cdot \mu_r \right),X\right).
\end{split}
\end{equation}
Putting everything together, and setting 
\begin{equation}
    L=-s=\Pi_{\mathfrak{a}^*_+}\left(\text{Ad}^*_{{a_+}}\cdot \Lambda_+ \right)-\Pi_{\mathfrak{a}^*_-}\left(\text{Ad}^*_{{a_-}}\cdot \Lambda_- \right),
\end{equation}
the effective Lagrangian is
\begin{equation}
    \Lag_{\text{eff}}=\left(L,\frac{\d a_+}{\d t}\,a_+^{-1} -\frac{\d a_-}{\d t}\,a_-^{-1}\right)-H\left(L\right).
\end{equation}
In the special case where $A=G$, $\mathfrak{a}=\g$, $\mathfrak{a}_\pm=\g_{\pm}$ with $\g_\pm=R_\pm(\g)$, this effective Lagrangian is exactly of the form of our Lagrangian coefficients $\Lag_k$. This alternative construction amounts to reducing a free system on $T^*G$ by acting with $G_+\times G_-\simeq G_R$. We refer the interested reader to \cite[Section 2.4]{STS} for a discussion of the connection between the reduction on $T^*G_R$ by left translations of $G_R$ and the reduction on $T^*G$ by left and right translations of $G_+\times G_-$. 

This brings us to the end of our presentation of the general framework of Lagrangian multiforms on coadjoint orbits. The next chapter will be dedicated to the application of this framework to some examples. These examples will demonstrate the scope of our framework while also providing, for some well-known integrable systems, explicit Lagrangian one-forms that were obtained for the first time in \cite{CDS} and \cite{CSV}.

\chapter{Examples}\label{chap:orbitexamples}
The theory of Lie dialgebras provides a general setup for constructing integrable models in a systematic manner. This setup is at the heart of the framework of geometric Lagrangian one-forms living on coadjoint orbits that we presented in Chapter \ref{chap:lm-orbits}. By fixing the algebraic data in the general expression for the geometric Lagrangian one-form in \eqref{our_Lag}, one can construct explicit examples of Lagrangian one-forms for specific integrable models. In this chapter, adapted from \cite{CDS} and \cite{CSV}, we demonstrate this for several well-known integrable models, thus obtaining Lagrangian one-forms describing the corresponding hierarchies.\footnote{Minor sign errors appear in some expressions in Sections 5 and 6 of \cite{CDS} (corresponding to Sections \ref{Flaschka} and \ref{Toda_pq} below) that have been corrected in this thesis. I would like to thank Caolfionn McLoughlin for pointing these out.}

In Section \ref{Flaschka}, we illustrate the construction for the open Toda chain associated with a Lie dialgebra via a non-skew-symmetric $r$-matrix. We present explicit expressions for the Lagrangian coefficients and relate our results to the well-known formulations of the Toda chain in Flaschka and canonical coordinates. In Section \ref{Toda_pq}, the same open Toda chain is used to illustrate our construction in the case of a skew-symmetric $r$-matrix. We also relate our results to the description in Flaschka and canonical coordinates. Section \ref{sec:noncycloGaudin} is concerned with the rational Gaudin model, while in Section \ref{sec:cycloGaudin}, we deal with a cyclotomic generalisation of this model. The two cases of Gaudin models will be opportunities for us to show how our construction works in the case of infinite-dimensional Lie algebras, which accounts for the presence of a spectral parameter in the Lax matrices. The case of the cyclotomic Gaudin model is also interesting because it admits two special realisations --- the periodic Toda chain and the discrete self-trapping (DST) model --- that are the subject of Section \ref{sec:realisations}. In this last section, we also demonstrate how to couple together the hierarchies of the periodic Toda chain and the DST model in a straightforward manner and derive a Lagrangian one-form for this coupled hierarchy.

The choice of examples in this chapter also allows us to illustrate the versatility of the theory of Lie dialgebras and, as a consequence, of our framework of geometric Lagrangian one-forms, since each of these examples requires a distinct algebraic setup, as will become clear over the course of this chapter.

\section{Open Toda chain in the AKS scheme}\label{Flaschka}

Toda chains, introduced in \cite{To1}, are among the most extensively studied integrable models. A Toda chain describes a system of particles with nearest-neighbour exponential interaction. The case with an open boundary condition is the subject of this section. The construction here is based on the AKS scheme \cite{A, K, Sy} and reproduces the approach of Flaschka \cite{F}. We start with the algebraic setup required for the construction and present the Lax description of the model before making the connection with our variational approach of geometric Lagrangian one-forms.

\subsection{Algebraic setup}

Let us choose $\g=\sl_{N+1}(\mathbb{C})$, the Lie algebra of $(N+1)\times (N+1)$ traceless real matrices, $\g_+$ the Lie subalgebra of skew-symmetric matrices and $\g_-$ the Lie subalgebra of upper triangular traceless matrices, yielding
	\be
	\g=\g_+\dotplus \g_-.
	\ee
	Here $R=P_+-P_-$ and $R_\pm=\pm P_\pm$ with $P_\pm$ the projector on $\g_\pm$ along $\g_\mp$. 
	The following $\text{Ad}$-invariant nondegenerate bilinear form
	\be
 \label{bilinear_form}
	\langle  X, Y\rangle={\Tr}(XY)
	\ee
	allows the identification $\g^*\simeq \g$, and it induces the decomposition  
	\be
	\g^*=\g_-^*\dotplus \g_+^*\simeq\g_+^\perp\dotplus \g_-^\perp,
	\ee
	where $\g_\pm^\perp$ is the orthogonal complement of $\g_{\pm}$ with respect to $\langle  ~~,~~ \rangle$: $\g_+^\perp$ is the subspace of traceless symmetric matrices and $\g_-^\perp$ the subspace of strictly upper triangular matrices. Let us choose
	\be \label{eq:Lambda}
	\Lambda =\begin{pmatrix}
		0 & 1 & 0 & 0 &\dots & 0\\
		1 & 0 & 1 & 0 &\dots & 0\\
		0 & 1 & 0 & 1 &\dots & 0\\
		0 & 0 & 1 &\ddots &\ddots & \vdots\\
		\vdots & & & \ddots & \ddots & 1\\
		0 & 0 & 0 & \dots &1 & 0\\
	\end{pmatrix}\in \g_-^*\simeq\g_+^\perp
	\ee
	and consider its orbit under the (co)adjoint action of $G_-$, the Lie subgroup associated to $\g_-$ consisting of upper triangular matrices with unit determinant. 
	
\subsection{Lax  description}

As explained in Section \ref{sec:AKS}, the AKS case corresponds to the particular case where $\varphi\in G_-$ so that
	\begin{equation}
		L=\text{Ad}^{R*}_{\varphi} \cdot \Lambda=-R_-^*(\text{Ad}^*_{\varphi_-}\cdot \Lambda),
	\end{equation}
	and the coadjoint orbit ${\cal O}_{\Lambda}$ lies in $\g_-^*$. 
	Using $\langle  ~~,~~  \rangle$, we can identify the adjoint and coadjoint actions. Also, we use it to identify the transpose $A^*:\g^*\to\g^*$ of any linear map $A:\g\to\g$ with the transpose of $A$ with respect to $\langle  ~~,~~  \rangle$ defined on $\g$. Writing $(\xi,X)=\langle Y,X\rangle$, this means that we have
	\begin{equation}
  (A^*(\xi),X)=(\xi,A(X))=\langle Y,A(X) \rangle=\langle A^*(Y),X \rangle.
  \end{equation}
	This allows us to work with 
	\begin{equation}
		L=-R_-^*(\varphi_- \Lambda \varphi_-^{-1})=-R_-^*(\varphi \Lambda \varphi^{-1}) ,
	\end{equation}
	where we have dropped the redundant subscript on $\varphi$ in the second equality with $\varphi=\varphi_-\in G_-$.
From the definitions $\langle  X,\,R_\pm Y   \rangle  =\langle  R^*_\pm X,\,Y   \rangle  $ and  $\langle  X,\,P_\pm Y   \rangle  =\langle  \Pi_\mp X,\,Y   \rangle  $, where we denote by $\Pi_\pm$ the projector onto $\g_\pm^\perp$ along $\g_\mp^\perp$, we find $R^*_\pm=\pm\Pi_\mp$. Note that this is an example of non-skew-symmetric $r$-matrix since
 \begin{equation}
     R^*=\Pi_--\Pi_+\neq -R=P_--P_+.
 \end{equation}
Now, $\varphi\,\Lambda \,\varphi^{-1}$ is the following traceless matrix:
	\be
	\varphi\,\Lambda \,\varphi^{-1}=\begin{pmatrix}
		~a_1~ & * & * & * &\dots & *\\[1ex]
		b_1 & a_2 & * & * &\dots & *\\[1ex]
		0 & b_2 & a_3 & * &\dots & *\\[1ex]
		0 & 0 & b_3 &\ddots &\ddots & \vdots\\[1ex]
		\vdots & & & \ddots & \ddots & *\\[1ex]
		0 & 0 & 0 & \dots &b_{N} & ~a_{N+1}~\\
	\end{pmatrix},
	\ee
from which we obtain $L$ as the symmetric tridiagonal matrix
	\be
	\label{form_L_Flaschka}
	L=\Pi_+(\varphi \Lambda  \varphi^{-1})=\begin{pmatrix}
		~a_1~ & b_1 & 0 & 0 &\dots & 0\\[1ex]
		b_1 & a_2 & b_2 & 0 &\dots & 0\\[1ex]
		0 & b_2 & a_3 & b_3 &\dots & 0\\[1ex]
		0 & 0 & b_3 &\ddots &\ddots & \vdots\\[1ex]
		\vdots & & & \ddots & \ddots & b_{N}\\[1ex]
		0 & 0 & 0 & \dots &b_{N} & ~a_{N+1}~\\
	\end{pmatrix}.
	\ee
Using the Hamiltonian 
	\begin{equation} \label{eq:Ham_1_Toda}
		H_1(L) = -\frac{1}{2} \,{\Tr}\, L^2, 
	\end{equation}
	we then find 
	\be
	R_+\nabla H_1(L)=P_+(-L)=\begin{pmatrix}
		0 & b_1 & 0 & 0 &\dots & 0\\[1ex]
		~-b_1~ & 0 & b_2 & 0 &\dots & 0\\[1ex]
		0 & -b_2 & 0 & b_3 &\dots & 0\\[1ex]
		0 & 0 & -b_3 &\ddots &\ddots & \vdots\\[1ex]
		\vdots & & & \ddots & \ddots & ~b_{N}~\\[1ex]
		0 & 0 & 0 & \dots &-b_{N} & 0\\
	\end{pmatrix}.
	\ee
	A direct substitution in \eqref{EL_Lax} with $k=1$, \ie
	\begin{equation}
  \partial_{t^1}L=[R_\pm \nabla H_1(L),L],
	\end{equation}
  reproduces the open Toda chain equations in Flaschka's coordinates $a_n$, $b_n$:
	\be \label{eq:Flaschka_coord}
	\begin{cases}
		\partial_{t^1}a_1=2b_1^2,\qquad \partial_{t^1}a_{N+1}=-2b_{N}^2,\\[1.5ex]		\partial_{t^1}a_j=2(b_{j}^2-b_{j-1}^2),\qquad j=2,\ldots,N, \\[1.5ex]
            \partial_{t^1}b_j=b_j(a_{j+1}-a_j),\qquad j=1,\ldots,N.
	\end{cases}
	\ee
	The next flow generated by the Hamiltonian 
	\begin{equation}\label{eq:Ham_2_Toda}
		H_2(L) = -\frac{1}{3}   {\Tr}   L^3, 
	\end{equation}
	with gradient $\nabla H_2(L)=-L^2$ yields $R_+\nabla H_2(L)=P_+(-L^2)$ as
	\begin{equation*}
    \resizebox{0.97\textwidth}{!}{$
		\begin{pmatrix}
			0 & b_1(a_1+a_2) & b_1  b_2 & 0 &\dots & 0\\[1ex]
			-b_1(a_1+a_2) & 0 & b_2(a_2+a_3) & b_2  b_3 &\dots & 0\\[1ex]
			-b_1  b_2 & -b_2(a_2+a_3) & 0 & b_3(a_3+a_4) &\dots & 0\\[1ex]
			0 & -b_2  b_3 & -b_3(a_3+a_4) &\ddots &\ddots & \vdots\\[1ex]
			\vdots & & & \ddots & \ddots & b_{N}(a_{N}+a_{N+1})\\[1ex]
			0 & 0 & 0 & \dots &-b_{N}(a_{N}+a_{N+1}) & 0\\
		\end{pmatrix}.
    $}
	\end{equation*}
The corresponding equations from \eqref{EL_Lax} with $k=2$ read
\be \label{eq:second_flow}
	\begin{cases}
\partial_{t^2}a_1=2  b_1^2(a_1+a_2),\qquad \partial_{t^2}a_{N+1}=-2  b_{N}^2(a_{N}+a_{N+1}),\\[1.8ex]
		\partial_{t^2}a_j=2b_j^2(a_j+a_{j+1})-2  b_{j-1}^2(a_{j-1}+a_j),\qquad j=2,\ldots,N, \\[1.8ex] 
		\partial_{t^2}b_1= b_1(a_2^2-a_1^2+b_2^2), \qquad \partial_{t^2}b_{N}=b_{N}(a_{N+1}^2-a_{N}^2-b_{N-1}^2),\\[1.8ex]
		\partial_{t^2}b_j=b_j(a_{j+1}^2-a_j^2+b_{j+1}^2-b_{j-1}^2),\qquad j=2,\ldots,N.
	\end{cases}
\ee

\subsection{Lagrangian description}

We need to choose a convenient parametrisation of $\varphi$ since this is the essential ingredient in the Lagrangians $\Lag_k$. We choose
	\begin{equation}
		\varphi=U  Y , 
	\end{equation}
	where $Y={\rm diag}(y_1,\ldots,y_{N+1})$ is the diagonal matrix of diagonal elements of $\varphi$ (\ie $y_i=\varphi_{ii}$) and $U=\varphi  Y^{-1}$ is the upper triangular matrix with $1$ on the diagonal and arbitrary elements $u_{ij}$, $1 \leq i < j \leq N$. Since $\varphi$ has non-zero determinant, $y_i\neq 0$, $i=1,\ldots,N+1$, and with this parametrisation, we find $L$ as in \eqref{form_L_Flaschka} with
	\begin{equation}\label{ab_uy_vars}
		\begin{cases} 
		a_1=\dfrac{y_{2}}{y_1}  u_{12},\qquad    
		a_{N+1}=-\dfrac{y_{N+1}}{y_{N}}  u_{N,N+1}, \\[2ex]
            a_i=\dfrac{y_{i+1}}{y_i}  u_{i,i+1}-\dfrac{y_{i}}{y_{i-1}}  u_{i-1,i}   ,\qquad i=2,\ldots,N,\\[2ex]  
            b_i=\dfrac{y_{i+1}}{y_i},\qquad i=1,\ldots,N    .
            \end{cases} 
	\end{equation}
	Note that $\displaystyle \sum_{j=1}^{N+1}a_j=0$, so we have $2N$ independent variables on the coadjoint orbit ${\cal O}_\Lambda$. We compute the kinetic part of $\Lag_k$ defined in \eqref{kin_part} as
	\begin{equation}
 \begin{split} 
		K_k&= \langle-R^*_-(\varphi   \Lambda  \varphi^{-1}),  \partial_{t^k}\varphi \cdot_R \varphi^{-1}\rangle  = -\langle\varphi   \Lambda  \varphi^{-1},  R_-(\partial_{t^k}\varphi \cdot_R \varphi^{-1})\rangle\\[1ex] 
		&=-\langle\varphi   \Lambda  \varphi^{-1},  \partial_{t^k}\varphi    \varphi^{-1}\rangle=-{\Tr}\left(\Lambda  \varphi^{-1}  \partial_{t^k}\varphi \right),
  \end{split} 
	\end{equation}
	where in the third step we have used the morphism property of $R_-$,
	\begin{equation}
      R_-(\partial_{t^k}\varphi \cdot_R \varphi^{-1})=\partial_{t^k}\varphi_-    \varphi_-^{-1},
	\end{equation} 
and $\varphi_-=\varphi\in G_-$.
	It remains to express it in terms of our chosen coordinates to get
	\begin{equation}
		K_k =-{\Tr}\left(\Lambda   Y^{-1}  U^{-1}  \partial_{t^k}(UY) \right) = - \sum_{j=1}^{N}\frac{y_{j+1}}{y_j}  \partial_{t^k}u_{j,j+1}.
	\end{equation}
	From these results, it becomes apparent that the convenient coordinates are $b_i$ as given in \eqref{ab_uy_vars} and $u_{i}\equiv u_{i,i+1}$, $i=1,\ldots,N$. 
	The first two Lagrangians involve the Hamiltonians \eqref{eq:Ham_1_Toda} and \eqref{eq:Ham_2_Toda} respectively, and can now be expressed in the $u_i,b_i$ coordinates as follows
  \begin{subequations}
	\begin{align}
		\Lag_1= K_1 - H_1 &= -\sum_{j=1}^{N}b_j  \partial_{t^1}u_{j}+ \frac{1}{2}\sum_{j=2}^{N}(b_j  u_j-b_{j-1}  u_{j-1})^2+\sum_{j=1}^{N}b_j^2+ \frac{1}{2}b_1^2  u_1^2+ \frac{1}{2}b_{N}^2  u_{N}^2, \\
		\Lag_2=K_2 - H_2 &= -\sum_{j=1}^{N}b_j  \partial_{t^2}u_{j}+\frac{1}{3}\sum_{j=2}^{N}(b_j  u_j-b_{j-1}  u_{j-1})^3+\frac{1}{3}(b_1  u_1)^3-\frac{1}{3}(b_{N}  u_{N})^3\nonumber\\
		&\quad+\sum_{j=2}^{N-1}b_j^2(b_{j+1}  u_{j+1}-b_{j-1}  u_{j-1})+b_1^2(b_2  u_2)-b_{N}^2(b_{N-1}  u_{N-1}) . 
	\end{align}
  \end{subequations}
	The variation of $\Lag_1$ reads
	\begin{equation}
    \begin{split}
		\delta\Lag_1&= -\sum_{j=1}^{N}  \partial_{t^1}u_{j}  \delta b_j+\sum_{j=1}^{N}  \partial_{t^1}b_j  \delta u_{j}-  \partial_{t^1} \sum_{j=1}^{N}b_j  \delta u_{j} \\
		&\quad+ \sum_{j=2}^{N}(b_ju_j-b_{j-1}u_{j-1})(u_j  \delta b_j+b_j  \delta u_j)- \sum_{j=1}^{N-1}(b_{j+1}u_{j+1}-b_{j}u_{j})(u_j  \delta b_j+b_j  \delta u_j)\\
		&\quad+2\sum_{j=1}^{N}b_j  \delta b_j+ b_1u_1^2  \delta b_1+b_1^2u_1  \delta u_1+ b_{N}u_{N}^2  \delta b_{N} + b_{N}^2u_{N}  \delta u_{N} , 
	  \end{split}
  \end{equation}
	and gives the following Euler--Lagrange equations
	\begin{equation}
		\begin{cases}
			\partial_{t^1}u_{1}=u_1^2  b_1-u_1  (b_{2}  u_{2}-b_{1}  u_{1})+2  b_1,\\[1.5ex]
			\partial_{t^1}u_{N}=u_N  (b_N  u_N-b_{N-1}  u_{N-1})+u_N^2  b_N+2  b_N,\\[1.5ex]
			\partial_{t^1}u_{j}=u_j(b_j  u_j-b_{j-1}  u_{j-1})-u_j(b_{j+1}  u_{j+1}-b_{j}  u_{j})+2  b_j,\\[1.5ex]
			\partial_{t^1}b_1=b_1(b_2u_2-b_{1}u_{1})-b_1^2  u_1       ,\quad \partial_{t^1}b_{N}=-b_{N}(b_{N}u_{N}-b_{N-1}u_{N-1})-b_{N}^2  u_{N},\\[1.5ex]
			\partial_{t^1}b_j=b_j(b_{j+1}u_{j+1}-b_{j}u_{j})-b_j(b_ju_j-b_{j-1}u_{j-1}),    
		\end{cases}
	\end{equation}
 for $j=2,\ldots,N-1$. It is easy to see that these equations give exactly \eqref{eq:Flaschka_coord} using the identification (see \eqref{ab_uy_vars})
	\begin{equation}	\label{identification}
 \begin{cases}
     a_1=b_1u_1,\qquad a_{N+1}=-b_{N}u_{N},\\[1ex] 
     a_j=b_ju_j-b_{j-1}u_{j-1},\qquad j=2,\ldots,N.
 \end{cases}
	\end{equation}
	This provides a very explicit check that our Lagrangians produce the corresponding Lax equations, in coordinates naturally dictated by the coadjoint orbit construction of the kinetic term, here $u_j$, $b_j$. As recalled in Section \ref{sec:closure-KK}, the kinetic part of a Lagrangian provides the (pullback of the) symplectic form of the model via the Cartan form $\theta_R$. Here, we have (see the total derivative term in $\delta \Lag_1$)
	\begin{equation}
		\label{omega_b_u}
		\theta_R=\sum_{j=1}^{N}b_j  \delta u_{j}~\Rightarrow ~\Omega_R=\sum_{j=1}^{N}\delta b_j\wedge\delta u_{j}.
	\end{equation}
	This shows that the coordinates $u_j$, $b_j$ are canonical. 
	In the present case, choosing $b_j$, $a_j$, for $j=1,\ldots,N$, as the coordinates on the coadjoint orbit ${\cal O}_\Lambda$, we can also express the Kostant--Kirillov--Souriau form explicitly using the formula
	\begin{equation}
  u_j=\frac{1}{b_j}\sum_{\ell=1}^j a_{\ell}
  \end{equation}
  to get
	\begin{equation}
		\omega_R=\sum_{j=1}^{N}\frac{1}{b_j}\sum_{\ell=1}^j\delta b_j\wedge\delta a_{\ell}.
	\end{equation}
	It is instructive to see how the usual Hamiltonian formulation of the open Toda chain in canonical coordinates $q^i,p_i$ is derived from our Lagrangian formulation. From the symplectic form \eqref{omega_b_u}, we deduce the following (canonical) Poisson brackets\footnote{Here we drop the subscript $R$ when referring to the Poisson bracket $\{\,~,~\}_R$ since there will be no confusion with another Poisson bracket.}
	\begin{equation}
		\label{PB_b_u}
		\{b_j,  u_k\}=\delta_{jk},\quad \{b_j,  b_k\}=0=\{u_j,  u_k\},\qquad j,k=1,\ldots,N.
	\end{equation}
	The Legendre transformation
	\begin{equation}
  \frac{\partial\Lag_1}{\partial (\partial_{t^1} u_j)} = - b_j    
\end{equation}
	reproduces, as it should, the Hamiltonian
	\begin{equation}
    \begin{split}
		&\sum_{j=1}^{N}\frac{\partial\Lag_1}{\partial (\partial_{t^1} u_j)}  \partial_{t^1}u_{j}-\Lag_1\\
    &\qquad \quad = -  \frac{1}{2}\sum_{j=2}^{N}(b_j  u_j-b_{j-1}  u_{j-1})^2-\sum_{j=1}^{N}b_j^2- \frac{1}{2}b_1^2  u_1^2- \frac{1}{2}b_{N}^2  u_{N}^2 = H_1(L) . 
    \end{split}
	\end{equation}
The matrix $L$ for $\sl_{N+1}(\mathbb{C})$ in canonical coordinates $(q^i,   p_i)$ is given by 
\begin{equation}\label{form_L_pq}
    L = \begin{pmatrix}
    p_1 & \text{e}^{q^1-q^2} & \hspace{2ex} 0 \hspace{2ex} & 0 & 0 & \dots \\[2ex]
    \text{e}^{q^1-q^2} & p_2 & \text{e}^{q^2-q^3} & 0 & 0 & \dots \\[2ex] 
    0 & \text{e}^{q^2-q^3} & p_3 & \text{e}^{q^3-q^4} & 0 & \dots\\[2ex] 
    0 & 0 & \ddots & \ddots & \ddots & & \\[2ex]
    \vdots & & & \text{e}^{q^{N-1}-q^{N}} & p_N &  \text{e}^{q^{N}-q^{N+1}}\!\!\!\! \\[2ex]
    0 & & &  &  \text{e}^{q^{N}-q^{N+1}} & p_{N+1}
    \end{pmatrix} ,
\end{equation}
and by comparison with \eqref{form_L_Flaschka}, we set the change of variables
	\begin{equation}
         \begin{cases} 
		      q^j=\displaystyle\sum_{k=j}^{N}\ln b_k,\qquad j=1,\ldots,N,\\[-.1ex]
		    p_j=b_j  u_j-b_{j-1}  u_{j-1},\qquad j=2,\ldots,N,\\[1.5ex]
		    p_1=b_1  u_1,\qquad p_{N+1}=-b_{N}  u_{N}.
         \end{cases}
	\end{equation}
From \eqref{PB_b_u}, we deduce by direct calculation:
	\begin{equation}
 \begin{split}
		\label{PB_q^p}
		&\{q^j,  p_k\}=\delta_{jk},\quad \{q^j,  q^k\}=0=\{p_j,  p_k\}, \qquad j,k=1,\ldots,N, \\[1ex]
		&\{q^j,  p_{N+1}\}=-1,\quad \{p_j,  p_{N+1}\}=0, \qquad j=1,\ldots,N.
  \end{split} 
	\end{equation}
Note that $p_{N+1}$ is redundant for the description of the dynamics since we only need the map $(u_j,b_j)\mapsto (q^j,p_j)$ for $j=1,\ldots,N$. This is captured by the fact that the previous relations imply that $C=\displaystyle \sum_{j=1}^{N+1} p_j$ is a Casimir on the $2N$-dimensional phase space with coordinates $(q^1,\ldots,q^{N},p_1,\ldots,p_{N+1})$ and we can work with $C=0$. The coordinate $p_{N+1}$ is still useful for writing the Hamiltonian in the familiar compact form as
\begin{equation} \label{eq:Ham_1_Toda_p_q}
		H_1= -\frac{1}{2}\sum_{j=1}^{N+1}p_j^2-\sum_{j=1}^{N-1}\text{e}^{2(q^j-q^{j+1})}-\text{e}^{2q^{N}}.
\end{equation}
Hamilton's equations $\partial_{t^1}q^j=\{q^j,H_1\}$, $\partial_{t^1}p_j=\{p_j,H_1\}$ yield
	\begin{equation}\label{eq:Toda_p_q}
		\begin{cases}
			\partial_{t^1}p_1=2\text{e}^{2(q^{1}-q^2)},\qquad \partial_{t^1}p_{N+1}=-2\text{e}^{2q^{N}},\\[1.2ex]
			\partial_{t^1}p_j=2\left(\text{e}^{2(q^j-q^{j+1})}-\text{e}^{2(q^{j-1}-q^j)}\right),\qquad j=2,\ldots,N-1,\\[1.2ex]
            \partial_{t^1}p_N = 2\left(\text{e}^{2q^N}-\text{e}^{2(q^{N-1}-q^N)}\right),\\[1.2ex]
			\partial_{t^1}q^j=p_{N+1}-p_j,\qquad j=1,\ldots,N.
		\end{cases}
	\end{equation}  
These can be seen to be equivalent to \eqref{eq:Flaschka_coord}, thus completing the Hamiltonian description of the first flow for the open Toda chain, from our Lagrangian formulation. The same analysis can be performed with $\Lag_2$, although the calculations are longer. We simply record here the Euler--Lagrange equations obtained from $\delta \Lag_2$:
	\begin{equation}
 \begin{cases}
		\partial_{t^2}u_1= u_1\left((b_{1}  u_{1})^2-(b_{2}  u_{2}-b_{1}  u_{1})^2-b_{2}^2  \right)  +2  b_1  b_{2}  u_{2}  ,\\[1.5ex]
		\partial_{t^2}u_{N}=u_N\left((b_{N}  u_{N}-b_{N-1}  u_{N-1})^2-(b_{N}  u_{N})^2+b_{N-1}^2  \right) -2  b_{N}  b_{N-1}  u_{N-1},\\[1.5ex]
		\partial_{t^2}u_j=u_j\left((b_{j}  u_{j}-b_{j-1}  u_{j-1})^2-(b_{j+1}  u_{j+1}-b_{j}  u_{j})^2\right)+u_j  (b_{j-1}^2-b_{j+1}^2) \\[1ex] 
        \hspace{10ex} +2  b_j  (b_{j+1}  u_{j+1}-b_{j-1}  u_{j-1}),\\[1.5ex]
		\partial_{t^2}b_1= b_1\left((b_{2}  u_{2}-b_{1}  u_{1})^2-(b_{1}  u_{1})^2+b_{2}^2  \right) ,\\[1.5ex]
		\partial_{t^2}b_{N}= b_N\left((b_{N}  u_{N})^2-(b_{N}  u_{N}-b_{N-1}  u_{N-1})^2-b_{N-1}^2  \right) ,\\[1.5ex]
		\partial_{t^2}b_j= b_j\left((b_{j+1}  u_{j+1}-b_{j}  u_{j})^2-(b_{j}  u_{j}-b_{j-1}  u_{j-1})^2\right)-b_j  (b_{j-1}^2-b_{j+1}^2)  ,
        \end{cases} 
	\end{equation}
 for $j=2,\ldots,N-1$. To conclude this example we establish the closure relation for the first two flows, \ie
 \begin{equation}
  \partial_{t^2}\Lag_1-\partial_{t^1}\Lag_2=0
 \end{equation}     
on shell. We know from our general results that this must hold, so this is simply an explicit check. The kinetic and potential contributions give zero separately, so we split the calculations accordingly. For the potential terms, it is more expedient to use the $a_j,b_j$ coordinates\footnote{Note that for conciseness, we treated the equations for $j=1$ and $j=N$ on the same level as for $j=2,\ldots,N-1$ by formally introducing $b_0=0$ and $b_{N+1}=0$.} and equations  \eqref{eq:Flaschka_coord} and \eqref{eq:second_flow}:
	\begin{equation}
    \begin{split}
		&\partial_{t^2}H_1-\partial_{t^1}H_2\\
    &= \partial_{t^1}\left(\sum_{j=1}^{N+1}\frac{a_j^3}{3}+\sum_{j=1}^{N}b_j^2(a_j+a_{j+1})  \right)-\partial_{t^2}\left(\sum_{j=1}^{N+1}\frac{a_j^2}{2}+\sum_{j=1}^{N}b_j^2  \right)\\
		&= \sum_{j=1}^{N+1} 2 a_j^2(b_{j-1}^2-b_j^2) +
  \sum_{j=1}^{N}2b_j^2(a_j^2-a_{j+1}^2+b_{j-1}^2-b_{j+1}^2)\\
		&\quad -\sum_{j=1}^{N+1} 2 a_j(b_{j-1}^2(a_{j-1}+a_j)-b_j^2(a_j+a_{j+1})) -\sum_{j=1}^{N}2b_j^2(a_j^2-a_{j+1}^2+b_{j-1}^2-b_{j+1}^2)\\
		&=\sum_{j=1}^{N+1} 2 (a_ja_{j+1}b_j^2-a_ja_{j-1}b_{j-1}^2 )\\
		&=0,
    \end{split}
	\end{equation}
	where in the last step we recognise a telescopic sum. For the kinetic terms, we also use the $a_j$, $b_j$ coordinates wherever possible to expedite the calculations:
	\begin{equation}
    \begin{split}
		&\partial_{t^1}K_2 - \partial_{t^2}K_1\\
    &= \sum_{j=1}^N(\partial_{t^1} (b_j\partial_{t^2}u_j)-\partial_{t^2}(b_j\partial_{t^1}u_j )) \\[-.5ex]
		&=\sum_{j=1}^N(\partial_{t^1}((a_{j+1}^2-a_j^2+b_{j+1}-b_{j-1}^2)u_jb_j-2b_j^2(a_j+a_{j+1}) )-\partial_{t^2}((a_{j+1}-a_j)u_jb_j-2b_j^2 ))\\
		&=\sum_{j=1}^N((\partial_{t^1}(a_{j+1}^2-a_j^2+b_{j+1}-b_{j-1}^2)- \partial_{t^2}(a_{j+1}-a_j)) u_jb_j  - 2b_j^2(b_{j+1}^2-b_{j-1}^2))\\
		&=0 , 
	  \end{split}
	\end{equation}
	where in the last step the first term gives zero for each $j$ upon using the equations of motion and the remaining terms form a telescopic sum adding up to zero. 

 A Lagrangian one-form for the Toda chain was first constructed in \cite{PS} using variational symmetries of a given starting Lagrangian, which would be $\Lag_1$ in our context, to construct higher Lagrangian coefficients which constitute a one-form when assembled together. The infinite Toda chain was studied more recently in \cite{SV} to illustrate the newly introduced theory of Lagrangian multiforms over semi-discrete multitime. In \cite{PS}, the analogue of our $\Lag_2$ and $\Lag_3$ were constructed. The Noether integrals $J_1$ and $J_2$ (equations (10.11) and (10.12) in \cite{PS}) which constitute the potential part of their Lagrangians are nothing but $H_2(L)$ and $H_3(L)$ with $L$ parametrised as in \eqref{form_L_pq}, up to an irrelevant change of convention $\text{e}^{q^i-q^{i+1}}\to \text{e}^{q^{i+1}-q^i}$ and setting $q^i=x^i$ and $p_i=\dot{x}_i$. The kinetic part of the higher Lagrangians in \cite{PS} involves the so-called alien derivatives which are symptomatic of constructing a Lagrangian multiform from a starting Lagrangian and building compatible higher Lagrangian coefficients. Our construction prevents the problem of alien derivatives altogether, putting all the Lagrangian coefficients on equal footing. This was also achieved previously in the context of IFTs in \cite{CS2,CStV}.

\section{Open Toda chain with a skew-symmetric \texorpdfstring{$r$}{r}-matrix}\label{Toda_pq}

We now present the open Toda chain for the same algebra, that is, $\g=\sl_{N+1}(\mathbb{C}),$ but endowed with a different Lie dialgebra structure. This is based on the Cartan decomposition of $\g$ and leads to a skew-symmetric $r$-matrix in contrast to the non-skew-symmetric $r$-matrix in the previous section. An additional interesting feature of this setup, which we only illustrate for $\sl_{N+1}(\mathbb{C})$ here, is that it allows for a generalisation to any finite semisimple Lie algebra. See, for instance, \cite[Chapter 4]{BBT}.

\subsection{Algebraic setup}

Consider the decomposition 
	\begin{equation}  
 \label{decomp_g}
		\mathfrak{g} =  \mathfrak{n}_+ \dotplus \mathfrak{h} \dotplus \mathfrak{n}_- ,
	\end{equation}
	where $\mathfrak{h}$ is the Cartan subalgebra of diagonal (traceless) matrices and $\mathfrak{n}_\pm$ the nilpotent subalgebra of strictly upper/lower triangular matrices. 
 Let $P_\pm$, $P_0$ be the projectors onto $\mathfrak{n}_\pm$ and $\mathfrak{h}$ respectively, relative to the decomposition \eqref{decomp_g} and set $R=P_+-P_-$. It can be verified that $R$ satisfies the mCYBE. Here $R_{\pm}=\pm(P_\pm+P_0/2)$ and 	
 \begin{equation}
		\mathfrak{g}_{\pm} = \text{Im}(R_{\pm}) = \mathfrak{b}_{\pm} = \mathfrak{h} \dotplus \mathfrak{n}_{\pm}. 
	\end{equation}
We have the following action of $R_\pm$ on the elements $y \in \mathfrak{h}$ and $w_{\pm} \in \mathfrak{n}_{\pm}$,
	\begin{equation} \label{eq:R_action_BBT}
		R_{\pm} (y) = \pm \frac{1}{2}   y ,\quad 
		R_{\pm} (w_{\pm}) = \pm    w_{\pm} ,\quad 
		R_{\pm} (w_{\mp}) = 0 .
	\end{equation}
 Taking the same bilinear form as in \eqref{bilinear_form}, \ie $\langle X, Y\rangle=\Tr(XY)$, we see that 
 \begin{equation}
 \label{adjoints}
     P_\pm^*=P_\mp,\quad P_0^*=P_0\qquad \text{so that}\,\, R^*=-R.
 \end{equation}
 Thus, we have a skew-symmetric $r$-matrix here. 
	For the related Lie groups, we have the following factorisations close to the identity, 
	\begin{equation} \label{eq:phi_BBT}
		\varphi =  \varphi_{+}  \varphi_-^{-1} , \quad \varphi_{\pm} =  W_{\pm}  Y^{\pm 1} , \qquad Y\in \text{exp}(\mathfrak{h}) ,\,\, W_{\pm}\in\text{exp}(\mathfrak{n}_{\pm}) . 
	\end{equation}

\subsection{Lax description}

For $\Lambda \in \mathfrak{g}^*\simeq \g$, the expression of $L$ as a coadjoint orbit of $\Lambda$ is given by
\begin{equation}
	\begin{split} \label{eq:L_mf_toda_complete}
		L &= \text{Ad}^{R*}_{\varphi} \cdot \Lambda = R^{*}_+( W_+  Y  \Lambda  Y^{-1}  W_+^{-1}) - R^{*}_-(W_-  Y^{-1}   \Lambda  Y  W_-^{-1}) .
	\end{split}
\end{equation}
	We choose $\Lambda$ as in \eqref{eq:Lambda}, emphasising that in this case it is an element of the full $\mathfrak{g}^* \simeq \mathfrak{g}$, and $Y \in \text{exp}(\mathfrak{h})$, $W_{\pm}\in \text{exp}(\mathfrak{n}_{\pm})$ given by
	\begin{subequations}\label{eq:YW_BBT}
  \begin{align} 
	Y &= \text{diag} \left( \eta_1 , \eta_2 \dots, \eta_{N+1} \right) , \quad \det Y = 1, \\[2ex]
	W_- &= \begin{pmatrix} 
			1 & 0  & 0 &\dots & 0\\[1.5ex]
			\omega^-_{2,1} & 1  & 0 &\dots & 0\\[1.5ex]
			\omega^-_{3,1} & \omega^-_{3,2} & 1 &\dots & 0\\[1.5ex]
			\vdots & \ddots & \ddots & \ddots & ~~0~~\\[1.5ex]
			~\omega^-_{N,1}~ & \omega^-_{N,2}  & \dots &\omega^-_{N,N-1} & 1\\
		\end{pmatrix},\\[2ex]
    W_+ &= \begin{pmatrix}
			1 & \omega^+_{1, 2} & \omega^+_{1, 3}  &\dots & \omega^+_{1, N}\\[1.5ex]
			0 & 1 & \omega^+_{2, 3}  &\dots & \omega^+_{2 ,N}\\[1.5ex]
			0 & 0 & 1 &\ddots & \vdots\\[1.5ex]
			\vdots & &  & \ddots & ~\omega^+_{N-1, N}~\\[1.5ex]
			~~0~~ & 0  & \dots & 0 & 1\\
		\end{pmatrix}.
	\end{align}
\end{subequations}
	From \eqref{adjoints}, we deduce that $R^*_{\pm}=\pm(P_\mp+P_0/2)$ so that
	\begin{equation}
			R_{\pm}^* (y) = \pm \frac{1}{2}   y ,\quad 
			R_{\pm}^* (w_{\pm}) = 0 ,\quad  
			R_{\pm}^* (w_{\mp}) =  \pm w_{\mp} ,
	\end{equation}
 for $y \in \mathfrak{h},~w_{\pm} \in \mathfrak{n}_{\pm}$.  Let us introduce the variables $(w_i,z_i)$, defined as  
	\begin{equation} \label{eq:variables_w_z}
		w_i = \frac{\omega^+_{i,i+1} - \omega^-_{i+1,i}}{2}, \quad z_i=2  \frac{\eta_{i+1}}{\eta_i} ,
	\end{equation}
from which we determine the Flaschka coordinates as 
	\begin{equation} \label{eq:BBT_Flash_coord}
 \begin{cases}
     	a_i = \dfrac{w_iz_i-w_{i-1} z_{i-1}}{2} , \qquad i = 2, \ldots, N-1, \\[2ex]
      a_1 = \dfrac{w_1z_1}{2} , \quad a_{N+1} = -\dfrac{w_Nz_N}{2} ,\\[2ex]
      b_i = \dfrac{z_i}{2} ,\qquad  i=1, \ldots, N.  
 \end{cases}
\end{equation}
The evaluation of \eqref{eq:L_mf_toda_complete} in those coordinates reproduces the tridiagonal form as in \eqref{form_L_Flaschka}. One can then check that the equations for the first two flows \eqref{eq:Flaschka_coord} and \eqref{eq:second_flow} in the previous section derive from the Lax equation
\begin{equation}
    \partial_{t^k} L = \left[ R_{+}(\nabla H_k(L)),L \right] , \qquad k = 1,2 ,
\end{equation}
where the Hamiltonians are taken as 
\begin{equation}
    H_1(L) = \Tr(L^2), \quad  H_2(L) = \frac{2}{3} \Tr(L^3),
\end{equation}
and we recall that $R_+=P_+ +P_0/2$ here.

\subsection{Lagrangian description}

The Lagrangian one-form takes the form
	\begin{equation} \label{eq:Lagrangian_Toda_a_la_BBT}
		\Lag = \Lag_{k}  dt^k = \left( {K}_k(L)  - {H}_k(L) \right) dt^k , 
	\end{equation}
for $L \in \mathcal{O}_{\Lambda} , \varphi \in G_R$, with the kinetic and the potential terms given by
	\begin{equation} \label{eq:kinetic_potential_terms_BBT}
		{K}_k(L) = \text{Tr} ( L   \partial_{t^k} \varphi \cdot_R \varphi^{-1} ), \quad  {H}_k(L) = \frac{2}{k+1} \text{Tr}(L^{k+1}), 
	\end{equation} 
 respectively. As in the previous section, the kinetic term will allow us to recognise natural canonical variables of the system in this description. Recalling \eqref{eq:phi_BBT}, \eqref{eq:YW_BBT}, and \eqref{eq:variables_w_z}, we find
	\begin{equation}
 \label{eq:BBT_kinetic_term}
 \begin{split} 
			{K}_k(L)  &=  \text{Tr}( \Lambda  \varphi^{-1} \cdot_R \partial_{t^k} \varphi  ) \\[1.2ex]
			&=   \text{Tr}( \Lambda  \varphi^{-1}_+ \cdot \partial_{t^k} \varphi_+   ) -  \text{Tr}( \Lambda  \varphi^{-1}_- \cdot \partial_{t^k} \varphi_-   ) \\
   &=\sum_{i=1}^N \dfrac{\eta_{i+1}}{\eta_i}  \partial_{t^k} \omega_{i,i+1}^+ -\sum_{i=1}^N \dfrac{\eta_{i+1}}{\eta_i}  \partial_{t^k} \omega_{i+1,i}^-  \\
   &=\sum_{i=1}^N z_i  \partial_{t^k} w_i .
   \end{split} 
	\end{equation} 
	The $k$-th Lagrangian coefficient expressed in terms of the coordinates $(w_i,z_i)$ reads
	\begin{equation}
		\Lag_k = {K}_k - {H}_k = \sum_{i=1}^N     z_i \partial_{t^k}w_i - {H}_k  ,
	\end{equation}
	with $(w_i,z_i)$ being canonical coordinates, and for $k=1,2$,
	\begin{subequations}
		\begin{align}
			{H}_1(L) &=  \text{Tr}(L^2) = \sum_{i=1}^{N} \frac{1}{2}\left(z_i^2+w_i^2z_i^2 \right)- \sum_{i=1}^{N-1} \frac{1}{2}w_iz_iw_{i+1}z_{i+1}    , \\
			{H}_2(L) &= \dfrac{2}{3} \text{Tr}(L^3) = \sum_{i=1}^{N} \frac{1}{4}\left(z_i^2w_{i+1}z_{i+1}-z_{i+1}^2w_iz_i + w_i^2z_i^2 w_{i+1}z_{i+1} -w_iz_iw_{i+1}^2z_{i+1}^2 \right)  .
		\end{align}
	\end{subequations}
To obtain the latter expressions of the Hamiltonians, it suffices to use the following expression for $L$ in the $w_i,z_i$ coordinates:
\begin{equation} 
		L =  \frac{1}{2}\begin{pmatrix}
			~w_1z_1 & \,z_1 & 0 & 0 &\dots & 0\\[2ex]
			\,z_1 & w_2z_2-w_1z_1 & \,z_2 & 0 &\dots & 0\\[2ex]
			0 & \,z_2 & w_3z_3-w_2z_2 & \,z_3 &\dots & 0\\[2ex]
			0 & 0 & \,z_3 &\ddots &\ddots & \vdots\\[2ex]
			\vdots & & & \ddots & \ddots & \,z_N\\[2ex]
			0 & 0 & 0 & \dots & \,z_N & -w_Nz_N ~
		\end{pmatrix} .
  \label{eq:L_BBT_R_skew}
	\end{equation}
 Note that we can just as easily determine the higher Hamiltonians and hence the higher Lagrangian coefficients $\Lag_k$, although the expressions become long. 
The variation $\delta \Lag_1$ yields the following Euler--Lagrange equations
	\begin{equation}
		\begin{cases} \label{eq:BBT_t1_w_z_R}
  \partial_{t^1} w_1 = z_1 - \dfrac{w_1}{2}   \left((w_{2} z_{2} - w_{1} z_{1}) -  w_{1} z_{1}  \right)  , \\[2ex] 
  \partial_{t^1} w_N =   z_N - \dfrac{w_N}{2}  \left( - w_{N} z_{N} - ( w_{N} z_{N} - w_{N-1} z_{N-1}   ) \right)  , \\[2ex]
\partial_{t^1} w_i =  z_i - \dfrac{w_i}{2}  \left( (w_{i+1} z_{i+1} -w_{i} z_{i}  ) -(w_{i} z_{i} - w_{i-1} z_{i-1} ) \right),  \\[2ex]
			\partial_{t^1} z_1 = \dfrac{z_1}{2}   \left((w_{2} z_{2} - w_{1} z_{1}) - w_{1} z_{1} \right), \\[2ex] 
   \partial_{t^1} z_N =  \dfrac{z_N}{2}   \left( -2 w_{N} z_{N} + w_{N-1} z_{N-1}  \right),\\[2ex]
        \partial_{t^1} z_i =  \dfrac{z_i}{2}   \left((w_{i+1} z_{i+1}-  w_{i} z_{i}) - ( w_{i} z_{i} - w_{i-1} z_{i-1} )  \right),
		\end{cases}
	\end{equation}
for $i=2,\ldots,N-1$, while the variation of $\Lag_2$ gives the Euler--Lagrange equations for the second flow
\begin{equation} 
    \begin{cases}
        \partial_{t^2} z_1 = \dfrac{z_1}{4} \big( (w_{2}z_{2}-w_1z_1)^2 -(w_{1}z_{1})^2 +  z_{2}^2  \big) ,  \\[2ex]
        \partial_{t^2} z_N = \dfrac{z_N}{4} \left( (w_Nz_N)^2 - (w_Nz_N - w_{N-1}z_{N-1})^2 -z_{N-1}^2 \right) , \\[2ex]
        \partial_{t^2} z_i = \dfrac{z_i}{4}\big( (w_{i+1}z_{i+1}-w_iz_i)^2 -(w_{i}z_{i}-w_{i-1}z_{i-1})^2 +  z_{i+1}^2- z_{i-1}^2  \big),\\[2ex] 
        \partial_{t^2} w_1 =  \dfrac{z_1}{2}\big(w_{2}z_{2}\big) -\dfrac{w_1}{4} \Big( (w_{2}z_{2}-w_1z_1)^2 -(w_{1}z_{1})^2 +  z_{2}^2  \Big) ,\\[2ex]
        \partial_{t^2} w_N =  - \dfrac{z_N}{2}\big(w_{N-1}z_{N-1}\big)  -\dfrac{w_N}{4} \Big( (w_Nz_N)^2 -(w_{N}z_{N}-w_{N-1}z_{N-1})^2- z_{N-1}^2  \Big) ,\\[2ex]
        \partial_{t^2} w_i = \dfrac{z_i}{2}\big((w_{i+1}z_{i+1}-w_iz_i)+(w_iz_i-w_{i-1}z_{i-1})\big)  -\dfrac{w_i}{4} \Big( (w_{i+1}z_{i+1}-w_iz_i)^2 \\[2ex] 
        \hspace{15ex} -(w_{i}z_{i}-w_{i-1}z_{i-1})^2 +  z_{i+1}^2- z_{i-1}^2  \Big) ,
    \end{cases}
\end{equation}
with $i=2,\ldots,N-1$. One can check that these reproduce the more familiar equations \eqref{eq:Flaschka_coord}--\eqref{eq:second_flow} in Flaschka coordinates, using \eqref{eq:BBT_Flash_coord}.
As in the previous section, we can relate our results with the Hamiltonian formulation of the Toda chain in traditional canonical coordinates $(q^i,p_i)$. With 
\begin{equation}
    \theta_R = \sum_{i=1}^N z_i\delta w_i \quad \implies \quad \{z_i,w_j\} = \delta_{ij} , \quad \{w_i,w_j\} = 0 = \{z_i,z_j\} ,
\end{equation}
we see that it suffices to set
\begin{equation}
\begin{cases}
    q^i = \displaystyle \sum_{\ell=i}^N \ln \dfrac{z_i}{2}, \qquad i = 1, \ldots, N \\[4ex]
    p_i = \dfrac{w_iz_i-w_{i-1}z_{i-1}}{2}, \qquad  i = 2, \ldots, N \\[2ex]
    p_1 = \dfrac{w_1z_1}{2} , \quad p_{N+1} = -\dfrac{w_Nz_N}{2} .
\end{cases}
\end{equation}

The explicit verification of the closure relation in the first two flows is completely analogous to that given at the end of the previous section.

\section{Rational Gaudin model}\label{sec:noncycloGaudin}

Gaudin models are a general class of integrable systems associated with Lie algebras with a nondegenerate invariant bilinear form. They were first introduced for the Lie algebra $\mathfrak{sl}_2(\mathbb{C})$ by Gaudin in \cite{Ga1} as quantum integrable spin chains with long-range interactions, and then generalised for arbitrary semisimple Lie algebras in \cite{Ga2}. 
At both the classical and quantum levels, the integrable structure of this model, associated with a Lie algebra $\g$, is underpinned by the rational skew-symmetric solution 
\begin{equation}
\label{rational_r_mat}
  r_{\ti{12}}^0(\lda, \mu) = \frac{C_{\ti{12}}}{\mu - \lda}
\end{equation}
of the CYBE\footnote{Note that this is the general version of the CYBE applicable to non-skew-symmetric matrices.}
\begin{equation}
   [r_{\ti{12}}(\lda, \mu),r_{\ti{13}}(\lda, \nu)]+[r_{\ti{12}}(\lda, \mu),r_{\ti{23}}( \mu,\nu)]+[r_{\ti{32}}(\nu, \mu),r_{\ti{13}}(\lda, \nu)]=0.
\end{equation}
The rational classical $r$-matrix \eqref{rational_r_mat} takes values in $\g \otimes \g$ and depends on spectral parameters $\lda, \mu \in \mathbb{C}$, and the tensor Casimir $C_{\ti{12}} \in \g \otimes \g$ appearing in \eqref{rational_r_mat} is defined as
\begin{equation}
    C_{\ti{12}} = I_a \otimes I^a
\end{equation}
where $\{I_a\}$ and $\{I^a\}$ are dual bases of $\g$ with respect to a fixed nondegenerate invariant bilinear form. The main property of $C_{\ti{12}}$ which gives it its name is $[C_{\ti{12}}, X_{\ti 1}+X_{\ti 2}]=0$ for all $X\in\g$ and where we use the standard tensorial notation $X_{\ti 1} \equiv X \otimes \mathbb{1}$ and $X_{\ti 2} \equiv \mathbb{1} \otimes X$.

Unlike the case of the open Toda chain, the Lax matrix of a Gaudin model is a Lie algebra-valued rational function of a variable $\lambda$, the spectral parameter. We will only look at (classical) finite Gaudin models here, which describe certain spin chains and mechanical systems. To accommodate this, we need to extend our construction to certain infinite-dimensional Lie algebras.

Before diving into the required algebraic machinery, it is useful to recall the usual presentation of the equations of the model that we are aiming at describing variationally. We do so in the simplest case of a rational Lax matrix with simple poles. Many generalisations are known, including elliptic and non-skew-symmetric cases \cite{S}. The latter will be the subject of Section \ref{sec:cycloGaudin}. 

The Lax matrix of a (rational) Gaudin model associated with a finite Lie algebra $\g$ and a set of points $\zeta_r \in \mathbb{C}$, $r=1, \ldots, N$, and the point at infinity is given by the following $\g$-valued rational function
\begin{equation}
\label{Lax_matrix_Gaudin-1}
    L(\lda)=  \sum_{r=1}^N \frac{X_r}{\lambda - \zeta_r} + X_\infty, \qquad X_1,\ldots,X_N,X_\infty\in\g.
\end{equation}
The coefficients $H^n_{k, r}$ of $(\lda-\zeta_r)^{-n-1}$, $n\ge 0$, in $\Tr(L(\lda)^{k+1})/{(k+1)}$, $ k\geq 1$, 
are Hamiltonians in involution (with respect to the Sklyanin bracket). Of course, only a finite subset of them are independent and generate nontrivial flows. In the rest of this work, we will focus on the coefficients corresponding to $n=0$ and drop the extra label by simply writing $H^0_{k, r}=H_{k, r}$. 
The most famous ones are the quadratic Gaudin Hamiltonians which are the coefficients $H_{1, r}$ in
\begin{equation}
    \frac{1}{2}\Tr(L(\lda)^2) = \frac{1}{2}\sum_{r=1}^N\frac{\Tr (X_r^2)}{(\lambda - \zeta_r)^2} + \sum_{r=1}^N\frac{H_{1, r}}{\lambda - \zeta_r} + \frac{1}{2}\Tr(X_\infty^2),
\end{equation}
and read
\begin{equation}
\label{Gaudin_Ham1}
H_{1, r}=\sum_{s\neq r}\frac{\Tr (X_rX_s)}{\zeta_r-\zeta_s}+\Tr(X_rX_\infty), \qquad r=1,\ldots,N.
\end{equation}
The functions $H_{k, r}$ give rise to a hierarchy of compatible equations in Lax form
\begin{equation}\label{generating_Gaudin_eqs}
    \partial_{t^{k, r}}L(\lda)=\left[M_{k, r}(\lda),L(\lda)\right].
\end{equation}
For $k=1$, we have
\begin{equation}
\label{first_M}
    M_{1, r} = -\frac{X_r}{\lda - \zeta_r},
\end{equation}
and the Lax equation \eqref{generating_Gaudin_eqs} leads to the following equations of motion for the degrees of freedom in $X_1,\ldots,X_N, X_\infty$:
\begin{subequations}
\begin{align}
\label{Gaudin_1}
   & \partial_{t^{1, r}}X_s =  \frac{[X_r,X_s]}{\zeta_r-\zeta_s},\qquad s\neq r,\\[1ex]
   \label{Gaudin_2}
   & \partial_{t^{1, r}}X_r =  -\sum_{s\neq r}\frac{[X_r,X_s]}{\zeta_r-\zeta_s}- [X_r,X_\infty] ,\\[1ex]
\label{Gaudin_3}   & \partial_{t^{1, r}}X_\infty = 0.
\end{align}
\end{subequations}
We proceed to derive a Lagrangian one-form description of the set of equations \eqref{Gaudin_1}--\eqref{Gaudin_3}, as well as those corresponding to the next higher Hamiltonians with $k=2$. In principle, we could also include all higher Hamiltonians, but the first two levels are enough to illustrate our method. To do so, we need to be able to interpret $L(\lda)$ as living in a coadjoint orbit. The corresponding setup is described in \cite[Lecture 3]{STS} which we now review and adapt to our purposes.

\subsection{Algebraic setup}

Let $Q = \{\zeta_1, \ldots, \zeta_N, \infty\} \subset \mathbb{C}P^1$ be a finite set of points in $\mathbb{C}P^1$ including the point at infinity, and denote by $\mathcal{F}_Q(\g)$ the algebra of $\g$-valued rational functions in the formal variable $\lambda$ with poles in $Q$. Further, define the local parameters
\begin{equation}
  \lambda_r = \lambda - \zeta_r,\quad \zeta_r \neq \infty \qquad \text{and} \qquad  \lambda_\infty = \frac{1}{\lambda},
\end{equation}
and let $S=\{1,\ldots,N,\infty\}$. This is to be used as an index set, so $\infty$ is viewed here purely as a label for an index, not as the point at infinity. For each $r\in S$, consider the algebra $\Lg_r$ of formal Laurent series in variable $\lambda_r$ with coefficients in $\g$,
\begin{equation}
  \Lg_r = \g\otimes \mathbb{C}((\lambda_r)),
\end{equation}
with Lie bracket
\begin{equation}
    [X\lambda_r^i, Y\lambda_r^j] = [X, Y] \lambda_r^{i+j},\qquad X,Y \in \g.
\end{equation}
We have the vector space decomposition into Lie subalgebras
\begin{equation}
    \Lg_r = \Lg_{r+} \dotplus \Lg_{r-},
\end{equation}
where
\begin{equation}
    \Lg_{r+}=\g\otimes \mathbb{C}[[\lambda_r]],\quad r\neq \infty,\qquad \Lg_{\infty +}=\g\otimes \lda_\infty\mathbb{C}[[\lambda_\infty]],
\end{equation}
and 
\begin{equation}
    \Lg_{r-}=\g\otimes \lambda_r^{-1}\mathbb{C}[\lambda_r^{-1}],\quad r\neq \infty,\qquad \Lg_{\infty -}=\g\otimes \mathbb{C}[\lambda_\infty^{-1}].
\end{equation}
In other words, $\Lg_{r+}$ is the algebra of formal Taylor series in $\lda_r$ (without constant term when $r=\infty$) and $\Lg_{r-}$ is the algebra of polynomials in $\lda_r^{-1}$ without constant term (except when $r=\infty$). Associated with this decomposition, we have projectors $P_{r\pm}$ onto $\Lg_{r\pm}$ relative to $\Lg_{r\mp}$.
Let us now consider $\Lg_Q$ defined as the following direct sum of Lie algebras
\begin{equation}
  \Lg_Q = \bigoplus_{r\in S} \Lg_r.
\end{equation}
The above decompositions yield the decomposition of $\Lg_Q$ as
\begin{equation}
\label{decomp_Lg_Q} 
\Lg_{Q}=\Lg_{Q+} \dotplus \Lg_{Q-} \quad \text{with} \quad  \Lg_{Q+} = \bigoplus_{r\in S} \Lg_{r+}   \quad  \text{and}  \quad  \Lg_{Q-} = \bigoplus_{r\in S} \Lg_{r-},
\end{equation}
and the related projectors $P_\pm$. 
Although useful, as we will see below, the decomposition \eqref{decomp_Lg_Q} is not what we need to interpret \eqref{generating_Gaudin_eqs} within the Lie dialgebra setup. So, let us consider the map 
\begin{equation}\label{eq:embed-rat-loop}
  \iota_\lambda: \mathcal{F}_Q(\g) \to \Lg_Q, \quad  f \mapsto \left(\iota_{\lambda_1} f, \ldots, \iota_{\lambda_N} f, \iota_{\lambda_\infty} f \right),
\end{equation}
where $\iota_{\lambda_r} f \in \Lg_r$ is the formal Laurent series of $f \in \mathcal{F}_Q(\g)$ at $\zeta_r \in \mathbb{C}P^1$ and $\iota_{\lambda_\infty} f \in \Lg_r$ that of $f \in \mathcal{F}_Q(\g)$ at $\zeta_\infty$. This is an embedding of Lie algebras. In addition, we have the vector space decomposition 
\begin{equation}
\label{decomp2}
  \Lg_Q = \Lg_{Q+} \dotplus \iota_\lambda \mathcal{F}_Q(\g).
\end{equation}
This decomposition reflects the fact that there exists a unique rational function on the Riemann sphere with a prescribed set of singularities and principal parts at its poles, and conversely, any rational function on the Riemann sphere admits a unique decomposition into elementary functions.

Let us introduce the projectors $\Pi_\pm$ associated with this decomposition. They are different from $P_\pm$ related to \eqref{decomp_Lg_Q}. The following relation is useful in practical calculations (see below when computing gradients or in \eqref{formula_orbit_elem})
 \begin{equation}
    \Pi_-(X)=\iota_\lda \circ \pi_\lda \circ P_-(X),\qquad X\in\Lg_Q,
\end{equation}
where the map $\pi_\lda:\Lg_{Q-}\to   \mathcal{F}_Q(\g)$ given by
\begin{equation}
\pi_\lda(Y_1(\lda_1),\dots, Y_N(\lda_N),Y_\infty(\lda_\infty))  =\sum_{r\in S}Y_r(\lda_r)
\end{equation}
puts elements of $\Lg_{Q-}$ and $\mathcal{F}_Q(\g)$ in one-to-one correspondence.
This amounts to decomposing an $f\in \mathcal{F}_Q(\g)$ into the sum of its partial fractions $Y_r(\lda_r)$. 

We define the $r$-matrix we need as
\begin{equation}
    R=\Pi_+-\Pi_-
\end{equation}
and use it to define on $\Lg_Q$ the structure of a Lie dialgebra to which we will apply the results of that theory. Since we want to work with rational fractions which we have naturally embedded as  $\iota_\lambda \mathcal{F}_Q(\g)$ into $\Lg_Q$, we need to identify the dual space this corresponds to, so that we can identify the coadjoint action and its orbits appropriately. 
The nondegenerate invariant symmetric bilinear form on $\g$, given by $(X, Y) \mapsto \Tr(XY)$, can be used to define a nondegenerate invariant symmetric bilinear form on  $\Lg_Q$ by setting
\begin{equation}\label{eq:bf-lg-q}
  \langle X, Y \rangle = \sum_{r\in S} \mathrm{res}_{\lambda_r = 0} \Tr(X_r(\lambda_r) Y_r(\lambda_r)).
\end{equation}
Both $\Lg_{Q+}$ and $\iota_\lambda \mathcal{F}_Q(\g)$ are Lie subalgebras which are (maximally) isotropic with respect to the bilinear form $\langle ~~,~~ \rangle$ in \eqref{eq:bf-lg-q}. This tells us that 
\begin{equation}
\label{identification2}
  \Lg_{Q+}^{*} \simeq \iota_\lambda \mathcal{F}_Q(\g),
\end{equation}
so that elements of $\Lg_{Q+}^{*}$ are those we should work with if we want to deal with Lax matrices which are rational fractions of the spectral parameter. Accordingly, coadjoint orbits of $\LG_{Q+}$ in $\Lg_{Q+}^{*}$ are the natural arena for the description of Gaudin Lax matrices. Here $\LG_{Q+}$ is the group associated with the algebra $\Lg_{Q+}$, with elements of the form
\begin{equation}
    \varphi_+=\left(\varphi_{1+}(\lda_1),\dots,\varphi_{N+}(\lda_N),\varphi_{\infty+}(\lda_\infty) \right),
\end{equation}
where each component $\varphi_{r+}(\lda_r)$ is a Taylor series in the local parameter $\lda_r$ valued in $G$, the matrix Lie group associated with the Lie algebra $\g$, that is,
\begin{equation}
    \varphi_{r+}(\lda_r)=\sum_{n=0}^\infty \phi_r^{(n)}\lda_r^n,\quad r\neq \infty,\qquad \varphi_{\infty+}(\lda_\infty)=\1+\sum_{n=1}^\infty\phi_\infty^{(n)} \lda_\infty^n.
\end{equation}
As always, in practice we use the identification \eqref{identification2} (identifying the action and coadjoint actions accordingly) and the (co)adjoint orbit of an element $f\in \iota_\lambda \mathcal{F}_Q(\g)$ can be seen to be given by the elements
\begin{equation}
\label{formula_orbit_elem}
    F=\Pi_-(\text{Ad}_{\varphi_+}\cdot f)\equiv \iota_\lda L.
\end{equation}
In \eqref{formula_orbit_elem}, the adjoint action of $\varphi_+$ on $f$ is defined component-wise
\begin{equation}
    (\text{Ad}_{\varphi_+}\cdot f)_r(\lda_r)=\varphi_{r+}(\lda_r) f_r(\lda_r)\varphi_{r+}(\lda_r)^{-1},\qquad r\in S.
\end{equation}
Thus, we have a construction that allows us to interpret a rational Lax matrix $L(\lda)$ as an element of a (co)adjoint orbit and recast \eqref{generating_Gaudin_eqs} as the following Lax equation in $\iota_\lambda \mathcal{F}_Q(\g)$
\begin{equation}
\label{embedded_Lax}
    \partial_{t^{k, r}}\iota_\lda L=[R_\pm\nabla H_{k, r}(\iota_\lda L),\iota_\lda L],
\end{equation}
where $H_{k, r}$ are the following invariant functions on $\Lg_Q$
\begin{equation}
\label{Hamiltonians}
    H_{k, r}: X\in\Lg_Q\mapsto \Res_{\lda_r=0}\frac{\Tr(X_r(\lda_r)^{k+1})}{k+1},\qquad k\ge 1.
\end{equation}
Having set up the required framework, let us now show how \eqref{generating_Gaudin_eqs} is derived in this context.

\subsection{Lax description}

Let us choose 
\begin{equation}
\Lambda(\lda) =\sum_{r=1}^N\frac{\Lambda_r}{\lda-\zeta_r}+\Omega , 
\end{equation}
and apply \eqref{formula_orbit_elem} to $f=\iota_\lda \Lambda$
to get 
\begin{equation}\label{orbit_L}
\begin{split}
\iota_\lda L=\Pi_-\left(\text{Ad}_{\varphi_+}\cdot \iota_\lda \Lambda\right)&= \iota_\lda\circ \pi_\lda\circ P_-\left(\text{Ad}_{\varphi_+}\cdot \iota_\lda \Lambda\right)\\[1ex] 
&=   \iota_\lda\circ \pi_\lda\left(\frac{\phi_1^{(0)}\Lambda_1(\phi_1^{(0)})^{-1}}{\lda-\zeta_1},\dots, \frac{\phi_N^{(0)}\Lambda_N(\phi_N^{(0)})^{-1}}{\lda-\zeta_N},\Omega \right)  \\[.5ex] 
 &\equiv\iota_\lda\circ \pi_\lda \left(\frac{A_1}{\lda-\zeta_1},\dots, \frac{A_N}{\lda-\zeta_N},\Omega \right) \\
&=\iota_\lda \left(\sum_{r=1}^N\frac{A_r}{\lda-\zeta_r}+\Omega\right).
\end{split} 
\end{equation}
This is the desired form of \eqref{Lax_matrix_Gaudin} where now each $X_r$ is of the form $A_r=\phi_r^{(0)}\Lambda_r(\phi_r^{(0)})^{-1}$ with $\Lambda_r\in\g$ fixed and $\phi_r^{(0)}$ containing the dynamical degrees of freedom. This is the (co)adjoint description required to compute our Lagrangian coefficients, see below. 

Next, we derive the Lax equations in $\iota_\lambda \mathcal{F}_Q(\g)$ associated with the functions $H_{k, r}(\iota_\lda L)$ for $k=1, 2$. The gradient of $H_{k, r}$ at the point $\iota_\lda L$ is defined as the element of $\Lg_Q$ satisfying
\begin{equation}
\label{def_gradient}
    \lim_{\epsilon\to 0}\frac{H_{k, r}(\iota_\lda L+\epsilon \eta)-H_{k, r}(\iota_\lda L)}{\epsilon}=\langle \eta, \nabla H_{k, r}(\iota_\lda L)\rangle,
\end{equation}
for all $\eta \in \Lg_Q$. It is enough for our purposes to calculate $R_-(\nabla H_{k, r}(\iota_\lda L))$, therefore, we can restrict $\eta$ to $\Lg_{Q+}$. Thus, writing
\begin{equation}
    \nabla H_{k, r}(\iota_\lda L)=N^{(k)}+\iota_\lda h^{(k)},\qquad N^{(k)}\in\Lg_{Q+}, \quad h^{(k)}(\lda)\in \mathcal{F}_Q(\g),
\end{equation}
recalling that $\Lg_{Q+}$ and $\iota_\lda \mathcal{F}_Q(\g)$ are isotropic with respect to the bilinear form in \eqref{eq:bf-lg-q}, \eqref{def_gradient} becomes
\begin{equation}
    \Res_{\lda_r=0}\Tr\left( \eta_r \iota_{\lda_r} L^k \right)=\sum_{s \in S}\Res_{\lda_s=0}
    \Tr \left( \eta_s \iota_{\lda_s} h^{(k)} \right),
\end{equation}
for any $\eta_s \in \Lg_{s+}$, $s\in S$, implying 
\begin{subequations}
\begin{align}
    ( \iota_{\lda_s}h^{(k)})_- &= 0, \qquad \forall s\neq r,\\[1ex] 
    (\iota_{\lda_r}h^{(k)})_- &= (\iota_{\lda_r} L^k)_-.
\end{align}
\end{subequations}
This means that the rational function $h^{(k)}(\lda)$ has a non-zero principal only at $\zeta_r$ which equals $(\iota_{\lda_r} L^k)_-$, so
\begin{equation}
    h^{(k)}(\lda) = (\iota_{\lda_r} L^k)_-(\lda),
\end{equation}
and we find
\begin{equation}
    R_-(\nabla H_{k, r}(\iota_\lda L)) = -\Pi_-(\nabla H_{k, r}(\iota_\lda L))= -\iota_\lda h^{(k)} = -\iota_\lda \left( (\iota_{\lda_r} L^k)_- \right).
\end{equation}
    For $k=1, 2$, this gives us
\begin{equation}
    R_-(\nabla H_{1, r}(\iota_\lda L)) = -\iota_\lda \frac{A_r}{\lda-\zeta_r}
\end{equation}
and
\begin{equation}
    R_-(\nabla H_{2, r}(\iota_\lda L)) = -\iota_\lda \left( \frac{A_r^2}{(\lda-\zeta_r)^2} +\sum_{s \neq r} \frac{A_rA_s+A_sA_r}{(\lda - \zeta_r)(\zeta_r - \zeta_s)} + \frac{A_r\Omega +\Omega A_r}{\lda - \zeta_r} \right), 
\end{equation}
respectively.
As a consequence, we find the Lax equations for the two levels of flows as
\begin{equation}
    \partial_{t^{1, r}}\iota_\lda L=\left[-\iota_\lda \frac{A_r}{\lda-\zeta_r}, \iota_\lda L\right]
\end{equation}
and
\begin{equation}\label{eq:gaudin-Laxeqtwo}
    \partial_{t^{2, r}}\iota_\lda L=\left[-\iota_\lda \left(\frac{A_r^2}{(\lda-\zeta_r)^2} +\sum_{s \neq r} \frac{A_rA_s+A_sA_r}{(\lda - \zeta_r)(\zeta_r - \zeta_s)} + \frac{A_r\Omega +\Omega A_r}{\lda - \zeta_r} \right), \iota_\lda L\right].
\end{equation}
Explicitly, they yield the following equations on the $A_s$,
\begin{equation}
\label{flow1}
    \begin{split}
  &   \partial_{t^{1, r}}A_s=  \frac{[A_r, A_s]}{\zeta_r-\zeta_s}, \qquad s\neq r,\\
   & \partial_{t^{1, r}}A_r=  -\sum_{s\neq r}\frac{[A_r, A_s]}{\zeta_r-\zeta_s}- [A_r, \Omega],
        \end{split}
\end{equation}
thus reproducing \eqref{Gaudin_1}--\eqref{Gaudin_2} (\eqref{Gaudin_3} is automatic here since $\Omega$ is a constant element of $\g$), and 
\begin{equation}
\label{flow2}
    \begin{split}
        \partial_{t^{2, r}}A_s &= - \frac{[A_r^2, A_s]}{(\zeta_r-\zeta_s)^2} + \sum_{{s^\prime} \neq r} \frac{[A_r A_{s^\prime} + A_{s^\prime}A_r, A_s]}{(\zeta_r-\zeta_s)(\zeta_r-\zeta_{s^\prime})} + \frac{[A_r\Omega + \Omega A_r, A_s]}{\zeta_r-\zeta_s} , \qquad s \neq r,\\
        \partial_{t^{2, r}}A_r &= \sum_{s \neq r} \frac{[A_r^2, A_s]}{(\zeta_r-\zeta_s)^2} -\sum_{s\neq r}\sum_{{s^\prime}\neq r}\frac{[A_r, A_s A_{s^\prime}]}{(\zeta_r-\zeta_s)(\zeta_r-\zeta_{s^\prime})} - \sum_{s \neq r} \frac{[A_r,  A_s \Omega + \Omega A_s]}{\zeta_r-\zeta_s} - [A_r,  \Omega^2].
        \end{split}
\end{equation}

\subsection{Lagrangian description}

Applying our formula for the Lagrangian coefficients, we obtain the following one-form
on the orbit of $\Lambda(\lda)$, with elements $\iota_\lda L$ given in \eqref{orbit_L},
\begin{equation}
    \Lag = \Lag_{k, r}  dt^{k, r} , \qquad k = 1, \ldots, N,\quad r \in S,
\end{equation}
with
\begin{equation}
    \Lag_{k, r} = \sum_{s\in S} \Res_{\lambda_s = 0} \Tr\left(\iota_{\lda_s}L  \partial_{t^{k, r}}\varphi_{s+}(\lda_s) \varphi_{s+}(\lda_s)^{-1}\right)-H_{k, r}(\iota_\lda L),
\end{equation}
where $H_{k, r}(\iota_\lda L)$ is the restriction of $H_{k, r}$ to $\iota_{\lda}L$. For the kinetic part, we have
\begin{equation}
    \Res_{\lambda_s = 0} \Tr(\iota_{\lda_s}L  \partial_{t^{k, r}}\varphi_{s+}(\lda_s) \varphi_{s+}(\lda_s)^{-1})=\Tr\left(\Lambda_s (\phi_s^{(0)})^{-1}\partial_{t^{k, r}}\phi_s^{(0)}\right), \qquad s=1,\ldots,N,
\end{equation}
and
\begin{equation}
\begin{split}
    \Res_{\lambda_\infty = 0} \Tr(\iota_{\lda_\infty}L  \partial_{t^{k, r}}\varphi_{\infty+}(\lda_\infty) \varphi_{\infty+}(\lda_\infty)^{-1})&=\Tr\left(\Omega \partial_{t^{k, r}}\phi_\infty^{(1)} \phi_\infty^{(1)}\right)\\
    &=\frac{1}{2}\partial_{t^{k, r}}\Tr\left(\Omega (\phi_\infty^{(1)})^2\right).
\end{split}
\end{equation}
The contribution at $\infty$ is a total derivative, so it will not enter the Euler--Lagrange equations and hence we discard it. Thus, only the term $\phi_s^{(0)}$ in the Taylor series of $\varphi_{s+}(\lda_s)$ appears in the kinetic term. We will simply denote it by $\phi_s$ to lighten notations. 
The Lagrangian coefficients of the Gaudin one-form take the form
\begin{equation}
    \Lag_{k, r} = \sum_{s=1}^N\Tr\left(\Lambda_s \phi_s^{-1}\partial_{t^{k, r}}\phi_s\right) - H_{k, r}(\iota_\lda L).
\end{equation}
More explicitly, for $k=1,2$, we have
\begin{equation}
    H_{1, r}(\iota_{\lda}L) = \sum_{s\neq r}\frac{\Tr (A_rA_s)}{\zeta_r-\zeta_s}+\Tr(A_r\Omega),
\end{equation}
and 
\begin{equation}
      H_{2, r}(\iota_{\lda}L) = \Tr\left(A_r \left(\sum_{s\neq r}\frac{A_s}{\zeta_r-\zeta_s}  +\Omega\right)^2\right) - \Tr\left(A_r^2 \left(\sum_{s\neq r}\frac{A_s}{(\zeta_r-\zeta_s)^2}\right) \right).
\end{equation}
Varying $\Lag_{1, r}$ and $\Lag_{2, r}$ with respect to $\phi_s$, $s=1,\ldots,N$ (recalling that $A_s=\phi_s \Lambda_s \phi_s^{-1}$), one can check by direct calculations that the Euler--Lagrange equations give exactly \eqref{flow1}--\eqref{flow2}.

Further, we know from the general theory that the closure relation $d\Lag = 0$ holds on shell. This implies
\begin{equation}
\label{eq:gaudin-closure}
    \partial_{t^{j, s}} \Lag_{k, r} - \partial_{t^{k, r}} \Lag_{j, s} = 0,
\end{equation}
for all possible combinations of $j,k$ and $r,s$. As we know, the kinetic and potential contributions give zero separately in each case. Let us illustrate the main steps here for $k=1$, $j=2$ and $r\neq s$ in \eqref{eq:gaudin-closure}, the left-hand side of which will then read
\begin{equation*}
  \begin{split}
   &\sum_{s^\prime=1}^N \partial_{t^{2, s}} \Tr\left(\Lambda_{s^\prime} \phi_{s^\prime}^{-1}\partial_{t^{1, r}}\phi_{s^\prime}\right) - \sum_{s^\prime=1}^N \partial_{t^{1, r}} \Tr\left(\Lambda_{s^\prime} \phi_{s^\prime}^{-1}\partial_{t^{2, s}}\phi_{s^\prime}\right)\\
   &\qquad \qquad \qquad \qquad \qquad \qquad \qquad \qquad - \partial_{t^{2, s}} H_{1, r}(\iota_\lda L) + \partial_{t^{1, r}} H_{2, s}(\iota_\lda L).
  \end{split}
\end{equation*}
Using the equations of motion, we have
\begin{equation}
  \begin{split}
    &\partial_{t^{2, s}} H_{1, r}(\iota_{\lda}L)\\
    &= \sum_{s^\prime\neq r} \frac{1}{\zeta_r-\zeta_{s^\prime}} \Tr \Bigg(\Bigg(  - \frac{[A_s^2, A_r]}{(\zeta_s-\zeta_r)^2} + \sum_{{s^{\prime\prime}} \neq s} \frac{[A_s A_{s^{\prime\prime}} + A_{s^{\prime\prime}}A_s, A_r]}{(\zeta_s-\zeta_r)(\zeta_s-\zeta_{s^{\prime\prime}})} +  \frac{[A_s \Omega + \Omega A_s, A_r]}{\zeta_s-\zeta_r} \Bigg) A_{s^\prime} \Bigg)\\
    &\quad + \sum_{\substack{s^\prime \neq r\\s^\prime \neq s}} \frac{1}{\zeta_r-\zeta_{s^\prime}} \Tr \Bigg(A_r \Bigg(  - \frac{[A_s^2, A_{s^\prime}]}{(\zeta_s-\zeta_{s^\prime})^2} + \sum_{{s^{\prime\prime}} \neq s} \frac{[A_s A_{s^{\prime\prime}} + A_{s^{\prime\prime}}A_s, A_{s^\prime}]}{(\zeta_s-\zeta_{s^\prime})(\zeta_s-\zeta_{s^{\prime\prime}})}\\
    &\qquad \qquad \qquad \qquad \qquad \qquad \qquad \qquad \qquad \qquad \qquad \qquad \qquad \quad +  \frac{[A_s \Omega + \Omega A_s, A_{s^\prime}]}{\zeta_s-\zeta_{s^\prime}}  \Bigg) \Bigg)\\
    &\quad + \frac{1}{\zeta_r-\zeta_s} \Tr \Bigg(A_r \Bigg( \sum_{s^\prime \neq s} \frac{[A_s^2, A_{s^\prime}]}{(\zeta_s-\zeta_{s^\prime})^2} -\sum_{s^\prime \neq s}\sum_{{s^{\prime\prime}}\neq s}\frac{[A_s, A_{s^\prime} A_{s^{\prime\prime}}]}{(\zeta_s-\zeta_{s^\prime})(\zeta_s-\zeta_{s^{\prime\prime}})}\\
    &\qquad \qquad \qquad \qquad \qquad \qquad \qquad \qquad \qquad \qquad \quad - \sum_{s^\prime \neq s} \frac{[A_s,  A_{s^\prime} \Omega + \Omega A_{s^\prime}]}{\zeta_s-\zeta_{s^\prime}}  - [A_s,  \Omega^2] \Bigg) \Bigg)\\
    &\quad +\Tr \Bigg(\Bigg( - \frac{[A_s^2, A_r]}{(\zeta_s-\zeta_r)^2} +  \sum_{{s^{\prime}} \neq s} \frac{[A_s A_{s^{\prime}} +  A_{s^{\prime}}A_s, A_r]}{(\zeta_s-\zeta_r)(\zeta_s-\zeta_{s^{\prime}})}  + \frac{[A_s \Omega + \Omega A_s, A_r]}{\zeta_s-\zeta_r} \Bigg) \Omega \Bigg). 
    \end{split}
  \end{equation}
This is seen to add up to zero by assembling the terms of the same nature (quartic, cubic or quadratic in $A$), manipulating the sums, using the ${\rm ad}$-invariance property $\Tr([A,B]C)=\Tr(A[B,C])$ and the identity
\begin{equation}
\frac{1}{(\zeta_r-\zeta_s)(\zeta_r-\zeta_{s^\prime})}+\frac{1}{(\zeta_s-\zeta_{s^\prime})(\zeta_r-\zeta_{s^\prime})} +\frac{1}{(\zeta_s-\zeta_r)(\zeta_s-\zeta_{s^\prime})}=0 .
\end{equation}
Similar calculations give $\partial_{t^{1, r}} H_{2, s}(\iota_\lda L)=0$.
For the kinetic terms we have
\begin{equation}
  \begin{split}
    &\partial_{t^{2, s}} \sum_{s^\prime=1}^N \Tr\left(\Lambda_{s^\prime} \phi_{s^\prime}^{-1}\partial_{t^{1, r}}\phi_{s^\prime}\right) - \partial_{t^{1, r}} \sum_{s^\prime=1}^N \Tr\left(\Lambda_{s^\prime} \phi_{s^\prime}^{-1}\partial_{t^{2, s}}\phi_{s^\prime}\right)\\ 
    &= \sum_{s^\prime=1}^N \Tr \left((\partial_{t^{2, s}}A_{s^\prime}) (\partial_{t^{1, r}}\phi_{s^\prime}) \phi_{s^\prime}^{-1} \right) - \sum_{s^\prime=1}^N \Tr \left((\partial_{t^{1, r}}A_{s^\prime}) (\partial_{t^{2, s}}\phi_{s^\prime}) \phi_{s^\prime}^{-1} \right)\\
    & \quad + \sum_{s^\prime=1}^N \Tr \left( A_{s^\prime} [(\partial_{t^{2, s}}\phi_{s^\prime})\phi_{s^\prime}^{-1}, (\partial_{t^{1, r}}\phi_{s^\prime})\phi_{s^\prime}^{-1}] \right)\\
    & \quad + \sum_{s^\prime=1}^N \Tr \left( A_{s^\prime} \left((\partial_{t^{2, s}}\partial_{t^{1, r}}\phi_{s^\prime})\phi_{s^\prime}^{-1} - (\partial_{t^{1, r}}\partial_{t^{2, s}}\phi_{s^\prime})\phi_{s^\prime}^{-1}\right) \right).
  \end{split}
\end{equation}
The commutativity of flows ensures that the last term equals zero. Further, using the relation
\begin{equation}\label{eq:gaudin-flowrearrange}
    \partial_{t^{2, s}}A_{s^\prime} = [(\partial_{t^{2, s}} \phi_{s^\prime})\phi_{s^\prime}^{-1}, A_{s^\prime}], \qquad s^\prime = 1,\ldots,N,
\end{equation}
it is easy to see that the first and the third terms cancel each other. Finally, for the second term, using ${\rm ad}$-invariance, \eqref{eq:gaudin-flowrearrange} and the on-shell relations in \eqref{flow1} and \eqref{flow2}, we have
\begin{equation}
\begin{split}
    &\sum_{s^\prime=1}^N \Tr \left((\partial_{t^{1, r}}A_{s^\prime}) (\partial_{t^{2, s}}\phi_{s^\prime}) \phi_{s^\prime}^{-1} \right)\\
    &= \Tr \left((\partial_{t^{1, r}}A_r) (\partial_{t^{2, s}}\phi_r) \phi_r^{-1} \right) + \sum_{s^\prime \neq r} \Tr \left((\partial_{t^{1, r}}A_{s^\prime}) (\partial_{t^{2, s}}\phi_{s^\prime}) \phi_{s^\prime}^{-1} \right)\\
    &= -\sum_{s^\prime \neq r} 
    \Tr \left( \frac{[A_r, A_{s^\prime}]}{\zeta_r - \zeta_{s^\prime}} (\partial_{t^{2, s}}\phi_r) \phi_r^{-1} \right) - \Tr \left([A_r, \Omega] (\partial_{t^{2, s}}\phi_r) \phi_r^{-1} \right)
    \\    & \quad 
    + \sum_{s^\prime \neq r} \Tr \left(\frac{[A_r, A_{s^\prime}]}{\zeta_r - \zeta_{s^\prime}} (\partial_{t^{2, s}}\phi_{s^\prime}) \phi_{s^\prime}^{-1} \right)\\
    &= -\sum_{s^\prime \neq r} \frac{\Tr (A_{s^\prime} \partial_{t^{2, s}} A_r)}{\zeta_r - \zeta_{s^\prime}} - \Tr(\Omega \partial_{t^{2, s}}A_r) -\sum_{s^\prime \neq r} \frac{\Tr (A_r \partial_{t^{2, s}} A_{s^\prime})}{\zeta_r - \zeta_{s^\prime}}\\
    &=-\partial_{t^{2, s}} H_{1, r}(\iota_{\lda}L)
\end{split}
\end{equation}
which we previously showed to be zero.

\begin{remark}
    The algebraic framework we have used to obtain a Lagrangian one-form for the Gaudin model is to a very large extent similar to that used in \cite{CStV} to construct Lagrangian multiforms of Zakharov--Mikhailov type. Therefore, in hindsight, it is perhaps not very surprising that the Lagrangian 
    \begin{equation}
\label{Lag_1k}
    \Lag_{1, r} = \sum_{s=1}^N\Tr\left(\Lambda_s  \phi_s^{-1}\partial_{t^{1, r}}\phi_s\right) - \sum_{s\neq r}\frac{\Tr (A_rA_s)}{\zeta_r-\zeta_s} - \Tr(A_r\Omega),
\end{equation}
appears to be the direct analogue in the finite-dimensional case of the Zakharov--Mikhailov Lagrangians which describe integrable field theories with rational Lax matrices \cite{ZM}. It is a rather satisfying outcome that we have unravelled the unifying structure underlying such Lagrangians, whether in finite or infinite dimensions. They are all connected to Lie dialgebras which control the structure of their kinetic part and tell us which potentials to include. 
\end{remark}

\section{Cyclotomic Gaudin model}\label{sec:cycloGaudin}

The purpose of this section is to formulate a Lagrangian one-form for a cyclotomic generalisation of the rational Gaudin model presented in Section \ref{sec:noncycloGaudin}. The first immediate task is to introduce the relevant infinite-dimensional Lie algebra $\mathbb{g}$ and a suitable model $V$ for its dual space $\mathbb{g}^\ast$ by constructing a nondegenerate invariant bilinear pairing as in \eqref{pairing V g}. This will be achieved in Section \ref{sec:Lie dialgebra Gaudin}, which culminates in a description of the relevant coadjoint orbits $\mathcal O_\Lambda \subset \mathbb{g}^\ast$ in Lemma \ref{identification_orbits}. In Section \ref{sec:cgmlaxm} we specialise to the case of simple poles at $\omg^k \zeta_r$, $k \in \{0, \ldots, T-1\}$, $r \in \{1, \ldots, N\}$, and double poles at the origin and at infinity, and then describe the Lax matrix \eqref{eq:cglax} of the cyclotomic Gaudin model as an element of this coadjoint orbit. In Section \ref{sec:cgmlaxeq} we then describe the Lax equations as flows on this coadjoint orbit. Finally, in Section \ref{sec:cgmultiform} we put everything together to obtain the Lagrangian one-form for the cyclotomic Gaudin model.

The cyclotomic Gaudin model of interest here arises as a specialisation of a general procedure which can be traced back to reduction group ideas \cite{M}, first applied in the form of an averaging procedure \cite{RF} to the rational classical $r$-matrix \eqref{rational_r_mat} in order to produce the trigonometric and elliptic $r$-matrices. This was generalised in various ways, for instance, in \cite{AT,Av1,Av2} in the context of Sklyanin's (linear) Poisson algebra  
\begin{equation}
\label{Sklyanin_algebra}
    \{U_{\ti 1}(\lda), U_{\ti 2}(\mu)\}=\left[ r_{\ti{12}}^0(\lda, \mu), U_{\ti 1}(\lda)+U_{\ti 2}(\mu)\right],
\end{equation}
where, as above, the index denotes which factor in the tensor product $\g \otimes \g$ the $\g$-valued function $U$ sits in, that is, $U_{\ti 1}(\lda) \equiv U(\lda) \otimes \mathbb{1}$ and $U_{\ti 2}(\mu) \equiv \mathbb{1} \otimes U(\mu)$.

The idea is to use a Lie algebra automorphism appropriately extended to a loop algebra automorphism. For our purpose, we use an automorphism $\sigma$ of order $T$ on $\g$ and define
 \begin{equation}
     \phi: U(\lda)\mapsto \omega \sigma^{-1}(U(\omega \lda))
 \end{equation}
 where $\omega$ is a $T$th root of unity. Thus, $\phi$ is an automorphism of the Poisson algebra \eqref{Sklyanin_algebra}. This leads us to consider the fixed point subalgebra generated by
 \begin{equation}
 \label{def_averaged_L}
     L(\lda)=\frac{1}{T}  \sum_{k=0}^{T-1} \omega^{-k}\sgm^k U(\omega^{-k}\lda), \quad \sigma(L(\lda))=\omega L(\omega \lda).
 \end{equation}
The Poisson algebra for $L(\lda)$ closes into
 \begin{equation}
 \label{PB_L}
    \{L_{\ti 1}(\lda), L_{\ti 2}(\mu)\}=\left[ r_{\ti{12}}(\lda, \mu), L_{\ti 1}(\lda)\right]-\left[ r_{\ti{21}}(\mu,\lda), L_{\ti 2}(\mu)\right]
\end{equation}
where
\begin{equation}
\label{eq:nss-rmat}
  r_{\ti{12}}(\lda, \mu) = \frac{1}{T} \sum_{k=0}^{T-1} \frac{\sgm_{\ti{1}}^k C_{\ti{12}}}{\mu - \omg^{-k}\lda}
\end{equation}
is a non-skew-symmetric solution of CYBE. These facts are special cases of Propositions 2.2 and 4.1 in \cite{CC}. It is well-known \cite{BV} that the Poisson algebra \eqref{PB_L} ensures that the quantities ${\Tr} L(\lda)^p$ Poisson commute and the equations of motion generated by them with respect to \eqref{PB_L} take the Lax form. More precisely, the generating Hamiltonian 
\begin{equation}
    H(\mu)=\frac{1}{p+1} \Tr L(\mu)^{p+1}, \qquad p\in \ZZ_{\ge 0},
\end{equation}
satisfies 
\begin{equation}
    \{L(\lda), H(\mu)\} = \left[M(\lda,\mu), L(\lda) \right],
\end{equation}
where, by definition $\{L(\lda), H(\mu)\}=\{L^a(\lda), H(\mu)\}I_a$, and
\begin{equation}
    M_{\ti 1}(\lda,\mu)= \Tr_{\ti 2}r_{\ti {12}}(\lda,\mu)L_{\ti 2}(\mu)^p.
\end{equation}
This yields an infinite number of Hamiltonians by specifying $p$ and extracting coefficients at the poles of $L(\mu)$. Of course, only a finite number of these Hamiltonians are independent when acting on a finite-dimensional phase space. They generate mutually compatible flows forming an integrable hierarchy.

If one applies this construction to a Lax matrix of rational Gaudin model type, with poles in the finite set $D = \{0, \zeta_1, \ldots, \zeta_N, \infty\} \subset \mathbb{C}P^1$,
\begin{equation}
\label{Lax_matrix_Gaudin}
    U(\lda)=\sum_{n=0}^{N_0 - 1} \frac{U_0^{(n)}}{\lda^{n+1}} + \sum_{r=1}^{N}  \sum_{n=0}^{N_{r}-1} \frac{U_{r}^{(n)}}{(\lda - \zeta_r)^{n+1}} + \sum_{n=0}^{N_\infty} U_\infty^{(n)} \lda^n, \qquad U_{s}^{(n)}\in \g,
\end{equation}
then one obtains the so-called cyclotomic Gaudin model \cite{S, VY1, VY2} and its associated integrable hierarchy. Note that in \eqref{Lax_matrix_Gaudin}, we have anticipated that the pole at $0$ and at infinity behave differently in \eqref{def_averaged_L} compared to the poles at $\lda=\zeta_r\neq 0,\infty$. Although we will present the general algebraic setup, in our examples we will only consider the case where $U(\lda)$ has simple poles at all $\zeta_r$, $r \in \{1, \ldots, N\}$, and double poles at the origin and at 
infinity for simplicity.\footnote{This will be sufficient for our application to the periodic Toda chain and the discrete self–trapping (DST) model that we present in Section \ref{sec:realisations}.} The corresponding Lax matrix of the cyclotomic Gaudin model then takes the form
\begin{equation}\label{eq:cglax}
  L(\lda) = \frac{X_{0}^{(0)}}{\lda} + \frac{X_{0}^{(1)}}{\lda^2} + \frac{1}{T} \sum_{r=1}^N \sum_{k=0}^{T - 1} \frac{\sgm^{k} X_{r}}{\lda - \omg^k \zeta_r} + X_{\infty}
\end{equation}
with $\sgm X_{0}^{(j)}=\omega^{-j} X_{0}^{(j)}$ and $\sgm X_{\infty}=\omega X_{\infty}$.

\subsection{Algebraic setup}\label{sec:Lie dialgebra Gaudin}

Let us fix $T \in \mathbb{Z}_{\geq 1}$ and pick a primitive $T$th root of unity $\omg \in \mathbb{C}^{\times}$. We can then define
\begin{equation}
  \Gamma \coloneqq \{ 1, \omg, \omg^2, \ldots, \omg^{T-1} \},
\end{equation}
a copy of the cyclic group $\mathbb{Z}_T$ of order $T$ that acts on $\mathbb{C}$ by multiplication. Further, we introduce the finite set of points $D = \{0, \zeta_1, \ldots, \zeta_N, \infty\} \subset \mathbb{C}P^1$ including the point at infinity such that the $\Gamma$-orbits of the points $\zeta_1, \ldots, \zeta_N$ are pairwise disjoint, that is,
\begin{equation}
  \Gamma \zeta_r \cap \Gamma \zeta_s = \emptyset, \qquad \text{for all} \,\, 1 \leq r \neq s \leq N.
\end{equation}
Note that unlike the non-zero finite points $\zeta_1, \ldots, \zeta_N$, the origin and the point at infinity are fixed under the action of $\Gamma$.

Let $\g$ be a finite-dimensional Lie algebra over $\mathbb{C}$ with an automorphism $\sgm$ of order $T$. For simplicity, we will only consider matrix Lie algebras in this work, with the Lie bracket being the commutator, and a nondegenerate invariant bilinear pairing given by the trace. We also assume that the automorphism $\sgm$ preserves the trace, that is, $\Tr(\sgm (x) \sgm (y)) = \Tr(xy)$ for any $x, y \in \g$. The eigenspaces of $\sgm$,
\begin{equation}\label{eq:sgmeigenspace}
  \g^{(k)} = \{X \in \g : \sgm(X) = \omg^k X \}, \qquad k \in \{0, \ldots, T-1\},
\end{equation}
form a $\mathbb{Z}_T$-gradation of $\g$, namely 
\begin{equation}
\label{gradation}
  \g =  \g^{(0)} \dotplus \ldots \dotplus \g^{(T-1)}, \qquad \text{with} \quad [\g^{(k_1)}, \g^{(k_2)}] = \g^{(k_1 + k_2\, \text{mod}\, T)}.
\end{equation}
From now on, we will simply write $\g^{(n)}$ to mean $\g^{(n\, \text{mod}\, T)}$. Further, let us define the local parameters
\begin{equation} \label{lambda r def}
  \lda_0 = \lda, \qquad  \lda_r = \lda - \zeta_r \quad \text{for}\,\, r \in \{1, \ldots, N\}, \qquad \lda_\infty = \dfrac{1}{\lda}.
\end{equation}
It will be convenient to also introduce an additional set of local parameters $\lda_{r, k} = \lda - \omg^k \zeta_r$ for all $k \in \{0, \ldots, T-1\}$ and $r \in \{1, \ldots, N\}$. We also introduce an index set $S = \{0, 1, \ldots, N, \infty\}$ where $0$ and $\infty$ denote labels for indices rather than points in $\mathbb{C}P^1$.

Let us denote by $\mathcal{F}_{D^{\prime}}$ the algebra of rational functions in the formal variable $\lda$ with values in $\g$ that are regular outside $D^{\prime} = \big\{ \omg^k \zeta_r : k \in \{0, \ldots, T-1\}, r \in \{1, \ldots, N\}\big\} \cup \{0, \infty\}$. Two subspaces of this algebra will be of relevance here: the subspace $\mathcal{F}_{D^\prime}^\Gamma$ of \emph{equivariant functions} and the subspace $\Omega_{D^\prime}^\Gamma$ of \emph{equivariant one-forms} that we define as
\begin{equation}
  \begin{split}
    &\mathcal{F}_{D^\prime}^{\Gamma} \coloneqq \{ f \in \mathcal{F}_{D^\prime} : \sgm (f (\lda)) = f (\omg \lda) \},\\
    &\Omega_{D^\prime}^{\Gamma} \coloneqq \{ g \in \mathcal{F}_{D^\prime} : \sgm (g (\lda)) = \omg g (\omg \lda) \}.
  \end{split}
\end{equation}
In general, equivariant functions and equivariant one-forms take the forms
\begin{equation}
  \begin{split}
    &f = \sum_{n=0}^{M_0 - 1} \frac{X_0^{(n)}}{\lda^{n+1}} + \sum_{r=1}^{N} \sum_{k=0}^{T-1} \sum_{n=0}^{M_r - 1} \frac{\sgm^k X_r^{(n)}}{(\omg^{-k}\lda - \zeta_r)^{n+1}} + \sum_{n=0}^{M_\infty} \frac{X_\infty^{(n)}}{\lda_\infty^n},\\
    &g = \sum_{n=0}^{N_0 - 1} \frac{Y_0^{(n)}}{\lda^{n+1}} + \sum_{r=1}^N \sum_{k=0}^{T-1} \sum_{n=0}^{N_r - 1} \frac{\omg^{-k} \sgm^k Y_r^{(n)}}{(\omg^{-k} \lda - \zeta_r)^{n+1}} + \sum_{n=0}^{N_\infty} \frac{Y_\infty^{(n)}} {\lda_\infty^n},
  \end{split}
\end{equation}
respectively, with
\begin{equation}\label{eq:coeffcond}
  \begin{split}
    &X_0^{(n)} \in \g^{(-n-1)},\qquad X_r^{(n)} \in \g \quad \text{for}\,\, r \in \{1, \ldots, N\},\qquad X_\infty^{(n)} \in \g^{(n)},\\
    &Y_0^{(n)} \in \g^{(-n)},\qquad Y_r^{(n)} \in \g \quad \text{for}\,\, r \in \{1, \ldots, N\}, \qquad Y_\infty^{(n)} \in \g^{(n+1)}.
  \end{split}
\end{equation}

One can check that the subspace $\mathcal{F}_{D^\prime}^\Gamma$ is, in fact, a Lie subalgebra of $\mathcal{F}_{D^\prime}$. The subspace $\Omega_{D^\prime}^\Gamma$, on the other hand, contains the Lax matrix \eqref{eq:cglax} of the model. To construct the cyclotomic Gaudin one-form \`{a} la \cite{CDS}, we will realise this Lax matrix as an element of a certain coadjoint orbit. To do so, let us define the loop algebra of formal Laurent series in the local parameter $\lda_r$ with coefficients in $\g$, for each $r \in S$,
\begin{equation} \label{Lgr def}
  \Lg_r = \g \otimes \mathbb{C}((\lda_r)),
\end{equation}
with Lie bracket
\begin{equation} \label{Lie bracket Lg}
  [X\lda_r^i, Y\lda_r^j] = [X, Y]\lda_r^{i+j}, \qquad X, Y \in \g,
\end{equation}
and then consider the direct sum
\begin{equation}
  \Lg_D = \bigoplus_{r \in S} \Lg_r.
\end{equation}

The Lie bracket of two elements $X, Y \in \Lg_D$ is defined component-wise
\begin{equation}
   [X,Y]_r(\lda_r)= [X_r(\lda_r), Y_r(\lda_r)].
\end{equation}
Of interest to us here are two particular subspaces of $\Lg_D$,
\begin{equation}
\label{def_g_V}
  \Lg_{D}^{(0)} = \Lg_0^{\Gamma, 0} \oplus \bigoplus_{r=1}^N \Lg_r \oplus \Lg_\infty^{\Gamma, 0} \quad \text{and} \quad \Lg_{D}^{(1)} = \Lg_0^{\Gamma, 1} \oplus \bigoplus_{r=1}^N \Lg_r \oplus \Lg_\infty^{\Gamma, 1},
\end{equation}
where the twisted spaces attached to the origin and infinity (recalling that $\lda_0 = \lda$ and $\lda_\infty = 1/\lda$) are defined as
\begin{equation} \label{Lg Gamma spaces}
  \begin{split}
    &\Lg_0^{\Gamma, k} = \{X_0(\lda_0) \in \Lg_0 : \sgm (X_0(\lda_0)) = \omg^k X_0(\omg \lda_0)\},\\
    &\Lg_\infty^{\Gamma, k} = \{X_\infty(\lda_\infty) \in \Lg_\infty : \sgm (X_\infty(\lda_\infty)) = \omg^k X_\infty(\omg^{-1} \lda_\infty)\},
  \end{split}  
\end{equation}
for $k = 0, 1$. Elements of $\Lg_{D}^{(0)}$ and $\Lg_{D}^{(1)}$ are tuples of the form
\begin{equation} \label{X g0D explicit}
  X = \left(\sum_{n=-\infty}^{M_0 - 1} \frac{X_0^{(n)}}{\lda^{n+1}}, \sum_{n=-\infty}^{M_1 - 1} \frac{X_1^{(n)}}{(\lda - \zeta_1)^{n+1}}, \ldots, \sum_{n=-\infty}^{M_N - 1} \frac{X_N^{(n)}}{(\lda - \zeta_N)^{n+1}}, \sum_{n=-\infty}^{M_\infty} \frac{X_\infty^{(n)}}{\lda_\infty^n} \right),
\end{equation}
and
\begin{equation} \label{Y g1D explicit}
    Y = \left(\sum_{n=-\infty}^{N_0 - 1} \frac{Y_0^{(n)}}{\lda^{n+1}}, \sum_{n=-\infty}^{N_1 - 1} \frac{Y_1^{(n)}}{(\lda - \zeta_1)^{n+1}}, \ldots, \sum_{n=-\infty}^{N_N - 1} \frac{Y_N^{(n)}}{(\lda - \zeta_N)^{n+1}}, \sum_{n=-\infty}^{N_\infty} \frac{Y_\infty^{(n)}}{\lda_\infty^n} \right),
  \end{equation}
  respectively, where the coefficients $X_r^{(n)}$ and $Y_r^{(n)}$, $r \in S$, satisfy the conditions in \eqref{eq:coeffcond}. Note that each entry in \eqref{X g0D explicit} and \eqref{Y g1D explicit} is a $\g$-valued Laurent series, i.e., an element of \eqref{Lgr def} for $r \in S$. This is clear from changing variables in each sum from $n$ to $-n$. The reason for the unusual choice of range for the indices $n$ in each of the above sums is to make the expressions for the maps in \eqref{pi01 def} defined below slightly more transparent.
  
  The subspace $\Lg_{D}^{(0)}$ is a Lie subalgebra of $\Lg_D$ which defines for us the desired infinite-dimensional Lie algebra $\mathbb{g} = \Lg_D^{(0)}$ in the notation of Section \ref{sec:AKS}. Continuing the identification of the different ingredients from the Lie dialgebra framework listed in Section \ref{sec:AKS}, notice that $V = \Lg_D^{(1)}$ is clearly a representation of the Lie algebra $\mathbb{g} = \Lg_D^{(0)}$ since we have $[\Lg_r^{\Gamma, 0}, \Lg_r^{\Gamma, 1}] \subset \Lg_r^{\Gamma, 1}$ for $r \in \{ 0, \infty \}$. In the notation of Section \ref{sec:AKS}, the representation $\ad^\ast : \mathbb{g} \times V \to V$ is given explicitly by $(X, Y) \mapsto [X, Y]$. The next proposition identifies $V = \Lg_D^{(1)}$ as a suitable model for the dual space $\mathbb{g}^\ast$.
  
\begin{proposition}\label{prop:pairing}
  The bilinear pairing $\langle \cdot, \cdot \rangle : \Lg_D^{(1)} \times \Lg_D^{(0)} \to \CC$ defined by 
  \begin{equation}\label{eq:bmgd}
    \langle Y, X \rangle = T \sum_{r=1}^{N} \Res_{\lda_r =0} \Tr(Y_r(\lda_r) X_r(\lda_r)) d\lda + \sum_{r \in \{0, \infty\}} \Res_{\lda_r =0} \Tr(Y_r(\lda_r) X_r(\lda_r))d\lda ,
\end{equation}
for any $Y \in \Lg_{D}^{(1)}$ and $X \in \Lg_{D}^{(0)}$, is nondegenerate and invariant under the adjoint action of $\Lg_D^{(0)}$.
\end{proposition}

\begin{proof}
Recall that the trace $\Tr : \g \times \g \to \CC$, $(x,y) \mapsto \Tr(xy)$ is a nondegenerate invariant bilinear pairing on $\g$ and invariant under the action of $\sgm$. Given any $x \in \g^{(m)}$ and $y \in \g^{(n)}$, we have $\Tr(xy) = \Tr(\sgm (x) \sgm (y)) = \omg^{m+n} \Tr(xy)$. Therefore, $\Tr(xy) = 0$ for all $x \in \g^{(m)}$ and $y \in \g^{(n)}$ if $m+n \neq 0\,\, \text{mod}\,\, T$. Now, since $\Tr : \g \times \g \to \CC$ is nondegenerate on $\g$, given $y \in \g^{(n)}$, there is $x\in \g$ such that $\Tr(xy)\neq 0$. From \eqref{gradation}, we have $x=x^{(0)}+\dots+x^{(T-1)}$, and from the previous result $\Tr(xy)=\Tr(x^{(m)}y)$ with $m+n= 0\,\, \text{mod}\,\, T$. Hence, $\Tr(x^{(m)}y)\neq 0$ and we get a nondegenerate pairing between the subspaces $\g^{(m)}$ and $\g^{(n)}$ with $m+n = 0\,\, \text{mod}\,\, T$.

Now given any non-zero element $Y \in \Lg_{D}^{(1)}$, it has a non-zero component $Y_r^{(m)} \lambda_r^{-m-1}$ for some $m \in \ZZ$ and $r \in S$. If $r=0$ then $Y_0^{(m)} \in \g^{(-m)}$ by \eqref{eq:coeffcond}, and we can find an $X_0^{(-m-1)} \in \g^{(m)}$ which pairs non-trivially with it under the trace. So, $X_0^{(-m-1)} \lambda^m$ pairs non-trivially with $Y$. Likewise, if $r=\infty$ then $Y_\infty^{(m)} \in \g^{(m+1)}$ by \eqref{eq:coeffcond} and we can find an $X_\infty^{(-m-1)} \in \g^{(-m-1)}$ which pairs non-trivially with it under the trace so that $X_\infty^{(-m-1)} \lambda_\infty^{m+1}$ pairs non-trivially with $Y$. And if $r \in \{1,\ldots, N\}$ then pick any $X_r^{(-m-1)} \in \g$ which pairs non-trivially with $Y_r^{(m)} \in \g$ under the trace so that $X_r^{(-m-1)} \lambda_r^m$ pairs non-trivially with $Y$. Therefore, the bilinear pairing $\langle \cdot, \cdot \rangle$ is nondegenerate in the first argument. A similar argument establishes the nondegeneracy in the second argument.

Finally, the invariance of the bilinear pairing $\langle \cdot, \cdot \rangle$ under the adjoint action of $\Lg_D^{(0)}$ follows from the definition \eqref{Lie bracket Lg} of the Lie bracket in $\Lg_D$ and the invariance of the trace under the adjoint action of $\g$.
\end{proof}

Next, we turn to the identification of the subalgebras $\mathbb{g}_\pm \subset \mathbb{g}$ and the corresponding subspaces $V_\pm \subset V$ in the notation of Section \ref{sec:AKS}.
Crucially for us, the spaces $\mathcal{F}_{D^\prime}^\Gamma$ and $\Omega_{D^\prime}^\Gamma$ embed into $\Lg_{D}^{(0)}$ and $\Lg_{D}^{(1)}$ respectively,
\begin{equation} \label{iota map def}
  \begin{split}
    &\iota_\lda \colon \mathcal{F}_{D^\prime}^\Gamma \longhookrightarrow \Lg_D^{(0)}, \quad f \longmapsto (\iota_{\lda_0}f, \iota_{\lda_1}f, \ldots, \iota_{\lda_N}f, \iota_{\lda_\infty}f),\\
    &\iota_\lda \colon \Omega_{D^\prime}^\Gamma \longhookrightarrow \Lg_D^{(1)}, \quad g \longmapsto (\iota_{\lda_0}g, \iota_{\lda_1}g, \ldots, \iota_{\lda_N}g, \iota_{\lda_\infty}g),
  \end{split}
\end{equation}
where, for each $r \in S$, $\iota_{\lda_r}f$ and $\iota_{\lda_r}g$ respectively denote the formal Laurent expansion of $f$ and $g$ about the point $\zeta_r$.
It will also be useful to define left inverses for these embeddings:
\begin{equation} \label{pi01 def}
  \begin{split}
    &\pi_\lda^{(0)} \colon \Lg_D^{(0)}  \longrightarrow \mathcal{F}_{D^\prime}^\Gamma,\\
    &\pi_\lda^{(1)} \colon \Lg_D^{(1)}  \longrightarrow \Omega_{D^\prime}^\Gamma,
  \end{split}
\end{equation}
defined explicitly as
\begin{equation}\label{eq:maptoequivfunc}
  \begin{split}
  &\pi_\lda^{(0)} \left(\sum_{n=-\infty}^{M_0 - 1} \frac{X_0^{(n)}}{\lda^{n+1}}, \sum_{n=-\infty}^{M_1 - 1} \frac{X_1^{(n)}}{(\lda - \zeta_1)^{n+1}}, \ldots, \sum_{n=-\infty}^{M_N - 1} \frac{X_N^{(n)}}{(\lda - \zeta_N)^{n+1}}, \sum_{n=-\infty}^{M_\infty} \frac{X_\infty^{(n)}}{\lda_\infty^n} \right)\\
  & \qquad \qquad \qquad \qquad \qquad = \sum_{n=0}^{M_0 - 1} \frac{X_0^{(n)}}{\lda^{n+1}} + \sum_{r=1}^N \sum_{k=0}^{T-1} \sum_{n = 0}^{M_r - 1} \frac{\sgm^k X_r^{(n)}}{(\omg^{-k}\lda - \zeta_r)^{n+1}} + \sum_{n=0}^{M_\infty} \frac{X_\infty^{(n)}}{\lda_\infty^n},
\end{split}
\end{equation}
and
\begin{equation}
  \begin{split}
    &\pi_\lda^{(1)} \left(\sum_{n=-\infty}^{N_0 - 1} \frac{Y_0^{(n)}}{\lda^{n+1}}, \sum_{n=-\infty}^{N_1 - 1} \frac{Y_1^{(n)}}{(\lda - \zeta_1)^{n+1}}, \ldots, \sum_{n=-\infty}^{N_N - 1} \frac{Y_N^{(n)}}{(\lda - \zeta_N)^{n+1}}, \sum_{n=-\infty}^{N_\infty} \frac{Y_\infty^{(n)}}{\lda_\infty^n} \right) \\
    & \qquad \qquad \qquad \qquad \qquad = \sum_{n=0}^{N_0 - 1} \frac{Y_0^{(n)}}{\lda^{n+1}} + \sum_{r=1}^N \sum_{k=0}^{T-1} \sum_{n=0}^{N_r - 1} \frac{\omg^{-k} \sgm^k Y_r^{(n)}}{(\omg^{-k} \lda - \zeta_r)^{n+1}} + \sum_{n=0}^{N_\infty} \frac{Y_\infty^{(n)}}{\lda_\infty^n}.
\end{split}
\end{equation}
We will also need to define the following subspaces of $\Lg_D^{(0)}$ and $\Lg_D^{(1)}$ respectively:
\begin{equation}
    \Lg_{D+}^{(0)} = \Lg_{0+}^{\Gamma, 0} \oplus \bigoplus_{r=1}^N \Lg_{r+} \oplus \Lg_{\infty+}^{\Gamma, 1} \quad \text{and} \quad \Lg_{D+}^{(1)} = \Lg_{0+}^{\Gamma, 1} \oplus \bigoplus_{r=1}^N \Lg_{r+} \oplus \Lg_{\infty+}^{\Gamma, 1}
\end{equation}
where, for $k = 0, 1$, we introduced (cf. \eqref{Lgr def} and \eqref{Lg Gamma spaces})
\begin{equation}
\begin{split}
  &\Lg_{0+}^{\Gamma, k} = \Lg_{0}^{\Gamma, k} \cap \g \otimes \mathbb{C}[[\lda_0]],\\
  &\Lg_{r+} = \g \otimes \mathbb{C}[[\lda_r]], \qquad r \in \{1, \ldots, N \},\\
  &\Lg_{\infty+}^{\Gamma, k} = \Lg_{\infty}^{\Gamma, k} \cap \g \otimes \lda_\infty \mathbb{C}[[\lda_\infty]].
\end{split}
\end{equation}
Here we denoted by $\g \otimes \mathbb{C}[[\lda_r]]$ the algebra of formal Taylor series in $\lda_r$ with coefficients in $\g$, for $r \neq \infty$, and by $\g \otimes \lda_\infty \mathbb{C}[[\lda_\infty]]$ the algebra of formal Taylor series in $\lda_\infty$ with coefficients in $\g$ without constant term.

Coming back to the identification of the Lie dialgebra ingredients from Section \ref{sec:AKS}, the next proposition identifies the desired decomposition of $\mathbb{g} = \Lg_D^{(0)}$ into complementary subalgebras $\mathbb{g}_\pm \subset \mathbb{g}$. Explicitly, we have the identifications $\mathbb{g}_+ = \Lg_{D+}^{(0)}$ and $\mathbb{g}_- = \iota_\lda \mathcal{F}_{D^\prime}^\Gamma$. We also identify a decomposition of our model $V = \Lg_D^{(1)}$ for the dual space $\mathbb{g}^\ast$ into complementary subspaces $V_\pm \subset V$, where explicitly $V_+ = \Lg_{D+}^{(1)}$ and $V_- = \iota_\lda \Omega_{D^\prime}^\Gamma$, but will show only later in Proposition \ref{prop:pairequivsp} that this decomposition of $V$ is the desired one induced by that of $\mathbb{g}$.

\begin{proposition}\label{prop:gddecomp}
  The spaces $\Lg_D^{(0)}$ and $\Lg_D^{(1)}$ admit the vector space decompositions
  \begin{subequations}
    \begin{align}
      \label{eq:gdzerodecomp}
      &\Lg_D^{(0)} = \Lg_{D+}^{(0)} \dotplus \iota_\lda \mathcal{F}_{D^\prime}^\Gamma,\\
      \label{eq:gdonedecomp}
      &\Lg_D^{(1)} = \Lg_{D+}^{(1)} \dotplus \iota_\lda \Omega_{D^\prime}^\Gamma,
    \end{align}    
\end{subequations}      
  respectively. Moreover, the subspaces $\Lg_{D+}^{(0)}$ and $\iota_\lda \mathcal{F}_{D^\prime}^\Gamma$ are Lie subalgebras of $\Lg_D^{(0)}$.
\end{proposition}
\begin{proof}
  To any $X = (X_0, X_1, \ldots, X_N, X_\infty) \in \Lg_D^{(0)}$, we associate an equivariant function $f = \pi_\lda^{(0)}X$. Notice that for all $r \in S$, $X_r - \iota_{\lda_r} f$ is a formal Taylor series in $\lda_r$. It follows that $X$ splits uniquely as the direct sum of the tuples $\left(X_0 - \iota_{\lda_0} f, X_1 - \iota_{\lda_1} f, \ldots, X_N - \iota_{\lda_N}f, \right.$ $\left. X_\infty - \iota_{\lda_\infty} f\right) \in \Lg_{D+}^{(0)}$ and $\left(\iota_{\lda_0} f, \iota_{\lda_1} f, \ldots, \iota_{\lda_N}f, \iota_{\lda_\infty} f\right) \in \iota_\lda \mathcal{F}_{D^\prime}^\Gamma$, as required.  

  One can repeat the above steps (with the map $\pi_\lda^{(1)}$ acting on some $Y \in \Lg_D^{(1)}$) to prove that \eqref{eq:gdonedecomp} defines a vector space decomposition as well.
\end{proof}

Let $P_+$ and $P_-$ denote the projectors onto the subspaces $\Lg_{D+}^{(0)}$ and $\iota_\lda \mathcal{F}_{D^\prime}^\Gamma$ respectively, relative to the decomposition in \eqref{eq:gdzerodecomp}, and $\P_+$ and $\P_-$ the projectors onto $\Lg_{D+}^{(1)}$ and $\iota_\lda \Omega_{D^\prime}^\Gamma$ respectively, relative to the decomposition in \eqref{eq:gdonedecomp}, in line with the notation from Section \ref{sec:AKS}. As recalled in Section \ref{sec:AKS}, the linear map
\begin{equation} \label{R def}
  R = P_+ - P_-,
\end{equation}
is a solution of the mCYBE on $\Lg_{D}^{(0)}$. It is also useful to define the maps
\begin{equation}
    R_\pm = \frac{1}{2}(R \pm \Id)
\end{equation}
which can be related to the projectors onto $\Lg_{D+}^{(0)}$ and $\iota_\lda \mathcal{F}_{D^\prime}^\Gamma$ relative to direct sum decomposition \eqref{eq:gdzerodecomp} as
\begin{equation}
    R_\pm = \pm P_\pm. 
\end{equation}
The linear map $R$ is related to the non-skew-symmetric $r$-matrix \eqref{eq:nss-rmat} underlying the cyclotomic Gaudin model, as one would naturally expect from the construction. More precisely, the expression \eqref{eq:nss-rmat} provides the kernel of the linear map \eqref{R def} with respect to the bilinear pairing \eqref{eq:bmgd}. To show this, we will use standard tensor product space notation as follows. Given any rational function $U_{\ti{12}}(\lambda, \mu)$ in the parameters $\lda$ and $\mu$ such that $\iota_\mu \iota_\lda U_{\ti{12}}, \iota_\lda \iota_\mu U_{\ti{12}} \in \Lg_D^{(0)} \,\widetilde{\otimes}\, \Lg_D^{(1)}$, where the first tensor factor in the formally completed tensor product is the loop algebra $\Lg_D^{(0)}$ with the loop parameter $\lambda$, and the second tensor factor is the space $\Lg_D^{(1)}$ with the loop parameter $\mu$, we introduce the following shorthand notations:
\begin{subequations}
\begin{align}\label{eq:bmgdtensorone}
  {\langle \iota_\mu \iota_\lda U_{\ti{12}},  X_{\ti{2}} \rangle}_{\ti{2}} &= T \sum_{r=1}^N \Res_{\mu_r = 0} \Tr_{\ti{2}} \left( \iota_{\mu_r} \iota_\lda U_{\ti{12}}(\lda, \mu) X_{r \ti{2}}(\mu_r)\right)d\mu \notag \\
  &\quad +  \sum_{r \in \{0, \infty\}} \Res_{\mu_r = 0} \Tr_{\ti{2}} \left( \iota_{\mu_r} \iota_\lda U_{\ti{12}}(\lda, \mu) X_{r \ti{2}}(\mu_r)\right)d\mu,
\end{align}  
\end{subequations}
and
\begin{subequations}
\begin{align}\label{eq:bmgdtensortwo}
  {\langle \iota_\lda \iota_\mu U_{\ti{12}},  X_{\ti{2}} \rangle}_{\ti{2}} &= T \sum_{r=1}^N \Res_{\mu_r = 0} \Tr_{\ti{2}} \left( \iota_\lda \iota_{\mu_r} U_{\ti{12}}(\lda, \mu) X_{r \ti{2}}(\mu_r)\right)d\mu \notag \\
  &\quad +  \sum_{r \in \{0, \infty\}} \Res_{\mu_r = 0} \Tr_{\ti{2}} \left( \iota_\lda \iota_{\mu_r} U_{\ti{12}}(\lda, \mu) X_{r \ti{2}}(\mu_r)\right)d\mu,
\end{align}  
\end{subequations}
for any $X \in \Lg_D^{(0)}$, where the parameters $\mu_r$ are defined analogously to the parameters $\lda_r$ in \eqref{lambda r def}, and the linear map $\iota_\mu$ is defined as in \eqref{iota map def}, returning the tuple of formal Laurent expansions in $\mu_r$, $r \in S$.
\begin{proposition}\label{prop:projrel}
For all $X \in \Lg_D^{(0)}$, the linear maps $R_+$ and $R_-$ satisfy
\begin{equation}\label{eq:projrel}
  R_+(X) = {\langle \iota_\mu \iota_\lda r_{\ti{12}},  X_{\ti{2}} \rangle}_{\ti{2}} \quad \text{and} \quad R_-(X) = {\langle \iota_\lda \iota_\mu r_{\ti{12}},  X_{\ti{2}} \rangle}_{\ti{2}}
\end{equation}
respectively, where
\begin{equation}
  r_{\ti{12}}(\lda, \mu) = \frac{1}{T} \sum_{k=0}^{T-1} \frac{\sgm_{\ti{1}}^k C_{\ti{12}}}{\mu - \omg^{-k}\lda}.
\end{equation}
\end{proposition}
\begin{proof}
In what follows, to treat the origin on the same footing as the points $\zeta_r$, $r \in \{1, \ldots, N \}$, it will be convenient to introduce the notation $\zeta_0 = 0$. We have
  \begin{equation}\label{eq:expldar}
    \iota_{\lda_s} \left( \frac{1}{T} \sum_{k=0}^{T-1} \frac{\sgm_{\ti{1}}^k C_{\ti{12}}}{\mu - \omg^{-k}\lda} \right) = \begin{dcases}
      \frac{1}{T} \sum_{k=0}^{T-1} \sum_{m=0}^{\infty} \frac{\omg^{-km}(\lda - \zeta_s)^m \sgm_{\ti{2}}^{-k} C_{\ti{12}}}{(\mu - \omg^{-k}\zeta_s)^{m+1}} &\text{for}\,\, s \in \{0, 1, \ldots, N \}\\
      - \frac{1}{T} \sum_{k=0}^{T-1} \sum_{m=0}^{\infty} \frac{\omg^{k(m+1)} \mu^m \sgm_{\ti{2}}^{-k} C_{\ti{12}}}{\lda^{m+1}} &\text{for}\,\, s = \infty,\\
    \end{dcases}
  \end{equation}
  and
  \begin{equation}\label{eq:expmur}
    \iota_{\mu_r} \left( \frac{1}{T} \sum_{k=0}^{T-1} \frac{\sgm_{\ti{1}}^k C_{\ti{12}}}{\mu - \omg^{-k}\lda} \right) = \begin{dcases}
      - \frac{1}{T} \sum_{k=0}^{T-1} \sum_{m=0}^{\infty} \frac{\omg^{k(m+1)}(\mu - \zeta_r)^m \sgm_{\ti{1}}^k C_{\ti{12}}}{(\lda - \omg^k\zeta_r)^{m+1}} &\text{for}\,\, r \in \{0, 1, \ldots, N \}\\
      \frac{1}{T} \sum_{k=0}^{T-1} \sum_{m=0}^{\infty} \frac{\omg^{-km} \lda^m \sgm_{\ti{1}}^k C_{\ti{12}}}{\mu^{m+1}} &\text{for}\,\, r = \infty.\\
    \end{dcases}
  \end{equation}
It follows that both $\iota_\mu \iota_\lda r_{\ti{12}}$ and $\iota_\lda \iota_\mu r_{\ti{12}}$ are elements of $\Lg_D^{(0)} \,\widetilde{\otimes}\, \Lg_D^{(1)}$ where the loop parameter $\lambda$ is in the first tensor factor and $\mu$ in the second. Now, pick an arbitrary $X \in \Lg_D^{(0)}$ and let $X = W+Z$ be its decomposition relative to \eqref{eq:gdzerodecomp} where the two components can be written explicitly as
  \begin{equation}
    W = \left( \sum_{n=0}^\infty W_0^{(n)} \lda^n, \sum_{n=0}^\infty W_1^{(n)} (\lda - \zeta_1)^n, \ldots, \sum_{n=0}^\infty W_N^{(n)} (\lda - \zeta_N)^n, \sum_{n=1}^\infty W_\infty^{(n)}\lda_\infty^n \right) \in \Lg_{D+}^{(0)},
  \end{equation}
  and
  \begin{equation}
    Z = \iota_\lda \left( \sum_{n=0}^{N_0 - 1} \frac{Z_0^{(n)}}{\lda^{n+1}} + \sum_{r=1}^{N} \sum_{k=0}^{T-1} \sum_{n=0}^{N_r - 1} \frac{\sgm^k Z_r^{(n)}}{(\omg^{-k}\lda - \zeta_r)^{n+1}} + \sum_{n=0}^{N_\infty} \frac{Z_\infty^{(n)}}{\lda_\infty^n} \right) \in \iota_\lda \mathcal{F}_{D^\prime}^\Gamma .
  \end{equation}
Using the expansion \eqref{eq:expldar}, we get
\begin{align}
    &T \sum_{r=1}^N \Res_{\mu_r = 0} \Tr_{\ti{2}} \left( \iota_{\mu_r} \iota_{\lda_s} r_{\ti{12}}(\lda, \mu) W_{r \ti{2}}(\mu_r)\right)d\mu \notag \\
    &+ \sum_{r \in \{0, \infty\}} \Res_{\mu_r = 0} \Tr_{\ti{2}} \left( \iota_{\mu_r} \iota_{\lda_s} r_{\ti{12}}(\lda, \mu) W_{r \ti{2}}(\mu_r)\right)d\mu \notag \\
    &= \sum_{m=0}^\infty W_s^{(m)} (\lda - \zeta_s)^m, \,\, \text{when} \,\, s \in \{0, 1, \ldots, N\},
\end{align}
and
\begin{align}
    &T \sum_{r=1}^N \Res_{\mu_r = 0} \Tr_{\ti{2}} \left( \iota_{\mu_r} \iota_{\lda_s} r_{\ti{12}}(\lda, \mu) W_{r \ti{2}}(\mu_r)\right)d\mu \notag \\
    &+ \sum_{r \in \{0, \infty\}} \Res_{\mu_r = 0} \Tr_{\ti{2}} \left( \iota_{\mu_r} \iota_{\lda_s} r_{\ti{12}}(\lda, \mu) W_{r \ti{2}}(\mu_r)\right)d\mu \notag\\
    &= \sum_{m=1}^\infty W_\infty^{(m)} \lda_\infty^m, \,\, \text{when} \,\, s = \infty.
\end{align}
So, we find that
  \begin{equation}\label{eq:rplusu}
    {\langle \iota_\mu \iota_\lda r_{\ti{12}},  W_{\ti{2}} \rangle}_{\ti{2}} = W.
\end{equation}
To evaluate ${\langle \iota_\mu \iota_\lda r_{\ti{12}},  Z_{\ti{2}} \rangle}_{\ti{2}}$, we note that for an arbitrary equivariant rational function $f \in \mathcal{F}_{D^\prime}^\Gamma$ and equivariant rational one-form $g \in \Omega_{D^\prime}^\Gamma$, we have the following relation:
\begin{align}\label{eq:respropallpoles}
    \Res_{\lda_r = 0} \Tr(g(\lda)f(\lda))d\lda &= \Res_{\lda_r = 0} \Tr(\sgm^k (g(\lda))\sgm^k (f(\lda)))d\lda \notag \\
    &= \Res_{\lda_r = 0} \Tr(\omg ^k g(\omg^k \lda)f(\omg^k \lda))d\lda \notag \\
    &= \Res_{\lda_{r,k} = 0} \Tr(g(\lda)f(\lda))d\lda,
\end{align}
for all $k \in \{0, \ldots, T-1\}$. This allows us to rewrite each entry of the tuple ${\langle \iota_\mu \iota_\lda r_{\ti{12}},  Z_{\ti{2}} \rangle}_{\ti{2}}$ as
\begin{equation}
  \begin{split}
    &T \sum_{r=1}^N \Res_{\mu_r = 0} \Tr_{\ti{2}} \left( \iota_{\mu_r} \iota_{\lda_s} r_{\ti{12}}(\lda, \mu) Z_{r \ti{2}}(\mu_r)\right)d\mu \\
    &+ \sum_{r \in \{0, \infty\}} \Res_{\mu_r = 0} \Tr_{\ti{2}} \left( \iota_{\mu_r} \iota_{\lda_s} r_{\ti{12}}(\lda, \mu) Z_{r \ti{2}}(\mu_r)\right)d\mu \notag\\
    &= \sum_{r=1}^N \sum_{k=0}^{T-1} \Res_{\mu_{r, k} = 0} \Tr_{\ti{2}} \left( \iota_{\mu_r} \iota_{\lda_s} r_{\ti{12}}(\lda, \mu) Z_{r \ti{2}}(\mu_r)\right)d\mu \notag\\
    &\quad+ \sum_{r \in \{0, \infty\}} \Res_{\mu_r = 0} \Tr_{\ti{2}} \left( \iota_{\mu_r} \iota_{\lda_s} r_{\ti{12}}(\lda, \mu) Z_{r \ti{2}}(\mu_r)\right)d\mu,\qquad s \in S,
  \end{split}
\end{equation}
which is a sum over all the residues of a meromorphic one-form on $\mathbb{C}P^1$. Hence, we deduce
\begin{equation}\label{eq:rplusv}
  {\langle \iota_\mu \iota_\lda r_{\ti{12}},  Z_{\ti{2}} \rangle}_{\ti{2}} = 0.
\end{equation}
From \eqref{eq:rplusu} and \eqref{eq:rplusv}, we then have
  \begin{equation}
    {\langle \iota_\mu \iota_\lda r_{\ti{12}},  X_{\ti{2}} \rangle}_{\ti{2}}=W = P_+(X) = R_+(X).
  \end{equation}
Since $X \in \Lg_D^{(0)}$ was arbitrary, this establishes the first relation in \eqref{eq:projrel}. Similarly, using the expansion \eqref{eq:expmur}, we find ${\langle \iota_\lda \iota_\mu r_{\ti{12}},  W_{\ti{2}} \rangle}_{\ti{2}} = 0$ and ${\langle \iota_\lda \iota_\mu r_{\ti{12}},  Z_{\ti{2}} \rangle}_{\ti{2}} =-Z$ from which we conclude that
\begin{equation}
  {\langle \iota_\lda \iota_\mu r_{\ti{12}},  X_{\ti{2}} \rangle}_{\ti{2}}=-Z = -P_-(X) = R_-(X) .
\end{equation}
Again, since $X \in \Lg_D^{(0)}$ was arbitrary, this establishes the second relation in \eqref{eq:projrel}.
\end{proof}

It follows from Proposition \ref{prop:projrel} that the kernel of the linear map $R = P_+ - P_-$ and that of the identity map $\Id = P_+ + P_-$ (with respect to the bilinear pairing of Proposition \ref{prop:pairing}) are
\begin{equation*}
  (\iota_\mu \iota_\lda + \iota_\lda \iota_\mu) r_{\ti{12}}(\lda, \mu)
  \quad \text{and} \quad (\iota_\mu \iota_\lda - \iota_\lda \iota_\mu)
  r_{\ti{12}}(\lda, \mu)
\end{equation*}
respectively.

Recall that in Proposition \ref{prop:gddecomp} we identified the Lax matrix \eqref{Lax_matrix_Gaudin} of the cyclotomic Gaudin model, or rather its image under the embedding $\iota_\lambda$ in \eqref{iota map def}, as living in the subspace $V_- = \iota_\lda \Omega_{D^\prime}^\Gamma$ of our model $V = \Lg_D^{(1)}$ for the dual space $\mathbb{g}^\ast$.
The next proposition establishes that this subspace $V_- \subset V$ is the orthogonal complement of $\iota_\lda \mathcal{F}_{D^\prime}^\Gamma \subset \Lg_D^{(0)}$, i.e., $\mathbb{g}_- \subset \mathbb{g}$, with respect to the nondegenerate bilinear pairing of Proposition \ref{prop:pairing}. This completes the proof of the claim that the decomposition $V = V_+ \dotplus V_-$ obtained in Proposition \ref{prop:gddecomp} is the one induced from the corresponding decomposition of the Lie algebra $\mathbb{g} = \mathbb{g}_+ \dotplus \mathbb{g}_-$.

\begin{proposition}\label{prop:pairequivsp}
An element $Y \in \Lg_D^{(1)}$ lies in $\iota_\lda \Omega_{D^\prime}^\Gamma$ if and only if $\langle Y, X \rangle = 0$ for all $X \in \iota_\lda \mathcal{F}_{D^\prime}^\Gamma$.
Moreover, $Y \in \Lg_D^{(1)}$ lies in $\Lg_{D+}^{(1)}$ if and only if $\langle Y, X \rangle = 0$ for all $X \in \Lg_{D+}^{(0)}$.
\end{proposition}

\begin{proof}
This is a particular case of the $\Gamma$-equivariant strong residue theorem \cite[Appendix A]{VY1}. We recall the proof here in the present setting for completeness.
  We start by proving that if $Y \in \iota_\lda \Omega_{D^\prime}^\Gamma$, then $\langle Y, X \rangle = 0$ for all $X \in \iota_\lda \mathcal{F}_{D^\prime}^\Gamma$. Let us pick elements $f \in \mathcal{F}_{D^\prime}^\Gamma$ and $g \in \Omega_{D^\prime}^\Gamma$, and let $X = \iota_\lda f$ and $Y = \iota_\lda g$. Since $g(\lambda) f(\lambda) d\lambda$ is a meromorphic one-form on $\CP$, by the residue theorem we have
  \begin{equation}\label{eq:sumresoneform}
      \sum_{r=1}^N \sum_{k=0}^{T-1} \Res_{\lda_{r, k} = 0} \Tr(\iota_{\lda_{r, k}}g(\lda)\iota_{\lda_{r, k}}f(\lda) )d\lda + \sum_{r \in \{0, \infty\}} \Res_{\lda_r = 0} \Tr(\iota_{\lda_r}g(\lda)\iota_{\lda_r}f(\lda) )d\lda = 0 .
  \end{equation}
Using the relation \eqref{eq:respropallpoles} valid for any $f \in \mathcal{F}_{D^\prime}^\Gamma$ and $g \in \Omega_{D^\prime}^\Gamma$, we may rewrite the first term on the left-hand side as $T$ times a sum over the residues at the points $\zeta_r$, for $r \in S$, so that
\begin{equation}\label{eq:orthspaceresone}
T \sum_{r=1}^N \Res_{\lda_r = 0} \Tr( \iota_{\lda_r} g(\lda)\iota_{\lda_r} f(\lda))d\lda + \sum_{r \in \{0, \infty\}} \Res_{\lda_r = 0} \Tr( \iota_{\lda_r} g(\lda)\iota_{\lda_r} f(\lda))d\lda = 0 .
\end{equation}
By the definition of the bilinear pairing from Proposition \ref{prop:pairing} we thus have $\langle Y, X \rangle = 0$, as desired.

Let us now prove the converse. Namely, let $Y \in \Lg_D^{(1)}$ be arbitrary and suppose that $\langle Y, X\rangle = 0$ for all $X \in \iota_\lda \mathcal{F}_{D^\prime}^\Gamma$. We must show that, in fact, $Y(\lda) \in \iota_\lda \Omega_{D^\prime}^\Gamma$. It follows from Proposition \ref{prop:gddecomp} that $Y = (Y_0, Y_1, \ldots, Y_N, Y_\infty) \in \Lg_D^{(1)}$ has a unique decomposition as a direct sum of the tuples $ \P_+(Y) = (\P_{0+}(Y_0), \P_{1+}(Y_1), \ldots, \P_{N+}(Y_N), \P_{\infty+}(Y_\infty)) \in \Lg_{D+}^{(1)} $ and $ \P_-(Y) = \iota_\lda g$ for some $g \in \Omega_{D^\prime}^\Gamma$. Now, let $X = \iota_\lda f$ for some $f \in \mathcal{F}_{D^\prime}^\Gamma$. Since $\langle Y, X \rangle = 0$, writing this out explicitly means
\begin{align}
    &T \sum_{r=1}^N \Res_{\lda_r = 0} \Tr(  \P_{r+}(Y_r(\lda_r)) \iota_{\lda_r}f(\lda) )d\lda + \sum_{r \in \{0, \infty\}} \Res_{\lda_r = 0} \Tr( \P_{r+}(Y_r(\lda_r))\iota_{\lda_r}f(\lda))d\lda \notag \\
    &+T \sum_{r=1}^N \Res_{\lda_r = 0} \Tr(\iota_{\lda_r}g(\lda)\iota_{\lda_r}f(\lda))d\lda + \sum_{r \in \{0, \infty\}} \Res_{\lda_r = 0} \Tr(\iota_{\lda_r}g(\lda)\iota_{\lda_r}f(\lda))d\lda = 0.
\end{align}
From \eqref{eq:orthspaceresone}, we have that the last two terms on the left-hand side vanish on their own. Therefore, we get
\begin{equation}
    T \sum_{r=1}^N \Res_{\lda_r = 0} \Tr( \P_{r+}(Y_r(\lda_r))\iota_{\lda_r}f(\lda))d\lda + \sum_{r \in \{0, \infty\}} \Res_{\lda_r = 0} \Tr( \P_{r+}(Y_r(\lda_r))\iota_{\lda_r}f(\lda))d\lda = 0.
\end{equation}
One can then show, along the same lines as the argument in the proof of Proposition \ref{prop:pairing}, that if $\mathcal P_+(Y) \neq 0$ then by picking a suitable $f \in \mathcal{F}_{D^\prime}^\Gamma$ adapted to $\mathcal{P}_+(Y) \in \Lg_{D+}^{(1)}$, one can ensure that the expression on the left-hand side is non-zero, which is a contradiction. Therefore, we conclude that $\mathcal P_+(Y) = 0$, and hence $Y \in \iota_\lambda \Omega_{D^\prime}^\Gamma$, as required.
The proof of the ``moreover'' part is completely analogous.
\end{proof}

The upshot of Proposition \ref{prop:pairequivsp} is that we are now exactly in the setting recalled in Section \ref{sec:AKS}. We will use this in Lemma \ref{identification_orbits} to give a concise description of the desired coadjoint orbit for the Lax matrix of the cyclotomic Gaudin model. Before stating the lemma, we first give an explicit description of the infinite-dimensional Lie group $\mathbb{G}_+ = \LG_{D+}^{(0)}$ associated with the Lie algebra $\mathbb{g}_+ = \Lg_{D+}^{(0)}$. Its elements are of the form
\begin{equation} \label{varphi+ def}
    \varphi_+=\left( \varphi_{0+}, \varphi_{1+}, \ldots, \varphi_{N+}, \varphi_{\infty+} \right),
\end{equation}
where $\varphi_{r+}(\lda_r)$ is a Taylor series in the local parameter $\lda_r$ with values in $G$, the matrix Lie group associated with the Lie algebra $\g$,
\begin{equation}\label{eq:groupelement}
    \varphi_{r+}(\lda_r) = \sum_{n=0}^\infty \phi_r^{(n)}\lda_r^n, \quad r \neq \infty,\qquad
    \varphi_{\infty+}(\lda_\infty) = \1 + \sum_{n=1}^\infty \phi_\infty^{(n)}\lda_\infty^n.
\end{equation}
The Lie group $\mathbb{G}_+ = \LG_{D+}^{(0)}$ has a natural action on $V = \Lg_{D}^{(1)}$ given by conjugation. This defines the coadjoint representation $\Ad^\ast : \mathbb{G}_+ \times V \to V$ from Section \ref{sec:AKS} explicitly as
\begin{equation}
(\varphi_+,Y) \longmapsto \Ad^\ast_{\varphi_+} Y=\left( \varphi_{0+}Y_0\varphi_{0+}^{-1}, \varphi_{1+}Y_1\varphi_{1+}^{-1}, \ldots, \varphi_{N+}Y_N\varphi_{N+}^{-1}, \varphi_{\infty+}Y_\infty\varphi_{\infty+}^{-1} \right).
\end{equation}
We are now in a position to give an explicit description of the $\mathbb{G}_R$-coadjoint orbit where the Lax matrix of the cyclotomic Gaudin model lives and which will act as our phase space.
\begin{lemma}\label{identification_orbits}
The orbit of the coadjoint action of $\mathbb{G}_R = (\LG_D^{(0)})_R$ on an element $\iota_\lda \Lambda \in V_- = \iota_\lda \Omega_{D^\prime}^\Gamma$ has the explicit form
    \begin{equation}
      \O_\Lambda = \big\{ \P_-(\Ad_{\varphi_+}^\ast \iota_\lda \Lambda ); \, \varphi_+ \in \LG_{D+}^{(0)} \big\}.
    \end{equation}
\end{lemma}

\begin{proof}
By definition, the coadjoint orbit of $(\LG_D^{(0)})_R$ on any $\iota_\lda \Lambda \in \iota_\lda \Omega_{D^\prime}^\Gamma$ is given by
\begin{equation}
\O_\Lambda =\big\{ \Ad_{\varphi}^{R\ast} \iota_\lda \Lambda ; \,\varphi \in (\LG_D^{(0)})_R \big\}.
\end{equation}
Using the explicit form for the coadjoint action of $\mathbb{G}_R$ we have
\begin{equation}\label{eq:rcoadorbdef}
\Ad^{R\ast}_\varphi \iota_\lda \Lambda =\mathcal{P}_-(\Ad^\ast_{\varphi_+} \iota_\lda \Lambda)+\mathcal{P}_+(\Ad^\ast_{\varphi_-}\iota_\lda \Lambda) = \mathcal{P}_-(\Ad^\ast_{\varphi_+} \iota_\lda \Lambda)
\end{equation}
where the last equality follows from the fact that $\iota_\lambda \Lambda \in V_- = \iota_\lambda \Omega_{D^\prime}^\Gamma$ so that $\Ad^\ast_{\varphi_-}\iota_\lda \Lambda \in V_-$ also and hence $\mathcal{P}_+(\Ad^\ast_{\varphi_-}\iota_\lda \Lambda) = 0$. 
\end{proof}

It will be useful in practice to express the action of the projector $\P_-$ on $Y \in \Lg_{D}^{(1)}$ as
\begin{equation}
\label{useful}
    \P_-(Y) = \iota_\lda \circ \pi_\lda^{(1)} (Y).
\end{equation}
In the remaining sections we will put this setup to use to describe the Lax matrix of the cyclotomic Gaudin model \eqref{Lax_matrix_Gaudin} as a point in a coadjoint orbit $\O_\Lambda$ for some suitable $\Lambda \in \Omega_{D^\prime}^\Gamma$ and then introduce the ingredients from Section \ref{sec:AKS} to derive the associated Lax equations.

\subsection{Lax description}\label{sec:cgmlaxm}
Our algebraic setup covers the case of the cyclotomic Gaudin model with arbitrary multiplicities. However, as mentioned at the start of this section, from now on we will restrict to the case with simple poles at all $\omg^k \zeta_r$, $k \in \{0, \ldots, T-1\}$, $r \in \{1, \ldots, N\}$, and double poles at the origin and at infinity, since this is the setting required for our examples in Section \ref{sec:realisations}. The discussion in the remaining sections is easily generalised to the case of arbitrary multiplicities.

To describe a coadjoint orbit $\O_\Lambda \in \iota_\lda \Omega_{D^\prime}^\Gamma$ from Lemma \ref{identification_orbits} corresponding to the Lax matrix \eqref{Lax_matrix_Gaudin} of the cyclotomic Gaudin model, we fix a non-dynamical element $\Lambda \in \Omega_{D^\prime}^\Gamma$ with the same pole structure as \eqref{Lax_matrix_Gaudin}, namely we introduce
\begin{equation}
  \Lambda(\lda) = \frac{\Lambda_0^{(0)}}{\lambda} + \frac{\Lambda_0^{(1)}}{\lambda^2} + \frac{1}{T} \sum_{r=1}^{N} \sum_{k=0}^{T-1} \frac{\sgm^k \Lambda_r}{\lda - \omg^k \zeta_r} + \Lambda_\infty \in \Omega_{D^\prime}^\Gamma .
\end{equation}
According to Lemma \ref{identification_orbits} and formula \eqref{useful}, the corresponding coadjoint orbit $\O_\Lambda$ then consists of elements of the form
\begin{align}\label{eq:cglaxembed}
    & \P_-\left(\varphi_+ \, \iota_\lda \left( \frac{\Lambda_0^{(0)}}{\lambda} + \frac{\Lambda_0^{(1)}}{\lambda^2} +  \frac{1}{T} \sum_{r=1}^{N} \sum_{k=0}^{T-1} \frac{\sgm^k \Lambda_r}{\lda - \omg^k \zeta_r} + \Lambda_\infty \right) \, \varphi_+^{-1}\right) \notag \\
    &= \iota_\lda \circ \pi_\lda^{(1)} \left( \frac{A_0^{(0)}}{\lda} + \frac{A_0^{(1)}}{\lda^2}, \frac{1}{T} \frac{A_1}{\lda - \zeta_1}, \ldots, \frac{1}{T} \frac{A_N}{\lda - \zeta_N}, A_\infty \right) \notag \\
    &= \iota_\lda \left( \frac{A_0^{(0)}}{\lda} + \frac{A_0^{(1)}}{\lda^2} + \frac{1}{T} \sum_{r=1}^{T-1} \sum_{k=0}^{T-1} \frac{ \sgm^k A_r }{\lda - \omg^k \zeta_r} + A_\infty \right)\notag\\
    &\equiv\iota_\lda L(\lda) ,
\end{align}
where, recalling the definitions \eqref{varphi+ def} and \eqref{eq:groupelement}, we have set
\begin{equation}\label{eq:cgcoadorb}
\begin{split}
    &A_0^{(0)} = \phi_0^{(0)} \Lambda_0^{(0)} \phi_0^{(0)\, -1} + \big[ \phi_0^{(1)} \phi_0^{(0)\, -1}, \phi_0^{(0)} \Lambda_0^{(1)} \phi_0^{(0)\, -1} \big], \\
    &A_0^{(1)} = \phi_0^{(0)} \Lambda_0^{(1)} \phi_0^{(0)\, -1}, \\
    &A_r = \phi_r \Lambda_r \phi_r^{-1}, \qquad r \in \{1, \ldots, N \}, \\
    &A_\infty = \Lambda_\infty,
\end{split}
\end{equation}
with $\phi_r^{(0)}$ denoted by $\phi_r$, for $r \in \{1, \ldots, N \}$, to simplify notation. Notice that we have the relations $\sgm\phi_{0}^{(0)} = \phi_{0}^{(0)}$ and $\sgm\phi_{0}^{(1)} = \omg\phi_{0}^{(1)}$ for the field elements which ensure that $A_{0}^{(0)} \in \g^{(0)}$ and $A_{0}^{(1)} \in \g^{(-1)}$. This gives us the desired parameterisation of the cyclotomic Gaudin Lax matrix
\begin{equation}\label{eq:cglm}
  L(\lda) = \frac{A_0^{(0)}}{\lda} + \frac{A_0^{(1)}}{\lda^2} + \frac{1}{T} \sum_{r=1}^{N} \sum_{k=0}^{T-1} \frac{ \sgm^k A_r }{\lda - \omg^k \zeta_r} + A_\infty
\end{equation}
viewed as an element of the coadjoint orbit $\O_\Lambda$.

\subsubsection{Lax equations}\label{sec:cgmlaxeq}
To derive Lax equations for the Lax matrix \eqref{eq:cglm} of the cyclotomic Gaudin model, let us return to the hierarchy of Lax equations \eqref{Lax_system} induced by the family of Hamiltonians in involution with respect to $\{\,,\,\}_R$. In our current setup, $\{\,,\,\}_R$ is the Lie--Poisson bracket on $\Lg_D^{(1)}$ associated with the linear map $R = P_+ - P_-$. Invariant functions on $\Lg_D^{(1)}$ take the form
\begin{equation}\label{eq:invfunc}
  H_{p, r} \colon Y \in \Lg_D^{(1)} \longmapsto \Res_{\lda_r = 0} \big(\ell_{p, r}(\lda_r) \Tr\big( Y_r(\lda_r)^{p+1} \big) \big)d\lda, \qquad p \geq 1, \quad r \in S,
\end{equation}
where $\ell_{p, r}(\lda_r) \in \mathbb{C}((\lda_r))$ is a collection of Laurent polynomials for $p \geq 1$ and $r \in S$. It follows from Proposition \ref{prop:pairing} that for these functions to be non-trivial, the Laurent polynomials $\ell_{p, r}(\lda_r)$, for $r \in \{0, \infty\}$, should be chosen such that $\ell_{p, r}(\lda_r) Y_r(\lda_r)^p \in \Lg_r^{\Gamma, 0}$, while $\ell_{p, r}(\lda_r)$, for $r \in \{1, \ldots, N \}$, can be any Laurent polynomials. Let us then choose
\begin{equation} \label{ell p r choice}
    \ell_{p, r}(\lambda_r) = \iota_{\lda_r} \frac{\lda^p}{p+1} \quad \text{for}\,\, r \in \{0, \infty\} \quad \text{and} \quad \ell_{p, r}(\lambda_r) = \iota_{\lda_r} \frac{T \lda^p}{p+1} \quad \text{for}\,\, r \in \{1, \ldots, N \}.
\end{equation}
The restriction of the functions $H_{p, r}$ to $\iota_\lda L$ are Hamiltonians (in involution) of the model and generate the elementary (pairwise-commuting) flows $\partial_{t^{p, r}}$. However, by virtue of the choice \eqref{ell p r choice} we made, it follows from \eqref{eq:respropallpoles} that $\sum_{r \in S} H_{p,r}(\iota_\lda L) = 0$ by the residue theorem, for each $p \geq 1$.

In what follows, it will thus be sufficient to focus on the Hamiltonians $H_{p, r}$, for $r \neq \infty$, and look at the associated equations of motion they produce through \eqref{Lax_system}.
Using $R_\pm = \frac{1}{2}(R \pm \Id)$, these equations can be rewritten as
\begin{equation}\label{eq:laxeqdialg}
  \partial_{t^{p, r}} \iota_\lda L = \big[ R_\pm \nabla H_{p, r}(\iota_\lda L), \iota_\lda L \big].
\end{equation}
Note that it suffices to calculate only one of the two expressions $R_+ \nabla H_{p, r}(\iota_\lda L)$ and $R_- \nabla H_{p, r}(\iota_\lda L)$. Let us compute the latter. The gradient of $H_{p, r}(\iota_\lda L)$ is an element of $\Lg_D^{(0)}$ and satisfies
\begin{equation} \label{eq:gradham}
  \lim_{\epsilon \rightarrow 0} \frac{H_{p, r}(\iota_\lda L + \epsilon \eta) - H_{p, r}(\iota_\lda L)}{\epsilon} = \big\langle \eta, \nabla H_{p, r} (\iota_\lda L) \big\rangle,
\end{equation}
for all $\eta \in \Lg_D^{(1)}$. Using Proposition \ref{prop:gddecomp}, we may decompose this gradient as follows
\begin{equation}
  \nabla H_{p, r} (\iota_\lda L) = N_r^{(p)} + \iota_\lda h_r^{(p)}, \qquad N_r^{(p)} \in \Lg_{D+}^{(0)}, \quad  h_r^{(p)} \in  \mathcal{F}_{D^\prime}^\Gamma,
\end{equation}
and rewrite \eqref{eq:laxeqdialg} as
\begin{equation}
  \partial_{t^{p, r}} \iota_\lda L = \big[ R_- \nabla H_{p, r}(\iota_\lda L), \iota_\lda L \big] = - \big[ P_-\big( \nabla H_{p, r} (\iota_\lda L) \big), \iota_\lda L \big] = - \big[\iota_\lda h_r^{(p)}, \iota_\lda L \big].
\end{equation}
Since $\iota_\lda$ is an embedding which also clearly commutes with the Lie brackets $[\cdot, \cdot] : \mathcal{F}_{D^\prime}^\Gamma \times \Omega_{D^\prime}^\Gamma \to \Omega_{D^\prime}^\Gamma$ and $[\cdot, \cdot] : \Lg_D^{(0)} \times \Lg_D^{(1)} \to \Lg_D^{(1)}$, the above equation implies
\begin{equation}\label{eq:laxeqdialgone}
  \partial_{t^{p, r}} L = - \big[ h_r^{(p)}, L \big].
\end{equation}
In order to calculate $h_r^{(p)} \in \mathcal{F}_{D^\prime}^\Gamma$, by virtue of Proposition \ref{prop:pairequivsp} it is sufficient to restrict $\eta$ in \eqref{eq:gradham} to live in $\Lg_{D+}^{(1)}$. We then have
\begin{equation}
  \big\langle \eta, N_r^{(p)} \big\rangle = 0\qquad \text{for all}\, \, \eta \in  \Lg_{D+}^{(1)},\,\, N_r^{(p)} \in \Lg_{D+}^{(0)},
\end{equation}
and the right-hand side of \eqref{eq:gradham} becomes
\begin{equation*}
  T \sum_{s=1}^N \Res_{\lda_s = 0} \Tr\big(\eta_s(\lambda_s) \iota_{\lda_s} h_r^{(p)}\big) + \sum_{s \in \{0, \infty\}} \Res_{\lda_s = 0} \Tr\big(\eta_s(\lambda_s) \iota_{\lda_s} h_r^{(p)}\big),
\end{equation*}
while for the left-hand side we find
\begin{equation}
  \lim_{\epsilon \rightarrow 0} \frac{H_{p, r}(\iota_\lda L + \epsilon \eta) - H_{p, r}(\iota_\lda L)}{\epsilon} = (p+1) \Res_{\lda_r = 0} \Tr\big(\ell_{p, r}(\lambda_r)\eta_r(\lambda_r) \iota_{\lda_r} L^p\big), \qquad r \in S,
\end{equation}
for any $\eta_s(\lambda_s) \in \Lg_{s+}$, $s \in \{1, \ldots, N\}$, and $\eta_s(\lambda_s) \in \Lg_{s+}^{\Gamma, 1}$, $s \in \{0, \infty\}$. By definition of $\ell_{p,r}(\lambda_r)$ in \eqref{ell p r choice}, this implies
\begin{equation}
    \big(\iota_{\lda_s} h_r^{(p)}\big)_- = \begin{dcases}
        0 &\text{for}\,\, s \neq r\\
        \big( \iota_{\lda_r} \lda^p \iota_{\lda_r} L^p \big)_- &\text{for}\,\, s = r,
    \end{dcases}
\end{equation}
for all $r \in S$, where $X_-$ denotes the principal part of a Laurent series $X$. Plugging the equivariant functions $h_r^{(p)}$ obtained from the above conditions into \eqref{eq:laxeqdialgone} gives a hierarchy of Lax equations corresponding to our choice of invariant functions $H_{p, r}$ in \eqref{eq:invfunc}. 

Explicitly, for $p = 1$, we get the Lax equations
\begin{equation}\label{eq:laxeqfirstflow}
  \partial_{t^{1, r}} L = - \big[ h_r^{(1)}, L \big] \quad \text{with} \quad h_r^{(1)} =
  \begin{dcases}
    \dfrac{A_0^{(1)}}{\lda} &\text{for}\,\, r = 0\\
    \dfrac{1}{T} \sum_{k=0}^{T-1} \dfrac{\omg^k \zeta_r \sgm^k A_r}{\lda - \omg^k \zeta_r} & \text{for}\,\, r \in \{1, \ldots, N \}.
  \end{dcases}
\end{equation}

Finally, we turn to the primary goal of this work which is to give a Lagrangian one-form for the cyclotomic Gaudin model that will provide a variational description of the hierarchy of Lax equations in \eqref{eq:laxeqdialgone}. Having just described the cyclotomic Gaudin model within the Lie dialgebra framework, we now have all the necessary ingredients to achieve this goal: the non-dynamical element $\Lambda \in \Omega_{D^\prime}^\Gamma$ that fixes the phase space, the field element $\varphi_+ \in \LG_{D+}^{(0)}$ containing the dynamical degrees of freedom, the linear map $R$ that equips the phase space with the required Poisson structure, and the invariant functions $H_{p, r}$ which induce non-trivial equations of motion with respect to this Poisson structure.

\subsection{Lagrangian description}\label{sec:cgmultiform}  
We can now define a Lagrangian one-form on the coadjoint orbit ${\cal O}_\Lambda$ of the cyclotomic Gaudin model as
\begin{equation}\label{eq:cgmultiform}
  \Lag = \Lag_{p, r} \mathrm{d}t^{p, r}, \qquad p = 1, \ldots, M, \quad r \in S \setminus \{\infty\},
\end{equation}
using the expression \eqref{our_Lag} for the Lagrangian coefficients. It will be useful to recall the notation in \eqref{eq:cgcoadorb} associated with the parameterisation of the cyclotomic Gaudin Lax matrix. The elementary times $t^k$ that appear in \eqref{our_Lag} are now naturally labelled by a pair of indices $p \geq 1$ and $r \in S$, namely we now have elementary times $t^{p, r}$, associated with the corresponding Hamiltonians \eqref{eq:invfunc}. Explicitly,

\begin{theorem}\label{th:cgmultiform}
The Lagrangian coefficients of the cyclotomic Gaudin one-form $\Lag$ take the form
\begin{equation}\label{eq:cglagcoeff}
    \Lag_{p, r} = \sum_{s = 1}^N \Tr \big( A_s \partial_{t^{p, r}} \phi_s \phi_s^{-1} \big) + \Tr \big( A_0^{(0)} \partial_{t^{p, r}} \phi_0^{(0)} \phi_0^{(0)\, -1} \big) - H_{p, r}(\iota_\lda L),
\end{equation}
with $A_s$ and $A_0^{(0)}$ given by \eqref{eq:cgcoadorb}, and the potential part given by
\begin{equation}\label{eq:cgpot}
    H_{p, r}(\iota_\lda L) = \Res_{\lda_r = 0} \big( \ell_{p, r}(\lda_r) \Tr(\iota_{\lda_r}L^{p+1}) \big)d\lda, \quad r \in \{0, 1, \ldots, N\},
\end{equation}
where $\ell_{p, r}(\lda_r)$ are the Laurent polynomials 
\begin{equation}
    \ell_{p, 0} = \iota_{\lda_0} \frac{\lda^p}{p+1} \quad \text{and} \quad \ell_{p, r} = \iota_{\lda_r} \frac{T \lda^p}{p+1} \quad \text{for}\,\, r \in \{1, \ldots, N \}.
\end{equation}
The Lagrangian one-form $\Lag$ satisfies the corner equations \eqref{simple_multitime_EL2}--\eqref{simple_multitime_EL3} of the multitime Euler--Lagrange equations, while the standard Euler--Lagrange equations for $\Lag_{p, r}$ give the hierarchy of Lax equations in \eqref{eq:laxeqdialgone}. Further, on solutions of \eqref{eq:laxeqdialgone}, we have the closure relation
\begin{equation}
  \partial_{t^{q, s}}\Lag_{p, r}-\partial_{t^{p, r}}\Lag_{q, s} = 0,
\end{equation}
for all possible combinations of $(p, r)$ and $(q, s)$ in $\ZZ_{\geq 1} \times S$.
\end{theorem}
\begin{proof}
Let us start by reinterpreting the formula in \eqref{our_Lag} in the present context of the cyclotomic Gaudin model. First, note that on the coadjoint orbit $\mathcal{O}_\Lambda$, where the Lagrangian one-form $\Lag$ lives, the role of the Lax matrix is played by the image of $L$ in $\Lg_{D}^{(1)}$ under the embedding $\iota_\lda$, given by \eqref{eq:cglaxembed}. Next, the bilinear pairing used to define the kinetic part is the one constructed in Proposition \ref{prop:pairing}. Furthermore, in the Adler--Kostant--Symes scheme where (locally) $\mathbb{G}_R \simeq \mathbb{G}_+ \times \mathbb{G}_-$, we explicitly have $\partial_{t^{p, r}} \varphi \cdot_R \varphi^{-1} = \partial_{t^{p, r}} \varphi_+ \varphi_+^{-1} + \partial_{t^{p, r}} \varphi_- \varphi_-^{-1}$, and since $\iota_\lambda L \in V_- = \iota_\lambda \Omega_{D^\prime}^\Gamma$, it follows from Proposition \ref{prop:pairequivsp} that only the $\partial_{t^{p, r}} \varphi_+ \varphi_+^{-1}$ piece contributes to the kinetic term. Finally, the potential part is simply the restriction of the invariant functions $H_{p, r}$ in \eqref{eq:invfunc} to $\iota_\lda L$. Therefore, in the present setup, the Lagrangian coefficients in \eqref{our_Lag} can be expressed as
\begin{align}\label{eq:cglagcoeffinit}
    \Lag_{p, r} &= \big\langle   \iota_\lda L, \partial_{t^{p, r}} \varphi_+ \varphi_+^{-1} \,\big\rangle - H_{p, r}(\iota_\lda L) \notag \\
    &= \big\langle   \iota_\lda \Lambda, \varphi_+^{-1} \partial_{t^{p, r}} \varphi_+ \big\rangle - H_{p, r}(\iota_\lda L), \quad r \in \{0, 1, \ldots, N\}
\end{align}
where in the second step we used the fact \eqref{eq:cglaxembed} that $\iota_\lambda L = \mathcal P_-\big( \varphi_+ (\iota_\lambda \Lambda) \varphi_+^{-1} \big)$. The kinetic term can be written out more explicitly in terms of the component fields \eqref{eq:groupelement} as
\begin{align}
    \big\langle   \iota_\lda \Lambda, \varphi_+^{-1} \partial_{t^{p, r}} \varphi_+ \,\big\rangle &= T \sum_{s=1}^N \Res_{\lda_s = 0} \Tr \big( \iota_{\lda_s} \Lambda \, \varphi_{s+}(\lda_s)^{-1} \, \partial_{t^{p, r}} \varphi_{s+}(\lda_s) \big)d\lda \notag \\
  &\quad + \sum_{s \in \{0, \infty\} } \Res_{\lda_s = 0} \Tr \big( \iota_{\lda_s} \Lambda \, \varphi_{s+}(\lda_s)^{-1} \, \partial_{t^{p, r}} \varphi_{s+}(\lda_s) \big)d\lda \notag \\
  &= \sum_{s = 1}^N \Tr \big( A_s \partial_{t^{p, r}} \phi_s \phi_s^{-1} \big) + \Tr \big( A_0^{(0)} \partial_{t^{p, r}} \phi_0^{(0)} \phi_0^{(0)\, -1} \big) \notag \\
  &\quad + \partial_{t^{p, r}} \Tr \big( \phi_0^{(1)} \phi_0^{(0)\, -1} A_0^{(1)} \big) + \frac{1}{2} \partial_{t^{p, r}} \Tr \big(A_\infty (\phi_\infty^{(1)})^2\big).
\end{align}
The last two terms associated to the poles at the origin and at infinity are total derivatives and do not contribute to the multitime Euler--Lagrange equations since dropping them amounts to changing the Lagrangian one-form $\Lag$ by $\mathrm{d} \Tr\left(  \phi_0^{(1)} \phi_0^{(0)\, -1} A_0^{(1)}  + \frac{1}{2}  A_\infty (\phi_\infty^{(1)})^2  \right)$ which is a total horizontal differential. Discarding these two terms, we are left with the required expression for the Lagrangian coefficients.

Since the Lagrangian coefficients $\Lag_{p, r}$ in \eqref{eq:cglagcoeffinit} are of the form \eqref{our_Lag}, it follows directly from Theorem \ref{Th_multi_EL} that the Lagrangian one-form $\Lag$ satisfies all the required conditions and the closure relation.
\end{proof}

To close this section, let us present the explicit expressions for the first set of Lagrangian coefficients $\Lag_{1, r}$ with $r \in \{0, 1, \ldots, N \}$. The kinetic terms are obtained by simply substituting $p=1$ in the kinetic part in \eqref{eq:cglagcoeff}, while the potential terms defined by \eqref{eq:cgpot} read
\begin{equation}
\begin{split}
    H_{1, 0} &= \frac{1}{2} \Tr \big( A_0^{(0)\, 2} \big) - \sum_{r=1}^N \frac{\Tr \big( A_0^{(1)} A_r \big)}{\zeta_r} + \Tr \big( A_0^{(1)}A_\infty \big),\\
    H_{1, r} &=  \Tr \big( A_0^{(0)}A_r \big) + \frac{\Tr\big( A_0^{(1)}A_r \big)}{\zeta_r} + \frac{1}{2T} \sum_{k=0}^{T-1} \Tr\big( A_r \sgm^k A_r \big) \\
    &\quad + \frac{1}{T} \sum_{s \neq r} \sum_{k=0}^{T-1} \frac{\Tr\big( A_r \sgm^k A_s \big) \zeta_r}{\zeta_r - \omg^k \zeta_s} + \Tr(A_r A_\infty) \zeta_r, \quad r \in \{1, \ldots, N\}.
\end{split}
\end{equation}
Upon varying $\Lag_{1, r}$ with respect to $\phi_s$, $s = 1, \ldots, N$, $\phi_0^{(0)}$, and $\phi_0^{(1)}$, it can be checked that the associated Euler--Lagrange equations correspond to the set of Lax equations for $p=1$ in \eqref{eq:laxeqfirstflow}, as it should be from Theorem \ref{th:cgmultiform}.

\section{Realisations of the cyclotomic Gaudin model}\label{sec:realisations}

In this section, we study two different realisations of the cyclotomic Gaudin model --- the periodic Toda chain and the discrete self-trapping (DST) model --- with the objective of describing their corresponding hierarchies variationally. The construction of the periodic Toda one-form, in particular, complements the result of Sections \ref{Flaschka} and \ref{Toda_pq} where we derived Lagrangian one-forms for the open Toda chain.\footnote{We also note the work \cite{SV}, where the infinite Toda chain was considered within the setup of semi-discrete Lagrangian one-forms.} Finally, in Section \ref{sec:todadst}, we illustrate how to couple the periodic Toda and DST hierarchies together in a straightforward manner and derive a Lagrangian one-form for this coupled hierarchy. We will then go on to show how our framework allows for a straightforward coupling of integrable hierarchies, by using the approach devised in \cite[Section 7]{CStV} to couple together hierarchies of integrable field theories. 

In what follows, we will work with the cyclotomic Gaudin model associated with the Lie algebra $\g \coloneqq \mathfrak{gl}_T(\mathbb{C})$ and the automorphism $\sgm \in \text{Aut}\, \g$ defined by $\sgm(E_{ij}) = \omg^{j-i}E_{ij}$, for every $i, j = 1, \ldots, T$. Here $\omg$ is a primitive $T$th root of unity, and by $E_{ij}$, $i, j = 1, \ldots, T$, we denote the standard basis of $\mathfrak{gl}_T(\mathbb{C})$ taking the indices $i$ and $j$ modulo $T$ by convention. The eigenspaces of $\sgm$ defined by \eqref{eq:sgmeigenspace} are then given by $\g^{(n)} = \text{span}\{E_{i, i+n}\}_{i=1}^T$. 

Let us also note the following useful identity that we shall frequently make use of:
\begin{equation}\label{eq:omgidentity}
    \frac{z_1^{T-1-[l]} z_2^{[l]}}{z_1^T - z_2^T} = \frac{1}{T} \sum_{k=0}^{T-1} \frac{\omg^{-kl}}{z_1 - \omg^k z_2}
\end{equation}
for any $z_1, z_2 \in \mathbb{C}$ and $l \in \mathbb{Z}$, where $[l] \in \{ 0, \ldots, T-1\}$ is such that $l = [l] \, \,\text{mod}\, \, T$. 

\subsection{Periodic Toda chain}\label{sec:toda}

The periodic Toda chain \cite{To2} describes a system of particles connected by ``exponential springs'' together with a periodic boundary condition (in contrast to the case with an open boundary condition that we saw in Sections \ref{Flaschka} and \ref{Toda_pq}) and has been extensively studied in the Hamiltonian formalism. See, for instance, \cite{F} for the widely used Flaschka change of coordinates, and \cite{AM} for a proof of its integrability. 

\paragraph{Lax matrix:} We will work with the Lax matrix
\begin{equation}
\label{Lax_Toda}
  L_{\text{Toda}}(\lda) =
  \left(\begin{matrix}
		~p_1 \lda^{-1}~ & 1 & 0 & & \dots & & a_T \lda^{-2}\\
		a_1 \lda^{-2} & p_2 \lda^{-1} & 1 & & \dots & & 0 &\\[1ex]
		\vdots & & \ddots & & & &\vdots\\
		0 & & a_{i-1} \lda^{-2} & p_i \lda^{-1} & 1 & & 0\\
    \vdots & & & & \ddots & &\vdots\\
    0 & & \dots & & a_{T-2} \lda^{-2} & p_{T-1} \lda^{-1} & 1\\
		1 & & \dots & & 0 & a_{T-1} \lda^{-2} & ~p_T \lda^{-1}
	\end{matrix}\!\!\!\!\right)
\end{equation}
where $a_i = e^{q^i - q^{i+1}}$, and $q^i, p_i$ are the canonical coordinates satisfying the canonical Poisson bracket relations $\{p_i, q^j\} = \delta_{ij}$, for $i, j = 1, \ldots, T$. We also have the periodic boundary conditions $(p_0, q^0) = (p_T, q^T)$ and $(p_{T+1}, q^{T+1}) = (p_1, q^1)$. The Lax matrix above is parametrised by $a_i, p_i$, $i = 1, \ldots, T$, satisfying $\prod_i^T a_i = 1$ and $\sum_i^T p_i = 0$, carrying $2(T-1)$ degrees of freedom. Equivalently, when describing the system in terms of the canonical coordinates $q^i, p_i$, we can eliminate the centre of mass motion by imposing the conditions $\sum_i^T q^i = 0$ and $\sum_i^T p_i = 0$, leaving us with a phase space of dimension $2(T-1)$.

The standard Lax matrix for the periodic Toda chain (see \cite[Chapter 6]{BBT}, for instance)
\begin{equation}
  \widetilde{L}_{\text{Toda}}(\lda) =
  \left(\;\begin{matrix}
		~p_1~ & a_1^{1/2} & 0 & & \dots & & a_T^{1/2} \lda^{-1}\\
		a_1^{1/2} & p_2 & a_2^{1/2} & & \dots & & 0 &\\[1ex]
		\vdots & & \ddots & & & &\vdots\\
		0 & & a_{i-1}^{1/2} & p_i & a_i^{1/2} & & 0\\
        \vdots & & & & \ddots & &\vdots\\
        0 & & \dots & & a_{T-2}^{1/2} & p_{T-1} & a_{T-1}^{1/2}\\
		a_T^{1/2} \lda & & \dots & & 0 & a_{T-1}^{1/2} & ~p_T~\\
	\end{matrix}\!\!\!\!\right)
\end{equation}
is related to $L_{\text{Toda}}(\lda)$ by conjugation by the diagonal matrix $\mathcal{Q} = \text{diag}\left(e^{-q^1/2}\lda^{-1}, \ldots, \right.$ $\left.e^{-q^T/2}\lda^{-T}\right)$ and multiplication by an overall factor of $\lda^{-1}$, together with a change of $\lda$-dependence, as follows
\begin{equation}\label{eq:todagaugetransform}
  L_{\text{Toda}}(\lda) = \lda^{-1} \mathcal{Q} \widetilde{L}_{\text{Toda}}(\lda^T) \mathcal{Q}^{-1}.
\end{equation}

The Poisson bracket of the Lax matrix $\widetilde{L}_{\text{Toda}}(\lda)$ can be written as
\begin{equation}
  \big\{ \widetilde{L}_{\text{Toda}\, \ti{1}}(\lda), \widetilde{L}_{\text{Toda}\, \ti{2}}(\mu) \big\} = \big[ \widetilde{r}_{\ti{12}}(\lda, \mu), \widetilde{L}_{\text{Toda}\, \ti{1}}(\lda) + \widetilde{L}_{\text{Toda}\, \ti{2}}(\mu) \big]
\end{equation}
where $\widetilde{r}_{\ti{12}}(\lda, \mu)$ is the skew-symmetric $r$-matrix
\begin{equation}
  \widetilde{r}_{\ti{12}}(\lda, \mu) = \frac{1}{2} \frac{\mu + \lda}{\mu - \lda} \sum_{i = 1}^T E_{ii} \otimes E_{ii} + \frac{1}{\mu - \lda} \bigg( \mu \sum_{i < j} + \lda \sum_{i > j} \bigg) E_{ij} \otimes E_{ji}.
\end{equation}
Under the gauge transformation \eqref{eq:todagaugetransform}, on using the identity \eqref{eq:omgidentity}, we find that the Lax matrix $L_{\text{Toda}}(\lda)$ satisfies the Poisson bracket
\begin{equation}
    \{L_{\text{Toda}\, \ti{1}}(\lda), L_{\text{Toda}\, \ti{2}}(\mu) \} = [r_{\ti{12}}(\lda, \mu), L_{\text{Toda}\, \ti{1}}(\lda) ] - [r_{\ti{21}}(\mu, \lda),  L_{\text{Toda}\, \ti{2}}(\mu)]
\end{equation}
where $r_{\ti{12}}(\lambda, \mu)$ is the non-skew-symmetric cyclotomic $r$-matrix
\begin{equation}\label{eq:nss-rmateij}
  r_{\ti{12}}(\lda, \mu) = \frac{1}{T} \sum_{k=0}^{T-1} \frac{\omg^{k(j-i)}}{\mu - \omg^{-k}\lda} E_{ij} \otimes E_{ji} \,
\end{equation}
that we have been working with. This explains our choice of Lax matrix \eqref{Lax_Toda} as opposed to the more traditional one. As just proved, it satisfies the Poisson algebra of the cyclotomic $\mathfrak{gl}_T(\mathbb{C})$-Gaudin model and therefore allows us to obtain the periodic Toda chain as a certain realisation of that model. Specifically, the Lax matrix $L_{\text{Toda}}(\lda)$ can be seen as a realisation of the cyclotomic Gaudin Lax matrix with double poles at the origin and at infinity, that is,
\begin{equation}\label{eq:todalax}
  L_{\text{Toda}}(\lda) = \frac{J_0^{(0)}}{\lda} + \frac{J_0^{(1)}}{\lda^2} + J_\infty
\end{equation}
where
\begin{equation}\label{eq:todacoeff}
  J_0^{(0)} = \sum_{i=1}^T p_i E_{ii}, \quad J_0^{(1)} = \sum_{i=1}^T e^{q^i - q^{i+1}} E_{i+1, i}, \quad J_\infty = \sum_{i=1}^T E_{i, i+1}.
\end{equation}

\paragraph{Orbit realisation:} Set $D = \{0, \infty\}$, and choose the non-dynamical element
\begin{equation}
    \Lambda_{\text{Toda}}(\lda)= \frac{\Lambda_0^{(1)}}{\lambda^2} + \Lambda_\infty \in \Omega_{D^\prime}^\Gamma 
\end{equation}
where
\begin{equation}
\Lambda_0^{(1)} = \sum_{i=1}^T E_{i+1, i} \in \g^{(-1)}, \quad \Lambda_\infty = \sum_{i=1}^T E_{i, i+1} \in \g^{(1)}.
\end{equation}
The group elements $\varphi_+ =\left( \varphi_{0+}, \varphi_{\infty+} \right)$, defined by \eqref{eq:groupelement}, contain the dynamical degrees of freedom. From \eqref{eq:cgcoadorb}, we know that the components of $L_{\text{Toda}}(\lda)$ can now be expressed as
\begin{equation}\label{eq:todacoadorb}
    J_0^{(0)} = [\phi_0^{(1)} \phi_0^{(0)\, -1}, J_0^{(1)}], \quad J_0^{(1)} = \phi_0^{(0)} \Lambda_0^{(1)} \phi_0^{(0)\, -1}, \quad J_\infty = \Lambda_\infty.
\end{equation}
This gives us a parametrisation of the Lax matrix $L_{\text{Toda}}(\lda)$ in \eqref{eq:todalax} as an element of the coadjoint orbit ${\cal O}_\Lambda^{\text{Toda}}$. Since $\sgm (\varphi_{0+}(\lda)) = \varphi_{0+}(\omg\lda)$, we have $\sgm \phi_0^{(n)} = \omg^n \phi_0^{(n)}$. In particular, $\phi_0^{(0)}$ and $\phi_0^{(1)}$ have the form
\begin{equation}
    \phi_0^{(0)} = \sum_{i=1}^T u_i E_{ii}, \quad \phi_0^{(1)} = \sum_{i=1}^T v_i E_{i, i+1}.
\end{equation}
For convenience, define $(u_0, v_0) = (u_T, v_T)$ and $(u_{T+1}, v_{T+1}) = (u_1, v_1)$ to encode the periodic boundary conditions. Then, from \eqref{eq:todacoadorb}, we get
\begin{equation}
    J_0^{(0)} = \sum_{i=1}^T \left(\frac{v_i}{u_i} - \frac{v_{i-1}}{u_{i-1}} \right) E_{ii}, \quad J_0^{(1)} = \sum_{i=1}^T \dfrac{u_{i+1}}{u_i} E_{i+1, i}, \quad J_\infty = \sum_{i=1}^T E_{i, i+1}.
\end{equation}
Defining
\begin{equation}
    p_i = \frac{v_i}{u_i} - \frac{v_{i-1}}{u_{i-1}}  \quad \text{and} \quad  q^i = -\ln{u_i}, \quad \text{for}\,\, i = 1, \ldots, T,
\end{equation}
we now have the desired realisation of the coefficients of the Lax matrix $L_{\text{Toda}}(\lda)$. The coadjoint orbit ${\cal O}_\Lambda^{\text{Toda}}$ where $L_{\text{Toda}}(\lda)$ lives is parameterised by $a_i, p_i$, $i = 1, \ldots, T$, satisfying $\prod_i^T a_i = 1$ and $\sum_i^T p_i = 0$.

\paragraph{Lax equations:} Let us now look at the Lax equations associated with the invariant functions on $\Lg_D^{(1)}$ defined in \eqref{eq:cgpot} for the general case. The only functions we need to consider are
\begin{equation}
    H_{p, 0} = \dfrac{1}{p+1} \Res_{\lda = 0} (\lda^p \Tr(\iota_{\lda_0}L_{\text{Toda}}^{p+1}))d\lda.
\end{equation}
The set of functions $H_{p, \infty}$ are not independent: we have $H_{p, \infty} = - H_{p, 0}$ for all $p \geq 1$. The Lax equations for the periodic Toda chain with respect to the elementary times $t^{p, 0}$ are given by 
\begin{equation}
  \partial_{t^{p, 0}} \iota_\lda L_{\text{Toda}} = [R_\pm \nabla H_{p, 0}(\iota_\lda L_{\text{Toda}}), \iota_\lda L_{\text{Toda}}].
\end{equation}
For $p=1$, these take the form of the set of Lax equations in \eqref{eq:laxeqfirstflow}:
\begin{equation}\label{eq:todalaxeq}
    \partial_{t^{1, 0}} L_{\text{Toda}} = [M_{1, 0}, L_{\text{Toda}}] \quad \text{with} \quad M_{1, 0} = - \dfrac{J_0^{(1)}}{\lda}.
\end{equation}
Taking residues on both sides in \eqref{eq:todalaxeq} gives the equations of motion for $p_i, q^i$, for $i = 1, \ldots, T$. To get the equations of motion for $q^i$ it is most convenient to ``undo'' the dressing and write the corresponding equation as 
\begin{equation}
[\partial_{t^{1, 0}}  \phi_0^{(0)} \phi_0^{(0)\, -1}-J_0^{(0)}, J_0^{(1)}]=0.
\end{equation}
This tells us that the diagonal matrix $\phi_0^{(0)\, -1}\left(\partial_{t^{1, 0}}  \phi_0^{(0)} \phi_0^{(0)\, -1}-J_0^{(0)}\right)\phi_0^{(0)}$ must commute with $\Lambda_0^{(1)}$, and is therefore equal to $\alpha\1$. Using the freedom $\phi_0^{(0)}\to\phi_0^{(0)} g$, where $g$ is diagonal, to set $\det\phi_0^{(0)}=1$, we see that $\alpha=0$ by taking the trace of $\partial_{t^{1, 0}}  \phi_0^{(0)} \phi_0^{(0)\, -1}-J_0^{(0)}=\alpha\1$. Thus, we have the desired Toda equations
\begin{equation}\label{eq:todaeom}
\begin{split}
  &\partial_{t^{1, 0}} q^i = - p_i,  \\
&\partial_{t^{1, 0}} p_i = e^{q^i - q^{i+1}} - e^{q^{i-1} - q^i}.  
\end{split}
\end{equation}

\paragraph{Lagrangian description:} Using Theorem \ref{th:cgmultiform}, we can now write a Lagrangian one-form for the periodic Toda hierarchy as
\begin{equation}
  \Lag_{\text{Toda}} = \Lag_{p, 0} \mathrm{d}t^{p, 0},\quad p = 1, \ldots, M,
\end{equation}
with
\begin{equation}
    \Lag_{p, 0} = - p_i \partial_{t^{p, 0}} q^i - \dfrac{1}{p+1} \Res_{\lda = 0} (\lda^p \Tr(L_{\text{Toda}}^{p+1}))d\lda,\quad i = 1, \ldots, T.
\end{equation}

For $p=1$, this gives us the Lagrangian coefficients for the periodic Toda chain with respect to the time $t^{1, 0}$:
\begin{equation}
  \Lag_{1, 0} = - p_i \partial_{t^{1, 0}} q^i - \frac{1}{2} \sum_{i=1}^T p_i^2 - \sum_{i=1}^T e^{q^i - q^{i+1}}.
\end{equation}
This is the expected {\it phase-space} Lagrangian that our method produces and which gives Hamilton's equations \eqref{eq:todaeom} for periodic Toda. It corresponds to the (tangent bundle) Lagrangian $\displaystyle\Lag_{1, 0} =  \frac{1}{2}\sum_{i=1}^T (\partial_{t^{1, 0}} q^i)^2  - \sum_{i=1}^T e^{q^i - q^{i+1}}$.

\subsection{DST model}\label{sec:dst}

The discrete self–trapping (DST) equation was introduced in \cite{ELS} to describe the dynamics of small molecules, which then led to detailed studies of the DST dimer using different methods. The DST model we cast into our framework here is a generalisation of the dimer case to arbitrary (finite) degrees of freedom. This general case first appeared in \cite{CJK} where its relationship with the periodic Toda chain was also hinted at. Our motivation here being different, we do not delve into this connection between the two theories. The interested reader is referred to \cite{KSS} where this aspect was explored further.

\paragraph{Lax matrix:} We work here with the following avatar of the Lax matrix of the DST model
\begin{equation}
    L_{\text{DST}}(\lda) = \dfrac{1}{\lda}\sum_{i=1}^T c_i E_{ii} + \frac{1}{T} \sum_{i, j = 1}^T \sum_{k=0}^{T-1} \frac{\omg^{k(j-i)}x^i X_j E_{ij}}{\lda - \omg^k \zeta_1} + \sum_{i=1}^T E_{i, i+1}
\end{equation}
where $c_i$, for $i = 1, \ldots, T$, are complex parameters, and $x^i, X_i$ are the canonical coordinates satisfying the canonical Poisson bracket relations $\{X_i, x^j\} = \delta_{ij}$, for $i, j = 1, \ldots, T$, and the periodic conditions $(X_0, x^0) = (X_T, x_T)$ and $(X_{T+1}, x^{T+1}) = (X_1, x^1)$. 

The DST Lax matrix in \cite[Equation 3.8]{KSS} is given as
\begin{equation}
    \widehat{L}_{\text{DST}}(\mu) = \sum_{i, j =1}^T \frac{b^{T+i-j} x^i X_j E_{ij}}{\mu - b^T} + \mu E_{T1} + \sum_{i \geq j} b^{i-j}x^i X_j E_{ij} + \sum_{i=1}^T c_i E_{ii} + \sum_{i=1}^{T-1} E_{i, i+1} 
\end{equation}
where $b, c_i \in \mathbb{C}$ are parameters of the model, and $\mu$ is the spectral parameter of the Lax matrix. In the present setup where we realise the DST model as a cyclotomic Gaudin model, the role of the parameter $b$ is played by the location of the pole $\zeta_1$ on $\mathbb{C}P^1 \setminus \{0, \infty\}$. 
One obtains the Lax matrix $L_{\text{DST}}(\lda)$ by conjugating $\widehat{L}_{\text{DST}}(\lda^T)$ by the diagonal matrix $\mathcal{D} = \text{diag}(\lda^{-1}, \ldots, \lda^{-T})$ followed by an overall multiplication by $\lda^{-1}$, together with a change of $\lda$-dependence, that is,
\begin{equation}\label{eq:dstgaugetransform}
  L_{\text{DST}}(\lda) = \lda^{-1} \mathcal{D} \widehat{L}_{\text{DST}}(\lda^T) \mathcal{D}^{-1},
\end{equation}
and then using the identity \eqref{eq:omgidentity}.

The Poisson bracket of the Lax matrix $\widehat{L}_{\text{DST}}(\lda)$ is given as
\begin{equation}
  \{\widehat{L}_{\text{DST}\, \ti{1}}(\lda), \widehat{L}_{\text{DST}\, \ti{2}}(\mu) \} = [\widehat{r}_{\ti{12}}(\lda, \mu), \widehat{L}_{\text{DST}\, \ti{1}}(\lda)] - [\widehat{r}_{\ti{21}}(\mu, \lda), \widehat{L}_{\text{DST}\, \ti{2}}(\mu)]
\end{equation}
where
\begin{equation}
  \widehat{r}_{\ti{12}}(\lda, \mu) = \frac{1}{\mu - \lda} \left( \mu \sum_{i \leq j} + \lda \sum_{i > j} \right) E_{ij} \otimes E_{ji}.
\end{equation}
Using the identity \eqref{eq:omgidentity} once again, one finds that the gauge transformation \eqref{eq:dstgaugetransform} gives for $L_{\text{DST}}(\lda)$ the $r$-matrix \eqref{eq:nss-rmateij} associated with the cyclotomic $\mathfrak{gl}_T(\mathbb{C})$-Gaudin model, as we would have anticipated. We then have the Poisson bracket
\begin{equation}
    \{L_{\text{DST}\, \ti{1}}(\lda), L_{\text{DST}\, \ti{2}}(\mu) \} = [r_{\ti{12}}(\lda, \mu), L_{\text{DST}\, \ti{1}}(\lda) ] - [r_{\ti{21}}(\mu, \lda),  L_{\text{DST}\, \ti{2}}(\mu)].
\end{equation}
Similar to the case of the periodic Toda chain, the Lax matrix $L_{\text{DST}}(\lda)$ can be seen as a realisation of the cyclotomic Gaudin Lax matrix, this time with simple poles at the origin and all $\omg^k \zeta_1$, $k \in \{0, \ldots, T-1\}$, for some $\zeta_1 \in \mathbb{C}^{\times}$, and a double pole at infinity, that is,
\begin{equation}\label{eq:dstlax}
  L_{\text{DST}}(\lda) = \frac{K_0^{(0)}}{\lda} + \frac{1}{T} \sum_{k=0}^{T-1} \frac{ \sgm^k K_1 }{\lda - \omg^k \zeta_1}  + K_\infty
\end{equation}
where
\begin{equation}\label{eq:dstcoeff}
  K_0^{(0)} = \sum_{i=1}^T c_i E_{ii},  \quad K_1 = \sum_{i, j = 1}^T x^i X_j E_{ij}, \quad K_\infty = \sum_{i=1}^T E_{i, i+1}.
\end{equation}

\paragraph{Orbit realisation:} Set $D = \{0, 1, \infty\}$, and choose the non-dynamical element
\begin{equation}
    \Lambda_{\text{DST}}(\lda) = \frac{\Lambda_0^{(0)}}{\lambda} + \frac{1}{T} \sum_{k=0}^{T-1} \frac{\sgm^k \Lambda_1}{\lda - \omg^k \zeta_1} + \Lambda_\infty \in \Omega_{D^\prime}^\Gamma
\end{equation}
where
\begin{equation}
\Lambda_0^{(0)} = \sum_{i=1}^T c_i E_{ii} \in \g^{(0)}, \quad \Lambda_1 = E_{11} \in \g, \quad \Lambda_\infty = \sum_{i=1}^T E_{i, i+1} \in \g^{(1)}.
\end{equation} 
The coadjoint action of the group elements $\varphi_+ =\left( \varphi_{0+}, \varphi_{1+}, \varphi_{\infty+} \right)$, defined by \eqref{eq:groupelement}, on $\Lambda_{\text{DST}}(\lda)$ gives the orbit where $L_{\text{DST}}(\lda)$ lives. From \eqref{eq:cgcoadorb}, we know that the components of $L_{\text{DST}}(\lda)$ can then be expressed as
\begin{equation}\label{eq:dstcoadorb}
    K_0^{(0)} = \phi_0^{(0)} \Lambda_0^{(0)} \phi_0^{(0)\, -1}, \quad K_1 = \phi_1 \Lambda_1 \phi_1^{-1}, \quad K_\infty = \Lambda_\infty,
\end{equation}
where we have denoted $\phi_1^{(0)}$ by $\phi_1$ for simplicity. 
This gives us a parameterisation of the Lax matrix $L_{\text{DST}}(\lda)$ in \eqref{eq:dsttodalax} as an element of the coadjoint orbit ${\cal O}_\Lambda^{\text{DST}}$. We parameterise $\phi_1$ as
\begin{equation}
    \phi_1 = \sum_{i,j=1}^T s_{ij} E_{ij},
\end{equation}
and denote by $\hat{s}_{ij}$, $i, j = 1, \ldots, T$, the entries of its inverse, that is,
\begin{equation}
    \phi_1^{-1} = \sum_{i,j=1}^T \hat{s}_{ij} E_{ij}.
\end{equation}
From \eqref{eq:dstcoadorb}, we then have
\begin{equation}
  K_0^{(0)} = \sum_{i=1}^T c_i E_{ii},  \quad K_1 = \sum_{i,j=1}^T  s_{i1} \hat{s}_{1j} E_{ij}, \quad K_\infty = \sum_{i=1}^T E_{i, i+1}.
\end{equation}
Defining
\begin{equation}
    x^i = s_{i1} \quad \text{and} \quad X_i = \hat{s}_{1i}, \quad \text{for}\,\, i = 1, \ldots, T,
\end{equation}
gives us the desired realisation of the Lax matrix $L_{\text{DST}}(\lda)$ as an element of the coadjoint orbit ${\cal O}_\Lambda^{\text{DST}}$. Notice that $\Tr K_1 = \sum_{i=1}^T x^i X_i = 1$ is an orbit invariant and can be seen as being generated by the symmetry $x^i \rightarrow a x^i$, $X_i \rightarrow a^{-1} X_i$.

\paragraph{Lax equations:} Let us choose invariant functions $H_{p, r}$ on $\Lg_D^{(1)} $ as defined in \eqref{eq:cgpot}, and treat $H_{p, 0}$ and $H_{p, 1}$ as the independent functions. The Lax equations for the DST model with respect to the elementary times $t^{p, r}$ are then given by 
\begin{equation}
  \partial_{t^{p, r}} \iota_\lda L_{\text{DST}} = [R_\pm \nabla H_{p, r}(\iota_\lda L_{\text{DST}}), \iota_\lda L_{\text{DST}}].
\end{equation}
For $p=1$, these take the form of the set of Lax equations in \eqref{eq:laxeqfirstflow}:
\begin{equation}\label{eq:dstlaxeq}
    \partial_{t^{1, r}} L_{\text{DST}} = [M_{1, r}, L_{\text{DST}}] \quad \text{with} \quad M_{1, r} =
    \begin{dcases}
      \quad 0 & \text{for}\,\, r = 0\\
      - \dfrac{1}{T} \sum_{k=0}^{T-1} \dfrac{\omg^k \zeta_1 \sgm^k K_1}{\lda - \omg^k \zeta_1} & \text{for}\,\, r = 1.
    \end{dcases}
\end{equation}
The $t^{1, 0}$ equations are trivial. Similar to the case of the periodic Toda chain, the easiest way to get the $t^{1, 1}$ equations of motion for $x^i$ and $X_i$, for $i = 1, \ldots, T$, is to undo the dressing. With $M=K_0^{(0)}+\zeta_1K_\infty+\dfrac{1}{T} \sum_{k=1}^{T-1} \sgm^k K_1$, we find that $ \phi_1^{-1}(\partial_{t^{1, 1}}\phi_1 \phi_1^{-1}-M)\phi_1$ must commute with $\Lambda_1$, and hence be block diagonal with a scalar ``block'', say $\rho$, and a $(T-1)\times (T-1)$ block which is irrelevant. Then we find, collecting $x^i$ in the vector ${\bf x}$, and $X_i$ in the vector ${\bf X}$,
\begin{equation}
\partial_{t^{1, 1}}{\bf x}-M{\bf x}=\rho {\bf x}, \quad \partial_{t^{1, 1}}{\bf X}^T-{\bf X}^TM=\rho {\bf X}^T.
\end{equation}
Using the freedom $x^i \rightarrow a x^i$, $X_i \rightarrow a^{-1} X_i$ mentioned above with $a=e^B$, $\partial_{t^{1, 1}}B=\rho$, we can set $\rho=0$. Explicitly, we obtain
\begin{equation}\label{eq:dsteom}
  \begin{cases}
  \partial_{t^{1, 0}} X_i = 0, \\[1.5ex]

  \partial_{t^{1, 0}} x^i = 0, \\[1.5ex]

  \partial_{t^{1, 1}} X_i = - c_i X_i - \zeta_1 X_{i-1} - \dfrac{1}{T} \sum\limits_{k=1}^{T-1} \sum\limits_{j=1}^T \omg^{k(j-i)}x^j X_j X_i, \\[1.5ex]
  
  \partial_{t^{1, 1}} x^i = c_i x^i + \zeta_1 x^{i+1} + \dfrac{1}{T} \sum\limits_{k=1}^{T-1} \sum\limits_{j=1}^T \omg^{k(j-i)} X_j x^j x^i.
  \end{cases}
\end{equation}

\paragraph{Lagrangian description:} We can now write a Lagrangian one-form for the DST hierarchy using Theorem \ref{th:cgmultiform} as
\begin{equation}
  \Lag_{\text{DST}} = \Lag_{p, 0} \mathrm{d}t^{p, 0} + \Lag_{p, 1} \mathrm{d}t^{p, 1},\qquad p =1 , \ldots, M, 
\end{equation}
with
\begin{equation}\label{eq:dstlagcoeff}
    \Lag_{p, r} = X_i \partial_{t^{p, r}} x^i - H_{p, r}(\iota_\lda L_{\text{DST}}),\qquad i = 1, \ldots, T,\quad r \in \{0, 1\},
\end{equation}
where the potential term $H_{p, r}(\iota_\lda L_{\text{DST}})$ is given by \eqref{eq:cgpot}, for $r \in \{0, 1\}$. Notice that we have dropped the kinetic contribution to $\Lag_{p, r}$ from the pole at origin since being a total derivative it will not enter the Euler--Lagrange equations.
For $p=1$, the Lagrangian coefficients explicitly read
\begin{equation}
  \begin{split}
    &\Lag_{1, 0} = X_i \partial_{t^{1, 0}} x^i - \dfrac{1}{2} \sum_{i=1}^T c_i^2,\\
    &\Lag_{1, 1} = X_i \partial_{t^{1, 1}} x^i - \dfrac{1}{2T} \sum_{k=0}^{T-1} \omg^{k(j-i)} x^i x^j X_i X_j - c_i x^i X_i - \zeta_1 x^{i+1} X_i,
  \end{split}
  \end{equation}
for $i, j = 1, \ldots, T$. It can be checked that the Euler--Lagrange equations obtained from varying $\Lag_{1, 0}$ and $\Lag_{1, 1}$ with respect to $X_i, x^i$ are exactly the equations in \eqref{eq:dsteom}.

\subsection{Coupled Toda--DST system}\label{sec:todadst}

Finally, we turn to the task of coupling together the two hierarchies we have described variationally in this section. The Lax matrix of the coupled hierarchy can be expressed as
\begin{equation}\label{eq:dsttodalax}
    L_{\text{Toda--DST}}(\lda) = L_{\text{Toda}}(\lda) + \beta L_{\text{DST}}(\lda),
\end{equation}
where $L_{\text{Toda}}(\lda)$ is the Lax matrix of the periodic Toda chain in \eqref{eq:todalax}, $L_{\text{DST}}(\lda)$ is the DST Lax matrix \eqref{eq:dstlax}, and the parameter $\beta$ is a real-valued scalar parameter dictating the strength of coupling between the two hierarchies. Naturally, the Lax matrix $L_{\text{Toda--DST}}(\lda)$ can be seen as a realisation of the cyclotomic Gaudin Lax matrix:
\begin{equation}
  L_{\text{Toda--DST}}(\lda) = \frac{J_0^{(0)} + \beta K_0^{(0)}}{\lda} + \frac{J_0^{(1)}}{\lda^2} + \frac{\beta}{T} \sum_{k=0}^{T-1} \frac{ \sgm^k K_1 }{\lda - \omg^k \zeta_1} + J_\infty + \beta K_\infty
\end{equation}
with the $\g$-valued coefficients defined in \eqref{eq:todacoadorb} and \eqref{eq:dstcoeff}.

As the number of finite (non-zero) poles in a cyclotomic Gaudin model is arbitrary, our construction allows us, in principle, to couple together an arbitrary number of copies of the DST model and a copy of the periodic Toda chain. Here we only illustrate it for a single copy each of the periodic Toda chain and the DST model. This simple mechanism for coupling together integrable hierarchies with Lax matrices living on coadjoint orbits can be applied if they have the same underlying Lie algebra. Readers interested in the application of this mechanism in the field-theoretic case are referred to \cite[Section 7]{CStV}.

\paragraph{Orbit realisation:} Set $D = \{0, 1, \infty\}$. Since we already have a parameterisation of the Lax matrices $L_{\text{Toda}}(\lda)$ and $L_{\text{DST}}(\lda)$ as orbit elements, we only need to check that the action of a generic group element $\varphi_+ =\left( \varphi_{0+}, \varphi_{1+}, \varphi_{\infty+}\right) \in \LG_{D+}^{(0)}$ defined by \eqref{eq:groupelement} on the non-dynamical element
\begin{equation}
    \Lambda_{\text{Toda--DST}}(\lda) = \Lambda_{\text{Toda}}(\lda) + \beta \Lambda_{\text{DST}}(\lda) \in \Omega_{D^\prime}^\Gamma
\end{equation}
where
\begin{equation}
    \Lambda_{\text{Toda}}(\lda)= \frac{\Lambda_0^{(1)}}{\lambda^2} + \Lambda_\infty, \quad \Lambda_{\text{DST}}(\lda) = \frac{\Lambda_0^{(0)}}{\lambda} + \frac{1}{T} \sum_{k=0}^{T-1} \frac{\sgm^k \Lambda_1}{\lda - \omg^k \zeta_1} + \Lambda_\infty
\end{equation}
results in the Lax matrix $L_{\text{Toda--DST}}(\lda)$. Indeed, using the result in \eqref{eq:cgcoadorb}, we have that the components of $L_{\text{Toda--DST}}(\lda)$ take the form
\begin{equation}\label{eq:todadstcoadorb}
  \begin{split}
    &J_0^{(0)} + \beta K_0^{(0)} = [\phi_0^{(1)} \phi_0^{(0)\, -1}, J_0^{(1)}] + \beta \phi_0^{(0)} \Lambda_0^{(0)} \phi_0^{(0)\, -1},\\
    &J_0^{(1)} = \phi_0^{(0)} \Lambda_0^{(1)} \phi_0^{(0)\, -1},\\
    &\beta K_1 = \beta \phi_1 \Lambda_1 \phi_1^{-1}, \quad \text{and} \\
    &J_\infty + \beta K_\infty = \Lambda_\infty + \beta \Lambda_\infty,
  \end{split}
\end{equation}
as desired. This gives us a parameterisation of the Lax matrix $L_{\text{Toda--DST}}(\lda)$ in \eqref{eq:dsttodalax} as an element of the coadjoint orbit ${\cal O}_\Lambda^{\text{Toda}} \times {\cal O}_\Lambda^{\text{DST}}$.

\paragraph{Lax equations:} As earlier, we will choose invariant functions $H_{p, r}$ on $\Lg_D^{(1)} $ as defined in \eqref{eq:cgpot}, for $r \in \{0, 1\}$. The Lax equations for the DST-Toda model with respect to the elementary times $t^{p, r}$ are given by 
\begin{equation}
  \partial_{t^{p, r}} \iota_\lda L_{\text{Toda--DST}} = [R_\pm \nabla H_{p, r}(\iota_\lda L_{\text{Toda--DST}}), \iota_\lda L_{\text{Toda--DST}}].
\end{equation}
For $p=1$, these take the form of the set of Lax equations in \eqref{eq:laxeqfirstflow}:
\begin{equation}\label{eq:todadstlaxeq}
    \partial_{t^{1, r}} L_{\text{Toda--DST}} = [M_{1, r}, L_{\text{Toda--DST}}]
\end{equation}
with
\begin{equation}
   M_{1, r} =
    \begin{dcases}
      - \dfrac{J_0^{(1)}}{\lda} &\text{for}\,\, r = 0\\
      - \dfrac{\beta}{T} \sum_{k=0}^{T-1} \dfrac{\omg^k \zeta_1 \sgm^k K_1}{\lda - \omg^k \zeta_1} & \text{for}\,\, r = 1.
    \end{dcases}
\end{equation}
Taking residues on both sides in \eqref{eq:todadstlaxeq} gives the equations of motion for $p_i, q^i, X_i, x^i$, for $i = 1, \ldots, T$. Like in the cases of the periodic Toda chain and the DST model, to get the equations of motion for $q^i, X_i, x^i$, one can ``undo'' the dressing in the corresponding equations. Explicitly, one gets the following equations for $p_i, q^i$, for $i = 1, \ldots, T$:
\begin{equation}\label{eq:todaeombeta}
  \begin{cases}
  \partial_{t^{1, 0}} p_i = (1 + \beta)(e^{q^i - q^{i+1}} - e^{q^{i-1} - q^i}) + \dfrac{\beta}{\zeta_1} (e^{q^{i-1} - q^i} x_{i-1} X_i - e^{q^i - q^{i+1}} x^i X_{i+1}), \\[1.5ex]
  
  \partial_{t^{1, 0}} q^i = - p_i - \beta c_i, \\[1.5ex]
  
  \partial_{t^{1, 1}} p_i = \dfrac{\beta}{\zeta_1} (e^{q^i - q^{i+1}} x^i X_{i+1} - e^{q^{i-1} - q^i} x_{i-1} X_i), \\[1.5ex]
  
  \partial_{t^{1, 1}} q^i = - \beta x^i X_i,
  \end{cases}
\end{equation}
and the following equations for $X_i, x^i$, for $i = 1, \ldots, T$:
\begin{equation}\label{eq:dsteombeta}
\begin{cases}
  \partial_{t^{1, 0}} X_i = \dfrac{1}{\zeta_1} e^{q^i - q^{i+1}} X_{i+1}, \\[1.5ex]
  
  \partial_{t^{1, 0}} x^i = - \dfrac{1}{\zeta_1} e^{q^{i-1} - q^i}x_{i-1}, \\[1.5ex]

  \partial_{t^{1, 1}} X_i = - p_i X_i \mkern-2mu - \mkern-2mu \beta c_i X_i \mkern-2mu - \mkern-2mu \dfrac{1}{\zeta_1} e^{q^i - q^{i+1}} X_{i+1} \mkern-2mu - \mkern-2mu \dfrac{\beta}{T} \sum\limits_{j=1}^T \sum\limits_{k=0}^{T-1} \omg^{k(j-i)}  x^j X_j X_i \mkern-2mu - \mkern-2mu (1 + \beta) \zeta_1 X_{i-1},\\[1.5ex]

  \partial_{t^{1, 1}} x^i = p_i x^i + \beta c_i x^i + \dfrac{1}{\zeta_1} e^{q^{i-1} - q^i}x_{i-1} + \dfrac{\beta}{T} \sum\limits_{j=1}^T \sum\limits_{k=0}^{T-1} \omg^{k(j-i)} X_j x^j x^i + (1 + \beta) \zeta_1 x^{i+1}.
\end{cases}
\end{equation}

Setting $\beta = 0$ in the above equations produces the equations of motion in \eqref{eq:todaeom} for the periodic Toda chain. In the limit $\beta \rightarrow \infty$, these reduce to the equations of motion we obtained for the DST model in \eqref{eq:dsteom}. To see this, note that the Lax matrix $L_{\text{DST}}$ comes as $\beta L_{\text{DST}}$ in the coupled system. Therefore, in the limit $\beta \rightarrow \infty$, the time flow $t^{p, r}$ is rescaled such that $\partial_{t^{p, r}}$ rescales to $\beta^p \partial_{t^{p, r}}$.

\paragraph{Lagrangian description:} Using Theorem \ref{th:cgmultiform}, we can now write a Lagrangian one-form for the Toda--DST hierarchy:
\begin{equation}
  \Lag_{\text{Toda--DST}} = \Lag_{p, 0} \mathrm{d}t^{p, 0} + \Lag_{p, 1} \mathrm{d}t_p^1,\qquad p =1 , \ldots, M,
\end{equation}
with
\begin{equation}
  \begin{split}
  &\Lag_{p, r} = \Tr (J_0^{(0)} \partial_{t^{p, r}} \phi_0^{(0)} \phi_0^{(0)\, -1}) + \beta \Tr (K_0^{(0)} \partial_{t^{p, r}} \phi_0^{(0)} \phi_0^{(0)\, -1})\\
  &\qquad \quad + \beta \Tr (K_1 \partial_{t^{p, r}} \phi_1 \phi_1^{-1}) - H_{p, r}(\iota_\lda L_{\text{Toda--DST}}),\quad r \in \{0, 1\},
  \end{split}
\end{equation}
where the potential term $H_{p, r}(\iota_\lda L_{\text{Toda--DST}})$ is given by \eqref{eq:cgpot} for $r \in \{0, 1\}$. Notice that $K_0^{(0)} = \sum_{i=1}^T c_i E_{ii}$. So, the second term on the right-hand side is, in fact, a total derivative and will not enter the Euler--Lagrange equations. So, we will simply drop it. In terms of the canonical coordinates, we then have
\begin{equation}\label{eq:dtlagcoeff}
    \Lag_{p, r} = - p_i \partial_{t^{p, r}} q^i + \beta X_i \partial_{t^{p, r}} x^i - H_{p, r}(\iota_\lda L_{\text{Toda--DST}}),\qquad i = 1, \ldots, T,\quad r \in \{0, 1\}.
\end{equation} 
The decoupled limits of the periodic Toda and the DST hierarchies are easily recovered by setting $\beta = 0$ and taking the limit $\beta \rightarrow \infty$ in \eqref{eq:dtlagcoeff} respectively.

For $p=1$, \eqref{eq:dtlagcoeff} gives the Lagrangian for the coupling of the periodic Toda chain and the DST model with the potential terms
\begin{equation}
\begin{split}
    &H_{1, 0} = \frac{1}{2} \sum_{i=1}^T p_i^2 + \beta \sum_{i=1}^T c_i p_i + \frac{\beta^2}{2} \sum_{i=1}^T c_i^2 + (1 + \beta) \sum_{i=1}^T e^{q^i - q^{i+1}} - \frac{\beta}{\zeta_1} e^{q^i - q^{i+1}} x^i X_{i+1}, \\
    &H_{1, 1} = \frac{\beta^2}{2T} \sum_{k=0}^{T-1} \omg^{k(j-i)} x^i x^j X_i X_j + \beta \sum_{i=1}^T p_i x^i X_i + \beta^2 c_i x^i X_i \\
    &\qquad \quad + \frac{\beta}{\zeta_1}  e^{q^i - q^{i+1}} x^i X_{i+1} + (\beta + \beta^2) \zeta_1 x^{i+1} X_i.
\end{split}
\end{equation}
The variation $\delta \Lag_{1, r}$ gives the Euler--Lagrange equations for the coupled system with respect to the time flow $t^{1, r}$, for each $r \in \{0, 1, \infty\}$. It can be checked that these are exactly the equations we obtained in \eqref{eq:todaeombeta}--\eqref{eq:dsteombeta}.

We have reached the end of Part \ref{part:coadjointmultiform}. We will now change tracks, and instead of the rather algebraic approach that allowed us to construct geometric Lagrangian one-forms living on coadjoint orbits, we will use a gauge-theoretic approach to cast a more general class of integrable models into the variational framework of Lagrangian multiforms.

\newpage
\thispagestyle{empty}
\mbox{}
\newpage

\part{Gauge theory and Lagrangian multiforms}\label{part:gaugemultiform}
\thispagestyle{empty}

\newpage
\thispagestyle{empty}
\mbox{}
\newpage

\chapter{3d mixed BF theory and Hitchin systems}\label{chap:bf-hitchin}
In 1987, Hitchin introduced a construction \cite{H} that provides integrable systems associated with the moduli space of stable holomorphic bundles over Riemann surfaces. This construction and its generalisations introduced in subsequent works (see, for instance, \cite{Ma, Bot, Ne, ER}) encompass a large class of classical integrable systems. Hitchin systems, as they have come to be known, have been studied extensively within the Hamiltonian framework leading to interesting discoveries that connect them to several areas of physics and mathematics.\footnote{We refer the interested reader to \cite{Schap1} for an overview of some of these connections.}

The shortcoming of \emph{traditional} Lagrangians in capturing the notion of integrable hierarchies --- as they only describe individual models --- becomes particularly detrimental in the case of Hitchin systems. Hitchin's construction provides a set of functionally independent Poisson-commuting conserved quantities on a suitable phase space: a Hitchin \emph{system} therefore represents an integrable \emph{hierarchy} by itself. This makes the traditional Lagrangian formalism ill-equipped for describing a Hitchin system variationally.

This part of the thesis, adapted from the joint work \cite{CHSV}, is dedicated to a solution to this problem through the construction of geometric Lagrangian one-forms for Hitchin systems associated with Riemann surfaces of arbitrary genus. The Lagrangian one-forms we construct in this part thus generalise the geometric Lagrangian one-forms on coadjoint orbits that were the subject of Part \ref{part:coadjointmultiform} of this thesis and first introduced in \cite{CDS}. In the process, we are also led to a Lagrangian multiform for \emph{$3$-dimensional mixed holomorphic-topological BF theory}\footnote{BF theory is named as such because its action is given in terms of a field $B$ and the curvature $F$ of a field $A$. One of the earliest occurrences of (topological) BF theory can be found in \cite{Hor}. In contrast to this earlier avatar of the theory, the BF theory of interest here looks topological in some directions and holomorphic in others, and is therefore $mixed$ in this sense.} (hereafter simply referred to as \emph{$3$d mixed BF theory}) introduced in \cite{VW}. Also see \cite{Ze, GW, GRW}. This connects us to another recent approach to integrability, also taking a Lagrangian perspective, based on so-called \emph{mixed holomorphic-topological (HT) gauge theories}.\footnote{In connection with combining the framework of Lagrangian multiforms with gauge-theoretic ideas, we note another recent work \cite{FNR}. However, it is based on an approach different from mixed HT gauge theories. In this work, Lagrangian multiforms for the Darboux--Kadomtsev--Petviashvili system were derived from a hierarchy of Chern--Simons actions in an infinite-dimensional space of Miwa variables.} It started with the work of Costello \cite{Cos}, which was further developed with Witten and Yamazaki in \cite{CosWY1, CosWY2}, demonstrating a relation between integrable lattice models and a $4$-dimensional semi-holomorphic variant of Chern--Simons theory in the presence of line defects. The main idea behind this gauge-theoretic approach to integrability is that the Lax formalism of an integrable model on a $d$-dimensional manifold $M$, depending on a spectral parameter living on a Riemann surface $C$, arises from the equations of motion of a suitable gauge theory on $M \times C$ which is holomorphic along $C$ and (possibly a mixture of holomorphic and) topological along $M$.

In the years since its inception, this approach has been extensively studied and generalised, most notably to the case of integrable field theories in $1+1$ dimensions. In \cite{CosY}, and many subsequent works (see, for instance, \cite{DLMV, BL, HL, LaV, CStV2}), $(1+1)$-dimensional integrable field theories were shown to arise from \emph{surface defects} in the same $4$-dimensional Chern--Simons theory. More recently, in \cite{SchV}, a $5$-dimensional semi-holomorphic \emph{higher} Chern--Simons theory was used to extend this framework to $3$-dimensional integrable field theories. This perspective has also been linked to an older gauge-theoretic approach to integrability in which integrable field theories arise as symmetry reduction of the anti-self-dual Yang--Mills (ASDYM) equations \cite{MasW} by deriving these integrable field theories from various defects in $6$-dimensional holomorphic Chern--Simons theory on twistor space \cite{BitS, CCHLT1, CCHLT2}. However, as alluded to earlier, the work most relevant here is \cite{VW} where a relationship between $3$-dimensional mixed BF theory with line defects and finite (rational) Gaudin model was established.

This chapter is structured as follows. In Section \ref{sec: geometric setup}, we discuss the geometric setup for Hitchin systems, thereby setting our conventions. Then, in Section \ref{sec: Lag for Hitchin}, we generalise the construction of Section \ref{gauged_univariational_principle} to an infinite-dimensional context and provide a variational description of the Hitchin system. In Section \ref{sec: Lag for Hitchin mod G}, we implement a symplectic reduction to obtain a Lagrangian one-form for the Hitchin system on a certain symplectic quotient. This turns out to be the Lagrangian one-form for $3$d mixed BF theory with type B defects\footnote{We use the terms ``type B defect'' and ``type A defect'' in line with \cite{VW} --- these defects depend on the $B$ field and the $A$ field respectively.}. Finally, in Section \ref{sec: adding punctures}, we go on to extend the construction to the case of Hitchin systems on compact Riemann surfaces with marked points, which leads to a Lagrangian one-form for $3$d mixed BF theory with type B and type A defects.

\section{Geometric setup for Hitchin systems}\label{sec: geometric setup}

To set the stage for the rest of the chapter, let us start by discussing the case of the Hitchin system on a Riemann surface $C$ of arbitrary genus $g \geq 2$ \emph{without} marked points.\footnote{Detailed discussions on the geometric setup for Hitchin systems in the case without marked points and related algebro-geometric notions can be found, for instance, in \cite{DonM}, \cite[Section III, Chapter 1]{KheWen}, and \cite{Schap2}.} The case with marked points is dealt with later in Section \ref{sec: adding punctures}.

The phase space of the Hitchin system for the case without marked points is given by the cotangent bundle $T^\ast \text{Bun}_G(C)$. A point in the base $\text{Bun}_G(C)$ is a stable holomorphic principal $G$-bundle $\mathcal P_{\rm hol}$, of a fixed topological type, which can always be described using a single holomorphic transition function $\gamma : U_0 \cap U_1 \to G$ relative to an open cover $\{ U_0, U_1 \}$ of $C$ with $U_0$ an open neighbourhood of a fixed point $\mathsf p \in C$ and $U_1 \coloneqq C \setminus \{ \mathsf p \}$. A point in the fibre $T^\ast_{\mathcal P_{\rm hol}} \text{Bun}_G(C)$ above $\mathcal P_{\rm hol}$ is called a \emph{holomorphic Higgs field} $L$. On identifying $\g$ with its dual $\g^\ast$ using a fixed nondegenerate invariant bilinear pairing on $\g$, it is given explicitly by a pair of $\g$-valued $(1,0)$-forms $L^0$ and $L^1$ on $U_0$ and $U_1$, respectively, which are related via the adjoint action $L^0 = \gamma L^1 \gamma^{-1}$ on the overlap $U_0 \cap U_1$. The pair $(\mathcal P_{\rm hol}, L)$ is an example of a \emph{stable Higgs bundle}. The \emph{Hitchin map}, also known as the \emph{Hitchin fibration}, provides a complete set of Poisson-commuting Hamiltonians $H_i$ for $i =1,\ldots, n$, where $n$ denotes half the dimension of the phase space, thereby encoding a finite-dimensional integrable hierarchy. 

These Hamiltonians, which we will now refer to as \emph{Hitchin Hamiltonians}, induce commuting time flows $\partial_{t^i}$, $i=1,\ldots, n$, on $T^\ast \text{Bun}_G(C)$, whose action on the pair $(L^0, \gamma)$ is given by
\begin{equation} \label{Hitchin flow intro}
\partial_{t^i} L^0 = [M^0_i, L^0], \qquad
M^0_i = \gamma M^1_i \gamma^{-1} + \partial_{t^i} \gamma \gamma^{-1},
\end{equation}
for $i =1, \ldots, n$, where $M^1_i$ are $\g$-valued meromorphic functions on $U_1$ each with a simple pole at a fixed marked point $\mathsf q_i \in U_1$ whose residue there is determined by the Hitchin Hamiltonian $H_i$, and $M^0_i$ are $\g$-valued holomorphic functions on $U_0$. 

To give a variational description of the hierarchy of commuting flows given by \eqref{Hitchin flow intro}, we will exploit the fact that the Hitchin phase space $T^\ast \text{Bun}_G(C)$ can be obtained as a symplectic reduction of the infinite-dimensional cotangent bundle $T^\ast \mathcal M$ of the space $\mathcal M$ of stable holomorphic structures on a fixed smooth principal $G$-bundle $\mathcal P$ by the action of the group $\mathcal G = \Aut\mathcal P$ of smooth bundle automorphisms of $\mathcal P$.\footnote{That this phase space is finite-dimensional is a consequence of a theorem of Narasimhan and Seshadri \cite{NarSesh}. See, for instance, \cite[Chapter 7]{BBT}.} A holomorphic structure on $\mathcal P$ can be specified in terms of a partial $(0,1)$-connection\footnote{A partial connection is a connection that can only take derivatives in certain directions.} $A''$ on $\mathcal P$ given in local coordinate patches by $\g$-valued $(0,1)$-forms on $C$. A smooth bundle automorphism $g \in \mathcal G$ acts on such a holomorphic structure $A''$ by gauge transformations $A'' \mapsto g A'' g^{-1} - \bar\partial g g^{-1}$. Two holomorphic structures related in this way define the same holomorphic principal $G$-bundle $\mathcal P_{\rm hol}$. In particular, the action of $\mathcal G$ on the space $\mathcal M$ of stable holomorphic structures is free, and we have an isomorphism $\mathcal M / \mathcal G \cong \text{Bun}_G(C)$. Moreover, the action of $\mathcal G$ on $\mathcal M$ lifts to a Hamiltonian action on $T^\ast \mathcal M$ with moment map given by $\mu(B, A'') = \bar\partial^{A''} B$, where $B$ denotes the smooth Higgs field\footnote{We use the symbol $B$ to denote the Higgs field instead of the symbol $\Phi$ which is more standard in works on Hitchin systems, as it will later be identified with the corresponding $B$ field of $3$d mixed BF theory.} parametrising the fibre of $T^\ast_{A''} \mathcal M$ over a fixed holomorphic structure $A'' \in \mathcal M$.

As the construction of geometric Lagrangian one-forms for Hitchin systems relies on some intricacies of the definition of $T^\ast \mathcal M$ and $\mathcal G$, we devote the rest of this section to the relevant details.

\subsection{Holomorphic structures on a principal \texorpdfstring{$G$}{G}-bundle} \label{sec: M and G details}

Let $G$ be a complex connected Lie group with Lie algebra $\g$. Denote by $C$ be a compact Riemann surface and fix a holomorphic atlas $\{ (U_I, z_I) \}_{I \in \mathcal I}$ of $C$ with $z_I : U_I \to \CC$ local holomorphic coordinates on each open subset $U_I \subset C$ and $\mathcal I$ some indexing set. 

We fix a smooth principal $G$-bundle $\pi :\mathcal P\to C$ which is specified relative to the open cover $\{ U_I \}$ of $C$ by local trivialisations $\psi_I : \pi^{-1}(U_I) \SimTo U_I \times G$, $p \mapsto (\pi(p), f_I(p))$. The principal bundle is equipped with a free right action $G \times \mathcal P \to\mathcal P$, $p \mapsto p \cdot g$, and the local trivialisations should be $G$-equivariant, that is, $f_I(p \cdot g) = f_I(p) g$ for any $g \in G$. The transition between local trivialisations $\psi_J$ and $\psi_I$ on overlapping charts $U_I \cap U_J \neq \emptyset$ is given by
\begin{equation}
\psi_I \circ \psi_J^{-1} : (U_I \cap U_J) \times G \, \longrightarrow \, (U_I \cap U_J) \times G ,\quad (x, g) \,\longmapsto\, \big( x, g_{IJ}(x) g \big)
\end{equation}
with smooth transition functions $g_{IJ} : U_I \cap U_J \to G$, given by $g_{IJ}(x) = f_I(p) f_J(p)^{-1}$ for any $p \in\mathcal P$ with $\pi(p) = x \in U_I \cap U_J$, satisfying the \v{C}ech cocycle condition
\begin{equation}
  g_{IJ} g_{JK} = g_{IK}
\end{equation}
on triple overlaps $U_I \cap U_J \cap U_K \neq \emptyset$. A \emph{change of local trivialisation} of $\mathcal P$ is specified by a family of smooth maps $h_I : U_I \to G$, that is, a \v{C}ech $0$-cochain $h = (h_I)_{I \in \mathcal I} \in \check{C}^0(C, G)$.
Indeed, given local trivialisations $\psi_I : \pi^{-1}(U_I) \SimTo U_I \times G$, $p \mapsto (\pi(p), f_I(p))$, we can define new local trivialisations by
\begin{equation} \label{new trivialisations}
\tilde \psi_I : \pi^{-1}(U_I) \overset{\cong}\longrightarrow U_I \times G , \quad p \longmapsto \big( \pi(p), h_I(\pi(p)) f_I(p) \big).
\end{equation}
The transition functions of $\mathcal P$ relative to these new local trivialisations are the smooth maps
\begin{equation} \label{gauge transf compat}
\tilde g_{IJ} = h_I g_{IJ} h_J^{-1} : U_I \cap U_J \to G .
\end{equation}
The change of local trivialisations from $\{ \psi_I \}_{I \in \mathcal I}$ to $\{ \tilde \psi_I \}_{I \in \mathcal I}$ on the fixed bundle $\mathcal P$ can equivalently be seen as producing a new principal $G$-bundle $\tilde {\mathcal P} \to C$ that is smoothly isomorphic to $\mathcal P$.

An automorphism of $\mathcal P$, or more precisely a \emph{fibre-preserving automorphism} of $\mathcal P$ which we will sometimes refer to as a \emph{gauge transformation}, is a \v{C}ech $0$-cochain $g = (g_I)_{I \in \mathcal I} \in \check{C}^0(C, G)$ which preserves the transition functions of $P$ in the sense that
\begin{equation} \label{bundle morphism compat}
g_I = g_{IJ} g_J g_{IJ}^{-1}
\end{equation}
on any overlap $U_I \cap U_J \neq \emptyset$.
We can describe the action of $g$ on $\mathcal P$ relative to a fixed choice of local trivialisations $\{\psi_I \}_{I \in \mathcal I}$ as sending $\psi_I(p) = (\pi(p), f_I(p))$ to $\psi_I(g \cdot p) \coloneqq \big( \pi(p), g_I(\pi(p)) f_I(p) \big)$. The compatibility condition \eqref{bundle morphism compat} ensures that this is well-defined on $\mathcal P$, in the sense that we can either perform the gauge transformation directly in the local trivialisation $\psi_I$, or we can first move to the local trivialisation $\psi_J$, perform the gauge transformation there and then move back to the local trivialisation $\psi_I$. Both give the same result. We let
\begin{equation} \label{cal G def}
\mathcal G \coloneqq \Aut \mathcal P \subset \check{C}^0(C, G)
\end{equation}
denote the infinite-dimensional group of automorphisms of the principal $G$-bundle $\mathcal P$. Note that, by the condition \eqref{bundle morphism compat}, we can equally describe automorphisms of $\mathcal P$ as sections of the fibre bundle $\mathcal P \times_{\Ad} G$ associated with the adjoint representation of $G$ on itself.

A \emph{holomorphic structure} on $\mathcal P$ is a choice of local trivialisations $\{ \psi_I \}_{I \in \mathcal I}$ with respect to which the transition functions $g_{IJ} : U_I \cap U_J \to G$ are holomorphic.
It can equally be described \cite{AB} as a family of $\g$-valued $(0,1)$-forms $A''_{\bar z_I}(z_I,\bar z_I)\d \bar z_I \in \Omega^{0,1}(U_I, \g)$ relative to a choice of local trivialisations $\{ \psi_I \}_{I \in \mathcal I}$, denoted collectively as $A''$, such that 
\begin{equation} \label{01-form compat}
A''_{\bar z_I}\d\bar z_I = g_{IJ} A''_{\bar z_J} g_{IJ}^{-1}\d\bar z_J - \bar\partial g_{IJ} g_{IJ}^{-1}
\end{equation}
on $U_I \cap U_J \neq \emptyset$. We let $\mathcal M$ denote the infinite-dimensional space of holomorphic structures on $\mathcal P$.

Under a change of local trivialisations $h \in \check{C}^0(C, G)$, the holomorphic structure $A''$ is described in the new local trivialisations \eqref{new trivialisations} by the family of $\g$-valued $(0,1)$-forms 
\begin{equation}
\label{change_triv_on_A}
\tilde A''_{\bar z_I}\d\bar z_I  = h_I A''_{\bar z_I} h_I^{-1}\d\bar z_I  - \bar\partial h_I h_I^{-1} \in \Omega^{0,1}(U_I, \g).
\end{equation}
In particular, by solving the equations $A''_{\bar{z}_I} = h_I^{-1} \partial h_I/\partial\bar{z}_I$, which is always possible locally \cite[Section 5]{AB}, we obtain smooth maps $h_I : U_I \to G$ which define a new local trivialisation where
\begin{equation} \label{A zbar 0 gauge}
\tilde A''_{\bar z_I}(z_I,\bar z_I) = 0.
\end{equation}
This represents the same holomorphic structure $A''$ of $\mathcal P$ but now in an adapted local trivialisation of $\mathcal P$ where its components vanish. In particular, it now follows from \eqref{01-form compat} that in this new local trivialisation the transition functions $\tilde g_{IJ} : U_I \cap U_J \to G$ of the bundle are holomorphic.

Under a gauge transformation by $g \in \mathcal G$, a holomorphic structure $A'' \in \mathcal M$ is transformed to a new holomorphic structure on $\mathcal P$ given by the family of $\g$-valued $(0,1)$-forms 
\begin{equation} \label{gauge transformed Aalpha}
{}^{g_I} A''_{\bar z_I}\d\bar z_I  \coloneqq g_I  A''_{\bar z_I} g_I^{-1} \d\bar z_I- \bar\partial g_I g_I^{-1} \in \Omega^{0,1}(U_I, \g).
\end{equation}
Let ${}^g A'' \coloneqq g A'' g^{-1} - \bar\partial g g^{-1} \in \mathcal M$ denote this transformed holomorphic structure. We have a left action of $\mathcal G$ on $\mathcal M$ given by
\begin{equation} \label{G action on M}
\mathcal G \times \mathcal M \,\longrightarrow\, \mathcal M, \qquad (g, A'') \,\longmapsto\, g\cdot A''\coloneq {}^g A''=g A'' g^{-1} - \bar\partial g g^{-1}.
\end{equation}
Any $A'' \in \mathcal M$ determines a Dolbeault operator $\bar\partial^{A''}$ on any vector bundle $V_{\mathcal P} \coloneqq \mathcal P \times_\rho V$ associated with $\mathcal P$ in some representation $\rho : G \to \Aut V$, which acts on local sections over $U_I$ as $\bar\partial + \rho( A''_{\bar z_I})\d\bar z_I$.
In terms of Dolbeault operators, the left action \eqref{G action on M} reads $(g, \bar\partial^{A''}) \mapsto g \bar\partial^{A''} g^{-1}$.

The Lie algebra $\mathfrak G \coloneqq \text{Lie}(\mathcal G)$ of $\mathcal{G}$ consists of sections $X$ of the vector bundle $\g_{\mathcal P} = {\mathcal P}\times_{{\rm ad}} \g$ associated with ${\mathcal P}$ in the adjoint representation. Explicitly, this is given by a family of $\g$-valued functions $X^I \in C^\infty(U_I, \g)$ in each local trivialisation such that on each overlap $U_I \cap U_J \neq \emptyset$ we have the relation $X^I = g_{IJ} X^J g_{IJ}^{-1}$. The left action \eqref{G action on M} of the group $\mathcal G$ induces an infinitesimal left action of a Lie algebra element $X \in \mathfrak G$ on $A'' \in \mathcal M$ given by
\begin{equation} \label{g left action on M}
\delta_X A'' = -\bar\partial^{A''} X = - \bar\partial X - [A'', X],
\end{equation}
or in local trivialisations by
\begin{equation}
  \delta_{X^I} A''^{I} = -\bar\partial X^I - [A''^{I}, X^I].
\end{equation}

\subsection{Cotangent bundle \texorpdfstring{$T^\ast \mathcal M$}{TM} and action of \texorpdfstring{$\mathcal{G}$}{G}} \label{sec: T*M and G Hitchin}

The tangent space $T_{A''} \mathcal M$ at any point $A'' \in \mathcal M$ is given by the space of sections of the bundle $\bigwedge^{0,1}C\otimes\g_{\mathcal P}$.  Similarly, the cotangent space $T^\ast_{A''} \mathcal M$ is the space of sections of $\bigwedge^{1,0}C\otimes\g_{\mathcal P}^\ast$, where $\g_{\mathcal P}^\ast \coloneqq {\mathcal P} \times_{{\rm ad}^\ast} \g^\ast$ is the vector bundle associated with ${\mathcal P}$ in the coadjoint representation. To put it concretely, any $X \in T_{A''} \mathcal M$ is described by a family of $\g$-valued $(0,1)$-forms $X^I = X_{\bar z_I}(z_I, \bar z_I) \d \bar z_I \in \Omega^{0,1}(U_I, \g)$, and any $Y \in T^\ast_{A''} \mathcal M$ by a family of $\g^\ast$-valued $(1,0)$-forms $Y^I = Y_{z_I}(z_I,\bar z_I)\d z_I \in \Omega^{1,0}(U_I, \g^\ast)$ such that on any overlap $U_I \cap U_J \neq \emptyset$, we have the relations
\begin{equation} \label{10 and 01-form compat}
X^I = g_{IJ} X^J g_{IJ}^{-1},\qquad Y^I = \Ad^\ast_{g_{IJ}} Y^J ,
\end{equation}
respectively. Using the canonical pairing $\langle~,~\rangle : \g^\ast \times \g \to \CC$, we obtain a family of local $(1,1)$-forms $\langle Y^I, X^I\rangle \in \Omega^{1,1}(U_I)$. Here we have suppressed a wedge product between the one-forms $Y^I$ and $X^I$. It follows from \eqref{10 and 01-form compat} that these local $(1,1)$-forms agree on overlaps, that is, $\langle Y^I, X^I\rangle = \langle Y^J, X^J\rangle$, and hence define a global $(1,1)$-form on $C$ which we denote by $\langle Y, X\rangle \in \Omega^{1,1}(C)$. In particular, the latter can be integrated over the compact Riemann surface $C$ to obtain a pairing
\begin{equation} \label{TM T*M pairing}
T_{A''}^\ast \mathcal M \times T_{A''} \mathcal M \longrightarrow \CC , \qquad  (Y,X)\,\longmapsto\, \frac{1}{2 \pi i}\int_C \langle Y, X\rangle .
\end{equation}

Now, a point in the cotangent bundle $T^\ast \mathcal M$ is given by a pair $(B, A'')$, with $A'' \in \mathcal M$ a holomorphic structure on ${\mathcal P}$ parametrising the base and $B$ a section of $\bigwedge^{1,0}C\otimes\g_{\mathcal P}^\ast$ parametrising the fibre. To describe vector fields on $T^\ast \mathcal M$, we note that we have the canonical isomorphism
\begin{equation} \label{T of T*M}
T_{(B, A'')} ( T^\ast \mathcal M ) \cong T_{A''}^\ast \mathcal M \oplus T_{A''} \mathcal M .
\end{equation}
The differential at $(B, A'') \in T^\ast \mathcal M$ of the projection
\begin{equation}
    \pi_{\mathcal M} : T^\ast \mathcal M \to \mathcal M, \qquad (B, A'') \mapsto A''
\end{equation}
is given by the projection onto the second summand under the isomorphism \eqref{T of T*M}, which we denote as
\begin{equation} \label{delta A def}
\delta_{(B, A'')} A'' : T_{(B, A'')}(T^\ast \mathcal M ) \,\longrightarrow\, T_{A''} \mathcal M , \qquad (Y,X) \,\longmapsto\, X.
\end{equation}
The tautological one-form on $T^\ast \mathcal M$ is then defined using the pairing \eqref{TM T*M pairing} as
\begin{equation} \label{tauto}
\alpha_{(B, A'')} \coloneqq \frac{1}{2 \pi i}\int_C \langle B, \delta A''\rangle .
\end{equation}
More explicitly, this can be described as a map $\alpha_{(B, A'')} : T_{(B, A'')}(T^\ast \mathcal M ) \to \CC$ given by $\alpha_{(B, A'')}(Y, X) \coloneqq \frac{1}{2 \pi i}\int_C\langle B,X\rangle$.
The corresponding symplectic form $\omega \coloneqq \delta \alpha$ is given by 
\begin{equation} \label{omega Hitchin}
\omega_{(B, A'')}\big( (Y_1,X_1), (Y_2,X_2) \big) = \frac{1}{2 \pi i}\int_C\langle Y_1,X_2\rangle - \frac{1}{2 \pi i}\int_C\langle Y_2,X_1\rangle.
\end{equation}

Let us recall the left action \eqref{G action on M} of the group of gauge transformations $\mathcal G$ on the space of holomorphic structures $\mathcal M$. We can lift this to an action of $\mathcal G$ on $T^\ast \mathcal M$ as follows. In a local trivialisation, an element $g \in \mathcal G$ is represented by smooth maps $g_I : U_I \to G$, and a section $B \in T^\ast_{A''} \mathcal M$ of the bundle $\bigwedge^{1,0} C \otimes \g^\ast_{\mathcal P}$ is described by a family of $\g^\ast$-valued $(1,0)$-forms $B^I \in \Omega^{1,0}(U_I, \g^\ast)$. Since
\begin{equation} \label{10-form compat}
\Ad^\ast_{g_I} B^I = \Ad^\ast_{g_I g_{IJ}} B^J = \Ad^\ast_{g_{IJ} g_J} B^J = \Ad^\ast_{g_{IJ}} \big( \Ad^\ast_{g_J} B^J \big) ,
    \end{equation}
we obtain a well-defined left action of $g = (g_I)_{I \in \mathcal I} \in \mathcal G$ on the fibres 
\begin{equation}\label{G act on B}
T^\ast_{A''} \mathcal M \,\longrightarrow\, T^\ast_{g \cdot A''} \mathcal M , \qquad B \,\longmapsto\, g \cdot B\coloneq \Ad^\ast_g B
\end{equation}
given explicitly in the local trivialisation over $U_I$ by $B^I \mapsto \Ad^\ast_{g_I} B^I$. Combining this with the left action of $\mathcal G$ on the base $\mathcal M$, we obtain the desired left action of $\mathcal G$ on $T^\ast \mathcal M$ given by
\begin{equation} \label{cal G action on T*M}
\mathcal G \times T^\ast \mathcal M \,\longrightarrow\, T^\ast \mathcal M , \qquad  \big(g, (B, A'') \big) \,\longmapsto\, g \cdot \big( B, A'' \big) \coloneqq \big( \Ad^\ast_g B, {}^g A'' \big) .
\end{equation}
This induces an infinitesimal left action of a Lie algebra element $X \in \mathfrak G$ on $(B, A'') \in T^\ast \mathcal M$ given by
\begin{equation} \label{global_inf_action}
\big( \delta_X B, \delta_X A'' \big) \coloneqq \big( {\rm ad}^*_X B, -\bar{\partial}^{A''} X \big) ,
\end{equation}
where $\delta_X B = {\rm ad}^*_X B$ is given in local trivialisations by $\delta_{X^I} B^I = {\rm ad}^*_{X^I} B^I \in \Omega^{1,0}(U_I, \g^\ast)$.

As in the finite-dimensional setup of Section \ref{gauged_univariational_principle}, we need a notion of freeness for the action of the group $\mathcal G$ on $T^\ast \mathcal M$.  We say that $(B, A'')\in T^\ast\mathcal{M}$ is \emph{stable} if
\begin{equation}\label{inf_freeness}
\bar\partial X+[A'',X]=0,\qquad {\rm ad}^*_X B=0\implies X=0,
\end{equation}
for all $X \in \mathfrak{G}$. This implies that the stabiliser of $(B, A'')$ is not a continuous subgroup of $\mathcal{G}$, making it an infinitesimal version of freeness.

\begin{remark}
  The notion of a stable principal Higgs bundle was introduced in \cite{BO}. We expect that any Higgs bundle that is stable in the sense of \cite{BO} satisfies \eqref{inf_freeness}. This was proved for semisimple irreducible reductive algebraic groups $G$ in the case $B=0$ in \cite[Proposition 3.2]{Ram}. A proof of this statement for $B\neq0$ and $G=SL_m(\CC)$ using the Kobayashi--Hitchin correspondence of Simpson can be found in \cite[Appendix A]{CHSV}.
\end{remark}

Having discussed the essential geometric ingredients, we now turn to our first main goal of this chapter: constructing a geometric Lagrangian one-form on the infinite-dimensional cotangent bundle $T^\ast \mathcal M$.

\section{Lagrangian one-form on \texorpdfstring{$T^\ast \mathcal{M}$}{TM}} \label{sec: Lag for Hitchin}

To construct a geometric Lagrangian one-form on $T^\ast \mathcal M$, we will generalise the finite-dimensional setup of geometric Lagrangian one-forms discussed in Chapter \ref{chap:lm-background} to the present infinite-dimensional case. The role of the underlying manifold $M$ there will now be played by the space $\M$ of holomorphic structures on a principle $G$-bundle $\mathcal P \to C$ over a compact Riemann surface $C$, for some connected Lie group $G$, and the role of the symmetry group $G$ will be played here by the group $\mathcal G = \Aut\mathcal P$ of (fibre-preserving) automorphisms of $\mathcal P$. 

The starting point for our construction is the fact that the Hitchin phase space is given by the symplectic quotient
\begin{equation} \label{Hitchin phase space intro}
T^\ast \text{Bun}_G(C) \;\cong\; \mu^{-1}(0) / \mathcal G .
\end{equation}
More specifically, we introduce a natural lift of the Hitchin map to the cotangent bundle $T^\ast \mathcal M$ which induces $n$ commuting flows on this infinite-dimensional symplectic manifold and that can be described variationally using a natural geometric multiform action $S_\Gamma[B, A'', t]$ on $T^\ast \mathcal M$. Then, upon performing the symplectic reduction to $\mu^{-1}(0) / \mathcal G$ at the level of the action, we obtain a multiform version of the action for $3$d mixed BF theory.

In the present infinite-dimensional setting, the analogue of the Lagrangian one-form \eqref{L} is
\begin{equation} \label{L_Phi_A}
\Lag = \alpha_{(B, A'')} - H_i( B,A'')\d t^i = \frac{1}{2 \pi i}\int_C \langle B,\delta A''\rangle - H_i( B,A'')\d t^i.
\end{equation}

To write down the corresponding action, that is, the analogue of \eqref{ungauged action}, let $\Sigma : \RR^n \to T^\ast \mathcal M \times \RR^n$ be an immersion. For any $u \in \RR^n$, we write $\Sigma(u) = \big( B(u), A''(u), t(u) \big)$. It is helpful to take a moment to describe each component of $\Sigma(u)$ in more detail. Firstly, $B(u)$ describes an $\RR^n$-dependent section of $\bigwedge^{1,0} C \otimes \g^\ast_{\mathcal P}$ given in each local trivialisation of $\mathcal P$ over $U_I$ by $\g^\ast$-valued $(1,0)$-forms $B_{z_I}(z_I, \bar z_I, u) \d z_I \in \Omega^{1,0}(U_I \times \RR^n, \g^\ast)$. These are related in chart overlaps $U_I \cap U_J \neq \emptyset$ by the second relation in \eqref{10 and 01-form compat}, explicitly
\begin{equation} \label{B(s) transition}
B_{ z_I}(z_I, \bar z_I,u)\d z_I = \Ad^\ast_{g_{IJ}} B_{ z_J}(z_J, \bar z_J,u)\d z_J.
\end{equation}
Secondly, $A''(u)$ describes an $\RR^n$-dependent element of $\mathcal M$ given in each local trivialisation of ${\mathcal P}$ over $U_I$ by $\g$-valued $(0,1)$-forms $A''_{\bar z_I}(z_I, \bar z_I, u) \d \bar z_I \in \Omega^{0,1}(U_I, \g)$. These are related in chart overlaps $U_I \cap U_J \neq \emptyset$ by \eqref{01-form compat}, explicitly
\begin{equation} \label{A(s) transition}
A''_{\bar z_I}(z_I, \bar z_I,u)\d\bar z_I = g_{IJ} A''_{\bar z_J}(z_J, \bar z_J,u) g_{IJ}^{-1}\d\bar z_J - \bar\partial g_{IJ} g_{IJ}^{-1} .
\end{equation}
Finally, $t(u)$ describes an $\RR^n$-dependent point in $\RR^n$ with components $t^i(u)$ for $i=1,\ldots, n$.

Note that since the transition functions $g_{IJ} : U_I \cap U_J \to G$ of $\mathcal P$ obviously do not depend on the parameter $u = (u^j) \in \RR^n$, differentiating the relation \eqref{A(s) transition} with respect to $u^j$, we obtain
\begin{equation} \label{dA(s) transition}
\partial_{u^j} A''_{\bar z_I}(z_I, \bar z_I,u)\d\bar z_I = g_{IJ} \partial_{u^j} A''_{\bar z_J}(z_J, \bar z_J,u) g_{IJ}^{-1}\d\bar z_J .
\end{equation}
Thus, for every $j = 1, \ldots, n$, the family of $\g$-valued $(0,1)$-forms $\partial_{u^j} A''_{\bar z_I}\d\bar z_I \in \Omega^{0,1}(U_I, \g)$ defines an $\RR^n$-dependent section of $\bigwedge^{0,1} C \otimes \g_{\mathcal P}$ which we denote by $\partial_{u^j} A''(u)$. Similarly, we denote by $\partial_{u^j} B(u)$ the family of $\g^\ast$-valued $(1,0)$-forms $\partial_{u^j} B_{ z_I}\d z_I \in \Omega^{1,0}(U_I, \g^\ast)$.
Now, let $\Sigma_1 : \RR^n \to T^\ast \mathcal M$, $u \mapsto \big( B(u), A''(u) \big)$ be the component of the map $\Sigma$ in $T^\ast \mathcal M$. Its differential at $u \in \RR^n$ is
\begin{equation}
\begin{split}
\d_u \Sigma_1 : T_u \RR^n \,&\longrightarrow\, T_{\Sigma_1(u)}(T^\ast \mathcal M ) \cong T^\ast_{A''(u)} \mathcal M \oplus T_{A''(u)} \mathcal M ,\\
\frac{\partial}{\partial u^j} \,&\longmapsto\, \big( \partial_{u^j} B(u), \partial_{u^j} A''(u) \big).
\end{split}
\end{equation}
The pullback of the differential $\delta A''$ defined in \eqref{delta A def} by the map $\Sigma_1$ is then given by the composition $\big( \Sigma^\ast (\delta A'') \big)(u) = \delta_{\Sigma_1(u)} A'' \circ \d_u \Sigma_1$ and hence $\Sigma^\ast (\delta A'') = \partial_{u^j} A''(u) \wedge\d u^j = -\d_{\RR^n} A''(u)$.
We now find that the pullback $\Sigma^\ast \Lag$ of the Lagrangian \eqref{L_Phi_A} by the map $\Sigma$ is given by
\begin{equation} \label{Sigma pull L}
\Sigma^\ast \Lag = \frac{1}{2 \pi i}\int_C \big\langle B(u), \d_{\RR^n} A''(u) \big\rangle - H_i\big( B(u), A''(u) \big) \d_{\RR^n} t^i .
\end{equation}
Given an arbitrary curve $\Gamma : (0,1) \to \RR^n$, $s \mapsto \big( u^j(s) \big)$ can now finally write down the analogue of the action \eqref{ungauged action} in the present case, which reads
\begin{equation} \label{ungauged action Phi A}
S_\Gamma[\Sigma] = \int_0^1 (\Sigma \circ \Gamma)^\ast \Lag = \int_0^1 \bigg(\! - \frac{1}{2 \pi i} \int_C \big\langle B(u), \partial_{u^j} A''(u) \big\rangle - H_i\big( B(u), A''(u) \big)\frac{\partial t^i}{\partial u^j} \bigg) \frac{\d u^j}{\d s} \d s .
\end{equation}
\begin{remark}
We make a few remarks on notation and terminology to avoid confusion. Note that the kinetic part $\frac{1}{2 \pi i}\int_C\langle B, \delta A''\rangle$ of \eqref{L_Phi_A} is the direct analogue of $p_\mu \d q^\mu$ in \eqref{L}. In particular, the integration over $C$ in the present setting is the analogue of the summation over $\mu \in \{ 1, \ldots, m \}$ in the finite-dimensional setting of Section \ref{sec: group actions}. Importantly, this means that, although the first term in the action \eqref{ungauged action Phi A} involves the integral of a $3$-form over $C \times (0,1)$, from the point of view of Lagrangian multiform theory we should really regard the whole action $S_\Gamma[\Sigma]$ as the integral of a $1$-form on the interval $(0,1)$, namely the pullback of \eqref{L_Phi_A} along $\Sigma \circ \Gamma : (0,1) \to T^\ast \mathcal M \times \RR^n$.
\end{remark}

\begin{remark}
The terminology ``gauge group'' for the group ${\cal G}$ of fibre-preserving automorphisms is standard in the geometric formulation of the Hitchin system and that is why we used it here. However, it is crucial to note that, at this stage of the construction, the group $\mathcal G$ is the analogue of what we called the {\it global} symmetry group $G$ in Section \ref{sec: group actions}, which we will later gauge by considering transformations parametrised by maps $g:\RR^n \to {\cal G}$ by analogy with the finite-dimensional setting of Section \ref{sec: gauging symmetry}; see Section \ref{sec: Lag for Hitchin mod G} for details. A crude way to say this is that we will ``gauge the gauge group $\mathcal{G}$''.
\end{remark}

We now prove the analogue of Proposition \ref{prop_global_symmetry} in the present infinite-dimensional context with this in mind. In fact, the setting of Section \ref{sec: group actions} was very generic and the infinitesimal action of a Lie algebra element $X \in \g$ on $M$ was specified only implicitly through the vector fields $X^\sharp$. By contrast, in the present context, we have an explicit description of the action of the Lie algebra $\mathfrak G$ on $T^\ast \mathcal M$ in \eqref{global_inf_action} and even of the action of the group $\mathcal G$ on $T^\ast \mathcal M$ in \eqref{cal G action on T*M}. Therefore, we can prove a stronger statement than Proposition \ref{prop_global_symmetry} in the present case. We first need to lift the action of $\mathcal G$ to $T^\ast \mathcal M \times \RR^n$ by letting it act trivially on $\RR^n$, that is, we set $g \cdot t = t$ for any $g \in \mathcal G$ and $t \in \RR^n$.

\begin{proposition} \label{prop: Hi invariance Hitchin}
The action \eqref{ungauged action Phi A} is invariant under the action of $\mathcal G$ on $T^\ast \mathcal M \times \RR^n$ given in \eqref{cal G action on T*M} if and only if each $H_i$, for $i=1, \ldots, n$, is invariant under the group action, that is,
\begin{equation} \label{Hi invariance prop}
H_i\big( \Ad^\ast_g B, {}^g A'' \big) = H_i(B, A'')
\end{equation}
for any $(B, A'') \in T^\ast \mathcal M$ and $g \in \mathcal G$. Moreover, the Noether charge associated with an infinitesimal bundle morphism $X \in \mathfrak G$ is given by
\begin{equation} \label{moment_map_Hitchin}
\mu_{(B, A'')}(X) = \frac{1}{2 \pi i}\int_C \big\langle B,\bar{\partial}^{A''} X \big\rangle .
\end{equation}
\begin{proof}
We closely follow the proof of Proposition \ref{prop_global_symmetry}. Let $\Sigma : \RR^n \to T^\ast \mathcal M \times \RR^n$ be given by $\Sigma(u) = \big( B(u), A''(u), t(u) \big)$. By \eqref{cal G action on T*M}, its pointwise image under the left action of any $g \in \mathcal G$ is $g \cdot \Sigma : \RR^n \to T^\ast \mathcal M \times \RR^n$ given by
\begin{equation} \label{g act on gamma Hitchin}
(g \cdot \Sigma)(u) = \big( \Ad^\ast_g B(u), {}^g A''(u), t(u) \big).
\end{equation}
Note, in particular, that this implies $\partial_{u^j} (g \cdot \Sigma)(u) = \big( \Ad^\ast_g \partial_{u^j} B(u), g \partial_{u^j} A''(u) g^{-1}, \partial_{u^j} t(u) \big)$. The action for the transformed map $g \cdot \Sigma$ therefore reads
\begin{equation}
\begin{split}
S_\Gamma&[g \cdot \Sigma]\\
&= \int_0^1\left(-\frac{1}{2 \pi i}\int_C \big\langle \Ad^\ast_g B(u), g \partial_{u^j} A''(u) g^{-1} \big\rangle - H_i\big( \Ad^\ast_g B(u), {}^g A''(u) \big)\frac{\partial t^i}{\partial u^j}\right) \frac{\d u^j}{\d s} \d s\\
&= \int_0^1\left(-\frac{1}{2 \pi i}\int_C \big\langle B(u), \partial_{u^j} A''(u) \big\rangle - H_i\big( \Ad^\ast_g B(u), {}^g A''(u) \big)\frac{\partial t^i}{\partial u^j}\right) \frac{\d u^j}{\d s} \d s\\
&= S_\Gamma[\Sigma] + \int_0^1 \Big( H_i\big( B(u), A''(u) \big) - H_i\big( \Ad^\ast_g B(u), {}^g A''(u) \big) \Big) \frac{\d t^i}{\d s} \d s
\end{split}
\end{equation}
The result now follows since for $g \in \mathcal G$ to be a symmetry means that $S_\Gamma[g \cdot \Sigma] = S_\Gamma[\Sigma]$ for any map $\Sigma$ and any curve $\Gamma$ and hence the integral on the right-hand side must vanish for any curve $\Gamma$ but this, in turn, is equivalent to the condition \eqref{Hi invariance prop}. 

To work out the Noether charge associated with the infinitesimal symmetry generated by a Lie algebra element $X \in \mathfrak G$, we introduce an arbitrary smooth function $\lambda : \RR^n \to \CC$ and consider the pointwise variations of the map $\Sigma(u) = \big( B(u), A''(u), t(u) \big)$ given by 
\begin{subequations}
\begin{align}
\delta_{\lambda(u) X} B(u) &= \lambda(u) {\rm ad}^*_X B(u) ,\\
\delta_{\lambda(u) X} A''(u) &= - \lambda(u) \bar\partial^{A''(u)} X = - \lambda(u) \big( \bar\partial X + [A''(u), X] \big).
\end{align}
\end{subequations}
The variation of the action \eqref{ungauged action Phi A} then only has a contribution from the kinetic term, which reads
\begin{equation}
\begin{split}
\delta_{\lambda(u) X} S_\Gamma[\Sigma] &= \delta_{\lambda(u) X} \int_0^1 \frac{-1}{2 \pi i}\int_C \big\langle B(u), \partial_{u^j} A''(u) \big\rangle \frac{\d u^j}{\d s} \d s \\
&=  \int_0^1 \partial_{u^j} \lambda(u) \left(\frac{1}{2 \pi i}\int_C \big\langle B(u), \bar\partial^{A''(u)} X \big\rangle \right)\frac{\d u^j}{\d s} \d s ,
\end{split}
\end{equation}
cf. the end of the proof of Propostion \ref{prop_global_symmetry}. From this we read off the desired expression \eqref{moment_map_Hitchin}. 
\end{proof}
\end{proposition}

\begin{remark} \label{rem: mu well defined}
Note that the expression \eqref{moment_map_Hitchin} is well-defined since $\bar\partial^{A''} X$ is a section of $\bigwedge^{0,1} C \otimes \g_{\mathcal P}$. Indeed, using the relations \eqref{01-form compat} and $X^I = g_{IJ} X^J g_{IJ}^{-1}$, respectively, between the local expressions of $A'' \in \mathcal M$ and $X \in \mathfrak G$ in overlapping charts $U^I \cap U^J \neq \emptyset$, we find that
\begin{equation}
\bar\partial X^I + [A''_{\bar z_I}, X^I]\d\bar z_I = g_{IJ} \big( \bar\partial X_J + [ A''_{\bar z_J}, X_J]\d\bar{z}^J \big) g_{IJ}^{-1} .
\end{equation}
We thus have $\langle B^I, \bar\partial^{A''} X^I \rangle = \langle B^J, \bar\partial^{A''} X^J \rangle$ on overlaps $U_I \cap U_J \neq \emptyset$ so that these define a global $(1,1)$-form $\langle B, \bar\partial^{A''} X\rangle \in \Omega^{1,1}(C)$ which can be integrated over the compact Riemann surface $C$. 
Likewise, we have a well-defined global $(1,1)$-form $\langle\bar\partial^{A''} B, X\rangle \in \Omega^{1,1}(C)$ given by the expression $\langle \bar\partial^{A''} 
B^I, X^I \rangle$ in each local chart $U_I$. Moreover, these are related by
\begin{equation} \label{three well defined sections}
\langle B, \bar\partial^{A''} X\rangle - \langle\bar\partial^{A''} B, X\rangle = - \d_C \langle B, X\rangle
\end{equation}
where $\langle B,X\rangle \in \Omega^{1,0}(C)$ is a well-defined $(1,0)$-form on $C$. The relative sign on the left-hand side comes from the fact that $B$ is a one-form and the operator $\bar\partial^{A''}$ has cohomological degree $1$. Integrating both sides over $C$ and using Stokes' theorem on the right-hand side, noting that $C$ has no boundary, we deduce that the Noether charge \eqref{moment_map_Hitchin} can equivalently be rewritten as
\begin{equation} \label{moment_map_Hitchin 2}
\mu_{(B, A'')}(X) = \frac{1}{2 \pi i}\int_C \big\langle \bar{\partial}^{A''} B, X \big\rangle.
\end{equation}
In other words, the value of the corresponding moment map $\mu : T^\ast \mathcal M \to \mathfrak G^\ast$ at $(B, A'') \in T^\ast \mathcal M$ is the element of $\mathfrak G^\ast$ given by the linear map $X \mapsto \frac{1}{2 \pi i}\int_C ( \bar{\partial}^{A''} B, X )$ which takes in any vector $X \in \mathfrak G$, that is, a section of $\g_{\mathcal P}$, and integrates it against the section $\bar{\partial}^{A''} B$ of $\bigwedge^{1,1} C \otimes \g^\ast_{\mathcal P}$.
\end{remark}

The construction of the Hamiltonians $H_i : T^\ast \mathcal M \to \CC$ satisfying \eqref{Hi invariance prop} will be inspired by that of the Hitchin map \cite{H}. Recall that the latter is constructed from a choice of
\begin{itemize}
  \item[$(i)$] $G$-invariant homogeneous polynomials $P_r:\g^\ast \to\CC$ for $r = 1,\ldots, \text{rk}\,\g$ of degree $d_r+1$, where $E = \{ d_r \}_{r=1}^{\text{rk} \, \g}$ is the set of exponents of $\g$.
\end{itemize}
Given such data, the Hitchin map 
\begin{equation} \label{Hitchin map}
P \coloneqq (P_1, \ldots, P_{\text{rk}\, \g}) : H^0 \Big( C, {\textstyle \bigwedge^{1,0}} C \otimes \g^\ast_{\mathcal P} \Big) \,\longrightarrow\, \bigoplus_{r=1}^{\text{rk}\,\g} H^0 \Big( C, \big( {\textstyle \bigwedge^{1,0}} C \big)^{\otimes (d_r+1)} \Big),
\end{equation}
takes as input a holomorphic section $B$ of the bundle $\bigwedge^{1,0} C \otimes \g^\ast_{\mathcal P}$.
Recall that this is given by a family of $\g^\ast$-valued $(1,0)$-forms $B^I = B_{z_I}(z_I) \d z_I \in \Omega^{1,0}(U_I, \g^\ast)$ in the local trivialisation over the chart $(U_I, z_I)$, where here the function $B_{z_I}$ depends holomorphically on $z_I$, related by the second relation in \eqref{10 and 01-form compat}, explicitly $B^I = \Ad^\ast_{g_{IJ}} B^J$ on $U_I \cap U_J \neq \emptyset$.
Since $P_r$ is $G$-invariant, it follows that we have
\begin{equation}
P_r(B^I) = P_r\big( \Ad^\ast_{g_{IJ}} B^J \big) = P_r(B^J)
\end{equation}
on non-trivial overlaps $U_I \cap U_J \neq \emptyset$. In this way, we obtain a holomorphic section of the bundle $(\bigwedge^{1,0} C)^{\otimes (d_r+1)}$ for each $r =1, \ldots, \text{rk}\, \g$, which we denote by $P_r(B)$, and the Hitchin map returns a holomorphic section of $\bigoplus_{r =1}^{\text{rk}\,\g} (\bigwedge^{1,0} C)^{\otimes (d_r+1)}$. To obtain individual complex-valued Hamiltonians one can then expand each component $P_r(B)$ of the Hitchin map in a basis of holomorphic $(d_r+1,0)$-differentials on $C$.

However, in our present setting, $B \in T^\ast_{A''} \mathcal M$ is only a \emph{smooth} section of $\bigwedge^{1,0} C \otimes \g^\ast_{\mathcal P}$. We can still form a smooth section $P_r(B)$ of $(\bigwedge^{1,0} C)^{\otimes (d_r+1)}$ for every $r = 1, \ldots, \text{rk}\,\g$, however we can no longer expand it in a basis of holomorphic sections of $(\bigwedge^{1,0} C)^{\otimes (d_r+1)}$. Instead, in order to produce complex-valued Hamiltonians we will proceed along the lines of \cite{VW} by introducing a marked point on $C$ for each Hamiltonian $H_i$.\footnote{Only one marked point was needed in \cite{VW} since only a single Hamiltonian was considered in that setup.} We therefore introduce the following additional data:
\begin{itemize}
  \item[$(ii)$] Points $\mathsf q_{rl} \in C$ labelled by pairs $(r,l)$ with $r = 1,\ldots, \text{rk}\,\g$ and $l = 1, \ldots, m_r$, where
\begin{equation}
m_r \coloneqq \dim \Big( H^0 \Big( C, \big( {\textstyle \bigwedge^{1,0}} C \big)^{\otimes (d_r+1)} \Big) \Big) = \left\{
\begin{array}{ll}
(2 d_r+1) (g-1) , & \quad \text{for}\; g \geq 2 ,\\
g , & \quad \text{for} \; g=0,1
\end{array}
\right.
\end{equation}
and a set of holomorphic tangent vectors $V_{\mathsf q_{rl}} \in T^{1,0}_{\mathsf q_{rl}} C$.
\end{itemize}

We can now define $H_i : T^\ast \mathcal M \to \CC$ by evaluating the smooth section $P_r(B)$ of $(\bigwedge^{1,0} C)^{\otimes (d_r+1)}$ at $\mathsf q_{rl} \in C$ and pairing the resulting element $P_r\big( B(\mathsf q_{rl}) \big) \in (\bigwedge^{1,0}_{\mathsf q_{rl}} C)^{\otimes (d_r+1)}$ with $V_{\mathsf q_{rl}}^{d_r+1} \in (T^{1,0}_{\mathsf q_{rl}} C)^{\otimes (d_r+1)}$, that is,
\begin{equation} \label{pre Hitchin Hamiltonians}
H_i(B) \coloneqq \big\langle P_r\big( B(\mathsf q_{rl}) \big), V_{\mathsf q_{rl}}^{d_r+1} \big\rangle .
\end{equation}
Concretely, if $(U_I, z_I)$ is a local chart around one of the points $\mathsf q_{rl} \in C$ then we can pick $V_{\mathsf q_{rl}} = \partial_{z_I}$ and the above geometric construction amounts to writing $P_r(B) = P_r\big( B_{z_I}(z_I, \bar z_I) \big) \d z_I^{\otimes (d_r+1)}$ locally in the coordinate $z_I$ and then evaluating its component at $\mathsf q_{rl} \in \CC$, that is,
\begin{equation} \label{Hitchin Hamiltonians}
H_i(B) = P_r\big( B_{z_I}(\mathsf q_{rl}) \big).
\end{equation}
The understanding in \eqref{pre Hitchin Hamiltonians} and \eqref{Hitchin Hamiltonians} is that  the label $i$ on the Hamiltonians runs over pairs $(r, l)$ with $r = 1, \ldots, \text{rk}\, \g$ and $l = 1, \ldots, m_r$.
Note that $m_r$ being the dimension of the space of holomorphic sections of $(\bigwedge^{1,0} C)^{\otimes (d_r+1)}$ ensures that, when the points $\mathsf q_{rl}$ are generic, the number of Hamiltonians we produce coincides with the number of Hamiltonians obtained via the construction of the Hitchin map \eqref{Hitchin map} when the smooth section $B$ of the bundle $\bigwedge^{1,0} C \otimes \g^\ast_{\mathcal P}$ becomes holomorphic. Indeed, our evaluation prescription \eqref{pre Hitchin Hamiltonians} defines a bijection
\begin{equation}
H^0 \Big( C, \big( {\textstyle \bigwedge^{1,0}} C \big)^{\otimes (d_r+1)} \Big) \,\overset{\cong}\longrightarrow\, \CC^{m_r} ,
\end{equation}
for generic points $\mathsf q_{rl} \in C$, $l = 1, \ldots, m_r$. It is enough to show this is injective, which follows from the fact that divisors $D_r \coloneqq \sum_{l=1}^{m_r} \mathsf q_{rl}$ on $C$ for which $\deg(D_r) = m_r = \dim \big( H^0 \big( C, ( {\textstyle \bigwedge^{1,0}} C \big)^{\otimes (d_r+1)} \big) \big)$ and $\dim \big( H^0 \big( C, ( {\textstyle \bigwedge^{1,0}} C \big)^{\otimes (d_r+1)} \otimes \mathcal O(-D_r) \big) \big) = 0$ are generic.

Since the label $i$ on the Hamiltonians runs over pairs $(r, l)$, with $r = 1, \ldots, \text{rk}\, \g$ and $l = 1, \ldots, m_r$, it runs from $1$ to
\begin{equation}
n \coloneqq \sum_{r=1}^{\text{rk}\, \g} m_r = (g-1) \sum_{r=1}^{\text{rk}\, \g} (2 d_r + 1) = (g-1) \dim \g 
\end{equation}
when $g \geq 2$ and from $1$ to $n \coloneqq \text{rk}\, \g$ when $g=1$. This number coincides with half the dimension of the phase space of the Hitchin system when $g \geq 2$ and $g=1$, respectively.\footnote{This is a consequence of the Riemann--Roch theorem. See, for instance, \cite[Chapter 7]{BBT}.} When $g=0$, however, there are no Hamiltonians since the set of points $\mathsf q_{rl} \in C$ introduced in condition $(ii)$ above is empty.
Producing non-trivial integrable systems in the case of genus $g=0$ will require introducing additional marked points on $C$ which we will turn to in Section \ref{sec: adding punctures} below.

\begin{remark}
    From now on, for notational convenience, we will simply use the common label $i$ for the Hamiltonians $H_i$, the polynomials $P_r$, the points $\mathsf q_{rl}$ and the times $t^i$ associated to the Hamiltonians. This amounts to relabelling $P_r$ as $P_{rl}$, with the understanding that $P_{rl} = P_{rl'}$ for any $l, l' =1, \ldots, m_r$, so that we can simply write $H_i(B) = P_i\big( B_{z_I}(\mathsf q_i) \big)$. Accordingly, we can keep denoting by $t^i$ the time associated to $H_i$, rather than the cumbersome $t^{rl}$ or $t^{(r,l)}$.
\end{remark}

\section{Lagrangian one-form for the Hitchin system on \texorpdfstring{$\mu^{-1}(0)/\mathcal{G}$}{mu G}} \label{sec: Lag for Hitchin mod G}

So far we have introduced the action $S_\Gamma[\Sigma]$ in \eqref{ungauged action Phi A} for an immersion $\Sigma : \RR^n \to T^\ast \mathcal M \times \RR^n$ and an arbitrary curve $\Gamma : (0,1) \to \RR^n$, and showed in Proposition \ref{prop: Hi invariance Hitchin} that it is invariant under the left action of the group $\mathcal G$, provided that the Hamiltonians $H_i$ themselves are $\mathcal G$-invariant in the sense that \eqref{Hi invariance prop} holds. Note that the latter condition clearly holds for the Hamiltonians introduced in \eqref{Hitchin Hamiltonians} by virtue of the $G$-invariance of the polynomials $P_i : \g^\ast \to \CC$. Moreover, we identified the moment map $\mu : T^\ast \mathcal M \to \mathfrak G^\ast$ associated with this symmetry as given by \eqref{moment_map_Hitchin}.

We now turn to generalising the gauging procedure of Section \ref{gauged_univariational_principle} to the present infinite-dimensional setting.  This requires introducing two additional elements: gauge transformations $g$ and a gauge field $\mathcal{A}$.  

Let us start with the gauge transformations. For the Hitchin system, the group by which we need to quotient is the group $\mathcal{G}$ of automorphisms of the bundle ${\mathcal P}\to C$. In the language of Section \ref{gauged_univariational_principle}, the group $\mathcal{G}$ corresponds to the \textit{global} symmetry group $G$. Thus, in order to gauge $\mathcal G$ here, we consider the group of local transformations $g:\RR^n\to \mathcal{G}$, \ie automorphisms of the bundle ${\mathcal P}\to C$ that depend smoothly on $u\in\RR^n$. In a local trivialisation of ${\mathcal P}$, this is represented by $G$-valued functions $g_I(z_I,\bar{z}_I,u)$. On overlaps $U_I \cap U_J \neq \emptyset$, they satisfy \eqref{bundle morphism compat}, in which $g_{IJ}$ are the transition functions relative to the chosen trivialisation of ${\mathcal P}\to C$.

The action of a local gauge transformation $g$ on the map $\Sigma:u\mapsto(A''(u),B(u),t(u))$ is exactly as in Section  \ref{sec: geometric setup}, except that $g$, $A''$, $B$, and $t$ depend on the parameters $u$ (in addition to depending on local coordinates $z_I$).  Explicitly, we write
\begin{equation}
g\cdot\Sigma :u\longmapsto \big( {}^{g(u)}A''(u),\Ad^\ast_{g(u)}B(u),t(u) \big),
\end{equation}
where ${}^{g(u)}A''(u)$ and $\Ad^\ast_{g(u)}B(u)$ are defined as in \eqref{G action on M} and \eqref{G act on B} for each $u$.

Next, we introduce the gauge field $\mathcal{A}=\mathcal{A}_i \d u^i$.  This consists of functions $\mathcal{A}_i$ from $\RR^n$ to the Lie algebra $\mathfrak{G}$ of $\mathcal{G}$.  The Lie algebra $\mathfrak{G}$ is the space of sections of the bundle $\mathfrak{g}_{\mathcal P}$, so each $\mathcal{A}_i$ is a section of $\mathfrak{g}_{\mathcal P}$ that depends smoothly on $u\in\mathbb{R}^n$.  In any local trivialisation, $\mathcal{A}_i$ is represented by functions $\mathcal{A}^I_{i}(z_I,\bar{z}_I,u)$ that take values in $\mathfrak{g}$. Further, on the overlap $U_I \cap U_J \neq \emptyset$, these are related by
\begin{equation}\label{Ai compat}
\mathcal{A}^I_{i}=g_{IJ}\mathcal{A}^J_{i}g_{IJ}^{-1} .
\end{equation}
Under a local transformation $g:\RR^n\to\mathcal{G}$, these transform as
\begin{equation}\label{G act on Ai}
\mathcal{A}^I_{i}\longmapsto g_I\mathcal{A}^I_{i}g_{I}^{-1} - \frac{\partial g_{I}}{\partial u^i}g_I^{-1}.
\end{equation}

Now that we have introduced the local transformations and the gauge field, let us write down the analogue of the gauged action in the finite-dimensional setup \eqref{uni_gauged action}. Recall that to do so we need to add to the ungauged action \eqref{ungauged action Phi A} a term coupling the moment map \eqref{moment_map_Hitchin} to the gauge field as
\begin{equation}\label{gauged Hitchin action}
S_\Gamma[\Sigma,\mathcal{A},t] = 
\int_0^1 \left( -\frac{1}{2 \pi i}\int_C \left\langle B(u), \frac{\partial A''}{\partial u^j}- \bar{\partial}^{A''}\mathcal{A}_j \right\rangle - H_i\big( B(u) \big)\frac{\partial t^i}{\partial u^j} \right) \frac{\d u^j}{\d s} \d s .
\end{equation}
The first term in this action is exactly a multiform action of $3$d mixed BF theory, as we now show. Let
\begin{equation}
A \coloneqq A'' + \mathcal{A}.
\end{equation}
This is a partial connection on the pullback bundle $\pi_C^\ast {\mathcal P}={\mathcal P}\times\RR^n$ over $C\times\mathbb{R}^n$ along the projection $\pi_C:C\times\RR^n\to C$.  By definition, the bundle $\pi_C^\ast {\mathcal P}$ is trivialised over open sets $U_I\times \RR^n$, and the transition functions between these sets are the transition functions $g_{IJ}$ of ${\mathcal P}$, which obviously satisfy
\begin{equation}\label{dgdu}
\frac{\partial g_{IJ}}{\partial u^j}=0 .
\end{equation}
In these local trivialisations, $A$ takes the form
\begin{equation}\label{A definition}
A^I = A''_{\bar{z}_I}(z_I,\bar{z}_I,u)\d\bar{z}_I + \mathcal{A}^I_{i}(z_I,\bar{z}_I,u)\d u^i.
\end{equation}
This looks like the local expression for a connection, except that it is missing a $\d z_I$-component, which is why we refer to it as a partial connection. To show that $A$ is a well-defined partial connection we must check that it satisfies
\begin{equation}\label{partial connection compat}
A^I = g_{IJ}A^Jg_{IJ}^{-1} - \bar{\partial}g_{IJ}g_{IJ}^{-1} - \d_{\RR^n} g_{IJ} g_{IJ}^{-1}
\end{equation}
on overlaps $U_I \cap U_J \neq \emptyset$, and that it is independent of the choice of local trivialisations of ${\mathcal P}\to C$. From \eqref{01-form compat}, \eqref{Ai compat} and \eqref{dgdu}, it follows that \eqref{partial connection compat} is satisfied. Independence of the choice of local trivialisation follows from \eqref{partial connection compat}: if we use different local trivialisations of ${\mathcal P}\to C$ over open sets $\{U_I\}_{I\in\mathcal{I}'}$ to construct a connection $A'$, then $A$ and $A'$ will be related as in \eqref{partial connection compat} on the overlaps $U_I\cap U_J$ for $I\in\mathcal{I}$ and $J\in\mathcal{I}'$. Therefore, $A'$ and $A$ represent the same connection.

The curvature $F_A$ of the partial connection $A$ is defined in local trivalisations by
\begin{equation}\label{FA def}
F_A^{I}=\left(\frac{\partial \mathcal A^I_{i}}{\partial \bar{z}_I}-\frac{\partial A''_{\bar{z}_I}}{\partial u^i}+[A''_{\bar{z}_I},\mathcal A^I_{i}]\right)\d\bar{z}_I\wedge \d u^i + \frac12\left(\frac{\partial \mathcal A^I_{j}}{\partial u^i}-\frac{\partial \mathcal A^I_{i}}{\partial u^j}+[\mathcal A^I_{i},\mathcal A^I_{j}]\right)\d u^i \wedge \d u^j.
\end{equation}
The family $B(u)$ of sections of $\bigwedge^{1,0}C\otimes\g_{\mathcal P}^\ast$ naturally determines a section of $\pi_C^\ast\bigwedge^{1,0}C\otimes\g_{\pi_C^\ast \mathcal P}^\ast$, which we will denote by the same symbol $B$.
The following theorem is a direct consequence of our definitions of $B$ and $A$, with corresponding curvature $F_A$.
\begin{theorem} \label{thm: BF Lagrangian}
The gauged multiform action \eqref{gauged Hitchin action} for the Hitchin system is equivalent to the multiform action for $3$d mixed BF theory on $C \times \RR^n$ for the pair $(B, A)$ with a type B line defect along each coordinate $t^i$ determined by the Hitchin Hamiltonian $H_i$ given in \eqref{Hitchin Hamiltonians}, namely
\begin{equation} \label{3d BF action}
S_\Gamma[B, A, t] = \frac{1}{2 \pi i}\int_{C\times\Gamma} \big\langle B, F_A \big\rangle - \int_0^1 H_i\big( B(u(s)) \big) \frac{\d t^i}{\d s} \d s
\end{equation}
for an arbitrary curve $\Gamma : (0,1) \to \RR^n$, $s \mapsto u(s)$.
\end{theorem}
The proof relies on the simple observation that, when pulled back along the one-dimensional curve $\Gamma$, only the first term in the expression \eqref{FA def} of the curvature is non-zero. That being said, the significance of Theorem \ref{thm: BF Lagrangian} is that it shows how the multiform version of $3$d mixed BF theory with type B defect is derived from our procedure of gauging a natural Lagrangian multiform on a cotangent bundle, applied to the Hitchin setup. As we show in Section \ref{sec: adding punctures}, a similar derivation with the inclusion of marked points in the Hitchin picture corresponds to the inclusion of so-called type A defects to the multiform action of $3$d mixed BF theory.

In this new interpretation of the action \eqref{gauged Hitchin action}, local (gauge) transformations $g:\RR^n\to\mathcal{G}$ play the role of bundle automorphisms of $\pi_C^\ast {\mathcal P}$, with local expressions $g_I(z_I,\bar{z}_I,u)$ in the trivialisations over $U_I\times\RR^n$.  From \eqref{G action on M}, \eqref{G act on B}, \eqref{G act on Ai} and \eqref{A definition}, it follows that these bundle automorphisms act on $A$ and $B$ in the expected way:
\begin{subequations}
\begin{align}
A^I &\mapsto {}^{g_I}A^I=g_IA^Ig_I^{-1}- \bar\partial g_{I} g_{I}^{-1} - \d_{\RR^n} g_{I} g_{I}^{-1} ,\\
B^I &\mapsto \Ad_{g_I}B^I.
\end{align}
\end{subequations}

\begin{remark}
The partial connection $A$ on $\pi_C^\ast {\mathcal P}$ was defined using the local trivialisations of $\pi_C^\ast {\mathcal P}$ obtained canonically from the local trivialisations of ${\mathcal P}$. In particular, in such a local trivialisation, the transition functions $g_{IJ}$ are independent of $u^j$ and $A$ transforms as in \eqref{partial connection compat} but without the last term. However, the resulting action \eqref{3d BF action} is well-defined in any local trivialisation of $\pi_C^\ast {\mathcal P}$, including those for which the transitions functions do depend on $u^j$, in which case $A$ transforms as in \eqref{partial connection compat} with the last term present. Indeed, with $A$ transforming as in \eqref{partial connection compat}, the curvature $F_A$ defines a section of $\bigwedge^2(C \times \RR^n) \otimes \g_{\pi_C^\ast {\mathcal P}}$ with $\g_{\pi_C^\ast {\mathcal P}}$ the vector bundle associated to $\pi_C^\ast {\mathcal P}$ in the adjoint representation, and hence $(B, F_A)$ is a well-defined global $3$-form on $C \times \RR^n$. We will exploit such a change of local trivialisation leading to $u^j$-dependent transition functions later on in Section \ref{sec:unifyingmultiform} to prove Theorem \ref{thm: Lag 1-form for Hitchin}.
\end{remark}

To derive the equations of motion for \eqref{3d BF action}, we first introduce some notation.
Let $(U_I, z_I)$ be a coordinate chart containing the point $\mathsf q \in U_I$ with coordinate $w = z_I(\mathsf q)$. We denote by $\delta(z_I-w)$ the $2$-dimensional Dirac $\delta$-distribution on $U_I$ at the point $\mathsf q \in U_I$ in the local coordinate $z_I$. It has the defining property
\begin{equation}\label{eq:Dirac-rel}
\int_{U_I} f(z_I) \delta(z_I - w) \d z_I \wedge \d \bar z_I = f(w)
\end{equation}
for any function $f : U_I \to \CC$. Note that the distribution-valued $2$-form $\delta(z_I - w) \d z_I \wedge \d \bar z_I$ is invariant under coordinate transformations. Specifically, if $(U_J, z_J)$ is another chart with $\mathsf q \in U_I \cap U_J \neq \emptyset$ and $w' \coloneqq z_J(\mathsf q)$, then we have
\begin{equation}
    \delta(z_I - w) \d z_I \wedge \d \bar z_I = \delta(z_J - w') \d z_J \wedge \d \bar z_J.
\end{equation}
Therefore, for notational convenience, we introduce
\begin{equation} \label{delta coord indep}
\delta_{\mathsf q} \coloneqq \delta\big( z_I - z_I(\mathsf q) \big) \d z_I \wedge \d \bar z_I
\end{equation}
for any $\mathsf q \in C$, which is independent of the coordinate chart $(U_I, z_I)$ used, so long as it contains the point $\mathsf q$.
We will also make use of the following useful fact
\begin{equation}\label{eq:del-Dirac-rel}
\partial_{\bar z_I} \frac{1}{z_I - w} = - 2 \pi i \, \delta (z_I - w).
\end{equation}

Given any polynomial $P : \g^\ast \to \CC$, its gradient $\nabla P : \g^\ast \to \g$ is defined through the first order term in the variation of $P(\phi)$ at any $\phi \in \g^\ast$, namely
\begin{equation} \label{gradient def}
P(\phi + \delta \phi) = P(\phi) + \big\langle \delta\phi, \nabla P(\phi) \big\rangle + O(\delta \phi^2) .
\end{equation}
In particular, for a coadjoint-invariant polynomial $P$, that is, a polynomial $P$ for which $P(\Ad^\ast_g \phi) = P(\phi)$ for any $g \in G$, it follows from considering $g = e^{h X}$ with $X \in \g$ and $h$ small such that $\big\langle {\rm ad}^\ast_X \phi, \nabla P(\phi) \big\rangle = 0$.

\begin{theorem}\label{th_one_form_Hitchin}
The gauged univariational principle applied to the $3$d mixed BF multiform action $S_\Gamma[B,A,t]$ in \eqref{3d BF action} gives rise to a set of equations for the fields $B$, $A''$ and $\mathcal A$. By working in any chart $(U_I, z_I)$ of $C$ in terms of the components $B_{z_I}(z_I, \bar z_I, u) $, $A''_{\bar z_I}(z_I, \bar z_I, u) $, and $\mathcal A^I_j(z_I, \bar z_I, u)$ of these fields, these equations take the following form:
\begin{subequations} \label{UV set}
\begin{align}
\partial_{\bar z_I} \widetilde{\mathcal A}^I_i - \partial_{t^i} A''_{\bar z_I} + [A''_{\bar z_I}, \widetilde{\mathcal A}^I_{i}] &= 2 \pi i\,\nabla P_i\big( B_{z_I}(\mathsf q_i) \big) \delta\big( z_I-z_I(\mathsf q_i) \big), \label{UV F}\\
\partial_{\bar z_I} B_{z_I} + {\rm ad}^*_{A''_{\bar z_I}} B_{z_I} &= 0 \label{UV Phi z},\\
\partial_{t^j} B_{z_I}+{\rm ad}^*_{\widetilde{\mathcal A}^I_{j}} B_{z_I} &= 0 \label{UV Phi i},\\
\partial_{t^j} H_i(B)&=0 \label{UV P}
\end{align}
\end{subequations}
where we used the invertibility of the map $(u^j) \mapsto \big( t^i(u) \big)$ to define $\widetilde{\mathcal A}^I_{i} = \frac{\partial u^j}{\partial t^i} \mathcal A^I_{j}$.
The equations associated to any pair of overlapping charts $(U_I, z_I)$ and $(U_J, z_J)$ are compatible on $U_I \cap U_J \neq \emptyset$.

Finally, if $B$ and $A''$ satisfy the stability condition \eqref{inf_freeness}, then the following zero curvature equations also hold
\begin{equation}
\label{ZC_eqs}
\partial_{t^i} \widetilde{\mathcal A}^I_{j} - \partial_{t^j} \widetilde{\mathcal A}^I_{i} + \big[ \widetilde{\mathcal A}^I_{i}, \widetilde{\mathcal A}^I_{j} \big] =0.
\end{equation}
\end{theorem}
\begin{proof}
Since the action \eqref{3d BF action} is local in all the fields, and in particular in the fields $B$ and $A$, to work out the equations of motion for the latter at any point $\mathsf p \in C$, it is sufficient to restrict the integration over $C$ in the action to any chart $(U_I, z_I)$ for which $\mathsf p \in U_I$. In other words, to work out the equations of motion for the $\g$-valued fields $B^I = B_{z_I}(z_I, \bar z_I, u) \d z_I$, $A''^I = A''_{\bar z_I}(z_I, \bar z_I, u) \d\bar z_I$ and $\mathcal A^I = \mathcal A^I_{j}(z_I, \bar z_I, u) \d u^j$, it is sufficient to consider the variation of the following action
\begin{equation} \label{3d BF action explicit}
\begin{split}
S_{I, \Gamma}[B_{z_I}, A''_{\bar z_I}, \mathcal A^I_{j}, t] &= \int_0^1 \Bigg( \frac{1}{2 \pi i} \int_{U_I} \big\langle B_{z_I}(z_I, \bar z_I, u(s)), F^I_{\bar z_Ij}(z_I, \bar z_I, u(s)) \big\rangle\, \d z_I \wedge \d\bar z_I\\
&\qquad\qquad\qquad\qquad - {\sum_{\substack{i = 1\\\mathsf q_i \in U_I}}^n } P_i\big( B_{z_I}(\mathsf q_i, u(s)) \big) \frac{\partial t^i}{\partial u^j} \Bigg) \frac{\d u^j}{\d s} \d s
\end{split}
\end{equation}
where we used the fact that the pullback of the curvature $F^I \in \Omega^2(U_I \times \RR^n, \g)$ along the curve $\Gamma : (0,1) \to \RR^n$ is given by
\begin{equation}
F^I(u(s)) = F^I_{\bar z_Ij}(z_I, \bar z_I, u) \, \d\bar z_I \wedge\frac{\d u^j}{\d s} \d s \in \Omega^2\big( U_I \times (0,1), \g \big) ,
\end{equation}
noting that the pullback of $F^I_{ij}(z_I, \bar z_I, u) \d u^i \wedge \d u^j$ vanishes since $\d s \wedge \d s = 0$.

We now work out the variation of \eqref{3d BF action explicit} with respect to all four fields $B_{z_I}(z_I, \bar z_I, u)$, $A''_{\bar z_I}(z_I, \bar z_I, u)$, $\mathcal A^I_{j}(z_I, \bar z_I, u)$ and $t^i(u)$. Firstly, we have
\begin{equation}
\begin{split}
&\delta_{B_{z_I}} S_{I, \Gamma}[B_{z_I}, A''_{\bar z_I}, \mathcal A^I_{j}, t]\\
&=  \int_0^1 \int_{U_I} \Bigg( \frac{1}{2 \pi i}\Big\langle \delta B_{z_I}(z_I, \bar z_I, u(s)), F^I_{\bar z_Ij}(z_I, \bar z_I, u(s)) \Big\rangle \\
&- \sum_{\substack{i = 1\\\mathsf q_i \in U_I}}^n \Big\langle \delta B_{z_I}(z_I, \bar z_I, u(s)), \nabla P_i\big( B_{z_I}(z_I, \bar z_I, u(s)) \big) \Big\rangle \delta\big( z_I - z_I(\mathsf q_i) \big) \frac{\partial t^i}{\partial u^j} \Bigg) \frac{\d u^j}{\d s} \d z_I \wedge \d\bar z_I \wedge \d s
\end{split}
\end{equation}
where, in the second term on the right-hand side, we used both the defining properties \eqref{eq:Dirac-rel} of the Dirac $\delta$-distribution and \eqref{gradient def} of the gradient of $P_i$. Since the above variation should vanish for all curves $\Gamma : (0,1) \to \RR^n$, $s \mapsto \big( u^j(s) \big)$ and for all variations $\delta^I B_{z_I}(z_I, \bar z_I, u)$, we deduce that
\begin{subequations}
\begin{equation} \label{3dBF eq a}
\begin{split}
&\partial_{\bar z_I} \mathcal A^I_{j}(z_I, \bar z_I, u(s)) - \partial_{u^j} A''_{\bar z_I}(z_I, \bar z_I, u(s)) + \big[ A''_{\bar z_I}(z_I, \bar z_I, u(s)), \mathcal A^I_{j}(z_I, \bar z_I, u(s)) \big] \\
&\qquad\qquad\qquad\qquad\qquad\qquad = 2 \pi i\sum_{\substack{i = 1\\\mathsf q_i \in U_I}}^n \nabla P_i\big( B_{z_I}(z_I, \bar z_I, u(s)) \big) \delta\big( z_I - z_I(\mathsf q_i) \big) \frac{\partial t^i}{\partial u^j}.
\end{split}
\end{equation}
The first equation \eqref{UV F} now follows by multiplying both sides by $\frac{\partial u^j}{\partial t^k}$, summing over $j = 1, \ldots, n$, and then relabelling $k$ as $j$, and using the definition of $\widetilde{\mathcal A}^I_{j}$ given in the statement of the theorem. Next, for the variation of \eqref{3d BF action explicit} with respect to $A''_{\bar z_I}(z_I, \bar z_I, u)$, we find
\begin{equation}
\begin{split}
&\delta_{A''_{\bar z_I}} S_{I, \Gamma}[B_{z_I}, A''_{\bar z_I}, \mathcal A^I_{j}, t]\\
&\quad = - \int_0^1 \frac{1}{2 \pi i}\int_{U_I} \Big\langle B_{z_I}(z_I, \bar z_I, u(s)),\\
&\qquad\qquad\partial_{u^j} \delta A''_{\bar z_I}(z_I, \bar z_I, u(s)) + \big[ \mathcal A^I_{j}(z_I, \bar z_I, u(s)), \delta A''_{\bar z_I}(z_I, \bar z_I, u(s)) \big] \Big\rangle \frac{\d u^j}{\d s} \d z_I \wedge \d\bar z_I \wedge \d s.
\end{split}
\end{equation}
Since this should vanish for all curves $\Gamma : (0,1) \to \RR^n$, $s \mapsto \big( u^j(s) \big)$ and all variations $\delta A''_{\bar z_I}(z_I, \bar z_I, u)$, we deduce that \eqref{UV Phi i} must hold, again after multiplying by $\frac{\partial u^j}{\partial t^k}$, summing over $j = 1, \ldots, n$, and relabelling $k$ as $j$, and also using the definition of $\widetilde{\mathcal A}^I_{j}$. Similarly, by varying \eqref{3d BF action explicit} with respect to $\mathcal A^I_{j}(z_I, \bar z_I, u)$, we deduce that \eqref{UV Phi z} must hold.
Finally, varying \eqref{3d BF action explicit} with respect to $t^i(u)$ leads to
\begin{equation}
\delta_{t^i(u)} S_{I, \Gamma}[B_{z_I}, A''_{\bar z_I}, \mathcal A^I_{j}, t] = - \int_0^1 \bigg( H_i\big( B(u(s)) \big) \frac{\partial \delta t^i}{\partial u^j} \bigg) \frac{\d u^j}{\d s} \d s,
\end{equation}
\end{subequations}
and since this should vanish for all curves $\Gamma : (0,1) \to \RR^n$, $s \mapsto \big( u^j(s) \big)$ and all variations $\delta t^i(u)$, we deduce that \eqref{UV P} must hold.

The derivation of \eqref{ZC_eqs} follows the same reasoning as in the proof of Theorem \ref{thm: gauged univar} for \eqref{flatness}. Using \eqref{UV Phi i}, and with $\widetilde{F}_{ij}^I=\partial_{t^i} \widetilde{\mathcal A}^I_{j} - \partial_{t^j} \widetilde{\mathcal A}^I_{i} + \big[ \widetilde{\mathcal A}^I_{i}, \widetilde{\mathcal A}^I_{j} \big]$, we deduce 
\begin{equation}
\left[\partial_{t^i},\partial_{t^j}\right] B_{z_I}+{\rm ad}^*_{\widetilde{F}_{ij}^I} B_{z_I} = 0.
\end{equation}
The first term on the left-hand side vanishes since partial derivatives commute. Hence,
\begin{equation} \label{Stab 1 for F=0}
{\rm ad}^*_{\widetilde{F}_{ij}^I} B_{z_I} = 0.     
\end{equation}
Similarly, using \eqref{UV F}, we deduce
\begin{equation}\label{Deriving dF AF zero}
\begin{split}
&\left[\partial_{t^i},\partial_{t^j}\right] A''_{\bar z_I}+\partial_{\bar z_I}\widetilde{F}_{ji}^I+\big[ A''_{\bar z_I}, \widetilde{F}_{ji}^I\big]\\
&\qquad\qquad=
 2 \pi i\left(\partial_{t^j}\,\nabla P_i\big( B_{z_I}(\mathsf q_i) \big)+ \big[ \widetilde{\mathcal A}^I_{j},\nabla P_i\big( B_{z_I}(\mathsf q_i) \big) \big] \right) \delta\big( z_I-z_I(\mathsf q_i) \big)\\
&\qquad\qquad\qquad - 
 2 \pi i\left(\partial_{t^i}\,\nabla P_j\big( B_{z_I}(\mathsf q_j) \big)+ \big[ \widetilde{\mathcal A}^I_{i},\nabla P_j\big( B_{z_I}(\mathsf q_j) \big) \big] \right) \delta\big( z_I-z_I(\mathsf q_j) \big).
\end{split}
\end{equation}
Now, note that \eqref{UV Phi i} implies 
\begin{equation}
\partial_{t^j} \nabla P_i\big( B_{z_I}(\mathsf q_i) \big)+\Big[{\widetilde{\mathcal A}^I_{j}},\nabla P_i\big( B_{z_I}(\mathsf q_i) \big)\Big] = 0,
\end{equation}
so that both terms on the right-hand side of \eqref{Deriving dF AF zero} vanish. The first term on the left-hand side of \eqref{Deriving dF AF zero} also vanishes since partial derivatives commute and \eqref{Deriving dF AF zero} reduces to
\begin{equation} \label{Stab 2 for F=0}
\partial_{\bar z_I}\widetilde{F}_{ji}^I+[A''_{\bar z_I}, \widetilde{F}_{ji}^I]=0.
\end{equation}
Equation \eqref{ZC_eqs} now follows from \eqref{Stab 1 for F=0} and \eqref{Stab 2 for F=0} using the stability condition \eqref{inf_freeness}.
\end{proof}

\begin{remark}
In standard terminology of the theory of Lagrangian multiforms (see Chapter \ref{chap:lm-background}), the first three equations \eqref{UV F}--\eqref{UV Phi i} are the multitime Euler--Lagrange equations and the last equation \eqref{UV P} is (related to) the closure relation. Importantly, the latter holds automatically as a consequence of \eqref{UV Phi i} and $G$-invariance of $P_i$, thus confirming that we indeed have a Lagrangian multiform --- a Lagrangian one-form here --- for the Hitchin system. To see this, note that
\begin{equation}
\partial_{t^j} P_i\big( B_{z_I} \big) = \big\langle \partial_{t^j} B_{z_I},\nabla P_i( B_{z_I}) \big\rangle = \Big\langle \partial_{t^j} B_{z_I} + {\rm ad}^*_{{\mathcal A}^I_{j}} B_{z_I},\nabla P_i( B_{z_I}) \Big\rangle = 0,
\end{equation}    
where we used the infinitesimal coadjoint-invariance of $P_i$ in the second equality (see the argument right after \eqref{gradient def}) and \eqref{UV Phi i} in the last equality.
\end{remark}

\section{Adding type A defects at marked points on \texorpdfstring{$C$}{C}} \label{sec: adding punctures}

Having obtained a variational description of the Hitchin system for the case without marked points, we now turn to extending the above construction to the case \emph{with} marked points. As we will see later on in the thesis, namely in Chapters \ref{chap:lax-hitchin} and \ref{chap:hitchinexamples}, this enables us to incorporate a wider class of integrable systems into our framework. We will work with a Riemann surface $C$ with $N \in \ZZ_{\geq 1}$ distinct marked points $\{ \mathsf p_\alpha \}_{\alpha=1}^N$ and will allow the Higgs field $B \in T^\ast_{A''} \mathcal M$ to have simple poles at these points by introducing extra structure  encoding the residues of $B$. 

More specifically, to each marked point $\mathsf p_\alpha$ we associate a coadjoint orbit $\mathcal{O}_\alpha$ in the fibre of $\g^\ast_{\mathcal P}$ over $\mathsf p_\alpha$.  This coadjoint orbit $\mathcal{O}_\alpha$ is determined by an element $\Lambda_\alpha\in\g^\ast$. In a local trivialisation of ${\mathcal P}$ over $U_I\ni \mathsf p_\alpha$, an element $X\in \mathcal{O}_\alpha$ is represented by
\begin{equation}\label{coadjoint orbit}
    X^I=\mathrm{Ad}^\ast_{\varphi_\alpha^I}\Lambda_\alpha, \quad \text{ for some }\varphi_\alpha^I\in G.
\end{equation}
If $\mathsf p_\alpha\in U_I\cap U_J$, then we must have $X^I=\mathrm{Ad}^\ast_{g_{IJ}(\mathsf p_\alpha)}X^J$. So, we require $\varphi_\alpha^I=g_{IJ}(\mathsf p_\alpha)\varphi_\alpha^J$.  This means that $\varphi_\alpha^I$ represents an element $\varphi_\alpha$ of the fibre of ${\mathcal P}$ over ${\mathsf p}_\alpha$.

The Kostant--Kirillov--Souriau form on $\mathcal{O}_\alpha$ is given by $\omega_\alpha(X,Y) \coloneqq \langle\Lambda_\alpha,[X,Y]\rangle$ for every $X,Y\in\mathfrak{g}$ representing tangent vectors in $T_{\Lambda_\alpha} \mathcal{O}_\alpha$.  The Kostant--Kirillov--Souriau form $\omega_\alpha$ can equally be described as follows. Consider the map
\begin{equation}
G\longrightarrow \mathcal O_\alpha , \qquad \varphi_\alpha^I \longmapsto \Ad^*_{\varphi_\alpha^I} \Lambda_\alpha .
\end{equation}
The pullback of $\omega_\alpha$ under this map, which, by abuse of notation, we also denote by $\omega_\alpha$, is the $2$-form given by $\omega_\alpha = \d\big\langle \Lambda_\alpha,(\varphi^I_\alpha)^{-1} \d\varphi^I_\alpha \big\rangle$. Note that this is exact, and that both $\omega_\alpha$ and $\big\langle \Lambda_\alpha,(\varphi^I_\alpha)^{-1} \d\varphi^I_\alpha \big\rangle$ are invariant under changes of local trivialisation. We will denote the latter by $\big\langle \Lambda_\alpha,(\varphi_\alpha)^{-1} \d\varphi_\alpha \big\rangle$ to emphasise this.

Instead of working on the phase space $(T^\ast \mathcal M, \omega)$ as in Section \ref{sec: Lag for Hitchin} and Section \ref{sec: Lag for Hitchin mod G}, we will now work with the extended phase space
\begin{equation} \label{extended phase space}
\bigg( T^\ast \mathcal M \times \prod_{\alpha = 1}^N \mathcal{O}_\alpha, \; \omega + \sum_{\alpha = 1}^N \omega_\alpha \bigg).
\end{equation}
We extend the tautological one-form $\alpha$ on $T^\ast \mathcal M$, given in \eqref{tauto}, to a one-form on the extended phase space \eqref{extended phase space}, which we still denote by $\alpha$. It is given explicitly by
\begin{equation}
\label{ext_tauto}
\alpha_{(B, A'', (\varphi_\alpha))} \coloneqq \frac{1}{2 \pi i}\int_C\langle B,\delta A''\rangle + \sum_{\alpha = 1}^N \big\langle \Lambda_\alpha, (\varphi_\alpha)^{-1} \d \varphi_\alpha \big\rangle.
\end{equation}
To evaluate the second term, for each $\alpha$, we choose an open set $U_I$ containing ${\mathsf p}_\alpha$, and evaluate $\big\langle \Lambda_\alpha, (\varphi_\alpha^I)^{-1} \d \varphi_\alpha^I \big\rangle$, where $\varphi_\alpha^I\in G$ is the representative of $\varphi_\alpha$ in our chosen trivialisation.

The action of the group $\mathcal{G}$ of bundle automorphisms on $T^\ast \mathcal M \times \prod_{\alpha = 1}^N \mathcal{O}_\alpha$ is as follows, cf. \eqref{cal G action on T*M},
\begin{equation} \label{group_action_marked}
\begin{split}
\mathcal G \times T^\ast \mathcal M \times \prod_{\alpha = 1}^N \mathcal{O}_\alpha &\longrightarrow T^\ast \mathcal M \times \prod_{\alpha = 1}^N \mathcal{O}_\alpha \\
\Big( g, \big(B, A'', (\varphi_\alpha^I) \big) \Big) &\longmapsto g\cdot \big( B, A'', (\varphi^I_\alpha) \big) = \Big( \Ad^*_g B, {}^g A'', \big( g^I(\mathsf p_\alpha) \varphi^I_\alpha \big) \Big).
\end{split}
\end{equation} 
This induces the following infinitesimal left action
\begin{equation}
\label{global_inf_action_X}
\delta_X B = {\rm ad}^*_X B, \quad \delta_X A'' = - \bar{\partial}^{A''} X , \quad \delta_X \big( \Ad_{\varphi^I_\alpha}^\ast \Lambda_\alpha \big) = {\rm ad}^*_{X^I(\mathsf p_\alpha)} \big( \Ad_{\varphi_\alpha^I}^\ast \Lambda_\alpha \big)
\end{equation}
of $X \in \mathfrak G$ on $\big( B, A'', (\varphi_\alpha^I) \big) \in T^\ast \mathcal M \times \prod_{\alpha = 1}^N \mathcal{O}_\alpha$.
\begin{remark}
We can describe the phase space \eqref{extended phase space} more invariantly in the following manner. Let $G_{\Lambda_\alpha}$ be the stabiliser of $\Lambda_\alpha\in\g^\ast$.  Then, $\Ad^\ast_{\varphi_\alpha^I}\Lambda_\alpha$ corresponds to a left coset $\varphi_\alpha^IG_{\Lambda_\alpha}\in G/G_{\Lambda_\alpha}$. This represents an element of the quotient $\mathcal P_{{\mathsf p}_\alpha}/G_{\Lambda_\alpha}$ in our chosen local trivialisation. An element of $\mathcal P_{{\mathsf p}_\alpha}/G_{\Lambda_\alpha}$ is usually called a reduction of the structure group of $\mathcal P$ at ${\mathsf p}_\alpha$ to $G_{\Lambda_\alpha}$. Therefore, our phase space is the product of the cotangent bundle of the space of holomorphic structures and the space of reductions of structure group to $G_{\Lambda_\alpha}$ at the subset of marked points $\{ {\mathsf p}_\alpha \}_{\alpha = 1}^N$.
\end{remark}

We proceed with generalising the construction of Section \ref{sec: Lag for Hitchin}. In the present setup, the Lagrangian one-form \eqref{L} and action \eqref{ungauged action} take the form
\begin{equation}
\Lag =\frac{1}{2 \pi i}\int_C\langle B,\delta A''\rangle + \sum_{\alpha=1}^N\langle\Lambda_\alpha,\varphi_\alpha^{-1}\d \varphi_\alpha\rangle - H_i\big( B, A'', (\varphi_\alpha) \big) \d t^i,    
\end{equation}
for some Hamiltonians $H_i : T^\ast \mathcal M \times \prod_{\alpha=1}^N \mathcal O_\alpha \to \CC$.
Its pullback along an arbitrary immersion $\Sigma : \RR^n \to T^\ast \mathcal M \times \prod_{\alpha=1}^N \mathcal O_\alpha \times \RR^n$, $\Sigma(u) = \big( B(u), A''(u), (\varphi_\alpha(u)), t(u) \big)$ is then
\begin{equation}
\Sigma^\ast \Lag = \frac{1}{2 \pi i}\int_C \langle B,\d_{\RR^n} A'' \rangle + \sum_{\alpha=1}^N \langle \Lambda_\alpha,\varphi_\alpha^{-1}\d_{\RR^n} \varphi_\alpha \rangle - H_i\big( B, A'', (\varphi_\alpha) \big) \d_{\RR^n} t^i,
\end{equation}
so that for any parametrised curve $\Gamma : (0,1) \to \RR^n$, $s \mapsto \big( u^j(s) \big)$, the corresponding action is given by $S_\Gamma[\Sigma] \coloneqq \int_0^1 (\Sigma \circ \Gamma)^\ast \Lag$, which explicitly reads
\begin{equation} \label{ungauged action Phi A Lambda}
\begin{split}
S_\Gamma[\Sigma] &= \int_0^1\Bigg(\! -\frac{1}{2 \pi i}\int_C \big\langle B(u),\partial_{u^j} A''(u) \big\rangle + \sum_{\alpha=1}^N \big\langle \Lambda_\alpha,\varphi_\alpha(u)^{-1} \partial_{u^j} \varphi_\alpha(u) \big\rangle \\
&\qquad\qquad\qquad\qquad\qquad\qquad - H_i\Big( B(u(s)), A''(u(s)), \big( \varphi_\alpha(u(s)) \big) \Big) \frac{\partial t^i}{\partial u^j}\Bigg) \frac{\d u^j}{\d s} \d s.
\end{split}
\end{equation}

We have the following analogue of Proposition \ref{prop: Hi invariance Hitchin}.

\begin{proposition} \label{prop: Hi invariance Hitchin type A}
The action \eqref{ungauged action Phi A Lambda} is invariant under the action of $\mathcal G$ on $T^\ast \mathcal M \times \prod_{\alpha=1}^N \mathcal O_\alpha \times \RR^n$ given by \eqref{group_action_marked} and extended trivially to the $\RR^n$-factor if and only if each $H_i$, for $i=1, \ldots, n$, is invariant under the group action, that is,
\begin{equation} \label{Hi invariance prop 2}
H_i\Big( \Ad^*_g B, {}^g A'', \big( g(\mathsf p_\alpha) \varphi_\alpha \big) \Big) = H_i\big( B, A'', (\varphi_\alpha) \big)
\end{equation}
for any $\big( B, A'', (\varphi_\alpha) \big) \in T^\ast \mathcal M \times \prod_{\alpha=1}^N \mathcal O_\alpha$ and $g \in \mathcal G$. Moreover,
the value of the associated moment map $\mu : T^\ast \mathcal M \times \prod_{\alpha=1}^N \mathcal O_\alpha \to \mathfrak{G}^\ast$ on an infinitesimal bundle morphism $X \in \mathfrak G$ is given by
\begin{equation} \label{moment_map_Hitchin type A}
\mu_{(B, A'',(\varphi_\alpha))}(X) = \frac{1}{2 \pi i}\int_C \big\langle B, \bar\partial^{A''} X \big\rangle  +\sum_{\alpha=1}^N \int_C\big\langle \Ad^\ast_{\varphi_\alpha} \Lambda_\alpha, X \big\rangle\delta_{\mathsf p_\alpha} .
\end{equation}
\begin{proof}
For the invariance of the action under the group action \eqref{group_action_marked}, by Proposition \ref{prop: Hi invariance Hitchin} we only need to check the invariance of the new kinetic terms for the $\varphi_\alpha$ but this is immediate.

The extra term in the moment map \eqref{moment_map_Hitchin type A} compared to the expression in \eqref{moment_map_Hitchin} comes from generalising the argument in the last part of the proof of Proposition \ref{prop: Hi invariance Hitchin} to the present case, and again, the new contribution comes from the coadjoint orbit terms in the kinetic part. 
\end{proof}
\end{proposition}

Following the discussion in Remark \ref{rem: mu well defined}, we can use Stokes' theorem to rewrite the expression \eqref{moment_map_Hitchin type A} for the moment map as
\begin{equation}
\mu_{(B, A'',(\varphi_\alpha))}(X) = \int_C \bigg\langle \frac{1}{2 \pi i}\bar\partial^{A''} B + \sum_{\alpha=1}^N \Ad^\ast_{\varphi_\alpha} \Lambda_\alpha \delta_{\mathsf p_\alpha} ,X \bigg\rangle.
\end{equation}
In particular, we see that the vanishing of $\mu$ at any $\big( B, A'', (\varphi_\alpha) \big) \in T^\ast \mathcal M \times \prod_{\alpha=1}^N \mathcal O_\alpha$ corresponds to the condition
\begin{equation} \label{mu_with_punctures}
\bar{\partial}^{A''} B = -2 \pi i\sum_{\alpha=1}^N \Ad^\ast_{\varphi_\alpha} \Lambda_\alpha \delta_{\mathsf p_\alpha},
\end{equation}
which reads locally as 
\begin{equation}
\label{local_mu_constraint}
    \partial_{\bar z_I} B_{z_I} + {\rm ad}^*_{A''_{\bar z_I}} B_{z_I} = 2 \pi i\sum_{\substack{\alpha=1\\ \mathsf p_\alpha \in U_I}}^N \Ad^\ast_{\varphi^I_\alpha} \Lambda_\alpha \delta\big( z_I - z_I(\mathsf p_\alpha) \big).
\end{equation}

To produce a (non-trivial) dynamical theory, we must choose Hamiltonians $H_i$ on the symplectic manifold $T^\ast\M\times \prod_\alpha\mathcal{O}_\alpha$. We do this in a manner similar to the case without marked points, with Hamiltonians given in \eqref{Hitchin Hamiltonians} in terms of invariant polynomials $P_i$ on $\g^\ast$ and points ${\mathsf q}_i$. This ensures, in particular, their invariance as in Proposition \ref{prop: Hi invariance Hitchin type A}. We will specify the number of points ${\mathsf q}_i$ more precisely when we look at specific examples in Chapter \ref{chap:hitchinexamples}.

Let us now proceed with generalising the construction of Section \ref{sec: Lag for Hitchin mod G} to the case with marked points.
Recall that we introduced the $\g$-valued gauge field ${\cal A}$ along $\RR^n$ and $\langle \Sigma^\ast\mu,\mathcal A \rangle$ was added to the pullback $\Sigma^\ast \Lag$ of the Lagrangian one-form \eqref{L_Phi_A} to eventually produce the gauged action \eqref{gauged Hitchin action}.
In the present case, we have the gauged action
\begin{multline}\label{gauged Hitchin action marked points}
S_\Gamma[\Sigma,\mathcal{A},t] = 
\int_0^1 \Bigg( -\frac{1}{2 \pi i}
\int_C \left\langle B(u), \frac{\partial A''}{\partial u^j}- \bar{\partial}^{A''}\mathcal{A}_j \right\rangle+\sum_{\alpha=1}^N \Big\langle \Lambda_\alpha, \varphi_\alpha^{-1} \big(\partial_{u^j} + \mathcal A_j(\mathsf p_\alpha) \big) \varphi_\alpha \Big\rangle\\
- H_i\big( B(u) \big)\frac{\partial t^i}{\partial u^j} \Bigg) \frac{\d u^j}{\d s} \d s .
\end{multline}
Rewriting this in terms of the connection $A=A''+\mathcal{A}$ over $C\times\RR^n$ yields the following analogue of Theorem \ref{thm: BF Lagrangian}.

\begin{theorem} \label{thm: 3d BF with type A defects 1}
The gauged multiform action \eqref{gauged Hitchin action marked points} for the Hitchin system with marked points is given by the multiform action for $3$d mixed BF theory on $C \times \RR^n$ for the collection of fields $\big( B, A, (\varphi_\alpha) \big)$ with a type B line defect along each coordinate $t^i$ determined by the Hitchin Hamiltonian $H_i$ defined as in \eqref{Hitchin Hamiltonians}, and a type A line defect at each marked point $\mathsf p_\alpha$, namely
\begin{equation}\label{S Gamma 2}
\begin{split}
S_\Gamma[B, A, (\varphi_\alpha), t] &= \frac{1}{2 \pi i}\int_{C\times \Gamma} \big\langle B,F_A \big\rangle + \sum_{\alpha=1}^N \int_0^1\Big\langle\Lambda_\alpha,\varphi_\alpha^{-1}\big(\partial_{u^j} + \mathcal A_j(\mathsf p_\alpha) \big)\varphi_\alpha \Big\rangle \frac{\d u^j}{\d s}\d s \\
&\qquad\qquad\qquad\qquad\qquad\qquad - \int_0^1H_i\big(B(u(s)) \big) \frac{\d t^i}{\d s} \d s ,
\end{split}
\end{equation}
for an arbitrary curve $\Gamma : (0,1) \to \RR^n$, $s \mapsto u(s)$, where $F_A$ is the curvature introduced in \eqref{FA def}.
\end{theorem}

The action \eqref{S Gamma 2} will appear again in Chapter \ref{chap:lax-hitchin}, where, following \cite{VW}, we perform a reduction to obtain a \emph{unifying action} for Hitchin systems associated with Riemann surfaces of arbitrary genus with marked points. Notably, the construction we have presented here extends the work of \cite{VW} in a key aspect: we have \emph{derived} the $3$d mixed BF action as a variational description of the Hitchin system rather than taking it as given as was done in \cite{VW}. Before we move to discussing this reduction, let us present the following analogue of Theorem \ref{th_one_form_Hitchin} that we need for what follows in the next chapter.

\begin{theorem} \label{thm: 3d BF with type A defects 2}
The gauged univariational principle applied to the $3$d mixed BF multiform action $S_\Gamma[B,A, (\varphi_\alpha), t]$ in \eqref{S Gamma 2} yields a set of equations for the fields $B$, $A$ and  $\varphi_\alpha$, for $\alpha =1,\ldots, N$. Working in any chart $(U_I, z_I)$ of $C$ and in terms of the components $B_{z_I}(z_I, \bar z_I, u) $, $A''_{\bar z_I}(z_I, \bar z_I, u)$, $\mathcal A^I_{j}(z_I, \bar z_I, u)$ of the various fields, these equations take the following form:
\begin{subequations}
\label{eqs_gauged_action}
\begin{align}
\partial_{\bar z_I} \widetilde{\mathcal A}^I_{i} - \partial_{t^i} A''_{\bar z_I} + [A''_{\bar z_I}, \widetilde{\mathcal A}^I_{i}] &= 2 \pi i\,\nabla P_i\big( B_{z_I}(\mathsf q_i) \big) \delta\big( z_I-z_I(\mathsf q_i) \big), \label{F_B}\\
\partial_{\bar z_I} B_{z_I} + {\rm ad}^*_{A''_{\bar z_I}} B_{z_I} &= 2 \pi i\sum_{\substack{\alpha=1\\ \mathsf p_\alpha \in U_I}}^N \Ad^\ast_{\varphi^I_\alpha} \Lambda_\alpha \delta\big( z_I - z_I(\mathsf p_\alpha) \big) \label{A_defect_modif},\\
\partial_{t^j} B_{z_I}+{\rm ad}^*_{\widetilde{\mathcal A}^I_{j}} B_{z_I} &= 0 \label{eq:B_z_punctures},\\
\partial_{t^j} H_i(B)&=0 ,\\
\partial_{t^j} \big( \Ad^\ast_{\varphi^I_\alpha} \Lambda_\alpha \big) + {\rm ad}^*_{\widetilde{\mathcal A}^I_{ j}(\mathsf p_\alpha)} \Ad^\ast_{\varphi^I_\alpha} \Lambda_\alpha &=0\label{eq:L_beta},
\end{align}
\end{subequations}
where we have used the invertibility of the map $(u^j) \mapsto \big( t^i(u) \big)$ to define $\widetilde{\mathcal A}^I_{i} = \frac{\partial u^j}{\partial t^i} \mathcal A^I_{j}$.
The equations associated to any pair of overlapping charts $(U_I, z_I)$ and $(U_J, z_J)$ are compatible on $U_I \cap U_J \neq \emptyset$.
The following zero curvature equations also hold
\begin{equation}\label{eq:zce}
\partial_{t^i} \widetilde{\mathcal A}^I_{j} - \partial_{t^j} \widetilde{\mathcal A}^I_{i} + \big[ \widetilde{\mathcal A}^I_{i}, \widetilde{\mathcal A}^I_{j} \big] =0.
\end{equation}
\end{theorem}
\begin{proof}
The computation for the variation of the action with respect to the fields $B^I$, $A^I$,  and $t^i$ is exactly as in the proof of Theorem \ref{th_one_form_Hitchin} and so is the derivation of \eqref{eq:zce}. The new terms in the action only modify the equation obtained by varying ${\cal A}^I$ and this gives the right-hand side of \eqref{A_defect_modif}. Finally, the variation with respect to $\varphi_\beta$ gives
\begin{equation}
\begin{split}
\delta_{\varphi_\beta} S_\Gamma[B, A, (\varphi_\alpha), t] &= \delta_{\varphi_\beta} \int_0^1\sum_{\alpha=1}^N \Big\langle\Lambda_\alpha,\varphi_\alpha^{-1}\big(\partial_{u^j} + \mathcal A_{j}(\mathsf p_\alpha) \big)\varphi_\alpha \Big\rangle \frac{\d u^j}{\d s} \d s\\
&= \int_0^1\Big\langle\Lambda_\beta,-(\varphi^I_\beta)^{-1}\delta \varphi^I_\beta(\varphi^I_\beta)^{-1}\big( \partial_{u^j} + \mathcal A^I_{j}(\mathsf p_\alpha) \big) \varphi^I_\beta\\
&\qquad\qquad\qquad\qquad\qquad +(\varphi^I_\beta)^{-1}\big( \partial_{u^j}+ \mathcal A^I_{j}(\mathsf p_\alpha) \big) \delta\varphi^I_\beta \Big\rangle \frac{\d u^j}{\d s} \d s\\
&= \int_0^1\Big\langle\Ad^\ast_{\varphi^I_\beta} \Lambda_\beta, \big[ \partial_{u^j} + \mathcal A^I_{j}(\mathsf p_\beta),\delta\varphi^I_\beta(\varphi^I_\beta)^{-1}\big] \Big\rangle \frac{\d u^j}{\d s} \d s\\
&= - \int_0^1\Big\langle \big( \partial_{u^j} + {\rm ad}^\ast_{ \mathcal A^I_{j}(\mathsf p_\beta)} \big)\big( \Ad^\ast_{\varphi^I_\beta} \Lambda_\beta \big), \delta\varphi^I_\beta(\varphi^I_\beta)^{-1} \Big\rangle \frac{\d u^j}{\d s}\d s
\end{split}
\end{equation}
so that we obtain \eqref{eq:L_beta}.
\end{proof}

To summarise this chapter, we have shown that in the case without marked points, the Hitchin system on the symplectic quotient $\mu^{-1}(0) / \mathcal G$ is described variationally by the action for $3$d mixed BF theory with type B line defects associated with each of the Hamiltonians $H_i$ and corresponding times $t^i$ for $i=1,\ldots, n$ (Theorem \ref{thm: BF Lagrangian}), and in the case with marked points by adding type A line defects at each marked point in $C$ to the action (Theorem \ref{thm: 3d BF with type A defects 1}). The latter case will be the subject of the next chapter where we obtain a \emph{unifying Lagrangian one-form} for Hitchin systems associated with Riemann surfaces of arbitrary genus with marked points.

\chapter{Hitchin system in Lax form and its variational formulation}\label{chap:lax-hitchin}
Hitchin's original construction \cite{H} provides integrable systems associated with Riemann surfaces without marked points. The generalisation of this construction to the case of Riemann surfaces with marked points allows for a wider class of integrable models --- even certain models with spectral parameters lying on curves of genus zero and genus one --- to be seen as realisations of the Hitchin system. 

The focus of this chapter, adapted from \cite[Section 4]{CHSV}, is this more general form of Hitchin systems: those associated with Riemann surfaces of any arbitrary genus \emph{with} marked points. In Section \ref{sec: adding punctures}, we obtained a variational description for these systems in terms of a multiform action \eqref{S Gamma 2} for $3$d mixed BF theory with type A and type B line defects. Now, in this chapter, we work with this multiform action as our starting point. 

Central to what follows is the isomorphism \eqref{Hitchin phase space intro} which we use to pass from the description of the Hitchin system on $\mu^{-1}(0) / \mathcal G$ in terms of smooth $\g$-valued $(0,1)$-connections $A'' \in \mathcal M$ and smooth Higgs fields $B \in T_{A''}^\ast \mathcal M$, as in Section \ref{sec: Lag for Hitchin mod G} and Section \ref{sec: adding punctures}, to a description of the Hitchin system on $T^\ast \text{Bun}_G(C)$ in terms of holomorphic transition functions and holomorphic Higgs fields. We will thus establish the right-hand side of the commutative diagram in Figure \ref{fig:bfcommutdiag}.
\begin{figure}
\begin{equation*}
\begin{tikzcd}[row sep=30, column sep=60]
S_\Gamma[B,A, (\varphi_\alpha), t] \arrow[r, "\text{holomorphic}", "\text{description}"'] \arrow[d, "\text{univariational}"' pos=0.35, "\text{principle}"' pos=0.65] & \text{Unifying action} \arrow[d, "\text{univariational}" pos=0.35, "\text{principle}" pos=0.65] \\
\text{Variational equations \eqref{eqs_gauged_action}} \arrow[r, "\text{holomorphic}", "\text{description}"'] & \partial_{t^i}L=[M_i,L] 
\end{tikzcd}
\end{equation*}
\caption{Two descriptions of Hitchin systems and corresponding variational equations}
\label{fig:bfcommutdiag}
\end{figure}

This chapter has a twofold purpose. First, in Section \ref{sec: Lax matrices}, we show that the set of variational equations derived in Theorem \ref{thm: 3d BF with type A defects 2} encode the hierarchy of equations of the Hitchin system in the standard Lax form\footnote{To our knowledge, the earliest connection between the Hitchin system and the Lax formalism appeared in \cite{Kri, LeOZ1}.}, namely $\partial_{t^i}L=[M_i,L]$ for a meromorphic Lax matrix $L$ and a collection of meromorphic Lax matrices $M_i$ associated with each time $t^i$ in the hierarchy for $i=1, \ldots, n$. Second, in Section \ref{sec:unifyingmultiform}, we show that this hierarchy of Lax equations for the Hitchin system is itself variational and can be derived as the Euler--Lagrange equations of a new multiform action, which we refer to as the \emph{unifying action}, derived, in turn, from the multiform action \eqref{S Gamma 2}.

\section{Hitchin system in Lax form} \label{sec: Lax matrices}

This section is devoted to obtaining a description of the Hitchin system as a hierarchy of Lax equations. Following \cite{VW}, we make the observation that \eqref{eq:B_z_punctures} takes the form of a hierarchy of Lax equations for Lax matrices $B_{z_I}$ and $\widetilde{\mathcal A}^I_i$ defined on the coordinate patch $(U_I, z_I)$ that become meromorphic when solving \eqref{F_B}--\eqref{A_defect_modif} in a local trivialisation such that $A''_{\bar z_I} = 0$. The compatibility of this hierarchy of Lax equations is ensured by \eqref{eq:zce}. 

As usual, in order to bring \eqref{eq:B_z_punctures} to the standard Lax form, we will identify $\g^\ast$ with $\g$ using a nondegenerate invariant bilinear form on $\g$. We still denote the latter by $\langle~,~\rangle$ (like the pairing between $\g^\ast$ and $\g$) hoping that it will not confuse the reader. This allows us to identify adjoint and coadjoint actions. Moreover, from now on we only consider matrix Lie groups and Lie algebras.

Next, we pick a point $\mathsf p \in C$ distinct from all the marked points $\mathsf p_\alpha$ with $\alpha = 1,\ldots, N$, and $\mathsf q_i$ with $i = 1,\ldots, n$. Pick a neighbourhood $U_0$ of $\mathsf p$ not containing any of these other points and equipped with a local coordinate $z_0 : U_0 \to C$. We also let $U_1 \coloneqq C \setminus \{ \mathsf p \}$ so that $\{ U_0, U_1 \}$ forms an open cover of $C$. Let us stress that for $g \geq 1$ the open $U_1 \subset C$ is not a coordinate chart since in general it cannot be equipped with a holomorphic coordinate $U_1 \to \CC$. However, we can further refine the cover by choosing a holomorphic atlas $\{ (U_I, z_I) \}_{I \in \mathcal I}$ for $U_1$, where $\mathcal I$ is any indexing set not containing $0$ and $1$ so that we can also use the notation $U_I$ for $I \in \{ 0, 1 \}$. In other words, $U_I$ for $I \in \mathcal I$ are open subsets of $U_1$ such that $\bigcup_{I \in \mathcal I} U_I = U_1$ and equipped with coordinates $z_I : U_I \to \CC$. 

The open cover $\{U_0, U_1\}$ of $C$ suffices for the purpose of describing the smooth principal $G$-bundle $\pi : {\mathcal P} \to C$ with a holomorphic structure specified by $A'' \in \mathcal M$. Indeed, since $U_1$ is a non-compact Riemann surface and $G$ is semisimple, we can trivialise the holomorphic principal $G$-bundle $({\mathcal P}, A'')$ over $U_1$. Since we can also trivialise $({\mathcal P}, A'')$ over $U_0$, we obtain local trivialisations for the holomorphic principal $G$-bundle $({\mathcal P}, A'')$ relative to the open cover $\{ U_0, U_1 \}$ of $C$. More importantly for us, we also have local trivialisations of the pullback bundle $\pi_C^\ast {\mathcal P} \cong {\mathcal P} \times \RR^n$, see Section \ref{sec: Lag for Hitchin mod G}, which we write as
\begin{subequations} \label{local triv examples}
\begin{align}
\label{loc triv 0} \psi_0 : \pi^{-1}(U_0 \times \RR^n) &\overset{\cong}\longrightarrow U_0 \times \RR^n \times G , \quad p \longmapsto (\pi(p), f_0(p)) ,\\
\label{loc triv 1} \psi_1 : \pi^{-1}(U_1 \times \RR^n) &\overset{\cong}\longrightarrow U_1 \times \RR^n \times G , \quad p \longmapsto (\pi(p), f_1(p))
\end{align}
\end{subequations}
relative to the open cover $\{ U_0 \times \RR^n, U_1 \times \RR^n \}$ of $C \times \RR^n$. Let $g_{01} : (U_0 \cap U_1) \times \RR^n \to G$ be the smooth transition function on the overlap $(U_0 \cap U_1) \times \RR^n \neq \emptyset$ so that, with $(x,t)=\pi(p)\in (U_0 \cap U_1) \times \RR^n$, we have
\begin{equation}
\psi_0 \circ \psi_1^{-1} : (U_0 \cap U_1) \times \RR^n \times G \longrightarrow (U_0 \cap U_1) \times \RR^n \times G , \quad (x, t, g) \, \longmapsto\, \big( x, t, g_{01}(x,t) g \big) .
\end{equation}
Since $U_I \subset U_1$ for each $I \in \mathcal I$, we will use the restriction of the local trivialisation \eqref{loc triv 1} on each $\pi^{-1}(U_I \times \RR^n)$ so that the transition functions $g_{IJ} : (U_I \cap U_J) \times \RR^n \to G$ and $g_{1I} : (U_1 \cap U_I) \times \RR^n \to G$ are trivial for all $I,J \in \mathcal I$, and $g_{0I} : (U_0 \cap U_I) \times \RR^n \to G$ is given by $g_{0I} = g_{01}$ for all $I \in \mathcal I$.

Let us revisit the equations of motion \eqref{eq:B_z_punctures} derived in Theorem \ref{thm: 3d BF with type A defects 2}. Since we are now identifying $\g^\ast$ with $\g$ and the coadjoint action with the adjoint action, we can rewrite this set of equations as
\begin{equation}\label{eq:eom-B_zA_k-def-hitchin}
\partial_{t^i} B_{z_I} = \big[ \! - \widetilde{\mathcal A}^I_i, B_{z_I} \big]
\end{equation}
on the open chart $U_I \subset C$ for each $I \in \mathcal I \cup \{0\}$. We note following \cite{VW} that these clearly resemble a hierarchy of Lax equations. We can rewrite them without using coordinates by expressing them in terms of the $\g$-valued $(1,0)$-forms $B^I = B_{z_I}(z_I, \bar z_I, t) \d z_I \in \Omega^{1,0}(U_I \times \RR^n, \g)$ for $I \in \mathcal I \cup \{0\}$ as
\begin{subequations} \label{global Lax}
\begin{equation}\label{global Lax a}
\partial_{t^i} B^I = \big[ \! - \widetilde{\mathcal A}^I_{i}, B^I \big] .
\end{equation}
Since these equations for $I \in \mathcal I$ are compatible on overlaps $U_I \cap U_J \neq \emptyset$ and in fact $B^I = B^J$ and $\widetilde{\mathcal A}^I_{i} = \widetilde{\mathcal A}^J_{i}$ since the transition functions on these overlaps are trivial, we may consider the equations \eqref{global Lax a} simply for $I \in \{ 0, 1 \}$. In that case, $B^I \in \Omega^{1,0}(U_I \times \RR^n, \g)$ and $\widetilde{\mathcal A}^I_{i}$ is a $\g$-valued function on $U_I \times \RR^n$ for $I\in \{0, 1 \}$ which on the overlap $U_0 \cap U_1$ are related by
\begin{equation} \label{global Lax b}
B^0 = g_{01} B^1 g_{01}^{-1} , \qquad \widetilde{\mathcal A}^0_{i} = g_{01} \widetilde{\mathcal A}^1_{i} g_{01}^{-1} - \partial_{t^i} g_{01} g_{01}^{-1}.
\end{equation}
\end{subequations}
We could also further rewrite \eqref{global Lax a} more compactly as
\begin{equation}\label{eq:eom-B_zA_k-def-hitchin U1 geom}
\d_{\RR^n} B^I = \big[\! - \mathcal A^I, B^I \big]
\end{equation}
recalling that $\mathcal A^I = \mathcal A^I_{j} \d u^j = \widetilde{\mathcal A}^I_{i} \d t^i$, see Theorem \ref{thm: 3d BF with type A defects 2}, and correspondingly the second equation in \eqref{global Lax b} would read 
\begin{equation}
\label{transfo_A}
    \mathcal A^0 = g_{01} \mathcal A^1 g_{01}^{-1} - \d_{\RR^n} g_{01} g_{01}^{-1}.
\end{equation}

In order for \eqref{global Lax} to describe Lax equations on the Riemann surface $C$ for the hierarchy of the Hitchin system, however, we need $B^I$ and $\widetilde{\mathcal A}^I_{i}$ to be solutions of \eqref{F_B} and \eqref{A_defect_modif} (recall that the latter represent the moment map condition $\mu=0$).
In addition, to ensure that the resulting Lax matrices are meromorphic in the spectral parameter, we make a change of local trivialisation on $\mathcal P$, moving from \eqref{local triv examples} to new local trivialisations $\tilde \psi_0$ and $\tilde \psi_1$ with respect to which $A''^I = 0$ for $I \in \{ 0, 1\}$, or equivalently $A''_{\bar z_I} = 0$ in each local chart $(U_I, z_I)$ for $I \in \mathcal I \cup \{ 0 \}$, see Section \ref{sec: M and G details} for details. Let $h = (h_I) \in \check C^0(C,G)$ be the \v{C}ech $0$-cochain implementing this change of trivialisation and let
\begin{equation} \label{Lax definition}
L^I \coloneqq h_I B^I h_I^{-1} , \qquad - M^I_{i} \coloneqq h_I \widetilde{\mathcal A}^I_{i} h_I - \partial_{t^i} h_I h_I^{-1} 
\end{equation}
be the local expressions for $B^I$ and $\widetilde{\mathcal A}^I_{i}$ with respect to this new local trivialisation. The additional minus sign in the definition of $M^I_{i}$ is introduced so that the equations \eqref{global Lax a} take on the standard Lax form in this local trivialisation, namely
\begin{subequations} \label{global Lax def}
\begin{equation}\label{global Lax def a}
\partial_{t^i} L^I = [ M^I_{i}, L^I ]
\end{equation}
on $U_I$ for $I \in \{ 0, 1 \}$. On the overlap $(U_0 \cap U_1) \times \RR^n$ we have the same relations as in \eqref{global Lax b}, namely
\begin{equation} \label{global Lax def b}
L^0 = \gamma L^1 \gamma^{-1} , \qquad M^0_{i} = \gamma M^1_{i} \gamma^{-1} + \partial_{t^i} \gamma \gamma^{-1},
\end{equation}
\end{subequations}
where the new transition function 
\begin{equation}
\label{def_gamma}
\gamma \coloneqq h_0 g_{01} h_1^{-1} : (U_0 \cap U_1) \times \RR^n \to G    
\end{equation} 
is now holomorphic on $U_0 \cap U_1$, see Section \ref{sec: M and G details}. We note here that the plus sign in the last term of the second equation in \eqref{global Lax def b} stems from the minus sign introduced in the second definition in \eqref{Lax definition}. In other words, $-M^0_{i}$ and $-M^1_{i}$ are related by an ordinary gauge transformation by $\gamma$. If we introduce also the $\g$-valued $1$-forms $M^I \coloneqq M^I_i \d_{\RR^n} t^i \in \Omega^1(U_I \times \RR^n, \g)$ then the Lax hierarchy \eqref{global Lax def a} can be rewritten more compactly as in \eqref{eq:eom-B_zA_k-def-hitchin U1 geom}, namely
\begin{subequations} \label{geometric Lax}
\begin{equation}\label{geometric Lax eq}
\d_{\RR^n} L^I = [ M^I, L^I ]
\end{equation}
and also the relations \eqref{global Lax def b} become
\begin{equation} \label{geometric Lax overlap}
L^0 = \gamma L^1 \gamma^{-1} , \qquad M^0 = \gamma M^1 \gamma^{-1} + \d_{\RR^n} \gamma \gamma^{-1}.
\end{equation}
\end{subequations}
In this adapted local trivialisation, 
the equations of motion \eqref{F_B} and \eqref{A_defect_modif} become 
\begin{subequations}
\begin{align}
\partial_{\bar z_I} M^I_i &=- 2 \pi i \nabla P_i\big( L_{z_I}(\mathsf q_i) \big) \delta\big( z_I - z_I(\mathsf q_i) \big),\\
\label{moment map constraint L} \partial_{\bar z_I} L_{z_I} &=  2 \pi i \sum_{\substack{\alpha=1\\ \mathsf p_\alpha \in U_I}}^N \varphi_\alpha \Lambda_\alpha \varphi_\alpha^{-1} \delta\big( z_I - z_I(\mathsf p_\alpha) \big)
\end{align}
\end{subequations}
for each $I \in \mathcal I \cup \{ 0\}$, where $L^I = L_{z_I}(z_I, \bar z_I, t) \d z_I$ is the local expression of the Lax matrix $L^I$ in the coordinate chart $(U_I, z_I)$.
Equivalently, using the identity \eqref{eq:del-Dirac-rel}, we have
\begin{subequations}
\begin{align}
&\partial_{\bar z_I} \bigg( M^I_i - \frac{\nabla P_i \big( L_{z_I}(\mathsf q_i)\big)}{z_I - z_I(\mathsf q_i)} \bigg) = 0,\label{eq:a_k-gauged}\\
&\partial_{\bar z_I} \Bigg( L_{z_I} + \sum_{\substack{\alpha=1\\ \mathsf p_\alpha \in U_I}}^N \frac{\varphi_\alpha \Lambda_\alpha \varphi_\alpha^{-1}}{z_I - z_I(\mathsf p_\alpha)}\Bigg) = 0.\label{eq:b_z-gauged}
\end{align}
\end{subequations}
So $M^I_i$ and $L_{z_I}$ are meromorphic in the chosen local trivialisation, with a specific pole structure.

First, equation \eqref{eq:b_z-gauged} tells us that $L^1$ is a $\g$-valued meromorphic $(1,0)$-form on $U_1$ with simple poles at each $\mathsf p_\alpha$ with residue $-\varphi_\alpha \Lambda_\alpha \varphi_\alpha^{-1}$ there. Indeed, note that the residue of a meromorphic $(1,0)$-form is independent of the local coordinate used. Locally around the point $\mathsf p_\alpha$ for any $\alpha = 1, \ldots, N$, if $\mathsf p_\alpha \in U_I$ for some $I \in \mathcal I$ then in the coordinate chart $(U_I, z_I)$ we can write
\begin{subequations} \label{Lax behaviour}
\begin{equation}\label{Lax behaviour a}
L^1 = L^I = \bigg( -\frac{\varphi_\alpha \Lambda_\alpha \varphi_\alpha^{-1}}{z_I - z_I(\mathsf p_\alpha)} + J^I_\alpha \bigg) \d z_I
\end{equation}
where the first equality follows from the fact that the transition functions of the principal $G$-bundle ${\mathcal P}$ between the opens $U_1$ and $U_I \subset U_1$ were, by definition, trivial. In the last expression, $J^I_\alpha$ denotes the holomorphic part of $L_{z_I}$ in the neighbourhood of $\mathsf p_\alpha$. Also, since $\mathsf p_\alpha \not \in U_0$ by assumption, it follows that $L^0 = L_{z_0} \d z_0$ is holomorphic on $U_0$. Therefore, using the relation \eqref{global Lax def b} we have
\begin{equation} \label{Lax behaviour b}
\gamma L^1 \gamma^{-1} = L^0
\end{equation}
\end{subequations}
on $U_0 \cap U_1$, which expresses the fact that the $\g$-valued $(1,0)$-form $\gamma L^1 \gamma^{-1} \in \Omega^{1,0} \big( (U_1 \cap U_0) \times \RR^n, \g \big)$ extends holomorphically to $U_0 \times \RR^n$. We will refer to $L$ as the \emph{Lax matrix} of the Hitchin system.

Next, equation \eqref{eq:a_k-gauged} tells us that $M^1_i$ is a $\g$-valued meromorphic function on $U_1$ with a simple pole at the point $\mathsf q_i$, with the local expression
\begin{equation} \label{MIi explicit}
M^1_i = M^I_i =  \frac{\nabla P_i \big( L_{z_I}(\mathsf q_i)\big)}{z_I - z_I(\mathsf q_i)} + K^I_i
\end{equation}
in the chart $(U_I, z_I)$ which is such that $\mathsf q_i \in U_I$. Here $K^I_i$ denotes the holomorphic part of $M^I_i$ in the neighbourhood of $\mathsf q_i$. Moreover, since we have $\mathsf q_i \not\in U_0$ by assumption, $M^0_i$ is a holomorphic $\g$-valued function on $U_0$ which is related to $M^1_i$ on the overlap $(U_0 \cap U_1) \times \RR^n$ by the second relation in \eqref{global Lax def b}. That is, $\gamma M^1_i \gamma^{-1} + \partial_{t^i} \gamma \gamma^{-1} = M^0_i$ extends holomorphically from $U_0 \cap U_1$ to $U_0$.

To summarise the above discussion, the Hitchin system can indeed be presented as the hierarchy of Lax equations \eqref{global Lax def a}, as a direct consequence expressing our variational equations \eqref{eqs_gauged_action} in a local trivialisation with $A''^I = 0$, in terms of:
\begin{enumerate}
    \item the Lax matrix $L$ which is a meromorphic section of $\pi_C^\ast \bigwedge^{1,0} C \otimes \g_{\pi_C^\ast {\mathcal P}}$ fixed in terms of the following degrees of freedom:
\begin{itemize}
  \item[$(a)$] Maps $\varphi_\alpha : \RR^n \to G / G_{\Lambda_\alpha}$ into the coadjoint orbits $\mathcal O_\alpha \cong G / G_{\Lambda_\alpha}$ for each $\alpha =1,\ldots, N$,
  \item[$(b)$] The transition function $\gamma : (U_0 \cap U_1) \times \RR^n \to G$ holomorphic in $U_0 \cap U_1$ and encoding the holomorphic structure on the principal $G$-bundle ${\mathcal P}$. In particular, $L$ satisfies \eqref{Lax behaviour b}.
\end{itemize}

\item The $\g$-valued meromorphic functions $M^0_i$ and $M^1_i$ for each $i =1, \ldots, n$ with pole structure dictated by \eqref{MIi explicit} and satisfying the second equation in \eqref{geometric Lax overlap}.
\end{enumerate}

\section{Unifying Lagrangian one-form} \label{sec:unifyingmultiform}

We now turn to the last remaining task of this chapter, which is to derive the \emph{unifying action} referred to in the commutative diagram in Figure \ref{fig:bfcommutdiag} that directly produces the hierarchy of Lax equations \eqref{global Lax def a} as its Euler--Lagrange equations. 

Recall that in Section \ref{sec: Lax matrices}, we obtained a description of the Hitchin system associated with a Riemann surface with marked points as the hierarchy of Lax equations \eqref{global Lax def a} from the set of equations of Theorem \ref{thm: 3d BF with type A defects 2} by moving to a local trivialisation in which $A''^I = 0$ and solving the moment map condition \eqref{A_defect_modif}. 

In Theorem \ref{thm: Lag 1-form for Hitchin}, we construct the \textit{unifying action} (and the associated unifying Lagrangian one-form) by writing the $3$d mixed BF action \eqref{S Gamma 2} in a local trivialisation with $A''^I = 0$ and explicitly solving the moment map condition \eqref{A_defect_modif}. Then, in Theorem \ref{thm: eom for 1d action}, we prove that varying this unifying action obtained after the above step does produce precisely the hierarchy of Lax equations \eqref{global Lax def a} thus establishing the commutativity of the diagram in Figure \ref{fig:bfcommutdiag}.

\begin{theorem} \label{thm: Lag 1-form for Hitchin}
The $3$d mixed BF action \eqref{S Gamma 2} written in a local trivialisation where $A''^0 = A''^1 = 0$ and with the moment map condition \eqref{A_defect_modif} explicitly solved, leads to the unifying action
\begin{subequations} \label{eq:unifyingaction}
\begin{equation} \label{eq:unifyingaction a}
S_{{\rm H}, \Gamma}\big[ L, \gamma, (\varphi_\alpha), t \big] = \int_\Gamma \Lag_{\rm H} ,
\end{equation}
for any parametrised curve $\Gamma : (0,1) \to \RR^n$, where $\Lag_{\rm H} \in \Omega^1(\RR^n)$ is the Hitchin Lagrangian one-form defined using a small counter-clockwise oriented loop $c_{\mathsf p}$ in $U_0 \cap U_1$ around the point $\mathsf p \in C$ by
\begin{equation} \label{eq:unifyingaction b}
\Lag_{\rm H} \coloneqq  \frac{1}{2 \pi i} \int_{c_{\mathsf p}} \big\langle L^0, \d_{\RR^n} \gamma \gamma^{-1} \big\rangle + \sum_{\alpha=1}^N \big\langle \Lambda_\alpha, \varphi_\alpha^{-1} \d_{\RR^n} \varphi_\alpha \big\rangle - H_i \d_{\RR^n} t^i,\qquad H_i=P_i\big( L^1_{z_I}(\mathsf{q}_i) \big) .
\end{equation}
\end{subequations}

In particular, under a change of `residual' local trivialisation $h = (h_0, h_1) \in \check C^0(C, G)$ with $h_I$ holomorphic on $U_I$ for $I \in \{0,1\}$ so that the condition $A''^0 = A''^1 = 0$ is preserved, the action \eqref{eq:unifyingaction a} is invariant in the sense that
\begin{equation} \label{1d action invariance}
S_{{\rm H}, \Gamma}\big[ hLh^{-1}, h_0 \gamma h_1^{-1}, (h_1 \varphi_\alpha) \big] = S_{{\rm H}, \Gamma}\big[ L^0, \gamma, (\varphi_\alpha) \big] ,
\end{equation}
where $hLh^{-1}$ stands for the Lax matrix given by $h_0 L^0 h_0^{-1}$ on $U_0$ and $h_1 L^1 h_1^{-1}$ on $U_1$.
\end{theorem}
\begin{proof}
We choose local trivialisations over $U_0\times \RR^n$ and $U_1\times \RR^n$ in which the components $A''_{\bar z_0}$, $A''_{\bar z_1}$ of the partial connection $A$ vanish.  Then the local expression for the curvature \eqref{FA def} becomes
\begin{equation}
F^I_A = \partial_{\bar{z}_I}\A^I_j\d\bar{z}_I\wedge \d u^j+\frac12 (\partial_{u^i}\A^I_j-\partial_{u^j}\A^I_i+[\A^I_i,\A^I_j])\d u^i\wedge \d u^j.
\end{equation}
Unlike earlier trivialisations used in Section \ref{sec: Lag for Hitchin mod G} for which we had \eqref{dgdu}, these trivialisations will have a transition function $\gamma = g_{01} : U_0\cap U_1\times\mathbb{R}^n\to G$ that depends on $u\in\mathbb{R}^n$ (because our choice of trivialisation depends on $A''$, and $A''$ depends on $u$).  On the overlaps of coordinate charts we have
\begin{equation}\label{eq:4.1 A transition}
    \A_j^0=\gamma \A_j^1\gamma^{-1}-\partial_{u^j}\gamma\gamma^{-1},\qquad \bar{\partial}\gamma=0.
\end{equation}

Recall from Section \ref{sec: Lax matrices} that to emphasise that $B$ solves the moment map condition \eqref{A_defect_modif} we denote it by $L$ and refer to it as a Lax matrix. Now the moment map condition \eqref{eq:b_z-gauged} can be written concisely as
\begin{equation} \label{moment map condition L}
\bar\partial L^0 =  0 , \qquad \bar\partial L^1 = -2\pi i \sum_{\alpha=1}^N \varphi_\alpha \Lambda_\alpha \varphi_\alpha^{-1} \delta_{\mathsf p_\alpha} .
\end{equation}

In order to evaluate the action \eqref{S Gamma 2} in these trivialisations, we choose a circle $c_{\mathsf p}$ as described in the statement of the theorem, and let $R_0\subset U_0$, $R_1\subset U_1$ be the closure of the interior and exterior of this circle.  Then $R_0\cap R_1=c_{\mathsf p}$ and $R_0\cup R_1=C$.  The first term in the action \eqref{S Gamma 2} is
\begin{equation}\label{eq:4.1 integral}
\frac{1}{2\pi i}\int_{C\times \Gamma}\langle B,F_A\rangle =\sum_{I \in \{0,1\}} \frac{1}{2\pi i}\int_{R_I\times \Gamma}\langle L^I,\bar{\partial}\A^I_j\rangle\wedge \d u^j.
\end{equation}
We evaluate these terms separately.  For $I=1$ we find
\begin{equation}
\begin{split}
\frac{1}{2 \pi i} \int_{R_1} \langle L^1, \bar{\partial}\A_j^1\rangle  &=  \frac{1}{2 \pi i}\int_{R_1}\left(-\bar{\partial}\langle L^1,\A_j^1\rangle + \langle\bar{\partial}L^1,\A_j^1\rangle\right)\\
&=- \frac{1}{2 \pi i} \int_{\partial R_1} \langle L^1, \A_j^1 \rangle - \sum_{\alpha=1}^N \big\langle \varphi_\alpha \Lambda_\alpha \varphi_\alpha^{-1}, \A_j^1(\mathsf p_\alpha) \big\rangle\\
&= \frac{1}{2 \pi i} \int_{c_\mathsf{p}} \big\langle L^0, \gamma \A_j^1 \gamma^{-1} \big\rangle - \sum_{\alpha=1}^N \big\langle \varphi_\alpha \Lambda_\alpha \varphi_\alpha^{-1}, \A_j^1(\mathsf p_\alpha) \big\rangle,    
\end{split}
\end{equation}
where in the second equality we used the second relation in \eqref{moment map condition L} and in the last step we used the first relation in \eqref{geometric Lax overlap} and the fact that $\partial R_1$ is equal to $c_{\mathsf{p}}$ with reversed orientation. For $I=0$,
\begin{equation}
\begin{split}
\frac{1}{2 \pi i} \int_{R_0} \langle L^0, \bar\partial \A_j^0\rangle &= - \frac{1}{2 \pi i} \int_{R_0}\bar{\partial} \langle L^0, \A_j^0 \rangle = - \frac{1}{2 \pi i} \int_{\partial R_0} \langle L^0, \A_j^0 \rangle\\
&= -\frac{1}{2 \pi i} \int_{c_{\mathsf{p}}} \big\langle L^0, \gamma \A_j^1 \gamma^{-1} \big\rangle + \frac{1}{2 \pi i} \int_{c_{\mathsf{p}}} \big\langle L^0, \partial_{u^j} \gamma \gamma^{-1} \big\rangle ,
\end{split}
\end{equation}
where the first equality uses the first relation in \eqref{moment map condition L} and in the third step we used \eqref{eq:4.1 A transition} and the fact that $\partial R_0$ coincides with $c_{\mathsf{p}}$ and has the same orientation. It now follows from the above computation of both integrals on the right-hand side of \eqref{eq:4.1 integral} that
\begin{equation}
\frac{1}{2 \pi i}\int_{C\times \Gamma}\langle B,F_A\rangle =\frac{1}{2\pi i}\int_{c_{\mathsf p}\times\Gamma}\langle L^0,\d_{\RR^n}\gamma \gamma^{-1}\rangle- \sum_{\alpha=1}^N \int_\Gamma\big\langle \varphi_\alpha \Lambda_\alpha \varphi_\alpha^{-1}, \A_j^1(\mathsf p_\alpha) \big\rangle \d u^j.
\end{equation}
Substituting this into \eqref{S Gamma 2} and setting $B=L$ in the remaining terms gives the desired result.

To prove the last ``in particular'' statement, consider a change of ``residual'' local trivialisation $h = (h_0, h_1) \in \check C^0(C, G)$, with $h_I$ holomorphic on $U_I$ so that the condition $A''_{\bar z_I} = 0$ is preserved. We have $\gamma \mapsto \tilde\gamma = h_0 \gamma h_1^{-1}$, $\tilde L^0 = h_0 L^0 h_0^{-1}$ and $\tilde L^1 = h_1 L^1 h_1^{-1}$ so that $\tilde L^0 = \tilde \gamma \tilde L^1 \tilde \gamma^{-1}$. Then the integral in the first term on the right-hand side of \eqref{eq:unifyingaction b} transforms to
\begin{equation} \label{pre Lag 1-form change loc triv 1}
\int_{c_{\mathsf p}} \big\langle \tilde L^0, \d_{\RR^n} \tilde \gamma \tilde \gamma^{-1} \big\rangle = \int_{c_{\mathsf p}} \big\langle L^0, \d_{\RR^n} \gamma \gamma^{-1} \big\rangle + \int_{c_{\mathsf p}} \big\langle L^0, h_0^{-1} \d_{\RR^n} h_0 \big\rangle - \int_{c_{\mathsf p}} \big\langle L^1, h_1^{-1} \d_{\RR^n} h_1 \big\rangle .
\end{equation}
Note that the second term on the right-hand side vanishes by Cauchy's theorem since $h_0$ and $L_0$ are both holomorphic on $U_0$ and $c_{\mathsf p}$ is a small contour around $\mathsf p \in U_0$. To evaluate the last term on the right-hand side of \eqref{pre Lag 1-form change loc triv 1}, recall that the section $L^1$ has poles at the marked points $\mathsf p_\alpha$ given by \eqref{Lax behaviour a}. Since $h_1 : U_1 \to G$ is holomorphic on $U_1 = C \setminus \{ \mathsf p\}$, we then deduce using the residue theorem that
\begin{equation}
\frac{1}{2 \pi i} \int_{c_{\mathsf p}} \big\langle L^1, h_1^{-1} \d_{\RR^n} h_1 \big\rangle = - \sum_{\alpha = 1}^N \big\langle \! -\varphi_\alpha \Lambda_\alpha \varphi_\alpha^{-1}, h_1(\mathsf p_\alpha)^{-1} \d_{\RR^n} h_1(\mathsf p_\alpha) \big\rangle
\end{equation}
where the first sign comes from noting that $c_{\mathsf p}$ can be contracted down to a sum of small \emph{clockwise} circles around each point $\mathsf p_\alpha$. Substituting this into the right-hand side of \eqref{pre Lag 1-form change loc triv 1} we deduce the transformation property
\begin{equation} \label{Lag 1-form change loc triv 1}
 \frac{1}{2 \pi i} \int_{c_{\mathsf p}} \big\langle \tilde L^0, \d_{\RR^n} \tilde \gamma \tilde \gamma^{-1} \big\rangle = \frac{1}{2 \pi i} \int_{c_{\mathsf p}} \big\langle L^0, \d_{\RR^n} \gamma \gamma^{-1} \big\rangle - \sum_{\alpha = 1}^N \Big\langle \Lambda_\alpha, \varphi_\alpha^{-1} \big( h_1(\mathsf p_\alpha)^{-1} \d_{\RR^n} h_1(\mathsf p_\alpha) \big) \varphi_\alpha \Big\rangle .
\end{equation}
Consider now the second term in the Lagrangian one-form \eqref{eq:unifyingaction b}. Since $\varphi_\alpha$ transforms under the change of local trivialisation as $\varphi_\alpha \mapsto \tilde \varphi_\alpha = h_1(\mathsf p_\alpha) \varphi_\alpha$, see the discussion at the start of Section \ref{sec: adding punctures}, this term in the Lagrangian transforms as
\begin{equation} \label{Lag 1-form change loc triv 2}
\sum_{\alpha=1}^N \big\langle \Lambda_\alpha, \tilde \varphi_\alpha^{-1} \d_{\RR^n} \tilde \varphi_\alpha \big\rangle
= \sum_{\alpha=1}^N \big\langle \Lambda_\alpha, \varphi_\alpha^{-1} \d_{\RR^n} \varphi_\alpha \big\rangle + \sum_{\alpha = 1}^N \Big\langle \Lambda_\alpha, \varphi_\alpha^{-1} \big( h_1(\mathsf p_\alpha)^{-1} \d_{\RR^n} h_1(\mathsf p_\alpha) \big) \varphi_\alpha \Big\rangle .
\end{equation}
We see now from \eqref{Lag 1-form change loc triv 1} and \eqref{Lag 1-form change loc triv 2} that the Lagrangian one-form \eqref{eq:unifyingaction b} is invariant under residual changes of local trivialisations $h = (h_0, h_1) \in \check C^0(C, G)$ with $h_I$ holomorphic on $U_I$.
\end{proof}

\begin{remark} \label{rem: integration by parts}
The term $\int_{c_{\mathsf p}} \big\langle L^0, \d_{\RR^n} \gamma \gamma^{-1} \big\rangle$ in the kinetic part of the unifying Lagrangian \eqref{eq:unifyingaction b} arises from a subtle mechanism. 
In the proof of Theorem \ref{thm: Lag 1-form for Hitchin}, we work on the pullback bundle $\pi_C^\ast \mathcal P$ but in a local trivialisation in which $A''^I = 0$ for each $I \in \mathcal I$. This implies, in particular, that the transition function $g_{IJ} : (U_I \cap U_J) \times \RR^n \to G$ is holomorphic in $U_I \cap U_J$ and dependent on the coordinates $u^j$ of $\RR^n$. Thus, specialising the $\mathcal A$ component of the identity \eqref{partial connection compat} to this setting, we find that
\begin{equation}
\bar\partial \mathcal A^I = g_{IJ} \, \bar\partial \mathcal A^J \, g_{IJ}^{-1} .
\end{equation}
In particular, it follows that $\langle B^I, \bar\partial \mathcal A^I\rangle = \langle B^J, \bar\partial \mathcal A^J\rangle$ on overlaps $U_I \cap U_J \neq \emptyset$ so that we obtain a $\g$-valued $3$-form $\langle B, \bar\partial \mathcal A\rangle \in \Omega^3(C \times \RR^n, \g)$ which, in particular, is a global $(1,1)$-form along $C$ so that it can be integrated over $C$.\footnote{Strictly speaking, this is a fibre integration along the fibres of the projection $\pi_{\RR^n} : C \times \RR^n \to \RR^n$.} Now, by contrast, in this local trivialisation where the transition functions of $\pi_C^\ast {\mathcal P}$ explicitly depend on $\RR^n$, we find using the transformation property \eqref{partial connection compat} that
\begin{equation} \label{del B A bad transformation}
\langle \bar\partial B^I, \mathcal A^I\rangle = \langle \bar\partial B^J, \mathcal A^J\rangle - \big\langle \bar\partial B^J, g_{IJ}^{-1} \d_{\RR^n} g_{IJ} \big\rangle .
\end{equation}
In other words, the local expressions $\langle \bar\partial B^I, \mathcal A^I\rangle$ do not define a global $\g$-valued $3$-form on $C \times \RR^n$. Likewise, the local expressions $\langle B^I, \mathcal A^I\rangle$ transform in a similar manner to \eqref{del B A bad transformation} and therefore do not define a global $\g$-valued $2$-form on $C \times \RR^n$ either. This means that in the present context, ``naive'' integration by parts in an expression like $\int_C \langle B, \bar\partial \mathcal A\rangle$ is not possible, cf. Remark \ref{rem: mu well defined}. This is the source of the term $\int_{c_{\mathsf p}} \big\langle L^0, \d_{\RR^n} \gamma \gamma^{-1} \big\rangle$.
\end{remark}

We now derive the univariational equations for the unifying $1$d action in \eqref{eq:unifyingaction}.  To do so, it is helpful to recall the constraints on the various degrees of freedom.  We recall that $L^0$ and $L^1$ are holomorphic $\mathfrak{g}$-valued (1,0)-forms on $U_0$ and $U_1\setminus\{\mathsf{p}_\alpha\}$, $\gamma$ is a holomorphic $G$-valued function on $U_0\cap U_1$, and $\varphi_\alpha$ are elements of $G$.  All of these depend on $t\in\RR^n$, and are constrained by
\begin{equation}\label{constraint}
L^1=\gamma^{-1}L^0\gamma,\qquad \Res_{\mathsf p_\alpha}L^1_{z_I} = -\varphi_\alpha\Lambda_\alpha\varphi_\alpha^{-1}.
\end{equation}
We continue to assume that the Higgs bundle determined by $\gamma$ and $L$ satisfies the stability condition \eqref{inf_freeness}.  This means that there are no non-zero holomorphic functions $X^I:U_I\to\g$ for $I \in \{0,1\}$ with the property that $[L^I,X^I]=0$ on $U_I$ and $X^1=\gamma^{-1}X^0\gamma$ on $U_0\cap U_1$.

\begin{theorem}\label{thm: eom for 1d action}
The equations of motion of the unifying $1$d action $S_{{\rm H}, \Gamma}\big[ L, \gamma, (\varphi_\alpha), t \big]$ in \eqref{eq:unifyingaction} with respect to the variables $L$, $\gamma$ and $\varphi_\alpha$ take the form
\begin{equation}\label{Lax equations}
\partial_{t^i} L^0=[M_i^0,L^0],\qquad \partial_{t^i} L^1=[M_i^1,L^1].
\end{equation}
Here $M_i^0,M_i^1$ are holomorphic $\mathfrak{g}$-valued functions on $U_0$ and $U_1\setminus\{\mathsf{q}_i\}$ that are uniquely determined by the constraints
\begin{equation}\label{M constraints}
\gamma^{-1}\partial_{t^i} \gamma = \gamma^{-1}M_i^0\gamma-M_i^1,\qquad \Res_{\mathsf{q}_i} M_i^1 = \nabla P_i\big( L_{z_I}^1(\mathsf{q}_i) \big).
\end{equation}
Moreover, they satisfy the zero curvature equations
\begin{equation}\label{eq:zce Mi}
\partial_{t^i} M^0_j - \partial_{t^j} M^0_i - [ M^0_i, M^0_j ] =0, \qquad
\partial_{t^i} M^1_j - \partial_{t^j} M^1_i - [ M^1_i, M^1_j ] =0.
\end{equation}
The equation of motion with respect to the set of times $t^i$ for $i=1,\ldots, n$ describe the conservation equations $\partial_{t^i} H_j = 0$ for all $i,j =1,\ldots, n$.
\end{theorem}
\begin{proof} The last statement for the variation with respect to $t^i$ for $i=1,\ldots, n$ is straightforward, so we focus on deriving the equations of motions associated to variations of $L$, $\gamma$ and $\varphi_\alpha$.

First we consider variations of $L$ with $\gamma$ and $\varphi_\alpha$ fixed. It follows from the condition \eqref{constraint} that these are described by holomorphic $\g$-valued $(1,0)$-forms $\delta L^I$ on $U_I$ for $I \in \{0,1\}$ satisfying
\begin{equation}\label{delta L constraint}
\delta L^1 = \gamma^{-1}\delta L^0\gamma
\end{equation}
on the intersection $U_0 \cap U_1$.
In particular, since $\varphi_\alpha$ is held fixed, varying the second constraint in \eqref{constraint} gives $\Res_{\mathsf p_\alpha} \delta L^1_{z_I} = 0$ so that $\delta L^1$ is indeed holomorphic on $U_1$.
The variation of $S_{{\rm H}, \Gamma}$ is
\begin{equation}
\delta S_{{\rm H}, \Gamma}= \int_0^1\left(\frac{1}{2\pi i}\oint_{c_{\mathsf p}} \langle \delta L^1,\gamma^{-1}\partial_{u^j}\gamma\rangle - \big\langle \delta L^1_{z_I}(\mathsf{q}_i),\nabla P_i(L^1_{z_I}(\mathsf{q}_i))\big\rangle \frac{\partial t^i}{\partial u^j} \right)\frac{\d u^j}{\d s}\d s .
\end{equation}
This must vanish for all curves $\Gamma : (0,1) \to \RR^n$ so that for $i =1,\ldots, n$ we have
\begin{equation}\label{L variation}
0=\frac{1}{2\pi i}\oint_{c_{\mathsf p}} \langle \delta L^1,\gamma^{-1}\partial_{t^i}\gamma\rangle - \big\langle \delta L^1_{z_I}(\mathsf{q}_i),\nabla P_i(L^1_{z_I}(\mathsf{q}_i))\big\rangle .
\end{equation}

Consider, to begin with, a variation such that $\delta L^1_{z_I}(\mathsf{q}_i)=0$ for some fixed $i$.  We find that
\begin{equation}\label{Serre}
0 = \frac{1}{2\pi i}\oint_{c_{\mathsf p}} \langle \delta L^1,\gamma^{-1}\partial_{t^i}\gamma\rangle = \frac{1}{2\pi i}\oint_{c_{\mathsf p}} \langle \delta L^0, \partial_{t^i}\gamma \gamma^{-1} \rangle ,
\end{equation}
where the second equality just follows from the invariance of the bilinear form $\langle \cdot, \cdot \rangle : \g \otimes \g \to \CC$ and the constraint \eqref{delta L constraint}. Let us interpret the equation \eqref{Serre} using sheaf cohomology. For each $t \in \RR^n$ we denote by $E_t \to C$ the holomorphic vector bundle with transition function $\Ad_{\gamma(t)}$ between the two local trivialisations over $U_1$ and $U_0$.
The pair $\delta L = (\delta L^0, \delta L^1)$ of $\g$-valud $(1,0)$-forms on $U_0$ and $U_1$ related by the constraint \eqref{delta L constraint} describes a holomorphic section of $\bigwedge^{1,0} C \otimes E_t$ that vanishes at $\mathsf{q}_i$ and hence it determines an element of the \v{C}ech cohomology group $H^0\big( C,\Lambda^{1,0}C\otimes E_t(-\mathsf{q}_i) \big)$, in which $(-\mathsf{q}_i)$ indicates that the section vanishes at $\mathsf{q}_i$. On the other hand, the pair $(\partial_{t^i}\gamma \gamma^{-1}, \gamma^{-1}\partial_{t^i}\gamma)$ of $\g$-valued functions on $U_0 \cap U_1$ describe a holomorphic section of $E_t$ over $U_0\cap U_1$ defined relative to the local trivialisations over $U_0$ and $U_1$. It thus determines an element of the \v{C}ech cohomology group $H^1\big( C,E_t(\mathsf{q}_i) \big)$. The two integral expressions in \eqref{Serre} then correspond to the Serre duality pairing between $H^0\big( C,\Lambda^{1,0}C\otimes E_t(-\mathsf{q}_i) \big)$ and $H^1\big( C,E_t(\mathsf{q}_i) \big)$ described using the local trivialisation of $E_t$ over $U_1$ and $U_0$, respectively. This pairing is nondegenerate, so if \eqref{Serre} holds for all variation $\delta L$ then $(\partial_{t^i}\gamma \gamma^{-1}, \gamma^{-1}\partial_{t^i}\gamma)$ is zero in cohomology.  This means that
\begin{equation}\label{M}
\gamma^{-1}\partial_{t^i}\gamma = \gamma^{-1}M_i^0\gamma-M_i^1 ,
\end{equation}
where $M_i^0$ and $M_i^1$ are sections of $E_t(\mathsf{q}_i)$ over $U_0$ and $U_1$, respectively, and the right-hand side of \eqref{M} is the coboundary map in sheaf cohomology, expressed in the local trivialisation over $U_1$. More precisely, $M_i^0$ is a $\g$-valued holomorphic function on $U_0$ and $M_i^1$ is a $\g$-valued holomorphic on $U_1\setminus\{\mathsf{q}_i\}$ with a simple pole at $\mathsf{q}_i$. There is some freedom in the choice of $M_i^I$ solving \eqref{M}: if $N_i^0$ and $N_i^1$ are sections of $E_t$ over $U_0$ and $U_1\setminus\{\mathsf{q}_i\}$, respectively, such that $N_i^1$ has a simple pole at $\mathsf{q}_i$ and $\gamma^{-1}N_i^0\gamma=N_i^1$, then adding $N_i^I$ to $M_i^I$ produces a new solution of \eqref{M}. Thus, $M_i^0,M_i^1$ are unique up to the addition of elements of $H^0\big( C,E_t(\mathsf{q}_i) \big)$.

Now, we insert \eqref{M} into \eqref{L variation}, in which the variation $\delta L^1$ is no longer constrained to vanish at $\mathsf{q}_i$.  We find that
\begin{equation}
\begin{split}
0 &= \frac{1}{2\pi i}\oint_{c_{\mathsf p}} \langle \delta L^1,\gamma^{-1}M_i^0\gamma-M_i^1\rangle - \big\langle \delta L^1_{z_I}(\mathsf{q}_i),\nabla P_i\big( L^1_{z_I}(\mathsf{q}_i) \big) \big\rangle \\
&= \frac{1}{2\pi i}\oint_{c_{\mathsf p}} \langle \delta L^0,M_i^0\rangle -  \frac{1}{2\pi i}\oint_{c_{\mathsf p}} \langle \delta L^1,M_i^1\rangle- \big\langle \delta L^1_{z_I}(\mathsf{q}_i),\nabla P_i\big( L^1_{z_I}(\mathsf{q}_i) \big) \big\rangle .
\label{L variation 2}
\end{split}
\end{equation}
The contour integral of $\langle \delta L^0,M_i^0\rangle$ on the right-hand side vanishes because its integrand is holomorphic on $U_0$. Deforming the contour integral of $\langle\delta L^1,M_i^1\rangle$ from the anticlockwise contour $c_{\mathsf p}$ to a clockwise small circle around $\mathsf{q}_i$ yields 
\begin{equation}
\frac{1}{2\pi i}\oint_{c_{\mathsf p}} \langle \delta L^1,M_i^1\rangle=-\Big\langle \delta L^1_{z_I}(\mathsf{q}_i),\Res_{\mathsf{q}_i} M_i^1\Big\rangle .
\end{equation}
We may then rewrite \eqref{L variation 2} as
\begin{equation}\label{M residue constraint}
0=\Big\langle \delta L^1_{z_I}(\mathsf{q}_i),\Res_{\mathsf{q}_i} M_i^1 - \nabla P_i\big( L_{z_I}^1(\mathsf{q}_i) \big)\Big\rangle .
\end{equation}
We would like to conclude from this that
\begin{equation}\label{M residue}
\Res_{\mathsf{q}_i} M_i^1 = \nabla P_i\big( L_{z_I}^1(\mathsf{q}_i) \big) .
\end{equation}
However, this does not follow immediately.  The variation $\delta L^1$ is constrained by \eqref{delta L constraint}, and this constraint may mean that $\delta L^1_{z_I}(\mathsf{q}_i)$ takes values in a proper subspace of the fibre $E_t|_{\mathsf{q}_i}$ of $E_t$ at $\mathsf{q}_i$. If so, \eqref{M residue constraint} does not constrain all the components of $\Res_{\mathsf{q}_i}M_i^1$. So to find a solution $M_i^0,M_i^1$ of \eqref{M residue constraint} we use more sophisticated methods.

Consider the exact sequence of sheaves on $C$ given by
\begin{equation} \label{short exact}
0 \longrightarrow E_t \longrightarrow E_t(\mathsf{q}_i) \longrightarrow E_t|_{\mathsf{q}_i} \longrightarrow 0 .
\end{equation}
Here, by abuse of notation, $E_t$ denotes the sheaf of holomorphic sections of the vector bundle $E_t$, and $E_t(\mathsf{q}_i)$ denotes the sheaf of sections that are holomorphic on $C\setminus\{\mathsf{q}_i\}$ with a simple pole at $\mathsf{q}_i$. Moreover, $E_t|_{\mathsf{q}_i}$ is the fibre of $E_t$ at $\mathsf{q}_i$, which we regard as a skyscraper sheaf on $C$ supported at the point $\mathsf{q}_i$.  This exact sequence of sheaves induces a long exact sequence in \v{C}ech cohomology:
\begin{equation}\label{exact sequence}
\begin{tikzcd}
0 \arrow[r] & H^0(C,E_t) \arrow[r, "\iota"] & H^0(C,E_t(\mathsf{q}_i)) \arrow[r, "\Res_{\mathsf{q}_i}"] & E_t|_{\mathsf{q}_i} \arrow[r, "h"] & H^1(C,E_t) \arrow[r] & \ldots .
\end{tikzcd}
\end{equation}
The map $\iota$ is the inclusion of the space of holomorphic sections in the space of sections with a simple pole. The map $\Res_{\mathsf q_i}$ is given by evaluating the residue of a section at the point $\mathsf{q}_i$. The so-called connecting homomorphism $h$ acts as follows: given $X_i \in E_t|_{\mathsf{q}_i}$, we use the exactness of \eqref{short exact} at $E_t|_{\mathsf q_i}$ to choose a holomorphic section $h(X_i)$ of $E_t$ over $U_1 \setminus \{\mathsf q_i\}$ with a simple pole at $\mathsf{q}_i$ with residue $X_i$ there. The restriction of this holomorphic section to $U_0\cap U_1$ determines a cohomology class in $H^1(C,E_t)$ which is independent of the choice of holomorphic section.

With this notation and letting $X_i \coloneqq \Res_{\mathsf{q}_i}M_i^1-\nabla P_i\big( L^1_{z_I}(\mathsf{q}_i) \big)$, we can rewrite \eqref{M residue constraint} as follows:
\begin{equation} \label{Serre 2}
0 = \big\langle \delta L^1_{z_I}(\mathsf{q}_i),X_i \big\rangle = -\frac{1}{2\pi i}\oint_{c_{\mathsf p}} \big\langle \delta L^1,h(X_i) \big\rangle = -\frac{1}{2\pi i}\oint_{c_{\mathsf p}} \big\langle \delta L^0, \gamma h(X_i) \gamma^{-1} \big\rangle ,
\end{equation}
where the last step is as in \eqref{Serre}.
These two integral expressions describe the Serre duality pairing between $\delta L\in H^0(C,\Lambda^{1,0}\otimes E^\ast_t)$ and $\big( \gamma h(X_i) \gamma^{-1}, h(X_i) \big) \in H^1(C,E_t)$ in the local trivialisation of $E_t$ over $U_1$ and $U_0$, respectively.
Since \eqref{Serre 2} holds for all $\delta L\in H^0(C,\Lambda^{1,0}\otimes E_t)$, and since the pairing is nondegenerate, we conclude that $h(X_i)=0\in H^1(C,E_t)$.  Since the sequence \eqref{exact sequence} is exact, there exists an $N_i \in H^0\big( C,E_t(\mathsf{q}_i) \big)$ such that $X_i = \Res_{\mathsf q_i} N_i$.  We then have that $\tilde{M}_i \coloneqq M_i - N_i$ solves both \eqref{M} and \eqref{M residue}, as required.

There is still some freedom in the choice of $M_i$ solving \eqref{M}, \eqref{M residue}. The solution of \eqref{M} is unique up to the addition of $\g$-valued functions $N_i^0,N_i^1$ on $U_0,U_1$ satisfying $N_i^1=\gamma^{-1}N_i^0\gamma$.  Since the pole of $M_i^1$ is fixed by \eqref{M residue}, these functions $N_i^I$ are holomorphic, so they determine a holomorphic section of $E_t$.  Therefore, $M_i$ is unique up to addition of elements of $H^0(C,E_t)$.

Having considered variations of $L$ only, we now consider variations of $L$, $\varphi_\alpha$ and $\gamma$.  We assume that $M_i$ has been chosen solving \eqref{M} and \eqref{M residue}. From \eqref{constraint}, $\delta L^1$ is now meromorphic and the variations satisfy
\begin{subequations}
\begin{align}
\label{variation constraint}
[L^0,\delta \gamma\gamma^{-1}]&=\gamma \delta L^1\gamma^{-1}-\delta L^0 ,\\
\label{delta L residue}
\Res_{\mathsf{p}_\alpha}\delta L^1_{z_I}&=-[\delta\varphi_\alpha\varphi_\alpha^{-1},\varphi_{\alpha}\Lambda_\alpha\varphi_\alpha^{-1}] .
\end{align}
\end{subequations}
The variation of $S_{{\rm H}, \Gamma}$ is
\begin{multline}
\delta S_{{\rm H}, \Gamma} = \int_0^1\bigg(
\frac{1}{2\pi i}\oint_{c_{\mathsf p}} \langle \delta L^1,\gamma^{-1}\partial_{u^j}\gamma\rangle
+\frac{1}{2\pi i}\oint_{c_{\mathsf p}} \big\langle L^1,\gamma^{-1}\partial_{u^j}(\delta \gamma\gamma^{-1})\gamma \big\rangle\\
+ \sum_{\alpha=1}^N\big\langle\Lambda_\alpha,\varphi_\alpha^{-1}\partial_{u^j}(\delta\varphi_\alpha\varphi_\alpha^{-1})\varphi_\alpha\big\rangle
- \big\langle\delta L^1_{z_I}(\mathsf{q}_i),\nabla P_i\big( L^1_{z_I}(\mathsf{q}_i) \big) \big\rangle \frac{\partial t^i}{\partial u^j}\bigg)\frac{\d u^j}{\d s}\d s .
\end{multline}
We will rewrite this by substituting \eqref{M} in the first term and by using the following contour integral, which is evaluated using \eqref{M residue} and \eqref{delta L residue}:
\begin{equation}
\frac{1}{2\pi i}\oint_{c_{\mathsf p}} \langle\delta L^1,M_i^1\rangle=-\big\langle \delta L^1_{z_I}(\mathsf{q}_i),\nabla P_i\big( L^1_{z_I}(\mathsf{q}_i) \big)\big\rangle+\sum_{\alpha=1}^N \big\langle [\delta\varphi_\alpha\varphi_\alpha^{-1},\varphi_\alpha\Lambda_\alpha\varphi_\alpha^{-1}],M_i^1(\mathsf{p}_\alpha)\big\rangle .
\end{equation}
We obtain 
\begin{multline}
\delta S_{{\rm H}, \Gamma} = \int_0^1\bigg(
\frac{1}{2\pi i}\oint_{c_{\mathsf p}} \langle \gamma\,\delta L^1\,\gamma^{-1},M_i^0\rangle\frac{\partial t^i}{\partial u^j}
+\frac{1}{2\pi i}\oint_{c_{\mathsf p}} \big\langle \gamma\,L^1\,\gamma^{-1},\partial_{u^j}(\delta \gamma\gamma^{-1}) \big\rangle\\
+ \sum_{\alpha=1}^N \big\langle\Lambda_\alpha,\varphi_\alpha^{-1}\partial_{u^j}(\delta\varphi_\alpha\varphi_\alpha^{-1})\varphi_\alpha \big\rangle
-
\sum_{\alpha=1}^N \big\langle [\delta\varphi_\alpha\varphi_\alpha^{-1},\varphi_\alpha\Lambda_\alpha\varphi_\alpha^{-1}],M_i^1(\mathsf{p}_\alpha)\big\rangle\frac{\partial t^i}{\partial u^j}\bigg)\frac{\d u^j}{\d s}\d s .
\end{multline}
Let us now rewrite this using \eqref{constraint} and \eqref{variation constraint}:
\begin{multline}
\delta S_{{\rm H}, \Gamma} = \int_0^1\bigg(
\frac{1}{2\pi i}\oint_{c_{\mathsf p}} \big\langle \delta L^0+[L^0,\delta\gamma\gamma^{-1}],M_i^0 \big\rangle\frac{\partial t^i}{\partial u^j}
+\frac{1}{2\pi i}\oint_{c_{\mathsf p}} \big\langle L^0,\partial_{u^j}(\delta \gamma\gamma^{-1}) \big\rangle\\
+ \sum_\alpha \big\langle \varphi_\alpha\,\Lambda_\alpha\,\varphi_\alpha^{-1},\partial_{u^j}(\delta\varphi_\alpha\varphi_\alpha^{-1})\big\rangle
-
\sum_\alpha \big\langle [\delta\varphi_\alpha\varphi_\alpha^{-1},\varphi_\alpha\Lambda_\alpha\varphi_\alpha^{-1}],M_i^1(\mathsf{p}_\alpha) \big\rangle\frac{\partial t^i}{\partial u^j}\bigg)\frac{\d u^j}{\d s}\d s .
\end{multline}
The contour integral of $\langle \delta L^0,M_i^0\rangle$ vanishes because it is holomorphic on $U_0$.  After integration by parts, the variation becomes
\begin{multline}\label{SH variation}
\delta S_{{\rm H}, \Gamma} = \int_0^1\bigg(
\frac{1}{2\pi i}\oint_{c_{\mathsf p}} \big\langle [M_i^0,L^0],\delta\gamma\gamma^{-1}\big\rangle\frac{\partial t^i}{\partial u^j}
-\frac{1}{2\pi i}\oint_{c_{\mathsf p}} \big\langle \partial_{u^j} L^0,\delta \gamma\gamma^{-1}\big\rangle\\
- \sum_\alpha \big\langle \partial_{u^j}(\varphi_\alpha\,\Lambda_\alpha\,\varphi_\alpha^{-1}),\delta\varphi_\alpha\varphi_\alpha^{-1} \big\rangle
+
\sum_\alpha \big\langle [M_i^1(\mathsf{p}_\alpha),\varphi_\alpha\Lambda_\alpha\varphi_\alpha^{-1}],\delta\varphi_\alpha\varphi_\alpha^{-1}\big\rangle\frac{\partial t^i}{\partial u^j}\bigg)\frac{\d u^j}{\d s}\d s .
\end{multline}

Let us restrict attention to variations for which $\delta\varphi_\alpha=0$.  Then $\delta S_{{\rm H}, \Gamma}$ vanishes for all curves $\Gamma$ if and only if
\begin{equation}\label{gamma variation}
0 = \oint_{c_{\mathsf p}} \big\langle \partial_{t^i} L^0-[M_i^0,L^0],\delta \gamma\gamma^{-1} \big\rangle = \oint_{c_{\mathsf p}} \big\langle \partial_{t^i} L^1-[M_i^1,L^1], \gamma^{-1}\delta \gamma \big\rangle
\end{equation}
for all variations $\delta\gamma\gamma^{-1}$ satisfying \eqref{variation constraint}, where the second equality follows from noting that
\begin{equation}
\gamma^{-1}\big( \partial_{t^i} L^0-[M_i^0,L^0] \big) \gamma=\partial_{t^i} L^1-[M_i^1,L^1]
\end{equation}
as a consequence of \eqref{M}. In particular, it follows that $\partial_{t^i}L^I-[M_i^I,L^I]$ can be interpreted as a section of $E_t$. It could have poles at the point $\mathsf{q}_i$ and any of the points $\mathsf{p}_\alpha$, because $M_i^1$ and $L^1$ have a simple poles there. However, from \eqref{M residue} the residue at $\mathsf{q}_i$ is given by 
\begin{equation}
-\Big[\Res_{\mathsf{q}_i}M_i^1,L_{z_I}^1(\mathsf{q}_i) \Big]=-\Big[\nabla P_i \big( L^1_{z_I}(\mathsf{q}_i) \big), L_{z_I}^1(\mathsf{q}_i) \Big]
\end{equation}
and this vanishes because $P_i$ is an invariant polynomial.  So $\partial_{t^i} L^1-[M_i^1,L^1]$ only has poles at the points $\mathsf{p}_\alpha$, and it determines a holomorphic section in  $H^0\big( C,\Lambda^{1,0}C\otimes E_t^\ast(\sum_\alpha \mathsf{p}_\alpha) \big)$.

To interpret \eqref{gamma variation} using Serre duality, we note using the constraint \eqref{variation constraint} that the components of the holomorphic section $Y \coloneqq (\delta\gamma\gamma^{-1}, \gamma^{-1} \delta \gamma)$ of $E_t$ over $U_0 \cap U_1$ defined relative to the local trivialisations over $U_0$ and $U_1$, respectively, satisfy
\begin{equation}
[L^0,\delta \gamma\gamma^{-1}] = \gamma \delta L^1\gamma^{-1}-\delta L^0, \qquad
[L^1,\gamma^{-1}\delta \gamma] = \delta L^1 - \gamma^{-1} \delta L^0 \gamma.
\end{equation}
Since $\delta L^0,\delta L^1$ are holomorphic on $U_0,U_1$ (because we are assuming that $\delta\varphi_\alpha=0$), these equations say that $[L, Y]$ lies in the image of the \v{C}ech coboundary map, so determines a trivial element of $H^1\big( C,\Lambda^{1,0}C\otimes E_t \big)$. In other words, $Y$ is in the kernel of the map
\begin{equation}\label{ad L}
[L,\cdot]:H^1\big(C,E_t ( -\textstyle\sum_\alpha \mathsf{p}_\alpha )\big) \longrightarrow H^1\big( C,\Lambda^{1,0}C\otimes E_t \big).
\end{equation}
(The notation $(-\sum_\alpha \mathsf{p}_\alpha)$ in the domain of this map denotes the sheaf of sections vanishing at the points $\mathsf{p}_\alpha$, and appears because $L$ has poles at these points). Thus, we may interpret the integrals in \eqref{gamma variation} as the Serre duality pairing between $\partial_{t^i} L - [M_i, L] \in H^0\big( C,\Lambda^{1,0}C\otimes E_t^\ast(\sum_\alpha \mathsf{p}_\alpha) \big)$ and $Y = (\delta\gamma\gamma^{-1}, \gamma^{-1} \delta \gamma) \in H^1\big(C,E_t(-\sum_\alpha \mathsf{p}_\alpha )\big)$. The pairing is required to vanish for all $Y$ in the kernel of the map \eqref{ad L}.  This happens if and only if $\partial_{t^i}L-[M_i,L]$ lies in the image of the adjoint of \eqref{ad L}.  The adjoint of \eqref{ad L} is the map
\begin{equation}\label{ad L adjoint}
-[L,\cdot]:H^0\left(C,E_t\right) \longrightarrow H^0\big( C,\Lambda^{1,0}C\otimes E_t(\textstyle\sum_\alpha \mathsf{p}_\alpha) \big)
\end{equation}
because, by Serre duality, the dual space of $H^1(C,V)$ is $H^0(C,\Lambda^{1,0}C\otimes V^\ast)$, and 
\begin{equation}
\oint_{c_{\mathsf p}}\langle [L,u],v\rangle =- \oint_{c_{\mathsf p}}\langle u,[L,v]\rangle
\end{equation}
for every $u\in H^1\big(C,E_t(-\textstyle\sum_\alpha \mathsf{p}_\alpha)\big)$ and $v\in H^0\left(C,E_t\right)$.
Since $\partial_{t^i}L-[M_i,L]$ is in the image of the map \eqref{ad L adjoint}, there exist holomorphic functions $N_i^I:U_I\to\g$ for $I=0,1$ satisfying
\begin{equation}
\partial_{t^i}L^I-[M^I_i,L^I]=[N^I_i,L^I],\qquad \gamma^{-1}N^0_i\gamma = N^1_i.
\end{equation}
Recall from earlier that the solution of \eqref{M} and \eqref{M residue} was unique up to addition of elements of $H^0(C,E_t)$.  Therefore, we are free to define $\tilde{M}^I_i \coloneqq M^I_i+N^I_i$, and this choice of $M_i$ satisfies the Lax equations \eqref{Lax equations} and the constraints \eqref{M constraints}.  The choice of $M_i$ is now unique up to addition of elements of $\ker [L,\cdot]\subset H^0(C,E_t)$.  But the stability condition \eqref{inf_freeness} means that there are no holomorphic sections of $C$ that commute with $L$. So $M_i$ is unique, as claimed.

We have shown that $S_{{\rm H}, \Gamma}$ is critical with respect to variations of $L,\gamma$ with $\varphi_\alpha$ fixed if and only if $L$ solves the Lax equation \eqref{Lax equations}.  It remains to check that solutions of these equations are also critical points of $S_{{\rm H}, \Gamma}$ with respect to variations of $\varphi_{\alpha}$.  The variation of $S_{{\rm H}, \Gamma}$ is given in \eqref{SH variation}.  The term involving $\delta\gamma$ vanishes because $L$ satisfies the Lax equations \eqref{Lax equations}.  The term involving $\delta\varphi_\alpha$ also vanishes because taking residues of \eqref{Lax equations} gives
\begin{equation}
\partial_{t^i} (\varphi_\alpha\Lambda_\alpha\varphi_\alpha^{-1}) = \big[ M_i^1(\mathsf{p}_\alpha),\varphi_\alpha\Lambda_\alpha\varphi_\alpha^{-1} \big].
\end{equation}
So the variation vanishes and $S_{{\rm H}, \Gamma}$ is critical with respect to variations of $L,\gamma,\varphi_\alpha$.

Finally, to show the zero curvature equation \eqref{eq:zce Mi} we note that the Lax equations \eqref{Lax equations} imply
\begin{equation}\label{ZCE Mi com L}
\big[ \partial_{t^i} M^I_j - \partial_{t^j} M^I_i - [ M^I_i, M^I_j ], L^I \big] =0
\end{equation}
for $I \in \{ 0, 1\}$. On the other hand, it also follows from the first equation in \eqref{M constraints} that
\begin{equation}
\gamma^{-1} \big( \partial_{t^i} M^0_j - \partial_{t^j} M^0_i - [ M^0_i, M^0_j ] \big) \gamma = \partial_{t^i} M^1_j - \partial_{t^j} M^1_i - [ M^1_i, M^1_j ]
\end{equation}
on $U_0 \cap U_1$. The zero curvature equation \eqref{eq:zce Mi} now follows from the stability condition.
\end{proof}

As we have shown, the unifying action \eqref{eq:unifyingaction} indeed produces the hierarchy of Lax equations \eqref{global Lax def a}. Notably, this action generalises the $1$d action obtained in \cite[Section 2.4]{VW} in two fundamental aspects. First, it describes Hitchin systems associated with Riemann surfaces of \emph{arbitrary} genus with marked points rather than just for the case of $\CP$. Second, the unifying action \eqref{eq:unifyingaction} encodes the entire hierarchy of such Hitchin systems rather than just one individual flow in this hierarchy. The ``unifying'' aspect of this construction lies in the fact that it incorporates a large class of integrable hierarchies within a single multiform action, in which fixing algebraic and topological data produces explicit variational descriptions of integrable hierarchies in Lax form. In the next chapter, we will put this result to use in some special cases, thereby producing explicit Lagrangian one-forms for hierarchies of certain well-known integrable models.

\chapter{Examples}\label{chap:hitchinexamples}
In this final chapter of Part \ref{part:gaugemultiform}, adapted from \cite[Section 5]{CHSV}, we will look at the specific cases of Hitchin systems associated with the Riemann surfaces of genus zero and genus one with marked points.\footnote{We refer the interested reader to \cite[Section 5]{GawT} for an explicit construction of a Hitchin system (in the Hamiltonian framework) associated with a genus-two Riemann surface.} We explicitly compute our unifying Lagrangian one-form $\Lag_{\rm H}$, or its multiform action $S_{{\rm H},\Gamma}[L^0, \gamma, (\varphi_\alpha), t]$ from Section \ref{sec:unifyingmultiform}, in these cases by exploiting the invariance of the action under changes of holomorphic local trivialisations to fix particularly nice representatives for the transition function $\gamma$. In the $g=0$ case, we recover the Lagrangian one-form for the rational Gaudin hierarchy first obtained in \cite{CDS} (and discussed in Section \ref{sec:noncycloGaudin}) by a completely different method. In the $g=1$ case, we obtain a novel Lagrangian one-form for the elliptic Gaudin hierarchy --- and the elliptic spin Calogero--Moser hierarchy as a special case --- thus filling a gap in the landscape of Lagrangian multiforms. 

The unifying Lagrangian one-form \eqref{eq:unifyingaction b} has the typical structure of a Lagrangian one-form in phase space coordinates as first used in \cite{CDS} for finite-dimensional integrable systems on coadjoint orbits, and further developed in \cite{CH} in relation to the univariational principle. It is the difference of a kinetic part and a potential part involving all the invariant Hamiltonians $H_i$.
The essential novelty compared to all Lagrangian one-forms considered in Part \ref{part:coadjointmultiform} is that the kinetic part involves not only the group coordinates $\varphi_\alpha$ of the coadjoint orbits at the marked points $\mathsf p_\alpha$ for $\alpha = 1,\ldots, N$, but also the transition function $\gamma : (U_0 \cap U_1) \times \RR^n \to G$ of the holomorphic principal $G$-bundle. This introduces a new type of kinetic term in the action which is crucial for generating the complete set of kinetic contributions. We will see explicit examples of this in the genus-$1$ case considered in Section \ref{sec:ellipticgaudin}, where we derive a Lagrangian one-form for the elliptic Gaudin model, and the elliptic spin Calogero--Moser model as its special case. In the genus-$0$ case considered in Section \ref{sec:rationalgaudin}, this new kinetic term is absent, and we recover the familiar geometric Lagrangian one-form living on a coadjoint orbit from Sections \ref{sec:noncycloGaudin} and \ref{sec:cycloGaudin}, first introduced in \cite{CDS,CSV}.

Since this section deals with explicit examples and we only work with a matrix Lie algebra $\g$, we fix the nondegenerate invariant bilinear pairing $\langle \cdot, \cdot \rangle : \g \otimes \g \to \CC$ to be the usual one given in terms of the trace by $\langle A,B \rangle = \Tr(AB)$. Also, throughout this section, we identify $\g^\ast$ with $\g$ and the coadjoint action with the adjoint action. Further, we fix a Cartan subalgebra $\h \subset \g$ with basis $\{\sfH_\mu\}_{\mu = 1}^{\text{rk}\, \g}$ and the corresponding root generators $\{\sfE_\varrho\}$ of $\g$.

\section{Rational Gaudin model}\label{sec:rationalgaudin}

We begin with the simplest case of genus $0$ so that $C = \CP$ is the Riemann sphere. We will show that when the unifying Lagrangian one-form \eqref{eq:unifyingaction b} is specialised to the case $\gamma = 1$, we obtain a Lagrangian one-form for the rational Gaudin model, which we have also encountered earlier in Section \ref{sec:noncycloGaudin}. This model provides the simplest example of a Hitchin system with marked points and has been extensively studied from this perspective; see, for instance, \cite{Ne, ER}.

We choose $N \in \ZZ_{\geq 1}$ distinct marked points $\{ \sfp_\alpha \}_{\alpha=1}^N$. We also need sufficiently many points $\{\sfq_i\}_{i=1}^n$ which are distinct from the $\{ \sfp_\alpha \}_{\alpha=1}^N$, where $n$ will be half the dimension of the phase space, to obtain a sufficient number of Hamiltonians. Let $U_0$ be a neighbourhood of a point $\sfp \in C$ distinct from $\{ \sfp_\alpha \}_{\alpha=1}^N$ and $\{\sfq_i\}_{i=1}^n$, such that $U_0$ does not contain any of these other points. Further, let $U_1 \coloneqq C \setminus \{\sfp\}$ which we equip with a holomorphic coordinate $z \colon U_1 \rightarrow \mathbb{C}$, where, in contrast to the general case of Chapter \ref{chap:lax-hitchin}, we have dropped the index on $z$ for notational simplicity.

\subsection{Lax description}

Following the general setup in Section  \ref{sec: Lax matrices}, we obtain a Lax matrix for the rational Gaudin model as the solution of the moment map condition $\mu=0$ \eqref{eq:b_z-gauged} in a local trivialisation where $A''^I = 0$. Using \eqref{Lax behaviour a}, we can write
\begin{equation}\label{eq:L1-genuszero}
L^1 = L_{z} \d z = \left( \frac{L_\alpha}{z - z(\mathsf p_\alpha)} + J_\alpha^1 \right) \d z
\end{equation}
locally near the point $\mathsf p_\alpha$. Here $L_\alpha = - \varphi_\alpha \Lambda_\alpha \varphi_\alpha^{-1}$ denotes the residue at $\mathsf p_\alpha$ and $J^1_\alpha$ the holomorphic part of $L_z$ in the neighbourhood of $\mathsf p_\alpha$. Since $\gamma = 1$ we have $L^0 = L^1$ on $U_0 \cap U_1$ so that $L^1$ extends holomorphically to the point $\mathsf p \in C = \CP$ at infinity in the coordinate $z$.
We can therefore write the Lax matrix $L^1$ as a global $\g$-valued meromorphic $(1,0)$-form
\begin{equation}\label{eq:B-Gaudin}
L^1 = L_z \d z = \sum_{\alpha=1}^N \frac{L_\alpha}{z-z(\mathsf p_\alpha)} \d z
\end{equation}
together with the constraint $\sum_\alpha L_\alpha = 0$ coming from the fact that $L^1$ is regular at $z(\mathsf p) = \infty$.

\subsection{Lagrangian description}

Let us look at the unifying Lagrangian one-form \eqref{eq:unifyingaction b} in the present case. Since $\gamma = 1$, the kinetic term in the Lagrangian one-form \eqref{eq:unifyingaction b} involving the transition function drops out, and we are left with
\begin{equation}\label{eq:rgmultiform}
\Lag_{\rm RG} = \sum_{\alpha=1}^N \Tr \left( \Lambda_\alpha \varphi_\alpha^{-1} \d_{\RR^n} \varphi_\alpha \right) - H_i\, \d_{\RR^n} t^i
\end{equation}
where
\begin{equation}
H_i = P_i( L_{z}(\sfq_i))
\end{equation}
with $G$-invariant polynomials $P_i$ acting on $L_{z}$ given by \eqref{eq:B-Gaudin}. The Lagrangian one-form \eqref{eq:rgmultiform} describes the rational Gaudin hierarchy. A Lagrangian one-form for the rational Gaudin hierarchy was first obtained in \cite[Section 7]{CDS} using an algebraic approach, and has been presented in this thesis in Section \ref{sec:noncycloGaudin}. A Lagrangian describing the dynamics for a single Hamiltonian of the Gaudin model was also obtained previously in \cite{VW}. By the same derivation as in these references, one shows that the Euler--Lagrange equations associated to \eqref{eq:rgmultiform} give the following collection of equations
\begin{equation}
  \partial_{t^i} L_\alpha = \left[ \frac{\nabla P_i(L_{z}(\sfq_i))}{z(\mathsf p_\alpha) - z(\sfq_i)}, L_\alpha \right],
\end{equation}
for all $i$. One can check that one recovers a hierarchy of Lax equations from the above collection of equations using \eqref{MIi explicit}.

\section{Elliptic Gaudin and elliptic spin Calogero--Moser models}\label{sec:ellipticgaudin}

Let us now consider the $g=1$ case so that the Riemann surface $C$ is a complex torus $\mathbb{C}/(\mathbb{Z} + \tau \mathbb{Z})$ with $\text{Im}(\tau)>0$. Following the general setup of Chapter \ref{chap:lax-hitchin}, we fix a point $\sfp \in C$, take $U_0$ to be a small neighbourhood of $\sfp$, and let $U_1 \coloneqq C \setminus \{\sfp\}$. We fix a coordinate $z$ on $C$ with the identifications $z \sim z + 1$ and $z \sim z + \tau$ such that $z(\sfp) = 0$.
Our starting point is the Lagrangian one-form \eqref{eq:unifyingaction b} but now specialised to a holomorphic transition function $\gamma$ on the annulus $U_0 \cap U_1$ given by
\begin{equation}\label{eq:tranfun-genusone}
    \gamma = \exp\left(\frac{\Q}{z} \right) \quad \text{with}\quad \Q = q^\mu \sfH_\mu
\end{equation}
where $q^\mu \in \mathbb{C}$, for each $\mu \in \{1, \ldots, \text{rk}\, \g\}$, is constant in $z$. We will show that the resulting Lagrangian one-form describes the elliptic Gaudin hierarchy, and in the case with $N=1$ marked point the elliptic spin Calogero--Moser hierarchy. The choice of the holomorphic transition function $\gamma$ is dictated by the realisation of the Hitchin system we wish to obtain, with the only condition on $\gamma$ being that it has a sufficient number of degrees of freedom to describe the moduli space in the $g=1$ case. Here we have taken $\gamma$ to have $\text{rk}\, \g$ parameters, which is exactly the dimension of the moduli space in this case. A heuristic argument for the form of transition functions for arbitrary genus can be found in \cite[Chapter 7]{BBT}.

The elliptic Gaudin model we construct was obtained from a Hamiltonian reduction procedure in \cite{ER, Ne}. To avoid possible confusion, it is worth noting in passing that for $\g = \sl_m$ there is another integrable system that also goes by the name elliptic Gaudin model and was originally derived as a limit of the XYZ spin chain in \cite{ST1, ST2}. It is not clear if and how these two models are related, especially since the elliptic Gaudin model we consider is known to have a dynamical $r$-matrix \cite{ER} while the one of \cite{ST1, ST2} can be built from Belavin's elliptic solution \cite{Be} of the CYBE by following the procedure of \cite{J}. A spin generalisation of the Calogero--Moser model was first defined in \cite{GiH}, and its case with elliptic potential was realised as a Hitchin system in \cite{Ma, GoNe, ER, Ne}. The main goal for us in this section is to use the unifying Lagrangian one-form \eqref{eq:unifyingaction b} to obtain variational descriptions of these two hierarchies.\footnote{Recently, the Lagrangian one-form structure of certain Calogero--Moser-type systems was studied in \cite{KNY}. Starting from a general ansatz for the kinetic term of the Lagrangian one-form, the authors used the resulting generalised Euler--Lagrange equations to derive the integrable cases of these systems.}

We choose $N \in \ZZ_{\geq 1}$ distinct marked points $\{ \sfp_\alpha \}_{\alpha=1}^N$ in $C \setminus \overline{U_0}$. As in the general case, we need sufficiently many additional points $\sfq_i \in C \setminus \overline{U_0}$ distinct from $\{\sfp_\alpha\}_{\alpha=1}^N$ to obtain a sufficient number of Hamiltonians.
We will later specialise to the case of a single marked point $\sfp_1$, i.e. taking $N=1$.

\subsection{Lax description}

From \eqref{Lax behaviour a}, we have
\begin{equation}\label{eq:L1-eg}
L^1 = L_z \d z = \left( \frac{L_\alpha}{z - z(\mathsf p_\alpha)} + J^1_\alpha \right) \d z
\end{equation}
locally around the point $\mathsf p_\alpha$ in our choice of local trivialisation. Here $L_\alpha = -\varphi_\alpha \Lambda_\alpha \varphi_\alpha^{-1}$ denotes the residue at $\mathsf p_\alpha$ and $J^1_\alpha$ the holomorphic part of $L_z$ in the neighbourhood of $\mathsf p_\alpha$. We can express $L_\alpha$ and $L^1 = L_z \d z$ in the basis $(\sfH_\mu, \sfE_\varrho)$ as
\begin{equation}\label{eq:L1-cartanbasis}
  L_\alpha = (L_\alpha)^\mu \sfH_\mu + (L_\alpha)^ \varrho \sfE_\varrho \quad \text{and} \quad L_z = L^\mu \sfH_\mu + L^\varrho \sfE_\varrho.
\end{equation}
Since $L^1 = L_z \d z$ is a meromorphic one-form on the punctured torus $U_1$, it satisfies
\begin{equation}
    L^1(z+1) = L^1(z+\tau) = L^1(z). 
\end{equation}
Also, recall from Section  \ref{sec: Lax matrices} that $L^0$ is holomorphic on $U_0$ and we have the relation $L^0 = \gamma L^1 \gamma^{-1}$ on $U_0 \cap U_1$. Therefore, we can write
\begin{equation}\label{eq:L0-cartanbasis}
    L^0 = \bigg( L^\mu \sfH_\mu +  L^\varrho \exp\Big(\frac{\varrho(\Q)}{z} \Big) \sfE_\varrho \bigg) \d z.
\end{equation}
We can express $L^\mu$ and $L^\varrho$ in \eqref{eq:L1-cartanbasis} and \eqref{eq:L0-cartanbasis} in terms of the Weierstrass $\zeta$-function and the Weierstrass $\sgm$-function which are respectively defined by the relations
\begin{equation}\label{eq:zetafunction}
  \frac{\d \zeta(z)}{\d z} = - \wp(z), \quad \lim_{z \rightarrow 0} \left( \zeta(z) - \frac{1}{z} \right) = 0
\end{equation}
and
\begin{equation}\label{eq:sigmafunction}
  \frac{\d \log (\sgm(z))}{\d z} = \zeta(z), \quad \lim_{z \rightarrow 0} \frac{\sgm(z)}{z} = 1
\end{equation}
where $\wp$ is the Weierstrass $\wp$-function. While the Weierstrass $\wp$-function is doubly periodic, the $\sigma$-function and the $\zeta$-function satisfy
\begin{subequations}
\begin{align}
  &\zeta(z + 2 \omg_l) = \zeta(z) + 2 \eta_l ,\\
\label{quasi_period2}  &\sgm(z + 2 \omg_l) = - \sgm(z) e^{2 \eta_l (z + 2 \omg_l)},
\end{align}
\end{subequations}
where $2\omg_1$ and $2\omg_2$ are the periods of a generic torus and $\eta_l = \zeta(\omg_l)$, for $l = 1, 2$. Here we have $2\omg_1 = 1$ and $2\omg_2 = \tau$.

It follows from \eqref{eq:L1-eg} that the elliptic function $L^\mu$ is meromorphic with simple poles at $\{\sfp_\alpha\}_{\alpha=1}^N$ with residues $(L_\alpha)^\mu$, so it can be expressed as
\begin{subequations} \label{L mu and constraint}
\begin{equation}\label{eq:Lmu-eg}
  L^\mu =  \pi^\mu + \sum_{\alpha=1}^N (L_\alpha)^\mu \zeta(z-z(\sfp_\alpha)) 
\end{equation}
where $\pi^\mu$ is constant in $z$, and we have
\begin{equation}\label{eq:sum_residue}
    \sum_{\alpha=1}^N (L_\alpha)^\mu = 0
\end{equation}
\end{subequations}
since the sum of residues over an irreducible set of poles of an elliptic function vanishes. Next, for the function $L^\varrho$, \eqref{eq:L1-eg} tells us that it has poles at $\{\sfp_\alpha\}_{\alpha=1}^N$ with residues $(L_\alpha)^\varrho$ and \eqref{eq:L0-cartanbasis} implies that the function $L^\varrho e^{\frac{\varrho(\Q)}{z}}$ is holomorphic near $z = 0$. It then follows that
\begin{equation}\label{eq:Lvarrho-eg}
    L^\varrho = \sum_{\alpha=1}^{N} (L_\alpha)^\varrho \frac{\sgm(\varrho(\Q) + z - z(\sfp_\alpha))}{\sgm(\varrho(\Q)) \sgm(z-z(\sfp_\alpha)) } e^{- \varrho(\Q)\zeta(z)}.
\end{equation}
Then, we get $L^1 = (L^\mu \sfH_\mu + L^\varrho \sfE_\varrho) \d z$ with $L^\mu$ and $L^\varrho$ given by \eqref{L mu and constraint} and \eqref{eq:Lvarrho-eg} as the Lax matrix of the elliptic Gaudin model.

In the special case when $N = 1$ and $\sfp_1 = 0$, the components \eqref{L mu and constraint} simply reduce to $L^\mu =  \pi^\mu$, while the components \eqref{eq:Lvarrho-eg} read
\begin{equation}
L^\varrho = \sum_{\alpha=1}^{N} (L_\alpha)^\varrho \frac{\sgm(\varrho(\Q) + z)}{\sgm(\varrho(\Q)) \sgm(z) } e^{- \varrho(\Q)\zeta(z)}.
\end{equation} 
When $\g = \mathfrak{sl}_m(\mathbb{C})$, coincides with the Lax matrix of the elliptic spin Calogero--Moser model, see, e.g. \cite[Chapter 7]{BBT}.

\subsection{Lagrangian description}

We now specialise the unifying Lagrangian one-form \eqref{eq:unifyingaction b} to the present case. Since $\g$ is taken to be a matrix Lie algebra and we take the nondegenerate bilinear form on $\g$ to be given by the trace, the Lagrangian one-form \eqref{eq:unifyingaction b} now reads
\begin{equation}\label{eq:egmultiform-unsimp}
  \Lag_{\rm EG} = \frac{1}{2 \pi i} \int_{c_{\mathsf p}} \Tr \left( L^1 \wedge \gamma^{-1} \d_{\RR^n} \gamma \right) + \sum_{\alpha=1}^N \Tr \left( \Lambda_\alpha \varphi_\alpha^{-1} \d_{\RR^n} \varphi_\alpha \right) - H_i \, \d_{\RR^n} t^i
\end{equation}
where the transition function $\gamma$ is given by \eqref{eq:tranfun-genusone} and
\begin{equation}\label{eq:Ham-genusone}
H_i = P_i( L_{z}(\mathsf q_i)).
\end{equation}
Let us evaluate the integral in the first term on the right-hand side of \eqref{eq:egmultiform-unsimp}. We have
\begin{equation}\label{eq:integraleval}
  \begin{split}
    \frac{1}{2 \pi i} \int_{c_{\mathsf p}} \Tr \left( L_z \d z \wedge \gamma^{-1} \d_{\RR^n} \gamma \right) = \frac{1}{2 \pi i} \int_{c_{\mathsf p}} \Tr \left( L_z \d z \wedge \frac{\d_{\RR^n} \Q}{z}  \right) = p_\mu \d_{\RR^n} q^\mu
  \end{split}
\end{equation}
where the coordinates $q^\mu$ were introduced in \eqref{eq:tranfun-genusone}, and we defined
\begin{equation}
p_\mu \coloneqq \Tr \big( L_{z}(z(\sfp)) \sfH_\mu \big),~~\mu \in \{1, \ldots, \text{rk}\, \g\}.
\end{equation}
Plugging back \eqref{eq:integraleval} into \eqref{eq:egmultiform-unsimp}, we get a Lagrangian one-form for the elliptic Gaudin hierarchy
\begin{equation}\label{eq:egmultiform-simp}
  \Lag_{\rm EG} = p_\mu \d_{\RR^n} q^\mu + \sum_{\alpha=1}^N \Tr \left( \Lambda_\alpha \varphi_\alpha^{-1} \d_{\RR^n} \varphi_\alpha \right) - H_i\, \d_{\RR^n} t^i,
\end{equation}
with $H_i$ defined by \eqref{eq:Ham-genusone} depending on the elliptic Gaudin Lax matrix given by \eqref{L mu and constraint} and \eqref{eq:Lvarrho-eg}. Note that in contrast with the Lagrangian one-form \eqref{eq:rgmultiform} obtained in the genus $0$ case, the Lagrangian one-form above has an additional kinetic term corresponding to the cotangent bundle degrees of freedom $p_\mu,q^\mu$ arising from the non-triviality of the principal $G$-bundle in genus 1.

Writing a Lagrangian one-form for the elliptic spin Calogero--Moser hierarchy only requires us to set $N=1$ in the Lagrangian one-form for the elliptic Gaudin hierarchy. We then get
\begin{equation}\label{eq:escmmultiform}
  \Lag_{\rm ESCM} = p_\mu \d_{\RR^n} q^\mu + \Tr \left( \Lambda_1 \varphi_1^{-1} \d_{\RR^n} \varphi_1 \right) - H_i \, \d_{\RR^n} t^i
\end{equation}
for this special case. 

\begin{remark}
    Lax matrices for the elliptic Gaudin and elliptic spin Calogero--Moser models have appeared before in the literature. These descriptions are related to ours by a change of local trivialisation. We refer the interested reader to \cite[Section 5]{CHSV} for further details.
\end{remark}

By obtaining an explicit geometric Lagrangian one-form for the elliptic Gaudin hierarchy (and the elliptic spin Calogero--Moser hierarchy as a special case), we have extended the results of Sections \ref{sec:noncycloGaudin} and \ref{sec:cycloGaudin} in Part \ref{part:coadjointmultiform}, where we obtained explicit geometric Lagrangian one-forms for integrable models with spectral parameters lying on the Riemann sphere.

However, it is important to note that our approach in Part \ref{part:coadjointmultiform} has the additional advantage of providing an explicit connection between (non-dynamical) classical $r$-matrices and Lagrangian multiforms. It will be interesting to generalise this connection to the present setting, especially since the Hitchin system associated with an elliptic curve is known to have a dynamical $r$-matrix \cite{ER}. We end this chapter by noting the works \cite{BDOZ, Ber} which offer useful insights in this direction.

\newpage
\thispagestyle{empty}
\mbox{}
\newpage

\part{Conclusion}\label{part:conclusion}
\thispagestyle{empty}

\newpage
\thispagestyle{empty}
\mbox{}
\newpage

\chapter{Perspectives and open questions}\label{chap:perspectives}
In this work, we introduced two approaches for systematically constructing geometric Lagrangian one-forms for large classes of finite-dimensional integrable systems, thus addressing a central open problem in the theory of Lagrangian multiforms. 

In the first approach, presented in Part \ref{part:coadjointmultiform}, we constructed geometric Lagrangian one-forms living on coadjoint orbits while also making important connections between the theory of Lagrangian multiforms and the more traditional Hamiltonian aspects of integrability. The versatility of this approach was demonstrated through the construction of explicit Lagrangian one-forms for the open Toda chain, the (non-cyclotomic and cyclotomic) rational Gaudin models, and the periodic Toda chain and the DST model as special realisations of the cyclotomic Gaudin model. In Part \ref{part:gaugemultiform}, we introduced a variational description of Hitchin systems associated with compact Riemann surfaces of arbitrary genus using the framework of Lagrangian multiforms and developed it further by presenting a gauging procedure for multiforms. As an application, this allowed us to obtain explicit geometric Lagrangian one-forms for the elliptic Gaudin hierarchy (and the elliptic spin Calogero--Moser hierarchy as its special case) for the first time, as well as recovering the rational Gaudin one-form obtained in Part \ref{part:coadjointmultiform}. Apart from providing a systematic framework for constructing Lagrangian one-forms and thus filling large gaps in the theory of Lagrangian multiforms, we were also able to make useful connections with the Hamiltonian framework for integrability (in Part \ref{part:coadjointmultiform}) and with HT gauge theories (in Part \ref{part:gaugemultiform}).

As we mentioned in Section \ref{sec: univar principle CH}, our Lagrangian one-forms have a structure similar to the so-called geometric actions considered in \cite{AFS} in the context of path integral quantisation of coadjoint orbits that give rise to localisation formulae for characters of the underlying Lie group. In fact, a strong motivation behind the Lagrangian multiform programme itself is a path integral quantisation of integrable hierarchies.\footnote{In relation to this research direction, it is important to note the work \cite{KiN}, where a new quantum variational principle was proposed in terms of multiform path integrals.} Thus, this connection with geometric actions points towards the possibility of using results from equivariant localisation of path integrals (see, for instance, \cite{Sz}) to our advantage.

Further, given the role of ideas coming from Hamiltonian reduction in both the approaches, it will be worthwhile to attempt to connect our framework of geometric Lagrangian one-forms with other reduction-based approaches to integrable systems. In this regard, we note two recent interesting works \cite{Feh, FehFai} based on the Hamiltonian framework.

It will also be interesting to explore the generalisation of our approaches to incorporate other classes of integrable models and their connections with other areas of physics and mathematics. We conclude this thesis by briefly discussing some of these possible directions of research. 

\paragraph{Dihedral affine Gaudin models and non-ultralocal IFTs:} In \cite{FF}, Feigin and Frenkel obtained a field-theoretic generalisation of Gaudin models by replacing the underlying finite-dimensional Lie algebras with infinite-dimensional Lie algebras, which leads to affine Gaudin models (AGMs).\footnote{The interested reader is referred to \cite{Lac} for a detailed introduction to affine Gaudin models.} These models were also produced independently using a more geometric approach by Levin, Olshanetsky, and Zotov in \cite{LeOZ1}. More recently, it was shown in \cite{Vic} that a large family of classical \emph{non-ultralocal} IFTs can be seen as realisations of AGMs with dihedral symmetry. 

One of the motivations for choosing the finite rational Gaudin model and its cyclotomic generalisation as examples to cast in the Lagrangian one-form framework of Part \ref{part:coadjointmultiform} has been these so-called \emph{dihedral affine Gaudin models} introduced in \cite{Vic}.\footnote{In connection with Gaudin models, it is worth noting the recent work \cite{AMS} on the construction of solutions of the general CYBE based on more general Lie algebra decompositions. One can associate generalised notions of Gaudin models to these solutions, and it will be interesting to explore how to incorporate them into our framework of geometric Lagrangian one-forms.} Generalising the setup of \cite{CDS, CSV} to construct a Lagrangian multiform for dihedral AGMs\footnote{Constructing a Lagrangian multiform for dihedral AGMs will also provide a generalisation of the results of \cite{CStV}, where a systematic framework for constructing Lagrangian multiforms for $2$d \emph{ultralocal} IFTs was introduced.} and analysing the associated path integral can therefore provide crucial insights into the long-standing open problem of a systematic quantisation of non-ultralocal IFTs, which we briefly discussed in Remark \ref{rem:nonultralocal}.

\paragraph{$4$d semi-holomorphic Chern--Simons theory and affine Higgs bundles:} The construction of \cite{CHSV}, presented in Part \ref{part:gaugemultiform}, brings together the framework of Lagrangian multiforms with that of $3$d mixed HT BF theory through the study of Hitchin's completely integrable system in the Lagrangian framework. A natural generalisation of the work of \cite{CHSV} will be to similarly merge the framework of Lagrangian multiforms with that of $4$d semi-holomorphic Chern--Simons theory \cite{CosY} with the aim of obtaining a gauge-theoretic origin of \emph{hierarchies} of $2$d integrable field theories.

In connection with these ideas, it is important to also note the work \cite{LeOZ2} on a $2$d field-theoretic generalisation of Hitchin's integrable system based on affine Higgs bundles. It was shown to be closely related to the $4$d Chern--Simons setup. Examples of \emph{affine Hitchin systems} constructed in this manner include $2$d field-theoretic analogues of the elliptic Gaudin and Calogero--Moser\footnote{See \cite{AZ} and references therein concerning the field-theoretic analogue of the elliptic (spin) Calogero--Moser model.} models, both of which have been cast in the variational framework of geometric Lagrangian one-forms in \cite{CHSV}; see Section \ref{sec:ellipticgaudin}. A natural next step is then to use a higher-dimensional generalisation of the construction of \cite{CHSV} to obtain a variational analogue of the affine Higgs bundle setup.

\paragraph{Integrable systems, gauge theories, and Langlands correspondence:} The approach of \cite{CHSV} also unravels a connection between $3$d classical mixed HT BF theory and Hitchin systems. At the quantum level, this connection has been studied in \cite{GW}, where an identification between the Hitchin moduli space and the phase space of $3$d quantum mixed HT BF theory was made to demonstrate the existence of commuting quantum Hitchin Hamiltonians. 

In the last few years, it has been established that quantum Hitchin systems play a central role in the \emph{geometric Langlands correspondence}, as well as its more recent analytic version; see, for instance, \cite{EtFK1, EtFK2, GT}. This connection goes back to the work of Beilinson and Drinfel'd \cite{BeD} on the quantisation of Hitchin systems. In particular, for the genus-zero case of Gaudin models, interesting connections between the geometric Langlands correspondence and the Bethe ansatz and Sklyanin's separation of variables methods have also been found. See, for instance, \cite{Fren1}. In this context, it is also worth noting the relation of this correspondence with the Feigin--Frenkel--Reshetikhin (FFR) approach \cite{FFR} that describes the spectrum of (quantum) Gaudin Hamiltonians in terms of differential operators called opers.

By constructing Lagrangian multiforms for Gaudin models in Part \ref{part:coadjointmultiform} and for the more general case of Hitchin systems in Part \ref{part:gaugemultiform}, we have opened up the avenue for studying the quantisation of their hierarchies using a path integral approach (together with techniques from equivariant localisation). This also provides the exciting possibility of discovering novel connections of these models (as well as their $2$d analogues) to gauge theories in the quantum realm. Finally, a path integral quantisation of Lagrangian multiforms of these integrable models can lead to useful insights into the unification of these ideas through the construction of explicit examples of the geometric Langlands correspondence.

\newpage
\thispagestyle{empty}
\mbox{}
\newpage

\printbibliography[heading=bibintoc,title={References}]

\end{document}